\documentclass[twocolumn]{aastex631}
\usepackage{lineno}
\shorttitle{Systematic analysis of BH spins}
\shortauthors{Draghis et al.}

\begin{document}

\title{Systematically Revisiting All NuSTAR Spins of Black Holes in X-ray Binaries}
\correspondingauthor{Paul A. Draghis}
\author[0000-0002-2218-2306]{Paul A. Draghis}
\email{pdraghis@umich.edu}
\affiliation{Department of Astronomy, University of Michigan, 1085 South University Avenue, Ann Arbor, MI 48109, USA}

\author[0000-0003-2869-7682]{Jon M. Miller}
\affiliation{Department of Astronomy, University of Michigan, 1085 South University Avenue, Ann Arbor, MI 48109, USA}

\author[0000-0001-8470-749X]{Elisa Costantini}
\affiliation{SRON Netherlands Institute for Space Research, Niels Bohrweg 4, 2333 CA Leiden, The Netherlands}
\affiliation{Anton Pannekoek Astronomical Institute, University of Amsterdam, P.O. Box 94249, 1090 GE Amsterdam, the Netherlands}

\author{Luigi C. Gallo}
\affiliation{Department of Astronomy \& Physics, Saint Mary’s University, 923 Robie Street, Halifax, Nova Scotia, B3H 3C3, Canada}

\author[0000-0003-1621-9392]{Mark Reynolds}
\affiliation{Department of Astronomy, Ohio State University,  140 W 18th Avenue, Columbus, OH 43210, USA}

\author[0000-0001-5506-9855]{John A. Tomsick}
\affiliation{Space Sciences Laboratory, 7 Gauss Way, University of California, Berkeley, CA 94720-7450, USA}

\author[0000-0002-0572-9613]{Abderahmen Zoghbi}
\affiliation{Department of Astronomy, University of Maryland, College Park, MD 20742, USA}
\affiliation{HEASARC, Code 6601, NASA/GSFC, Greenbelt, MD 20771, USA}
\affiliation{CRESST II, NASA Goddard Space Flight Center, Greenbelt, MD 20771, USA}

\begin{abstract}

We extend our recent work on black hole spin in X-ray binary systems to include an analysis of 189 archival NuSTAR observations from 24 sources. Using self-consistent data reduction pipelines, spectral models, and statistical techniques, we report an unprecedented and uniform sample of 36 stellar-mass black hole spin measurements based on relativistic reflection. This treatment suggests that prior reports of low spins in a small number of sources were generally erroneous: our comprehensive treatment finds that those sources tend to harbor black holes with high spin values. Overall, within $1\sigma$ uncertainty, $\sim86\%$ of the sample are consistent with $a \geq 0.95$, $\sim94\%$ of the sample are consistent with $a\geq 0.9$, and $100\%$ is consistent with $a\geq 0.7$ (the theoretical maximum for neutron stars; $a = cJ/GM^{2}$). We also find that the high-mass X-ray binaries (those with A-, B-, or O-type companions) are consistent with $a\geq 0.9$ within the $1\sigma$ errors; this is in agreement with the low-mass X-ray binary population and may be especially important for comparisons to black holes discovered in gravitational wave events. In some cases, different spectra from the same source yield similar spin measurements but conflicting values for the inclination of the inner disk; we suggest that this is due to variable disk winds obscuring the blue wing of the relativistic Fe K emission line. We discuss the implications of our measurements, the unique view of systematic uncertainties enabled by our treatment, and future efforts to characterize black hole spins with new missions.

\end{abstract}

%\keywords{These are the keywords}

\section{Introduction} \label{sec:intro}
The ``relativistic reflection" spin measurement technique (\citealt{2000PASP..112.1145F}) has been extensively used to measure black hole (BH) spin\footnote{The BH spin $a$ is defined as  $-1<a=cJ/GM^2<1$, where c is the speed of light, G is the gravitational constant, and J \& M are the angular momentum and mass of the BH, respectively.} across the entire mass range, from stellar-mass BHs in X-ray binary (XB) systems (see, e.g. \citealt{2006ApJ...652.1028B, 2009ApJ...697..900M, 2023ApJ...947...39D}) to the BHs in active galactic nuclei (AGN; see, e.g., \citealt{2015MNRAS.446..633G, 2015ApJ...806..149K}). For a review of BH spin measurement techniques, see \cite{2021ARA&A..59..117R}. Furthermore, NuSTAR (\citealt{2013ApJ...770..103H}) observations taken over the past decade have enhanced the findings of the method, by providing spectra ideally suited for such studies, which cover the main features of relativistic reflection (the Fe K complex and the Compton "hump") with high sensitivity. Relativistic reflection studies on NuSTAR data have led to more than 30 BH spin measurements in XB systems, with more sources being discovered, observed, and analyzed every year. However, as the studies occurred over a decade and were performed by independent authors, as understanding of the physical mechanisms present in the analyzed systems grew and our models matured, and as our ability to quantify instrumental effects evolved, a comparison between independent measurements or a uniform analysis of the entire BH distribution would be unproductive.  

In contrast, gravitational wave (GW) observations have so far detected $\sim90$ binary black hole (BBH) merger candidates, translating to roughly twice as many BHs observed and analyzed in the third edition of the Gravitational Wave Transient Catalog (GWTC-3; \citealt{2023PhRvX..13d1039A}). The efforts of the GW community have produced a uniform description of the entire observed BH sample (\citealt{2023PhRvX..13a1048A}), leading to an inference on the population of merging BHs. This study suggests that the BHs observed to merge in BBH systems have preferentially low spins, in contrast with the BH spin distribution observed for the BHs in XBs, hinting at the fact that different populations of BHs are observed through GWs and electromagnetic radiation. In this regard, \cite{2022ApJ...938L..19G} concluded that at most 20\% of BBHs evolve from high-mass X-ray binary (HMXB) systems formed through Case-A mass transfer and fewer than 11\% of Case-A HMXB systems evolve into BBHs, further hinting at the different origins of the BH populations. In order to understand BH formation and evolution as a whole, any theory needs to be able to reproduce both the high spins observed in XBs and the low spins observed in BBHs. However, before taking such steps, a uniform analysis of the entire population of BH spins in XB systems is necessary, similarly to the case of BBHs observed through GWs.

A significant topic of discussion within the field of X-ray measurements of BH spins is the magnitude of the systematic uncertainties of such measurements, with existing studies reporting only the statistical uncertainties of the measurements. A few of the most significant possible sources of systematic uncertainty come from (1) disagreement between existing families of models, or between different flavors from the same family of models; (2) the validity of the assumptions regarding the physical processes occurring in the observed systems, and the accuracy with which the models capture the extent of the source behavior; (3) small-scale source variability, which is unaccounted for by the models used to fit the data; and (4) an incomplete understanding of the models used, and of possible correlations between sets of parameters which produce broad regions of the parameter space that are not properly explored. Furthermore, when attempting to describe the entire population of BH spins in XBs, it is important to acknowledge that so far, only a small subset of such systems in our Galaxy have been observed, and to consider the existence of possible selection effects and observational biases. Works such as \cite{2020ApJ...900...78D} and \cite{2021ApJ...920...88D} attempt to address the first point highlighted above, by exploring a large array of models when fitting the data, composed of multiple combinations of different model components. The second point in the list will be addressed in time, as more sources will be observed with instruments such as NuSTAR or NICER (\citealt{2016SPIE.9905E..1HG}), and in the more distant future with observations from proposed missions such as AXIS (\citealt{2018SPIE10699E..29M}), HEX-P (\citealt{2018SPIE10699E..6MM}), or STROBE-X (\citealt{2019arXiv190303035R}). In the near future, pairing NuSTAR and NICER observations with high-resolution spectra obtained using the Resolve instrument on board XRISM (\citealt{2020SPIE11444E..22T}) will further help in evolving our understanding of physical mechanisms that might be unaccounted for during a relativistic reflection study, and their importance on the precision of BH spin measurements. 

In \cite{Draghis23}, we began addressing the second possible source of systematic uncertainties listed above, and hinted at the possibility of taking steps in the direction to address the third point listed above. In this paper, we continue the effort of trying to incorporate the information available throughout all existing NuSTAR observations in BH spin measurements, and lay the groundwork for a large-scale analysis of the entire parameter space across a large number of observations. In that regard, we analyze 189 archival NuSTAR spectra of 24 sources that have previous spin measurements using the relativistic reflection method using NuSTAR data. For consistency with previous work, we performed this analysis in a manner that follows the pipeline highlighted in \cite{Draghis23}. The best-fit parameters from the array of spectral fits will be analyzed in a future paper to systematically investigate the behavior of the models used under a variety of observational cases. The work presented in this paper, together with previous results, creates a sample of 36 sources with spins measured using all the available data, by using fully consistent methods and making the same sets of assumptions. Through this large data-set of BH spins measured using a uniform treatment, we set the groundwork for properly deriving the BH spin distribution in XB systems. 

This paper is structured as follows. In Section \ref{sec:obs}, we briefly review the methods used to extract spectra from the NuSTAR observations, we explain the criteria used to select the observations analyzed, we discuss the assumptions made by our models, and briefly describe the analysis pipeline. In Section \ref{sec:measurements}, we present an analysis and results of the BH spin measurements for each of the 24 sources treated in this paper. Lastly, in Section \ref{sec:discussion} we discuss the findings of this large-scale project, their implications, and future directions for improving our understanding of BH spin and of BH formation as a whole. 

\section{Observations, Models, and Analysis} \label{sec:obs}

For most observations, we extracted spectra using the standard \texttt{nustardas} v2.1.1 routines in Heasoft v6.29c, using the NuSTAR CALDB version 20211103. Unless otherwise stated, we extracted the source spectra from circular regions of 120'' radius, centered at the position of the source, and background spectra from an annulus centered on the source, with inner radius of 200'' and outer radius of 300''. We rebinned the spectra using the ``optimal" binning scheme (\citealt{2016A&A...587A.151K}) through the \texttt{ftgrouppha} ftool, and only fit the spectra in the energy band where the source spectra are significantly above the background spectra, also ensuring that usage of the $\chi^2$ statistic is appropriate. Background spectra are subtracted from the source spectra during spectral fitting. We performed the spectral analysis using the \texttt{Xspec} v12.12.0g (\citealt{1996ASPC..101...17A}) software, ensuring \texttt{wilm} abundances (\citealt{2000ApJ...542..914W}) and \texttt{vern} photoelectric absorption cross sections (\citealt{1996ApJ...465..487V}). Instead of taking the customary approach of allowing a constant offset between the spectra from the two NuSTAR focal plane module (FPM) detectors, we allowed the normalizations of the components to vary freely. 

We only fit observations for which the Eddington fraction of the source during the observation was between $10^{-3}\leq L/L_{\rm Edd}\leq 0.3$, under the assumption that for sources within this Eddington fraction range, the accretion disk is not truncated and extends to the innermost stable circular orbit (ISCO). If measurements of BH mass and distance to the source were available in the literature, we used them to estimate the Eddington fraction of the source during the observation. For the remaining nine sources where estimates of BH mass and distance were unavailable, we assumed generic values: $M_{\rm BH}\sim8\pm 4 M_{\odot}$ and $d\sim 10\pm 5 \rm kpc$. We note that while the assumption of an accretion disk extending to the ISCO is backed by theoretical, numerical, and observational results, the lower limit of the Eddington fraction for which this behavior is expected is a topic of debate, with many works suggesting significant disk truncation, especially in hard spectral states (see reviews such as \citealt{2007A&ARv..15....1D, 2021SSRv..217...65B}). If disk truncation is present in the faint hard state, the choice of Eddington fraction for which this assumption holds is unlikely to significantly influence the final reported results for the measured spin, as low flux observations generally produce poor spin constraints, which contribute less to the overall spin probability distribution obtained by combining information from all available observations. Furthermore, it is likely that the lower limit of the Eddington fraction for which the disk extends to the ISCO is not a universal number, applicable for all sources. Therefore, for consistency with numerous amounts of existing work, we continue to employ this assumption, despite acknowledging its limitations and potential effects.

We first fit all observations with a ``baseline" model, which includes Galactic absorption along the line of sight, simplistic emission from a thermal accretion disk, and power-law emission from a compact corona: \texttt{TBabs*(diskbb+powerlaw)}. Then, we fit all spectra with an array of six models that replace the \texttt{powerlaw} component in our baseline model with different flavors of the \texttt{relxill} v.1.4.3 family of models (\citealt{2014MNRAS.444L.100D, 2014ApJ...782...76G}). We test the quality of the fits using the \texttt{relxill}, \texttt{relxillCp}, \texttt{relxilllp}, and the \texttt{relxillD} flavor, with the accretion disk density fixed to $\log(n)=10^{15},10^{17}$, and $10^{19}\;\rm cm^{-3}$. It is important to note that the \texttt{relxill} family of models only considers scattering by thermal electrons, and current-generation data are unable to constrain the need for scattering by a population of nonthermal electrons. Furthermore, fits to our current-generation data appear to be unable to constrain the need for a component describing the Comptonization of disk photons, which could impact the ability to disentangle the reflected radiation from the underlying continuum, potentially biasing  the spin measurements. In the future, pairing NuSTAR data with high-resolution XRISM spectra will enable probing the effects that these effects have on our ability to constrain BH spin.

For spectra that show the presence of additional narrow absorption and emission features, we introduce the multiplicative component \texttt{zxipcf} to the models, which describes the shape of the spectrum after being partially covered by ionized material. Furthermore, when informed by the data, we introduce the \texttt{gaussian} model component with negative (or positive, when describing narrow emission features) normalization to describe narrow absorption features around 7~keV, likely related to absorption by an ionized outflow (or possible residual distant refection when positive - for Cygnus X-1 and MAXI J1848-015). In a few instances, our models include both the \texttt{zxipcf} and \texttt{gaussian} components. When the \texttt{diskbb} component is not required by the data, the normalization of the component and/or the temperature of the disk take low values, which ensure that no significant contribution from the component is present in the total model. However, in a few situations, in order to reduce the complexity of the model, we remove the \texttt{diskbb} component from the models when it is not required by the data. This happens for a few of the spectra of sources in a hard state, especially when the models also require the \texttt{zxipcf} and \texttt{gaussian} components. 

In this analysis, we fit each observation individually. Upon obtaining the best-fit combination of parameters for the models for each spectrum, we run a Markov Chain Monte Carlo (MCMC) analysis as described in Section 2.2 in \cite{Draghis23}, and combine the individual 1D posterior probability distributions using a method similar to the one described in \cite{Draghis23}, but as expanded upon in \cite{2023ApJ...954...62D}. When running the combining algorithm, we weigh the independent measurements by the strength of reflection during the observation. We report the outputs of this Bayesian combining algorithm as our results, for the BH spin and the viewing inclination angle. As all the sources treated in this paper have been analyzed in previous works and as the goal of this paper is to quantify the information regarding the BH spin from all the available observations together, we do not report the full list of measured parameters for each observation. However, this entire data-set will be analyzed in a future paper and will be published at that time, and we note that it can be made available by request to the authors.

\section{New Measurements of Old Spins}\label{sec:measurements}

\subsection{AT 2019wey}\label{sec:AT_2019wey}
At the time of the analysis, three NuSTAR observations of AT 2019wey were public, all taken in 2022 (ObsIDs 90601315002, 90601315004, and 90601315006). \cite{2022SCPMA..6519512F} analyzed the third observation using the \texttt{relxill} family of models to find a BH spin of $a\sim0.96-0.97$ and an inclination $\theta<30^\circ$. Given that the distance estimate to the source is uncertain, $1\lesssim d \lesssim 10 \; \rm kpc$ (\citealt{2021ApJ...920..120Y}), and that the BH mass is unknown (we assumed $M=8\pm4\;M_\odot$), all three observations fall within the range of Eddington fractions for which we expect the accretion disk to extend to the ISCO. Therefore, we analyzed all three existing observations of the source.  The top panels in Figure \ref{fig:AT_2019wey_delchi} show the unfolded FPMA and FPMB NuSTAR spectra obtained during the three observations, while the middle and bottom panels show the residuals produced when fitting the spectra with a power-law model and with the best-fitting reflection models, respectively, together with the statistic produced.

Upon running the MCMC analysis on the best-performing reflection models, we obtained the posterior probability distributions shown by the colored lines in the left (for spin) and right (for inclination) panels of Figure \ref{fig:AT_2019wey_combined}. By combining the individual measurements using our prescription, which accounts for systematic variations between observations, we obtain a spin measurement of $a=0.906^{+0.084}_{-0.202}$ and an inclination of $\theta=14^{+12}_{-10}\;[^\circ]$. Our measurements formally agree with those of \cite{2022SCPMA..6519512F}, indicating that our spin constraint is accurate, but less precise. The increase in the uncertainty of the measurement comes from incorporating information from all existing observations in a way that better accounts for systematic variations. We note that if only the posterior distribution for the third observation was considered (red curve in the left panel of Figure \ref{fig:AT_2019wey_combined}), the spin measurement would be much more narrowly constrained and at nearly maximal values, similar to the result of \cite{2022SCPMA..6519512F}.

\subsection{LMC X-3} \label{sec:LMC_X-3}
At the time of the analysis, there were eight NuSTAR public observations of LMC X-3. As the source is relatively faint and persistently in a disk-dominated state, we only report the analysis of spectra which have shown a statistical improvement in fit statistic when including reflection features in the models: ObsIDs 10601308002, 10601308004, 30402035002, 30402035004, 30402035008, and 30902041002. We note that using a mass of $M=6.98\pm0.56\;M_\odot$ and a distance of $d=48.1\pm2.2\;\rm kpc$ (\citealt{2014ApJ...794..154O}), all observations fall within the acceptable Eddington limit regime. Figure \ref{fig:LMC_X3_delchi} shows the spectra from the six observations analyzed, together with the residuals produced when fitting with our baseline model and with the best-performing reflection models. We note that while our baseline model produces acceptable fits and the residuals do not indicate clear reflection features, for the six observations presented the fit statistic improves significantly when including relativistic reflection to the models.

The 1D posterior distributions obtained by running the MCMC analysis on the fits to the spectra from the six observations (shown in Figure \ref{fig:LMC_X3_combined}) show a general agreement for a high measured BH spin. However, the quality of the spectra prohibits a consentaneous inclination measurement. Using our combining algorithm, we obtain a spin of $a=0.928^{+0.058}_{-0.146}$ and a poorly constrained inclination of $\theta=38^{+14}_{-13}\;[^\circ]$. Interestingly, the measured spin is in strong disagreement with results obtained using continuum fitting, which indicate a spin $a\sim0.25$ (\citealt{2014ApJ...793L..29S, 2021MNRAS.507.4779J, 2024ApJ...960....3S, 2024MNRAS.527L..76M}). However, it is important to note that \cite{2024ApJ...962..101Z} showed that spin measurements using continuum fitting for LMC X-3 and Cygnus X-1 can be heavily model dependent, and rely on assumptions regarding the accretion disk. Furthermore, \cite{2024ApJ...960....3S} placed an upper limit on the spin of the source to be $a\leq0.7$ using an X-ray polarimetry study. This disagreement between the methods is of great concern, and should be used as a starting point for identifying potential issues with the methods used by these BH spin measurement techniques. Ultimately, a spectropolarimetric analysis of the source, in which continuum fitting is combined with a relativistic reflection analysis, should lead to consistent results of the BH spin.

\subsection{LMC X-1}\label{sec:LMC_X-1}
The BH in HMXB LMC X-1 has been well studied, and previous results strongly suggest that it is rapidly rotating. \cite{2012MNRAS.427.2552S} used relativistic reflection on Suzaku and RXTE data to measure the spin of the BH in LMC X-1, finding $a=0.97^{+0.02}_{-0.25}$. \cite{2021MNRAS.507.4779J} used three NuSTAR observations to measure a BH spin of $a=0.935\pm0.015$ using relativistic reflection, result confirmed by \cite{2022AdSpR..69..483B}, who measured $a=0.93\pm0.01$. Continuum fitting measurements find agreeing results, with \cite{2009ApJ...701.1076G} reporting $a=0.92^{+0.05}_{-0.07}$ based on RXTE data, result confirmed by \cite{2020MNRAS.498.4404M} who measure $a=0.93^{+0.04}_{-0.03}$. 

LMC X-1 was observed four times by NuSTAR, under ObsIDs 30001039002, 30001143002, 30201029002, and 90801324002. Using the mass and distance measurements of $M=10.91\pm1.41\;M_\odot$ and $d=48.1\pm2.2\;\rm kpc$ (\citealt{2009ApJ...697..573O}), we find that all four observations occurred while the source had an Eddington fraction $0.1\%\leq L/L_{\rm Edd}\leq 30\%$. We fit the spectra from the four observations in the 3-25, 3-30, 3-30, and 3-20 keV bands, respectively. The top panels in Figure \ref{fig:LMC_X1_delchi} show the unfolded spectra obtained from the four NuSTAR observations. The middle panels in Figure \ref{fig:LMC_X1_delchi} show the residuals produced when fitting the NuSTAR spectra of LMC X-3 using \texttt{TBabs*(diskbb+powerlaw)}, indicating clear signs of relativistic reflection. 

The bottom panels in Figure \ref{fig:LMC_X1_delchi} show the residuals produced when fitting the spectra with the best-performing reflection models for each observation. For three of the four observations, the preferred \texttt{relxill} variant is the one that allows probing higher accretion disk densities (\texttt{relxillD}), with the spectra from ObsIDs 30001143002 and 90801324002 preferring $\log(N)=19$ and the spectra from ObsID 30201029002 preferring the \texttt{relxillD} variant with $\log(N)=17$. For remaining observation (ObsID 30001039002), the preferred reflection model statistically favors the addition of a narrow Gaussian absorption component around $6.67\pm0.04~\rm keV$, likely consistent with a slightly blueshifted neutral Fe K$\alpha$ absorption caused by the stellar wind launched by the massive companion star in the system. The individual posterior distributions for the BH spin and inclination of the inner disk are shown in Figure \ref{fig:LMC_X1_combined}, with the thickness of the lines being proportional to the ratios of the reflected to the total flux in the 3-79~keV band, which were used as weights when combining the individual measurements. The combined distributions are represented in Figure \ref{fig:LMC_X1_combined} through the solid black curves, while the vertical solid and dashed lines indicate the mode and $1\sigma$ credible intervals of the distributions, respectively. We find a spin of $a=0.897^{+0.077}_{-0.176}$, formally consistent with previous results. The inclination was allowed to vary freely in our analysis, and we measured a value of $50^{+10}_{-13}[^\circ]$. This value is slightly higher but formally consistent with the binary inclination of the system, $\theta=36.38^\circ\pm1.92^\circ$ (\citealt{2009ApJ...697..573O}), suggesting that the inner accretion disk is likely close to being aligned with the binary orbit.

\subsection{MAXI J1348-630}\label{sec:MAXI_J1348-630}
There are nine archival NuSTAR observations of MAXI J1348-630 (ObsIDs 80402315002, 80402315004, 80402315006, 80402315008, 80402315010, 80402315012, 80502304002, 80502304004, and 80502304006). All nine observations were previously analyzed in \cite{2022MNRAS.511.3125J}, who obtained a spin of $a=0.78\pm0.04$ and an inclination of $\theta=29.2^{+0.3}_{-0.5}\;[^\circ]$. However, in their analysis, \cite{2022MNRAS.511.3125J} fixed the inner and outer emissivity indices $q_1=q_2=3$, which suggests that this measurement should be interpreted as a lower limit on the spin (\citealt{2014MNRAS.439.2307F}). Additionally, \cite{2022MNRAS.511.3125J} did not test for the possibility of the presence of narrow absorption features in the spectra. In our analysis, we find that narrow absorption \texttt{gaussian} features are statistically preferred in seven out of the nine observations. Using a mass of $M=11\pm2\;M_\odot$ and a distance of $d=3.4\pm0.4\rm kpc$ (\citealt{2021A&A...647A...7L}), we find that all observations are within the assumed constraints of Eddington range for an observation to be used in our analysis.

Panels (a) in Figure \ref{fig:MAXI_J1348-630_delchi} show the residuals produced when fitting the NuSTAR spectra from the nine observations with a power-law model, together with the statistic produced. panels (b) show the residuals produced when fitting the spectra with our best-performing models. The individual 1D posterior probability distributions for the BH spin and the viewing inclination are shown in the left and right panels of Figure \ref{fig:MAXI_J1348-630_combined}, respectively, using the colored curves. The combined distributions are shown through the solid black curves in Figure \ref{fig:MAXI_J1348-630_combined}. Through our analysis, we measure $a=0.977^{+0.017}_{-0.055}$ and $\theta=52^{+8}_{-11}\;[^\circ]$. The BH spin takes values higher than those measured by \cite{2022MNRAS.511.3125J}, as expected when allowing the emissivity parameters to vary freely. Through the introduction of the \texttt{gaussian} absorption features, the inclination generally takes higher values, leading to an inclination measurement higher than that reported by \cite{2022MNRAS.511.3125J}. 

It is worth focusing on the results from a few observations. For ObsID 80402315012, the fits do not require the addition of an absorption feature, with the statistical improvement being insignificant. This observation has a high reflection strength, leading to the results produced by the fits to the spectra from this observation being weighted more strongly when combining independent results. The fits to the spectra from ObsID 80402315012 produce spin and inclination constraints lower than the combined distributions, peaking around $a\sim0.84$ and $\theta\sim24^\circ$, respectively. However, when including a \texttt{gaussian} component around 7 keV, both the spin and the inclination distributions peak at higher values, more consistent with the other measurements: $a\sim0.94$ and $\theta\sim39^\circ$. Nevertheless, as the improvement in terms of $\chi^2$ was minimal and the deviance information criterion (DIC) computed based on the MCMC runs was worse due to the increased complexity of the model when including the \texttt{gaussian} component, we chose report the results of the analysis which did not include it, even if the inclusion would lead to results in better agreement with the ones from the other observations. In contrast, the fits to the spectra from ObsID 80502304006 produce a low inclination despite the inclusion of the absorption feature in the model. At the same time, the fits to the spectra from ObsID 80402315004 produce a high inclination measurement despite not including the absorption feature around 7 keV. At the same time, the MCMC analysis of the spectra from ObsID 80502304002 produces a posterior distribution for spin that has two peaks, one around $a=0.77$ and one for nearly maximal values of $a$. This is likely due to the inability of the data to distinguish between similarly good solutions. Despite the best fit in terms of $\chi^2$ being achieved for a nearly maximal spin, the solution with lower spin produces a similar statistic. As the parameter space is wider around the lower spin solution, the walkers converge to that solution and have trouble returning to the best-fit solution (with high spin) due to the parameter space being narrower. A similar phenomenon was described in \cite{2023ApJ...954...62D}. All these systematic variations are encapsulated in the uncertainties reported on our measurements.

\subsection{GS 1354-645}\label{sec:GS_1354-645}
Two of the existing three NuSTAR observations of GS 1354-645 were analyzed by \cite{2016ApJ...826L..12E}, where a BH spin of $a\geq0.98$ and a high inclination of $\theta=75^\circ\pm2^\circ$ were reported. We analyzed all three existing observations (ObsIDs 90101006002, 90101006004, and 90101006006), and fit the spectra in the entire 3-79 keV NuSTAR band. We note that the BH mass and distance to the system are highly uncertain. We used $M=7.6\pm0.7\;M_\odot$ and $d=43\pm18\;\rm kpc$ (\citealt{2009ApJS..181..238C}).

The top panels in Figure \ref{fig:GS_1354_delchi} show the unfolded spectra of the three observations, and the middle panels show the residuals produced when fitting the spectra with models that do not account for relativistic reflection. These residuals show clear features of relativistic reflection. The bottom panels of Figure \ref{fig:GS_1354_delchi} show the residuals produced by the best-fitting reflection models, together with the fit statistic. 

The posterior distributions produced by the MCMC analysis for the spin and inclination parameters are shown in the left and right panels of Figure \ref{fig:GS_1354_combined}, with the width of the lines being proportional to the weighting used when combining the measurements. In the two panels of Figure \ref{fig:GS_1354_combined}, the black dotted curves show the combined distribution obtained using the method highlighted in \cite{Draghis23}, and the black solid curves show the probability distribution of the combined measurements obtained using the method presented in \cite{2023ApJ...954...62D}. We measured a spin of $a=0.849^{+0.103}_{-0.221}$ and an inclination of $\theta=47^{+11}_{-10}\;[^\circ]$. It is interesting to note that the inclination measurement is lower and inconsistent with that of \cite{2016ApJ...826L..12E}, and that the spin distribution is driven to lower values by the measurements of two of the three available observations. However, the third observation (ObsID 90101006006) produces a posterior distribution that is strongly peaked at very high values and it does produce a measurement consistent with that previous results. When combining the independent measurements from the three observations in a way that more strongly accounts for systematic uncertainties between observations, we obtain a more loose constraint on the spin, which peaks at a lower value than previously reported.

\subsection{MAXI J1535-571}\label{sec:MAXI_J1535-571}
We fit 16 of the 17 available NuSTAR observations of MAXI J1535-571, excluding the observation during which the source flux was placing the observation outside the Eddington ratio range for which we expect the accretion disk to extend to the ISCO. The fluxes were calculated assuming a BH mass of $M=8.9\pm1\;M_\odot$ (\citealt{2019ApJ...875....4S}) and a distance of $d=4.1\pm0.6\rm kpc$ (\citealt{2019MNRAS.488L.129C}). The residuals produced when fitting the spectra with our baseline models are shown in the top panels of Figure \ref{fig:MAXI_J1535_delchi}, and the residuals produced when fitting the spectra with the best-performing reflection models are show in the bottom panels in Figure \ref{fig:MAXI_J1535_delchi}.

Similar to the case of a few other sources, the 1D posterior distributions obtained from the MCMC runs on the best-performing reflection models indicate that the inclination of the inner disk is not consistently measured (shown in the right panel of Figure \ref{fig:MAXI_J1535_combined}). The combining algorithm produces a poorly constrained inclination measurement of $\theta=44^{+17}_{-19}\;[^\circ]$. The large uncertainty on the measurement is caused by some independent measurements preferring very low inclination values, while others preferring nearly maximal inclinations. The fits for a handful of spectra statistically prefer the addition of \texttt{gaussian} absorption component around 7~keV, consistent with absorption caused by ionized winds, which would suggest a high viewing inclination of the source. However, there is no apparent trend that would suggest that the addition of the absorption component leads to preferentially low or high inclination measurements. This result suggests the need to better quantify the ability of the reflection models to constrain the viewing inclination. The spin measurements are all consistent with a high value, with the individual 1D posterior distributions being shown in the left panel of Figure \ref{fig:MAXI_J1535_combined}. However, a few measurements suggest a BH spin around $a\sim0.9$, while most prefer nearly maximal values. Interestingly, the same observations that seem to prefer a somewhat lower spin also prefer a low inclination, such as the ones produced by fits to the spectra from ObsIDs 80302309006 and 80402302008. Using the individual 1D posterior distributions, our combining algorithm produces a spin measurement of $a=0.979^{+0.015}_{-0.049}$, in agreement with the measurement of \cite{2022MNRAS.514.1422D}, who report a spin of $a=0.985^{+0.002}_{-0.004}$ and an inclination in the range $70^\circ\leq\theta\leq74^\circ$. Interestingly, our inclination measurement is in reasonable agreement with the inclination determined using ballistic motion in the observed radio jet of $\theta\leq45^\circ$ (\citealt{2019ApJ...883..198R}).

\subsection{MAXI J1631-479}\label{sec:MAXI_J1631-479}
We analyzed the four existing NuSTAR observations of MAXI J1631-479 (ObsIDs 80401316002, 80401316004, 90401372002, and 90501301001). Due to its proximity to the Sun during ObsID 90501301001, the source tracking was unstable at times, causing the position of the source on the detectors to drift. During one such episode of imperfect tracking, the source landed on detector (DET) 1 in FPMA (nominally the source was placed on DET 0), and on the gap between DET 0 and DET 1 in FPMB. The source being located on the gap between the detectors in FPMB for part of the observation led to improper source reconstruction while running the default \texttt{nupipeline} procedures, causing a significant difference between the spectra measured by the FPMA and FPMB NuSTAR detectors. Therefore, we split the Observing Mode 01 (SCIENCE) files produced during the observation based on the camera head unit (CHU) configuration used to reconstruct the science image, similarly to the procedure outlined in the suggestions regarding Mode 06 observations. Using the event files generated while using different CHU combinations, we identified the interval during the observation during which the source drifted into the gap between DET 0 and DET 1 on FPMB, and created user good-time intervals (GTIs) to exclude this segment of the observation, which were used to reextract Mode 01 science products from both the FPMA and FPMB detectors, leading to spectra in significantly better agreement. The spectra from the four observations, together with the residuals produced when fitting them with our baseline model and with the best-fitting reflection models, are shown in Figure \ref{fig:MAXI_J1631_delchi}. 

Upon running the MCMC analysis of the best-fitting spectral models for the four individual observations, we obtained the 1D posterior probability distributions shown in Figure \ref{fig:MAXI_J1631_combined}. We combined the independent measurements using our combining algorithm, and measured a spin of $a=0.951^{+0.039}_{-0.077}$ and an inclination of $\theta=22^{+10}_{-12}\;[^\circ]$. It is important to note that the independent probability distributions for inclination produced by the four observations are in disagreement, with the inclination measured using the spectra from ObsID 90501301001 taking a value of $\theta\sim75^\circ$. The discrepancy in inclination measurements likely comes from an inability of the data to constrain the presence and importance of absorption features around $\sim7$ keV, which land on the blue wing of the Fe K line complex, strongly influencing the ability to measure inclination. However, our measurements for both the BH spin and inclination are in good agreement with those of \cite{2020ApJ...893...30X}, who measured a spin $a\geq0.94$ and an inclination of $\theta=29^\circ\pm1^\circ$.

\subsection{4U 1630-472}\label{sec:4U_1630-472}
At the time of the analysis, six NuSTAR observations of 4U 1630-472 were publicly available (ObsIDs 40014009001, 80802313002, 80802313004, 80802313006, 90002004002, and 90002004004). Using the spectra from ObsID 40014009001, \cite{2014ApJ...784L...2K} measured a BH spin of $a=0.985_{-0.014}^{+0.005}$, and highlighted the presence of absorption features linked to ionized outflows in the source. We detected similar absorption features in all six observations analyzed. As the goal of the analysis is to describe the shape of the relativistically broadened reflection features and to measure the BH spin, we adopted the simplest models which properly account for the absorption features in the spectrum. Therefore, in all models, we added a \texttt{gaussian} absorption feature and included the \texttt{zxipcf} multiplicative component. The unfolded spectra from the six NuSTAR observations are shown in Figure \ref{fig:4U_1630_delchi}, together with the residuals produced using our baseline model, and the best-performing reflection models. 

Upon running the MCMC analysis of the best-performing models, we obtained the 1D posterior probability distributions shown in Figure \ref{fig:4U_1630_combined} for spin (left) and inclination (right). We combined the measurements using our combining algorithm and obtained a spin of $a=0.857^{+0.095}_{-0.211}$ and an inclination of $\theta=55^{+8}_{-11}\;[^\circ]$. The inclination and the spin measured are formally consistent with those measured by \cite{2014ApJ...784L...2K}, however taking lower values, and with larger uncertainties. This is a product of analyzing a larger number of observations, which independently lead to varying results. It is important to note that while all observations show a preference for high spin values, the measurement is strongly determined by the results of the analysis on the spectra from ObsIDs 90002004004 and 80802313006.

\subsection{Swift J1658.2-4242}\label{sec:Swift_J1658.2-4242}
\cite{2018ApJ...865...18X} used NuSTAR observation 90401307002 to place a limit on the spin of Swift J1658.2-4242 to be $a>0.94$, and measured the inclination of the inner disk to be $\theta=64^{+2}_{-3}\;[^\circ]$. In our analysis, we used the eight available NuSTAR observations: ObsIDs 80301301002, 80302302002, 80302302004, 80302302006, 80302302008, 80302302010, 90401307002, and 90401317002. In all observations, a clear absorption feature was detected around $\sim7.1$ keV, so we included a \texttt{gaussian} absorption feature in all the models that we tested. We note that this source shows long- and short-scale dips in the light curves. For consistency with the rest of the analysis, we extracted and fit the time-averaged spectra from all NuSTAR observations.

The top panels in Figure \ref{fig:Swift_J1658-4242_delchi} show the residuals produced when fitting the spectra from the eight observations with the \texttt{TBabs*(diskbb+powerlaw)} model. The bottom panels show the residuals in terms of $\sigma$ produced when fitting the spectra with the best-performing reflection models, which include the \texttt{gaussian} absorption feature around $\sim7.1$ keV. No strong residuals are apparent in the lower panels, suggesting that good fits were achieved for all spectra.

The individual 1D posterior distributions for spin and inclination are shown in the left and right panels of Figure \ref{fig:Swift_J1658-4242_combined}, respectively. Using our combining algorithm, we obtain $a=0.951^{+0.031}_{-0.069}$ and $\theta=50^{+9}_{-10}\;[^\circ]$. We note that for one single observation, the inclination measurement is systematically different from the rest. If we fix the inclination in the fits to the spectra from ObsID 90401317002 to $\theta=50^\circ$, we obtain a fit worse by $\Delta\chi^2=36$, but the spin measurement is unaffected. This again highlights the importance of analyzing all available data in order to encapsulate and characterize systematic uncertainties due to variations between observations. 

\subsection{GX 339-4}\label{sec:GX_339-4}
Of the 38 available archival NuSTAR observations, two occurred while the source was at an Eddington fraction below $10^{-3}$. We computed this using the values estimated by \cite{2016ApJ...821L...6P}: $M=9\pm1.5\;M_\odot$ and $d=8.4\pm0.9\;\rm kpc$. We fit the spectra from the other 36 observations, with ObsIDs listed in Appendix \ref{sec:obsids}. We show the residuals obtained through fits using our baseline model and the best-performing reflection models in Figure \ref{fig:GX_339_delchi}.

The 1D posterior distributions for BH spin and inclination are shown in the left and right panels of Figure \ref{fig:GX_339_combined}, respectively. We combined the measurements using our combining method, and obtained a spin of $a=0.970^{+0.026}_{-0.076}$, in good agreement with the result of \cite{2016ApJ...821L...6P} who find a spin of $a=0.95_{-0.08}^{+0.02}$. However, the independent 1D posteriors for inclination range throughout the entire parameter space, with our combining algorithm measuring a poorly constrained inclination of $\theta=49\pm14\;[^\circ]$. It is unclear what aspects of the spectra or the models lead to the inability to constrain the inclination of the source reliably between observations, and further investigation is required to understand this discrepancy. 

\subsection{IGR J17091-3624}\label{sec:IGR_J17091-3624}

We analyzed the 14 public NuSTAR observations available at the time. We assumed a mass of $M=12.15\pm3.45\;M_\odot$ and a distance of $d=12.6\pm2.0\;\rm kpc$ (\citealt{2015ApJ...807..108I}). A few observations produced fluxes in slight excess of the upper limit of Eddington fraction required by our analysis. However, as the flux estimates are model dependent and as the uncertainties of the BH properties are relatively large, we included all observations in the analysis. The residuals produced when fitting the spectra from the 14 observations with our baseline model are shown in the top panels in Figure \ref{fig:IGR_J17091_delchi}, and the residuals produced by the best-performing reflection models together with the statistic produced are shown in the bottom panels in Figure \ref{fig:IGR_J17091_delchi}.

Similarly to the case of LMC X-3 (described in Subsection \ref{sec:LMC_X-3}), the individual spin measurements agree on a high value, and the inclination measurements take values across the entire parameter space. The 1D posterior distributions obtained through the MCMC analysis of the best-performing reflection models are shown in Figure \ref{fig:IGR_J17091_combined}. Using our measurement combining algorithm, we obtain $a=0.963^{+0.027}_{-0.085}$ and $\theta=47^{+10}_{-11}\;[^\circ]$. The high spin measurement for this source is in disagreement with that measured by \cite{2018MNRAS.478.4837W}, who report a spin in the range $-0.13 \leq a \leq 0.27$. It is worth noting that in their analysis, \cite{2018MNRAS.478.4837W} reported that the source favors a low inner emissivity index, $q_1\sim3.7$, which could potentially lead to an underestimation of the BH spin (see \citealt{2023ApJ...954...62D} and Section \ref{sec:GRS_1739-278}). This source is particularly interesting, as it exhibits `heartbeat' variability (\citealt{2011ApJ...742L..17A}) similar to that observed in GRS 1915+105 (see, e.g. \citealt{1997ApJ...479L.145B, 2011ApJ...737...69N}), for which studies have shown that the BH spin is high.

\subsection{GRS 1716-249}\label{sec:GRS_1716-249}
NuSTAR observed GRS 1716-249 six times during its 2017 outburst. However, ObsID 80201034004 had an effective exposure of only 408 s, and the spectra are well fit using a power-law model. Therefore, we only analyzed the other five observations, with ObsIDs 80201034006, 80201034007, 90202055002, 90202055004, and 90301007002. All observations fall within the required Eddington range regime when assuming a mass of $M=6.5\pm1.5\;M_\odot$ (\citealt{2019ApJ...887..184T}) and a distance of $d=2.4\pm0.4\;\rm kpc$ (\citealt{1994A&A...290..803D}). The residuals produced when fitting the NuSTAR FPMA and FPMB spectra using a power-law model are shown in the panels (b) in Figure \ref{fig:GRS_1716_delchi}, clearly indicating signs of relativistic reflection. \cite{2019ApJ...887..184T} analyzed the last three observations mentioned above, jointly with simultaneous Swift observations. By only fitting the NuSTAR data in the 4.5-79 keV band, accounting for the thermal emission from the accretion disk using the \texttt{kerrbb} model, and by using only the default \texttt{relxill} model flavor while fixing the inner and outer emissivity indices $q_1=q_2=3$, they constrain the spin to $a>0.92$ and the inclination $\theta$ to $40-50^\circ$. 

We used our pipeline on all the existing NuSTAR data in the entire 3-79 keV band pass. The best-fit models that account for reflection features produce the residuals shown in panels (c) in Figure \ref{fig:GRS_1716_delchi}, while panels (a) show the unfolded NuSTAR spectra during the observations. The individual posterior probability distributions for BH spin and inner disk inclination produced by the MCMC analysis on the best-performing models are shown in the left and right panels of Figure \ref{fig:GRS_1716_combined}, respectively. Combining the individual measurements produces a spin measurement of $a=0.97_{-0.06}^{+0.02}$ and an inclination measurement of $\theta=59^{+7}_{-12}\;[^\circ]$.

\subsection{GRS 1739-278}\label{sec:GRS_1739-278}

At the time of the analysis, there were five public NuSTAR observations of GRS 1739-278. ObsID 80002018002 was taken in 2014, and is the only one of the five during which the source had a flux high enough to fall within the Eddington ratio limits for which we expect the accretion disk to extend to the ISCO. We computed this using estimated values for the BH mass, $M=6.75\pm2.75\;M_\odot$, and distance, $d=7.25\pm1.25\; \rm kpc$, in agreement with the findings of \cite{2018PASJ...70...67W}. During the other four archival observations, taken in 2015 and 2016, the source had a significantly lower flux. Additionally, we obtained a NuSTAR DDT observation of GRS 1739-278 during its 2023 outburst (ObsID 90901323002), during which the source was in a soft spectral state, and again within the preferred Eddington ratio regime. 

For ObsID 80002018002, we extracted spectra from circular source regions with a radius of 90'', in order to exclude the contribution from the nearby source 2CXO J174231.5-274349. The FPMA observation was contaminated by stay light, so we manually placed a circular background region of the same size in an uncontaminated region, and located the background extraction region for FPMB in a similar location to that in FPMA. For ObsID 90901323002, 2CXO J174231.5-274349 does not show any activity, so we used our default sized 120'' circular source region for the source. Similarly to the previous observation, the FMPB observation is contaminated by stray light, so we placed background regions of the same size as the source regions in appropriate locations to account for the effect. 

The spectra extracted are shown in the top panels of Figure \ref{fig:GRS_1739_delchi}. During the 2014 observation, GRS 1739-278 was in a hard state, while the 2023 observation captured the source in a soft state. The central panels show the residuals produced when fitting the spectra with the \texttt{TBabs*(diskbb+powerlaw)} model, indicating clear signs of relativistic reflection. The bottom panels of Figure \ref{fig:GRS_1739_delchi} show the residuals produced by our best-performing reflection models. 

Two aspects of the fits are important to note. The spectra obtained from ObsID 80002018002 do not statistically require the addition of a \texttt{diskbb} component, so our best fit is achieved using the \texttt{TBabs*relxill} model. In the fits to the spectra from ObsID 90901323002, the Galactic column density $N_{\rm H}$ goes to zero, indicating either that during the first observation, the measured column density is mostly due to intrinsic absorption in the system, or that the correlation between the \texttt{TBabs} and the \texttt{diskbb} components in the model cannot be broken given the low energy bound of the NuSTAR spectra of 3~keV. All the fits to the spectra from the two observations prefer high spins and inclinations. Using the spectra from ObsID 80002018002, \cite{2015ApJ...799L...6M} measured a high spin, $a=0.8\pm0.2$ and an intermediate inclination $\theta=43.2\pm0.5\;[^\circ]$. 

The posterior distributions for the spin and inclination obtained by running the MCMC analysis on the best-performing models are shown through the blue and red curves in the left and right panels of Figure \ref{fig:GRS_1739_combined}. Given the quality of the data from ObsID 90901323002, despite the best fit being achieved for a high spin, a large number of the posterior samples prefer a solution with similar but worse statistic in which the spin is negative and the inner emissivity index $q_1$ is low. This behavior is frequent when the quality of the data is not good enough for the fit to strongly distinguish between two similar families of solutions. This behavior is described in \cite{2023ApJ...954...62D}, and the measured low spin is to be interpreted as a lower limit due to the measured low inner emissivity index. Nevertheless, when combining the posterior distributions using our combining algorithm, we measure a spin of $a=0.968^{+0.022}_{-0.074}$ and an inclination $\theta=70^{+5}_{-11}\;[^\circ]$. The spin is formally in agreement with the value measured by \cite{2015ApJ...799L...6M}, but the measured inclination takes a higher value, consistent with the values measured by the fits to the two individual sets of NuSTAR spectra. Visually inspecting the left panel of Figure \ref{fig:GRS_1739_combined} indicates that the combining algorithm accounts for the contribution of the second independent measurement, which produces a low spin by bringing the lower bound of our uncertainty interval to lower values, suggesting that the method well includes the systematic uncertainties of the variation of the measurements caused by differences between observations. 

\subsection{1E 1740.7-2942}\label{sec:1E_1740.7-2942}
At the time of the analysis, there were four NuSTAR observations of 1E 1740.7-2942 available (ObsIDs 10002012001, 10002021001, 10002021003, 90701317002). \cite{2020MNRAS.493.2694S} analyzed the FPMA spectrum from ObsID 10002012001 in addition to XMM-Newton and INTEGRAL observations of the source, and using the relativistic reflection method they find a nearly maximal but poorly constrained BH spin, quoting $a\geq0.5$.  Additionally, \cite{2020MNRAS.493.2694S} estimated the BH mass to be around $M\sim5\;M_\odot$ and the distance to the system to be comparable to that to the Galactic center. Therefore, for estimating the Eddington fractions, we assumed a BH mass of $M=5\pm0.5\;M_\odot$ and a distance of $d=8.5\pm2\;\rm kpc$. All four available NuSTAR observations of the source happened while, given the above assumptions, the source was at an Eddington fraction within the range required for this analysis.

We extracted the spectra from the four NuSTAR observations and fit them with our baseline family of models. The reflection models perform significantly better than the simple \texttt{TBabs*(diskbb+powerlaw)} model, but similarly to models that describe the continuum as a thermally Comptonized continuum: \texttt{TBabs*(diskbb+nthcomp)}. This suggests that reflection is not statistically significant in the four NuSTAR observations. However, for the same observation as the one treated by \cite{2020MNRAS.493.2694S} (ObsID 10002012001), the \texttt{TBabs*(diskbb+relxill)} model does reach a solution that produces a statistically significant improvement through reflection, but some of the parameters are unphysical. The required accretion disk temperature is high, $T_{\rm in}\sim2.5~\rm keV$, with the normalization for the FPMB spectrum going to zero. In other words, the FPMA spectrum drives the \texttt{diskbb} component to a high temperature to account for a difference between the two NuSTAR FPM spectra. To test for a possible difference between the detectors artificially induced by our spectral extraction, we reextracted the spectra from this observation, this time manually placing circular background extraction regions instead of the annular background regions. The resulted spectra produce the same fits, suggesting that there is indeed a difference between the two NuSTAR spectra. When fitting the two spectra simultaneously and allowing the normalization of the \texttt{relxill} component and the power-law indices to differ for the two spectra, the quality of the fit is significantly improved and the \texttt{diskbb} component becomes no longer required. 

Inspecting the light curves of the observations shows that toward the end of the exposure, the count rate in the FPMB detector drops significantly, while the rate in the FPMA detector remained unchanged. We identified the source of this abnormal behavior to be a shift of the position of the source to the gap between two of the four detectors that make up the NuSTAR FPMs. We generated GTI files that excluded the last part of the observation, during which the decrease in FPMB count rate occurred. We reextracted the spectra using the GTI files, and fit them again with the same models. 

Fitting the spectra with \texttt{TBabs*(diskbb+powerlaw)} produces $\chi^2/\nu=387.43/344$, while fitting with \texttt{TBabs*(diskbb+nthcomp)} produces $\chi^2/\nu=371/343$. If we add reflection to the models, we obtain  $\chi^2/\nu=365.21/335$ when fitting with \texttt{TBabs*(diskbb+relxill)}. Further improvement is achieved if the power-law index is allowed to vary between the FPMA and FPMB spectra in this model, obtaining $\chi^2/\nu=356.22/337$, but this model no longer requires the presence of a \texttt{diskbb} component, making the model \texttt{TBabs*relxill}. Similarly, when fitting the spectra with \texttt{TBabs*relxillCp} with free power-law indices, the fit returns $\chi^2/\nu=356.25/337$. Note that given the high column density along the line of sight, the low energies are heavily obscured in the observed spectra. Therefore, it is unsurprising that the disk component is weakly required by the NuSTAR spectra. However, when the power-law indices are linked between the spectra from the two NuSTAR detectors, removing the \texttt{diskbb} component produces a significantly worse fit, with $\chi^2/\nu=378.07/338$. Therefore, we ran the MCMC analysis for the mentioned best-performing reflection models. 

The top panel in Figure \ref{fig:1E_1740_delchi} shows the unfolded spectra from ObsID 10002012001. The middle and bottom panels in Figure \ref{fig:1E_1740_delchi} show the residuals in terms of $\sigma$ produced when fitting the spectra with the \texttt{TBabs*(diskbb+powerlaw)} and the \texttt{TBabs*(diskbb+relxill)} models, together with the statistic produced. In this figure, we show the residuals produced by the reflection model in which the power-law indices $\Gamma$ are linked between the FPMA and FPMB spectra. Despite the models with free $\Gamma$ being slightly statistically favored in terms of DIC following the MCMC analysis, they both measure low inner disk inclinations, which are unlikely given the clearly observed radio jets in this source (\citealt{2015A&A...584A.122L}). Furthermore, the broadband fits performed by \cite{2020MNRAS.493.2694S}, which include XMM-Newton and Integral spectra, show the presence of a thermal disk component in the system. For these reasons, and in order to maintain consistency with the rest of the analyzed sample, we report the results of the \texttt{TBabs*(diskbb+relxill)} model with $\Gamma$ linked between the two FPM spectra, despite it being slightly statistically disfavored. However, as seen in Figure \ref{fig:1E_1740_combined}, the spin measurements are generally consistent between the fits.

The two panels in Figure \ref{fig:1E_1740_combined} show the posterior distributions for the spin (left) and inclination (right) for the \texttt{TBabs*(diskbb+relxill)} model with $\Gamma$ linked (blue line) and for the \texttt{TBabs*relxill} model with $\Gamma$ free (red line). The dotted and solid black curves show the results produced by our combining algorithm when ran \textit{only} on the blue posterior distributions. The values measured are $a=0.856^{+0.134}_{-0.443}$ and $\theta=31^{+29}_{-18}\;[^\circ]$. Note that the inclination of the system is unconstrained by the fit when linking the power-law indices between the spectra, and constrained to low values when allowing them to vary independently. This latter constraint is likely an artifact of an improper characterization of the underlying continuum caused by the lack of inclusion of the \texttt{diskbb} component, which given the quality of the data, is not statistically significant. The modes of the independent distributions for the BH spin are high and much more narrowly constrained, with $a=0.97^{+0.3}_{-0.32}$ when linking $\Gamma$, and $a=0.994^{+0.004}_{-0.059}$ when allowing $\Gamma$ to vary independently. The independent measurements likely strongly underestimate the size of the systematic uncertainty, but by applying our combining algorithm even on one fit alone, we obtain a much better estimate of the true size of the uncertainties of the measurement. 

Visually inspecting the residuals in Figure \ref{fig:1E_1740_delchi}, the presence of relativistic reflection is not immediately obvious. However, the statistics of the fits indicate that reflection is statistically significant. This raises the valuable point that it is possible that many observations with low signal-to-noise ratios that could still place constraints on the BH spin were simply not analyzed because the residuals did not indicate visually clear reflection features, but the quality of the data might still be good enough to obtain a statistically significant constraint. Lastly, given the shape of the residuals when fitting with \texttt{TBabs*(diskbb+powerlaw)} and \texttt{TBabs*(diskbb+relxill)}, another possible interpretation can be offered to the spin constraint: if the spin of the BH was small, the spectra would have shown narrower features, which would have been easier to distinguish from the underlying continuum. Ultimately, we encourage the readers to visit the more in-depth analysis of \cite{2020MNRAS.493.2694S}.

\subsection{Swift J174540.7-290015}\label{sec:T15}

\begin{figure*}[ht]
    \centering
    \includegraphics[width= 0.9\textwidth]{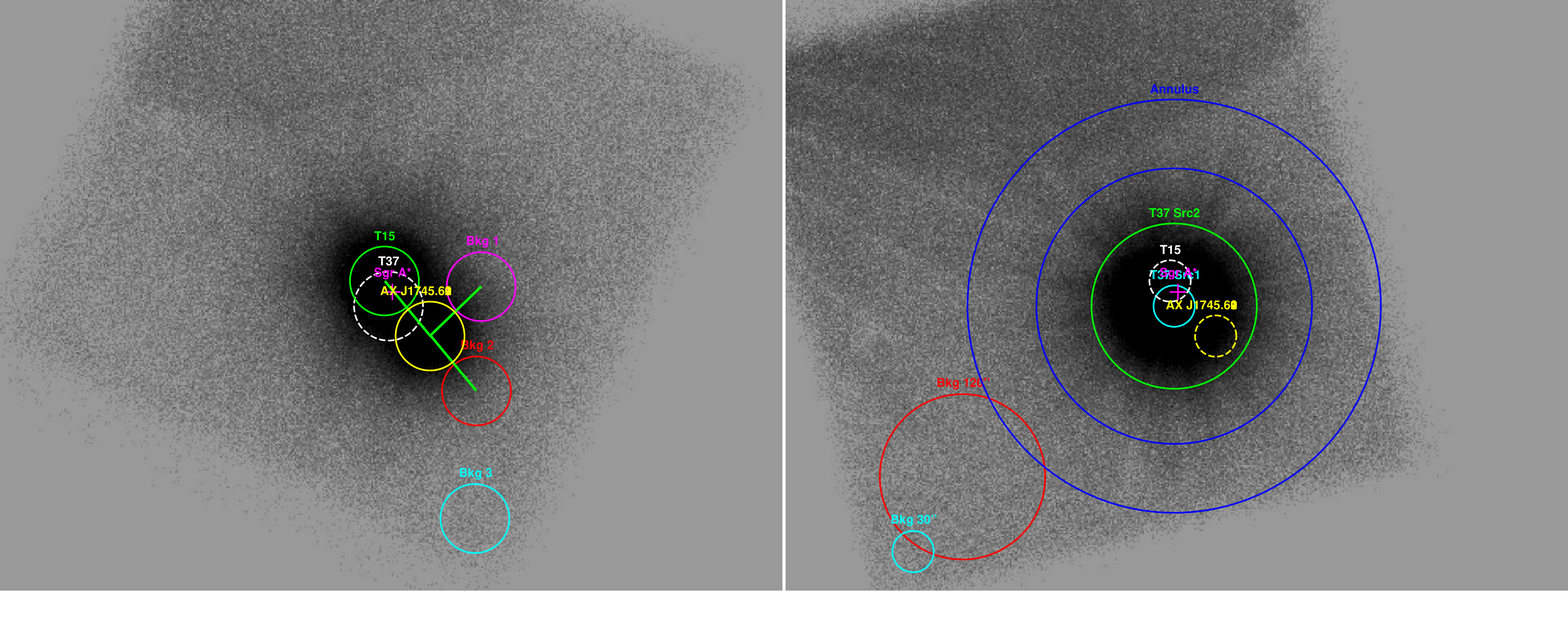}
    \caption{Source and background regions used for extracting spectra of T15 (left) and T37 (right). See text in Sections \ref{sec:T15} and \ref{sec:T37}. }
    \label{fig:GC_regions}
\end{figure*}

Swift J174540.7-290015 (hereafter T15) is an XB located near the line of sight to the Galactic center. It was observed to be in an outburst during the NuSTAR observation 90101022002. During the same observation, the source AX J1745.6-2901 was also above quiescence. The left panel in Figure \ref{fig:GC_regions} shows the exposure map of the NuSTAR observation. The green circle shows the 50'' region used for extraction of the source spectrum. The white dashed circle indicates a 50'' region at the position of Swift J174540.2-290037, the other source present in this field, also analyzed in this paper (see Section \ref{sec:T37}), which was in quiescence during this observation. The yellow circle shows a 50'' region around the source AX J1745.6-2901. The magenta cross represents the position of the Galactic center. The magenta, red, and cyan circles (labeled Bkg 1, Bkg 2, and Bkg 3, respectively) show three different background regions tested for extraction of background spectra. Regions 1 and 2 were placed around AX J1745.6-2901 at a distance equal to the distance between AX J1745.6-2901 and T15 (our target), with the aim of quantifying not only the contribution from the detector background, but also from AX J1745.6-2901 to our source region. Region 3 was placed away from the sources, and quantifies the detector background. The difference between the background spectra from regions 1 and 2 were minimal, with slightly increased rates in region 1 due to contribution from T15. In order to adopt the most conservative background choice, we continued our analysis with the background spectra from region 1. However, we note that our results are consistent when using the background spectrum obtained from region 2. 

T15 was analyzed by \cite{2019ApJ...885..142M}, who measured a BH spin of $a=0.94^{+0.03}_{-0.01}$ and an inclination of $\theta=64.2^{+0.9}_{-1.6}\;[^\circ]$. Of the models tested on the NuSTAR spectra of T15, the fit using \texttt{TBabs*(diskbb+relxillD\_19+gaussian)} perform best in terms of DIC. This model includes a \texttt{gaussian} absorption feature around 7.16~keV. The unfolded spectra of the observation are presented in the top panel of Figure \ref{fig:T15_delchi}. The middle and bottom panels of Figure \ref{fig:T15_delchi} show the residuals produced when fitting the spectra with the \texttt{TBabs*(diskbb+powerlaw)} model and the best-performing reflection model, respectively, together with the statistic produced. The posterior probability distributions for BH spin and inclination of the inner accretion disk are shown by the blue curves in the left and right panels of Figure \ref{fig:T15_combined}, respectively. Using only the posterior distributions from the MCMC analysis of this observation in our combining algorithm produces the black curves shown in Figure \ref{fig:T15_combined}. We measure $a=0.884^{+0.068}_{-0.109}$ and $\theta=63^{+10}_{-8}\;[^\circ]$. These results are formally consistent with those of \cite{2019ApJ...885..142M}, with larger uncertainties which better encapsulate the systematic uncertainties of the measurement.

\subsection{Swift J174540.2-290037}\label{sec:T37}

Similarly to T15 (see Section \ref{sec:T15}), Swift J174540.2-290037 (hereafter T37) is located near the Galactic center. The source was detected during the NuSTAR observation 90201026002. The right panel of Figure \ref{fig:GC_regions} shows the NuSTAR exposure map during the observation. The cyan and green circles show 30'' and 120'' regions centered at the position of T37, respectively. The magenta cross shows the position of the Galactic center, and the white and yellow dashed circles show the positions of T15 and AX J1745.6-2901, both in quiescence during the observation. The blue annulus shows the typical background region used throughout our analysis, with an inner radius of 200'' and an outer radius of 300''. The red and cyan circles respectively show the 120'' and 30'' circles that were used for extracting background regions. We extracted source and background spectra from the 30'' regions, from the 120'' source and background regions, and from the 120'' source region and the background annulus region. We chose to continue our analysis with the source spectra extracted from the 120'' source region and background spectra extracted from the annulus, as this is the most conservative pair of source and background regions, and also ensures consistency with the method used throughout the entire paper. We note that using different pairs of source and background regions produce consistent results.

T37 was analyzed by \cite{2019ApJ...885..142M}, who measured a BH spin of $a=0.92^{+0.07}_{-0.05}$ and an inclination of $\theta=21^{+2}_{-3}\;[^\circ]$. The top panel in Figure \ref{fig:T37_delchi} shows the unfolded spectra of the NuSTAR observation of T37, while the middle and bottom panels show the residuals in terms of $\sigma$ produced when fitting the spectra with a model that does not account for reflection, and with our best-performing reflection model, \texttt{TBabs*(diskbb+relxillCp)}, respectively. The 1D posterior probability distributions for the spin and inclination are shown through the red curves in the left and right panels of Figure \ref{fig:T37_combined}, respectively. The dotted and solid black curves show the results of applying the two methods of our combining algorithm on the single posterior distributions. We measure $a=0.774^{+0.082}_{-0.106}$ and $\theta=31^{+8}_{-9}\;[^\circ]$.
\\
\newline

\subsection{MAXI J1803-298}\label{sec:MAXI_J1803-298}

NuSTAR observed the 2021 outburst of MAXI J1803-298 four times (ObsIDs 80701332002, 90702316002, 90702318002, and 90702318003). We fit the spectra extracted from all four observations (shown in the top panels of Figure \ref{fig:MAXI_J1803_delchi}) in the entire 3-79 keV NuSTAR band pass, with the exception of the spectra from ObsID 90702318003, which we only fit in the 3-20 keV band, as they became background dominated at higher energies. \cite{2022MNRAS.516.2074F} used the spectra from ObsID 90702318002 paired with a NICER observation to measure a BH spin of $a=0.991\pm0.001$ and an inclination of $\theta=70.8\pm1.1\;[^\circ]$. Additionally, based on all the existing NuSTAR observations, \cite{2023ApJ...949...70C} measured a spin of $a=0.988^{+0.004}_{-0.010}$ and an inclination of $\theta=75\pm2\;[^\circ]$. 

The middle and bottom panels show the residuals produced when fitting the spectra with the model that ignores reflection and with our best-performing reflection models, which include narrow absorption features around 7~keV in all cases. Running the MCMC analysis on our best-performing models produces the 1D posterior probability distributions shown in Figure \ref{fig:MAXI_J1803_combined}, with the left panel showing the distributions for the spin parameter, and the right panel showing the distributions for the inclination parameter. By combining the individual posterior probability distributions using our combining algorithm, we measure a spin of $a=0.987^{+0.007}_{-0.037}$ and an inclination of $\theta=72_{-9}^{+6}\;[^\circ]$, in good agreement with the results of \cite{2022MNRAS.516.2074F} and \cite{2023ApJ...949...70C}.

\subsection{MAXI J1813-095} \label{sec:MAXI_J1813-095}
NuSTAR observed MAXI J1813-095 three times in 2018 (ObsIDs 80402303002, 80402303004, and 80402303006). \cite{2021RAA....21..125J} fit the three observations with a model that accounts for relativistic reflection to find a lower limit for the spin parameter of $a\geq0.76$ and an inclination in between $28^\circ<\theta<45^\circ$. Later, \cite{2022MNRAS.514.1952J} used an averaged spectrum obtained from the existing three NuSTAR observations to study the reflection component, and by freezing the spin parameter in their fits to the maximum possible value of $a=0.998$, they find an inner disk radius consistent with the ISCO and an inner disk inclination of $\theta=20^{+7}_{-10}\;[^\circ]$. 

Using the mass ($M=7.4\pm1.5\;M_{\odot}$) and distance ($d=6\pm1\;\rm kpc$) estimates for the BH in the system obtained by \cite{2021RAA....21..125J}, we find that all three existing NuSTAR observations occurred while the source was at an Eddington fraction for which we expect the inner disk radius to extend to the ISCO. Therefore, we continue analyzing all three existing observations of MAXI J1813-095 in the entire NuSTAR band pass of 3-79 keV band. The spectra from the three observations together with the residuals produced when fitting the spectra with a model not accounting for reflection are shown in the top and middle panels in Figure \ref{fig:MAXI_J1813-095_delchi}, indicating clear signs of relativistic reflection.

The residuals produced by fitting the spectra with our best-performing reflection models are shown in the bottom panels of Figure \ref{fig:MAXI_J1813-095_delchi}, together with the statistic produced. We note that for the third observation (80402303006), the \texttt{diskbb} component was not required by the data, but adding a narrow Gaussian absorption feature around $\sim 6.7$ keV produced a statistically significant improvement in the fit. In order to best assess the statistical significance of the improvement in statistic produced by the addition of the narrow absorption component, we removed the \texttt{diskbb} component for simplicity. We tested whether the addition of similar Gaussian features improves the quality of the fit for the other two observations, but the decrease in $\chi^2$ was not statistically significant.

The posterior distributions for the spin and inclination parameters obtained through the MCMC analysis are shown in the left and right panels of Figure \ref{fig:MAXI_J1813-095_combined}, respectively. The combined distributions for the two parameters are shown through the solid black curves in Figure \ref{fig:MAXI_J1813-095_combined}. We measure a spin of $a=0.88^{+0.10}_{-0.27}$ and an inclination of $\theta=42^{+11}_{-13}\;[^\circ]$. The three individual spin measurements agree on a high spin measurements, with the third observation producing a significantly narrower spin constraint, which drives the $1\sigma$ credible region of the combined distribution to produce a relatively precise spin constraint. The three individual inclination measurements are not in good agreement, hinting perhaps at the fact that the simplicity of the models does not fully capture the complexity of the data. However, this source highlights the benefit of using an averaging function when combining the individual measurements into a single posterior distribution which encapsulates systematic uncertainties not accounted for during the analysis. Our results are in agreement with those of \cite{2021RAA....21..125J}.

\subsection{MAXI J1848-015}\label{sec:MAXI_J1848-015}

The two existing NuSTAR observations of MAXI J1848-015 (ObsIDs 90601340002 and 90601341002) have been analyzed by \cite{2022ApJ...927..190P} who measured a spin of $a=0.967\pm0.013$ and an inclination of $\theta=26.4\pm0.5\;[^\circ]$. Due to the proximity of the source to the Sun during the observations, Mode 1 scientific data were unavailable. Therefore, we extracted spectra from the Mode 6 scientific data produced during the observation. The spectra taken during the two observations are shown in the top panels of Figure \ref{fig:MAXI_J1848_delchi}. Fits using the \texttt{TBabs*(diskbb+powerlaw)} model produce the residuals shown in the central panels of Figure \ref{fig:MAXI_J1848_delchi}. 

The residuals produced by the best-performing reflection models are shown in the bottom panels of Figure \ref{fig:MAXI_J1848_delchi}. For the spectra from ObsID 90601340002, the model that is statistically favored requires the addition of a narrow absorption line around 6.8~keV, while the spectra from ObsID 90601341002 were taken after the source had transitioned to a hard state and do not statistically require the presence of a \texttt{diskbb} component, but they do require the addition of a narrow emission line around 6.3~keV. Additionally, the fits measure a high hydrogen column density in the first observation ($N_{\rm H}\sim7\;\rm cm^{-2}$) and much lower in the second observation ($N_{\rm H}\sim2.6\;\rm cm^{-2}$). The change in measured column density and the presence of the narrow emission features were also pointed out and explained in-depth by \cite{2022ApJ...927..190P}, so we refrain from physically interpreting these effects.

The 1D posterior distributions for the BH spin and viewing inclination obtained from running the MCMC analysis are shown in Figure \ref{fig:MAXI_J1848_combined}. By combining the independent posterior distributions using our combining algorithm, we measure a spin of $a=0.753^{+0.191}_{-0.652}$ and an inclination of $\theta=29^{+13}_{-10}\;[^\circ]$. While the measured inclination agrees with that obtained by \cite{2022ApJ...927..190P}, the larger uncertainty better encapsulates the systematic uncertainties of the methods. Formally, our measured spin disagrees with that obtained by \cite{2022ApJ...927..190P}. However, visually inspecting the individual posterior probability distributions for the BH spin shown in the left panel of Figure \ref{fig:MAXI_J1848_combined}, we see that the distribution produced by the MCMC run on the observation taken during the hard state strongly prefers high BH spins, while the soft state observation places much weaker constraints on the spin. Combining the two results while weighting them by the strength of the reflection during the observation produces a relatively loose constraint on the BH spin. This apparent discrepancy should not be interpreted as a clear disagreement, but rather as proof that when encapsulating the variation between the constraining capabilities of multiple independent observations, the uncertainty of this particular measurement is significantly underestimated when not considering systematic effects.\\

\subsection{EXO 1846-031}\label{sec:EXO_1846-031}
NuSTAR observed EXO 1846-031 six times during its 2019 outburst (ObsIDs 80502303002, 80502303004, 80502303006, 80502303008, 90501334002, and 90501350002). The first observation was analyzed in \cite{2020ApJ...900...78D}, where a spin of $a=0.997^{+0.001}_{-0.002}$ and an inclination of $\theta=73_{-1}^{+2}\;[^\circ]$ were reported. \cite{2020ApJ...900...78D} also reported the presence of an absorption feature around 7 keV, which was modeled through the addition of a \texttt{gaussian} component with negative amplitude. Therefore, for the analysis of the six observations of EXO 1846-031, in addition to the models usually tested, we also verified whether the addition of \texttt{gaussian} components was statistically significantly. We find that for three of the six observations, the \texttt{gaussian} component improves the quality of the fit significantly. 

The top panels of Figure \ref{fig:EXO_1846_delchi} show the unfolded spectra of the six NuSTAR observations, while the middle supbanels show the residuals produced when fitting the spectra with a power-law model. The bottom panels of Figure \ref{fig:EXO_1846_delchi} show the residuals obtained when fitting the spectra with best-performing models in terms of statistic, together with the $\chi^2$ values produced. The individual posterior distributions for the spin and inclination obtained from the MCMC analysis of the six best-performing models are shown in the left and right panels of Figure \ref{fig:EXO_1846_combined} through the colored curves, respectively. The thickness of the lines is proportional to the reflection strength during the observations, which was used as weighting when combining the individual posterior distributions. The combined distributions for the spin and the inclination are shown in Figure \ref{fig:EXO_1846_combined} through the black solid curves, while the dotted curves show the distribution combined using the method described in \cite{Draghis23}. The vertical black solid and dashed lines show the mode of the combined distribution, together with the $1\sigma$ credible interval. Using the six NuSTAR observations, we measure a spin of $a=0.959^{+0.031}_{-0.077}$ and an inclination of $\theta=62^{+9}_{-10}\;[^\circ]$. These values are consistent with those reported in \cite{2020ApJ...900...78D}, but in addition to statistical uncertainties, they include the systematic variation originating from the independent measurements on the six different NuSTAR observations.

\subsection{XTE J1908+094}\label{sec:XTE_J1908+094}
The two existing NuSTAR observations of XTE J1908+094 were analyzed by \cite{2021ApJ...920...88D} (ObsIDs 90501317002 and 80001014002), who reported a spin of $a=0.55^{+0.29}_{-0.45}$ and an inclination of $\theta=27^{+2}_{-3}\;[^\circ]$. \cite{2021ApJ...920...88D} divided the spectra based on the hardness ratios during the observations and fit the four resulting pairs of spectra jointly. We extracted time-averaged spectra and fit them independently in the 3-50 keV band. The top panels of Figure \ref{fig:XTE_J1908_delchi} show the unfolded spectra of the observations. The center panels show the residuals of the models that do not account for reflection, while the lower panels show the residuals produced when accounting for reflection. 

We ran the MCMC analysis for the best-performing reflection models, and the posterior distributions for the BH spin and inclination are shown in the left and right panels of Figure \ref{fig:XTE_J1908_combined}, respectively. By combining the posterior distributions, we measure $a=0.466^{+0.368}_{-0.517}$ and $\theta=28\pm11\;[^\circ]$. These results are consistent with those found by \cite{2021ApJ...920...88D}, but with larger uncertainties owing to our joint treatment of individual observations. 

\subsection{GRS 1915+105}\label{sec:GRS_1915+105}
At the time of the analysis, there were 29 public archival NuSTAR observations of GRS 1915+105. Of the 29 observations, the source was at an Eddingon fraction $10^{-3}\leq L/L_{\rm Edd} \leq 0.3$ only in 18 of them. This was calculated using the mass and distance estimates from \cite{2014ApJ...796....2R}: $M=12.4\pm2\;M_\odot$ and $d=8.6\pm2\;\rm kpc$. Furthermore, due to the complexity of the obscuration of the central engine in this source, we were unable to obtain a good fit using our combination of simplistic models for the spectra from ObsID 30202033004. Therefore, we only fit the remaining 17 observations, listed in Appendix \ref{sec:obsids}. Figure \ref{fig:GRS_1915_delchi} shows the residuals produced by fits using our baseline model (top) and the best-performing reflection models (bottom). 

Using our combining algorithm on the individual 1D posterior probability distributions inferred using the MCMC analysis on the best-fit models (shown through the colored lines in the left and right panels of Figure \ref{fig:GRS_1915_combined} for spin and inclination, respectively), we measure a BH spin of $a=0.976^{+0.018}_{-0.056}$ and an inclination of $\theta=60\pm8\;[^\circ]$. Our spin measurement is in good agreement with that of \cite{2013ApJ...775L..45M}, who performed a reflection study on the NuSTAR spectrum from ObsID 10002004001 to measure $a=0.98\pm0.01$. However, our inclination measurement is lower than the value of $\theta=72\pm1\;[^\circ]$ determined by \cite{2013ApJ...775L..45M}. Interestingly, looking at the right panel in Figure \ref{fig:GRS_1915_combined}, we can see that the inclination that we measured from the same observation does indeed take larger values, consistent with that measured by \cite{2013ApJ...775L..45M}. This result suggests the importance of analyzing all the available observations in order to ensure minimal bias in our measurements originating from physical effects for which our models do not account. 

\subsection{Cygnus X-1}\label{sec:Cyg_X-1}
We analyzed all the 34 public NuSTAR observations of Cygnus X-1, as they all occur while the source was at an Eddington fraction for which it is expected that the accretion disk should extend to the ISCO. For this calculation, we used $M=21.2\pm2.2M_\odot$ and $d=2.22\pm0.18\;\rm kpc$ (\citealt{2021Sci...371.1046M}). We show the residuals produced when fitting the spectra from the individual observations in Figure \ref{fig:Cyg_X-1_delchi}, with the top panels indicating the residuals produced by our baseline model, and the bottom panels indicating the residuals produced by the best-performing reflection models. The source shows signs of variable narrow absorption and emission features, which we account for using the \texttt{zxipcf} multiplicative component, and through the addition of \texttt{gaussian} components with positive or negative normalizations, depending on whether the component is meant to account for an emission or absorption feature, respectively. The broad reflection features are well accounted for.

Figure \ref{fig:Cyg_X-1_combined} shows the 1D posterior distributions resulting from the MCMC analysis of the independent observations for spin (left) and inclination (right). The majority of the spin posterior distributions agree on a high spin. However, fits to a few observations prefer lower spin values, likely due to an incomplete or improper characterization of the narrow spectral features present. Furthermore, partial obscuration of the inner disk would likely also artificially weigh the emission from the outer regions of the accretion disk more, leading to a lower spin measurement. The origin and shape of the narrow spectral features, together with their influence on the spin measurement will be properly quantified using spectra from the \textit{Resolve} instrument on board the recently launched XRISM mission. Our combining algorithm finds a spin of $a=0.95^{+0.04}_{-0.08}$. Our measurement is consistent with archival spin measurements both through the relativistic reflection method on NuSTAR (see, e.g. \citealt{2015ApJ...808....9P} who find $a\geq0.97$) and through measurements made using continuum fitting (see, e.g. \citealt{2014ApJ...790...29G} who find $a\geq0.983$).

The inclination measurements are well concentrated around an intermediate value, with few outliers (right panel of Figure \ref{fig:Cyg_X-1_combined}. Our combining algorithm measures $\theta=47^{+9}_{-11}\;[^\circ]$, in good agreement with the value measured by \cite{2015ApJ...808....9P}, who find $\theta=45.3\pm0.4\;[^\circ]$, and in disagreement with the orbital inclination of $\theta=27.1 \pm 0.8\;[^\circ]$ found by \cite{2011ApJ...742...84O}.

\subsection{V404 Cygnus}\label{sec:V404_Cyg}
%https://ui.adsabs.harvard.edu/abs/2015ApJ...813L..37K/abstract
Given the mass estimate of $M=12\pm3\;M_\odot$ (\citealt{1994MNRAS.271L..10S}) and the distance of $d=2.39\pm0.14\;\rm kpc$ (\citealt{2009ApJ...706L.230M}), only two of the 11 available NuSTAR observations of V404 Cygnus happened while the source was at an Eddington ratio above $10^{-3}$: ObsIDs 90102007002 and 90102007003. The two observations were taken in succession and show significant flaring of the source, with count rates increasing by a factor of $\sim 5-10$. For both observations, we split the light curves into ``flare" and ``nonflare" segments, and extracted spectral products for both states. We fit the resulting spectra independently. Figure \ref{fig:V404_Cyg_delchi} shows the flare (left) and nonflare (right) spectra extracted from ObsIDs 90102007002 (top) and 90102007003 (bottom), together with the residuals produced when being fit with our baseline model and with the best-performing reflection models. The source was in a hard spectral state, and the preferred reflection models did not require the presence of a \texttt{diskbb} component, which was subsequently removed in order to reduce the size of the parameter space. All spectra required strong ionized obscuration, which we accounted for using \texttt{zxipcf}. Despite the best-fit statistics being poor, especially for the nonflaring state, we note that the residuals produced by our best-performing models indicate that the reflection features have well been accounted for. At this stage, the remaining residuals come from calibration uncertainties in the NuSTAR band, which under normal circumstances are negligible, but given the high signal-to-noise ratios of the observations are strongly accentuated. Furthermore, \cite{2015ApJ...813L..37K} found that the source showed a multitude of narrow spectral emission features which evolved on a timescale of a few kiloseconds. Our models do not account for any such features, as they are not resolvable given the NuSTAR energy resolution. Therefore, despite the formally poor statistic, we continue our analysis under the assumption that the relativistically broadened spectral features are properly accounted for.

Figure \ref{fig:V404_Cyg_combined} shows the individual 1D posterior distributions for spin (left) and inclination (right) produced by the MCMC analysis of the individual fits. The spin is generally high, and the inclination measurements are in broad agreement. Our combining algorithm finds $a=0.935^{+0.037}_{-0.075}$ and $\theta=37^{+9}_{-8}\;[^\circ]$. The spin measurement is consistent with that of \cite{2017ApJ...839..110W}, who report $a\geq0.92$, but our inclination measurement is lower than their reported value of $\theta\sim52^\circ$. 

Our models include significant obscuration of the central engine in this source, and both our results and the conclusion of \cite{2017ApJ...839..110W} suggest that the flaring episodes that this source experienced were in fact caused by a reduction in obscuration along the line of sight. If that is the case, it is likely that our spin inference should be treated as a lower limit only, as much of the emission from the innermost regions of the disk would have been attenuated, and larger contribution from outer regions of the disk would weigh more in the observed spectra, artificially biasing the measurement to lower values.

\section{Discussion}\label{sec:discussion}
In this paper, we reanalyzed all the existing archival NuSTAR observations of sources that had previous spin measurement using the relativistic reflection method. This analysis was performed in a way fully consistent with that in \cite{Draghis23}, adopting the same set of theoretical assumptions. In conjunction with the analysis in \cite{Draghis23}, we have compiled a sample of spin measurements of considerable size. 

For each source analyzed, we show the individual posterior probability distributions for the spin and inclination, obtained from the MCMC analysis of the spectra from individual observations. We report the values obtained by using a Bayesian combining algorithm to combine the individual measurements into a single result. This resulting combined distribution encapsulates the ignorance of the simplistic treatment of our models, which stems from using a single model to fit observations across various spectral states, at different accretion rates, when likely many more physical processes are present and relevant. Generally, the values that we report in this paper for the spin are in agreement with previous measurements, but the sizes of the uncertainties that we report are significantly larger. 

While the spin measurements generally agree between observations, the inclination measurements often tend to disagree, especially in sources where there are many observations available. This is likely due to difficulties in constraining the blue wing of the relativistically broadened Fe K line, owing to uncertainties in constraining the underlying continuum and possible ionized outflows. This suggests that the inclination values obtained through relativistic reflection measurements from a single observation should be taken with caution, and not be treated as definitive. However, when the analysis is performed systematically on a large number of observations, the underlying true value can be better estimated. It is important to note that for a few sources, timing analysis has suggested the presence of the Lense-Thirring effect (see, e.g., \citealt{2023arXiv230700867M,2023MNRAS.523.4394Z}), which could partly explain a change in the inferred viewing inclination of the inner disk.

To further probe the hypothesis regarding the effect of the ionized outflows, we ran a case study on ObsID 80402302008 of MAXI J1535-571 (Section \ref{sec:MAXI_J1535-571}). The best-fit solution obtained predicts a spin around $a\sim0.89$, and inclination of $\theta\sim5^\circ$, with $\chi^2_r=581/465$. If we fix the spin to the value derived using all observations, $a=0.98$, we obtain $\theta\sim7^\circ$, with a slightly worse $\chi^2_r=587/466$. If instead we fix the inclination to the one derived by averaging all observations, $\theta=45^\circ$, we obtain a spin of $a\sim0.97$, but with worse $\chi^2_r=643/466$. However, adding an absorption \texttt{gaussian} component at 7keV while maintaining the inclination fixed at $\theta=45^\circ$ maintains the spin high, in good agreement with the combined measurement, but the fit statistic does not represent a significant improvement ($\chi^2_r=579/463$) over the solution that does not fix the spin and inclination and does not include absorption features ($\chi^2_r=581/465$). However, the $\chi^2$ value does formally improve, suggesting that in observations with higher signal-to-noise ratios, the absorption feature could have been significant and produce results in agreement with the rest of the observations. As the spin and inclination measurements might be somewhat correlated, and as absorption features from ioinized outflows produce narrow spectral features that overlap with the blue wing of the relativistically broadened Fe K line, improperly characterizing the impact of the ionized absorption could directly impact our ability to accurately constrain BH spins and inclinations. In the future, pairing NuSTAR broadband spectra with high energy resolution XRISM spectra will help to definitively address this issue. 

%insert paragraph about GX 339 fixed inclination experiment
\begin{figure}[h]
    \centering
    \includegraphics[width= 0.48\textwidth]{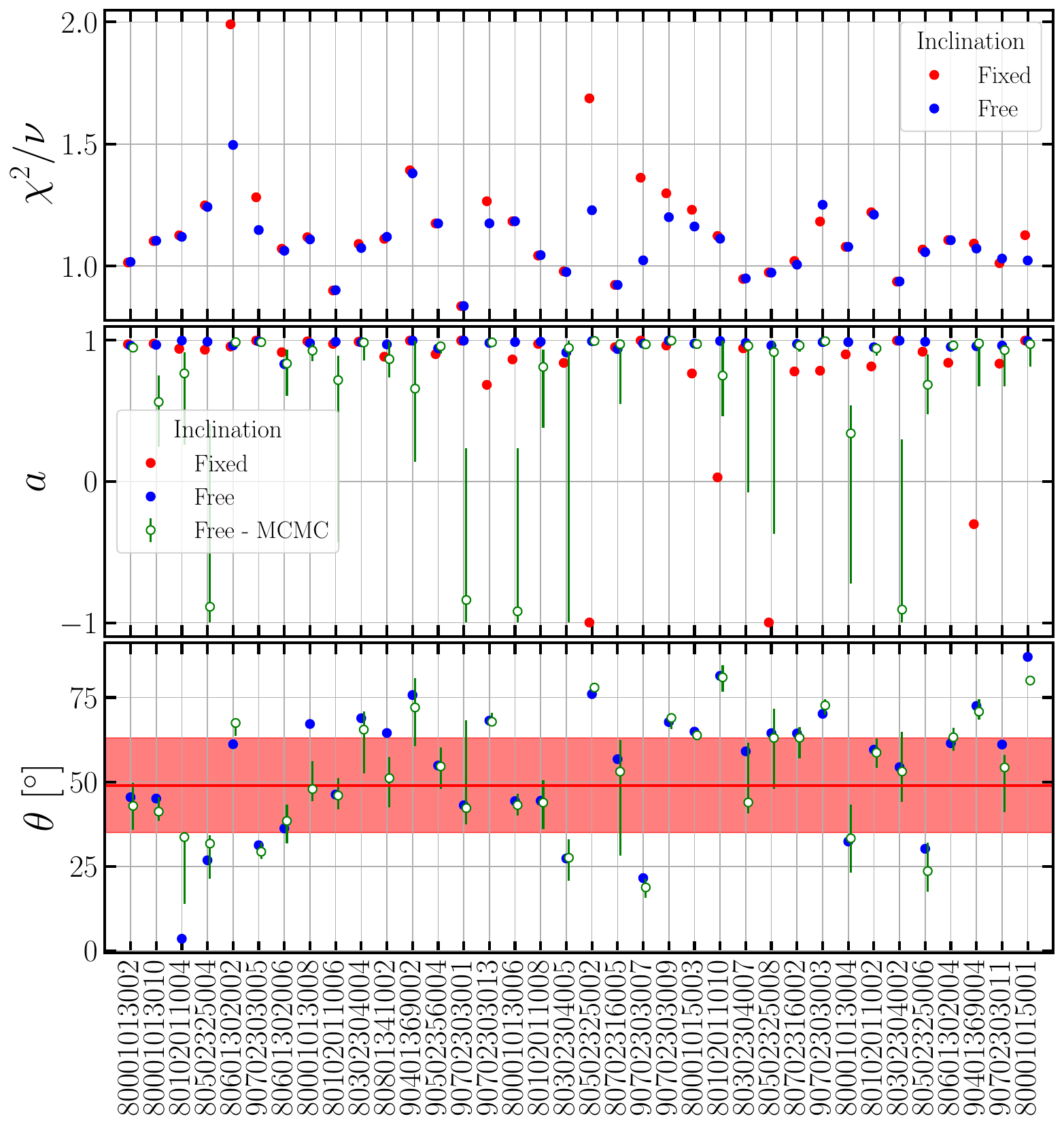}
    \caption{Top panel shows the reduced $\chi^2$ obtained when fitting all the spectra of GX 339-4 with inclination free (blue) and fixed at $49^\circ$ (red). The central panel shows the spin measured through direct fitting when the inclination is fixed at $49^\circ$ (red), free (blue), and measured through the MCMC analysis when the inclination is free (green). Note that for the measurements obtained through $\chi^2$ minimization in \texttt{xspec}, the uncertainties are not included. The bottom panel shows the inclinations measured through $\chi^2$ minimization in \texttt{xspec} (blue) and through MCMC analysis (green). The horizontal red line and red shaded region show, respectively, the mode and $\pm1\sigma$ credible interval on the inclination measurement obtained by combining all the independent measurements.}
    \label{fig:fixed_incl_results}
\end{figure}

We tested the impact of the ability to measure the viewing inclination on our ability to recover BH spin by reanalyzing all observations of GX 339-4 while fixing the inclination at the value determined through our combining algorithm, and comparing the results with the ones obtained with a free inclination. The top panel in Figure \ref{fig:fixed_incl_results} shows the fit statistic produced when fitting the spectra with a free inclination (blue points) and with the inclination fixed at $\theta=49^\circ$ (red points). Generally, the fits are very similar, or worse. Interestingly, in two cases, the statistic becomes slightly better, indicating that our automatic fitting algorithm might have not fully reached a global statistic minimum, but became stuck in a nearby local minimum.

The bottom panel in Figure \ref{fig:fixed_incl_results} shows the inclination measurements obtained when allowed to vary freely (in blue) by direct fitting within \texttt{xspec}, and the ones obtained following the MCMC analysis (in empty green circles with the associated error bars). The points showing the values from \texttt{xspec} fitting do not include error bars, as they are often uninformative, and as the errors reported in our work come from the MCMC analysis. The horizontal red line represents the value of $49^\circ$ to which the inclination was fixed during this experiment, and the red band shows the $\pm1\sigma$ uncertainty region on the combined measurement. The fact that the inclinations vary significantly despite the quality of the fits not being significantly different when compared to fits using a fixed inclination suggest that one needs to use caution when considering the inclination measurements obtained through this method, and again suggests that mischaracterizing disk winds which would imprint absorption features around 7keV could have a significant impact on our ability to recover the inclination. This hypothesis will be tested in the near future using XRISM data.

The central panel in Figure \ref{fig:fixed_incl_results} highlights the effect of fixing the inclination on the recovered spins. The blue points represent the spins inferred from fitting the spectra within \texttt{xspec} when the inclination is allowed to vary freely, while the red points show the spins measured when the inclination is fixed at the value determined by combining all independent measurements. The green circles show the spins determined after running the MCMC analysis, starting from the best-fit values with free inclination (blue points). Generally, the spins measured when the inclination is free agree with those measured when the inclination is fixed at $49^\circ$ (note that uncertainties are not shown), suggesting the ability of the data to separate the spin and inclination constraints. The spin generally determines the red wing of the Fe line, while the inclination impacts the blue wing of the line. Nevertheless, through this experiment, we have reconfirmed that the inclination measurements need to be taken with caution, while the spin measurements appear to be robust, especially when combining multiple independent observations.

However, a more interesting discrepancy, highlighted by Figure \ref{fig:fixed_incl_results}, is that obtained by running the MCMC analysis with free inclinations vs. results obtained from \texttt{xspec} $\chi^2$ minimization. Despite the MCMC runs starting with the walkers initialized around the value obtained from $\chi^2$ minimization, the posterior distributions sometimes cluster around a different value. This is likely because of the way in which the parameter space is explored by the walkers. When the quality of the data is reduced, there are often multiple families of solutions with similar statistics. Generally, when that happens, as highlighted previously in the paper, and as explained in \cite{2023ApJ...954...62D}, solutions with a high spin and a high inner coronal emissivity index produce similar statistics with solutions with a low (even negative) spin and a low inner emissivity index ($q_1$). Due to the nature of the shape of the parameter space, walkers preferentially navigate from the high spin solution (narrow region of the parameter space) to the low spin solution (wider region of the parameter space), and have difficulty navigating back to the narrow parameter space around the high spin solution, despite it being slightly preferred in terms of statistics. This ability to explore the parameter space also explains why when fixing the inner disk radius at the ISCO, we sometimes find low or negative spins, while for the same data, when allowing the inner disk radius to vary, the low spin solutions disappear: the families of solutions become more statistically distinct, and the walkers navigate more easily to the high spin solutions. Additionally, when the inner disk radius is free, the size of the parameter space increases, making it less likely for walkers to wander outside of the starting solution and become stuck in low spin solutions. (See the discussion regarding the effects of allowing the inner disk radius to vary below.) 

The figures illustrating the posterior distributions obtained from our spectral analysis contain two different ``combined" distributions. We report the mode and $\pm1\sigma$ credible interval of the black solid curves, which are obtained by averaging a large sample of beta distributions used to fit the individual measurements to a single distribution. This distribution encapsulates the information from all available measurements from individual observations. In contrast, the black dotted curves show the beta distributions obtained by the most frequently occurring shape of the beta function in the analysis, which combines the individual posterior distributions. While this value could be interpreted as the underlying, ``true" spin value from which the measurements are drawn, it does not fully incorporate the size of the uncertainties originating from using incomplete theoretical models to fit the spectra. In other words, the dotted distribution assumes that our models are correct, and the solid curves allow for a more relaxed interpretation of the independent results, leading to larger uncertainties. 

\begin{deluxetable*}{l|cc|ccr}
\tablecaption{All Current Spin Measurements in XBs Performed Using Our Pipeline}
\label{tab:all_spins}
\tablewidth{\textwidth} 
\tabletypesize{\scriptsize}
\tablehead{
\colhead{Source} & \colhead{Spin} & \colhead{Inclination $(^\circ)$} & \colhead{Previous Spin} & \colhead{Previous Inclination $(^\circ)$} & \colhead{Reference}}
\startdata\\
AT 2019wey & $0.906_{-0.202}^{+0.084}$ & $14_{-10}^{+12}$ & $0.97_{-0.03}^{+0.02}$ & $22.0_{-2.9}^{+2.6}$ & \cite{2022SCPMA..6519512F} \\
LMC X-3 & $0.928_{-0.146}^{+0.058}$ & $38_{-13}^{+14}$ & $0.24\pm0.05^\ddag$ & $69.24^*$ & \cite{2021MNRAS.507.4779J} \\
LMC X-1 & $0.897_{-0.176}^{+0.077}$ & $50_{-13}^{+10}$ & $0.9395\pm0.015$ & $36.38^*$ & \cite{2021MNRAS.507.4779J} \\
MAXI J0637-430 & $0.984_{-0.042}^{+0.012}$ & $63_{-10}^{+9}$ & $0.97\pm0.02$ & $62_{-4}^{+3}$ & \cite{Draghis23} \\
MAXI J1348-630 & $0.977_{-0.055}^{+0.017}$ & $52_{-11}^{+8}$ & $0.78\pm0.04$ & $29.2_{-0.5}^{+0.3}$ & \cite{2022MNRAS.511.3125J} \\
GS 1354-645 & $0.849_{-0.221}^{+0.103}$ & $47_{-10}^{+11}$ & $\geq0.98$ & $75\pm2$ & \cite{2016ApJ...826L..12E} \\
MAXI J1535-571 & $0.979_{-0.049}^{+0.015}$ & $44_{-19}^{+17}$ & $0.985_{-0.004}^{+0.002}$ & $72\pm2$ & \cite{2022MNRAS.514.1422D} \\
4U 1543-47 & $0.959_{-0.079}^{+0.031}$ & $67_{-8}^{+7}$ & $0.98_{-0.02}^{+0.01}$ & $68_{-4}^{+3}$ & \cite{Draghis23} \\
MAXI J1631-479 & $0.951_{-0.077}^{+0.039}$ & $22_{-12}^{+10}$ & $\geq0.94$ & $29\pm1$ & \cite{2020ApJ...893...30X} \\
4U 1630-472 & $0.857_{-0.211}^{+0.095}$ & $55_{-11}^{+8}$ & $0.985_{-0.014}^{+0.005}$ & $64_{-3}^{+2}$ & \cite{2014ApJ...784L...2K} \\
Swift J1658.2-4242 & $0.951_{-0.069}^{+0.031}$ & $50_{-10}^{+9}$ & $\geq0.96$ & $64_{-3}^{+2}$ & \cite{2018ApJ...865...18X} \\
GX 339-4 & $0.97_{-0.076}^{+0.026}$ & $49\pm14$ & $0.95_{-0.08}^{+0.02}$ & $30\pm1$ & \cite{2016ApJ...821L...6P} \\
IGR J17091-3624 & $0.963_{-0.085}^{+0.027}$ & $47_{-11}^{+10}$ & $0.07\pm0.2$ & $45.3\pm0.7$ & \cite{2018MNRAS.478.4837W} \\
GRS 1716-249 & $0.97_{-0.06}^{+0.022}$ & $59_{-12}^{+7}$ & $\geq0.92$ & $49.9_{-1.3}^{+1.0}$ & \cite{2019ApJ...887..184T} \\
MAXI J1727-203 & $0.962_{-0.414}^{+0.034}$ & $65_{-14}^{+11}$ & $0.986_{-0.159}^{+0.012}$ & $64_{-7}^{+10}$ & \cite{Draghis23} \\
Swift J1728.9-3613 & $0.868_{-0.088}^{+0.058}$ & $7_{-3}^{+8}$ & $0.86\pm0.02$ & $3.5_{-0.5}^{+6.2}$ & \cite{2023ApJ...947...39D} \\
GRS 1739-278 & $0.968_{-0.074}^{+0.022}$ & $70_{-11}^{+5}$ & $0.8\pm0.2$ & $43.2\pm0.5$ & \cite{2015ApJ...799L...6M} \\
1E 1740.7-2942 & $0.856_{-0.443}^{+0.134}$ & $31_{-18}^{+29}$ & $\geq0.5$ & $63.7_{-7.9}^{+4.6}$ & \cite{2020MNRAS.493.2694S} \\
IGR J17454-2919 & $0.932_{-0.363}^{+0.06}$ & $54_{-14}^{+15}$ & $0.97_{-0.17}^{+0.03}$ & $60_{-14}^{+8}$ & \cite{Draghis23} \\
Swift J174540.2-290037 & $0.774_{-0.106}^{+0.082}$ & $31_{-9}^{+8}$ & $0.92_{-0.07}^{+0.05}$ & $21_{-3}^{+2}$ & \cite{2019ApJ...885..142M} \\
Swift J174540.7-290015 & $0.884_{-0.109}^{+0.068}$ & $63_{-8}^{+10}$ & $0.94_{-0.1}^{+0.03}$ & $64.2_{-1.6}^{+0.9}$ & \cite{2019ApJ...885..142M} \\
H 1743-322 & $0.949_{-0.127}^{+0.039}$ & $54_{-13}^{+12}$ & $0.98_{-0.02}^{+0.01}$ & $56\pm4$ & \cite{Draghis23} \\
Swift J1753.5-0127 & $0.989_{-0.035}^{+0.007}$ & $73\pm8$ & $0.997_{-0.003}^{+0.001}$ & $74\pm3$ & \cite{Draghis23} \\
GRS 1758-258 & $0.98_{-0.058}^{+0.014}$ & $67_{-13}^{+8}$ & $0.991_{-0.019}^{+0.007}$ & $66_{-5}^{+8}$ & \cite{Draghis23} \\
MAXI J1803-298 & $0.987_{-0.037}^{+0.007}$ & $72_{-9}^{+6}$ & $0.991\pm0.001$ & $70.8\pm1.1$ & \cite{2022MNRAS.516.2074F} \\
MAXI J1813-095 & $0.88_{-0.27}^{+0.1}$ & $42_{-13}^{+11}$ & $\geq0.76$ & $36.5\pm8.5$ & \cite{2021RAA....21..125J} \\
V4641 Sgr & $0.701_{-0.272}^{+0.211}$ & $66_{-11}^{+7}$ & $0.86_{-0.06}^{+0.04}$ & $66_{-4}^{+3}$ & \cite{Draghis23} \\
MAXI J1820+070 & $0.967_{-0.061}^{+0.025}$ & $64_{-9}^{+8}$ & $0.988_{-0.028}^{+0.006}$ & $64_{-4}^{+3}$ & \cite{Draghis23} \\
MAXI J1848-015 & $0.753_{-0.652}^{+0.191}$ & $29_{-10}^{+13}$ & $0.967\pm0.013$ & $26.4\pm0.5$ & \cite{2022ApJ...927..190P} \\
EXO 1846-031 & $0.959_{-0.077}^{+0.031}$ & $62_{-10}^{+9}$ & $0.997_{-0.002}^{+0.001}$ & $73_{-1}^{+2}$ & \cite{2020ApJ...900...78D} \\
XTE J1908+094 & $0.466_{-0.517}^{+0.368}$ & $28\pm11$ & $0.55_{-0.45}^{+0.29}$ & $27_{-3}^{+2}$ & \cite{2021ApJ...920...88D} \\
GRS 1915+105 & $0.976_{-0.056}^{+0.018}$ & $60\pm8$ & $0.98\pm0.01$ & $72\pm1$ & \cite{2013ApJ...775L..45M} \\
Cyg X-1 & $0.95_{-0.084}^{+0.04}$ & $47_{-11}^{+9}$ & $\geq0.97$ & $45.3\pm0.4$ & \cite{2015ApJ...808....9P} \\
4U 1957+11 & $0.9_{-0.28}^{+0.08}$ & $52_{-13}^{+12}$ & $0.95_{-0.04}^{+0.02}$ & $52_{-5}^{+4}$ & \cite{Draghis23} \\
XTE J2012+381$^\dagger$ & $0.988_{-0.03}^{+0.008}$ & $68_{-11}^{+6}$ & \nodata & \nodata & \cite{2023ApJ...954...62D} \\
V404 Cyg & $0.935_{-0.075}^{+0.037}$ & $37_{-8}^{+9}$ & $\geq0.92$ & $52_{-3}^{+2}$ & \cite{2017ApJ...839..110W}
\enddata
\tablecomments{The sources are listed in order of increasing Right Ascension. All uncertainties reported are at the $1\sigma$ level. The values in the left columns are the measurements obtained using the pipeline highlighted in this paper, and represent the modes and $\pm1\sigma$ credible intervals of the combined posterior probability distributions obtained through the analysis (i.e. the solid black curves in the figures showing the posterior distributions). The values in the right columns show the previous measurements of spin and inclination for the sources, together with their respective references. $^\dagger$ XTE J2012+381 does not have a spin measurement prior to \cite{2023ApJ...954...62D}. $^\ddag$ The spin of LMC X-3 was measured in \citealt{2021MNRAS.507.4779J} using continuum fitting. $^*$ For LMC X-1 and LMC X-3, the inclination values were fixed to the orbital inclination.}
\end{deluxetable*}

In order to ensure consistency between the spins measured in this paper with those in \cite{Draghis23}, we ran the new version of the combining algorithm, which was established in \cite{2023ApJ...954...62D}, on the 10 measurements in \cite{Draghis23}. Additionally, we ran the same combining algorithm on the results from the analysis of single observation of Swift J1728.9-3613 (\citealt{2023ApJ...947...39D}). We combined the measurements of BH spin and inclination from this paper with the results from \cite{Draghis23}, the spin measurement of XTE J2012+381 (\citealt{2023ApJ...954...62D}), and Swift J1728.9-3613 (\citealt{2023ApJ...947...39D}) to compile a list of 36 BH spin measurements performed in a systematic way. We show the results in Table \ref{tab:all_spins}. We present the original measurements for the BH spin and the inner disk inclination, together with our updated measurements. All our reported values have been obtained through our measurement combining algorithm, even when a single observation of the sources is available. This ensures that the same treatment is applied to all sources, and that the reported uncertainties encapsulate part of the inherent systematic uncertainty that comes with spin measurements made using the relativistic reflection method.

\begin{figure}[ht]
    \centering
    \includegraphics[width= 0.48\textwidth]{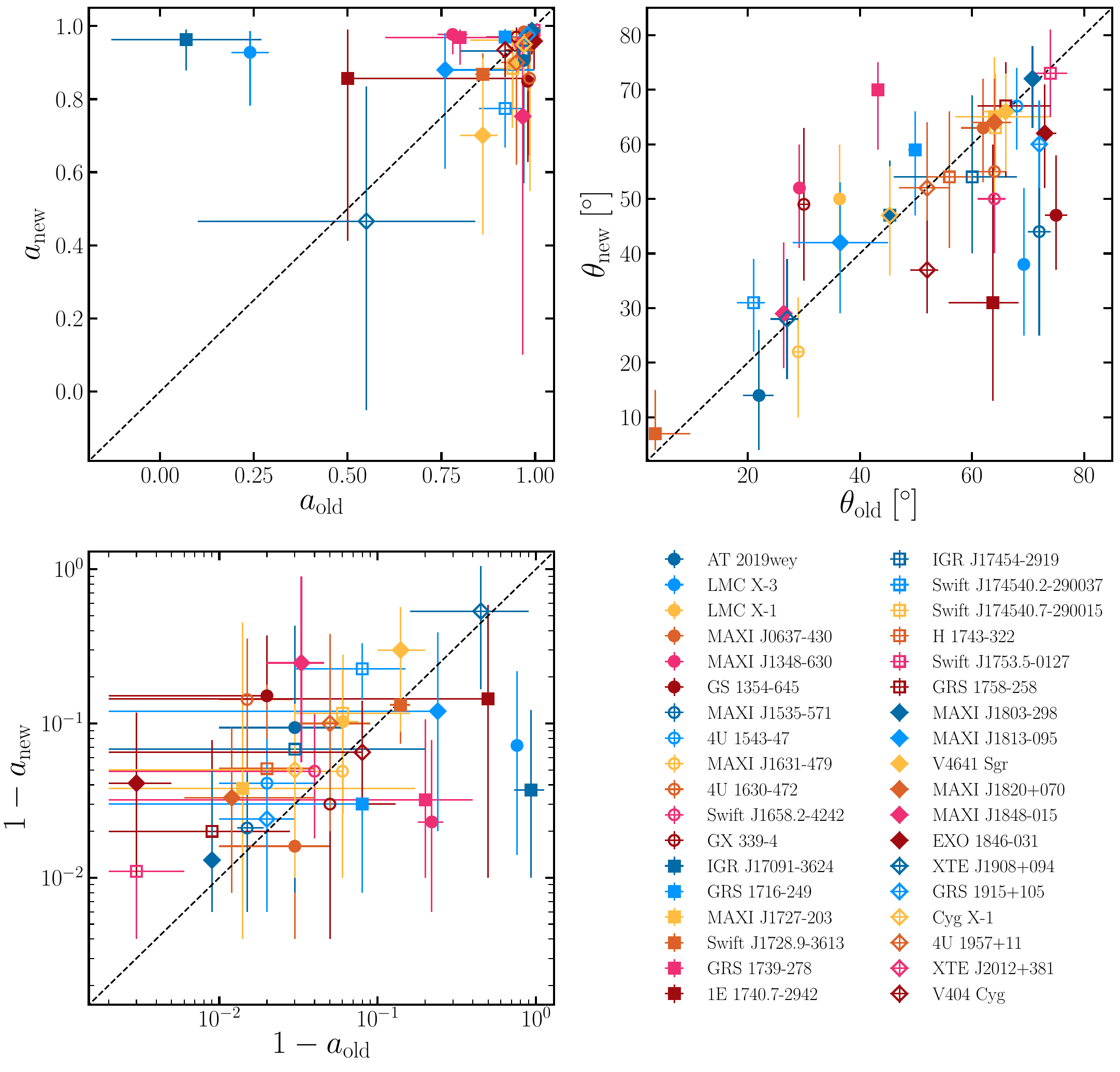}
    \caption{Comparison between the new measurements obtained in this paper and the old values reported in the literature for spin (top left) and inclination (top right). The bottom-left panel shows a comparison between $1-a_{\rm new}$ and $1-a_{\rm old}$ in logarithmic scale, to better highlight the differences in measurements where both the initial and final spin are nearly maximal. All values are taken from Table \ref{tab:all_spins}.}
    \label{fig:old_vs_new_1}
\end{figure}

To better illustrate the change induced by our new methods, we plotted all the values in Table \ref{tab:all_spins}. The top panels in Figure \ref{fig:old_vs_new_1} show the measurements updated using our pipeline versus the values in the literature for the spins (left) and inclinations (right). The bottom-left panel shows the complement to one of the updated and literature measurements, shown on a logarithmic scale. This panel is shown for visualization purpose only, and presents the same information as the top-left panel inverted across the origin. This was done to better highlight changes in the measured values at spins close to the maximal value. Generally, the measurements are consistent with their previous estimates. Three of the four spins with values less than $a=0.75$ are now closer to maximal, indicated by their position above the diagonal line in the top-left panel of Figure \ref{fig:old_vs_new_1}. Many updated spin measurements now take smaller values than previously, but with uncertainties that encapsulate the previous measurements. For the inclinations, the updated measurements are again in good agreement, but with significantly larger uncertainties. The many points below the diagonal line in the top-right panel of Figure \ref{fig:old_vs_new_1} indicate a decrease in the measured inclination compared to the literature values.

\begin{figure}[ht]
    \centering
    \includegraphics[width= 0.48\textwidth]{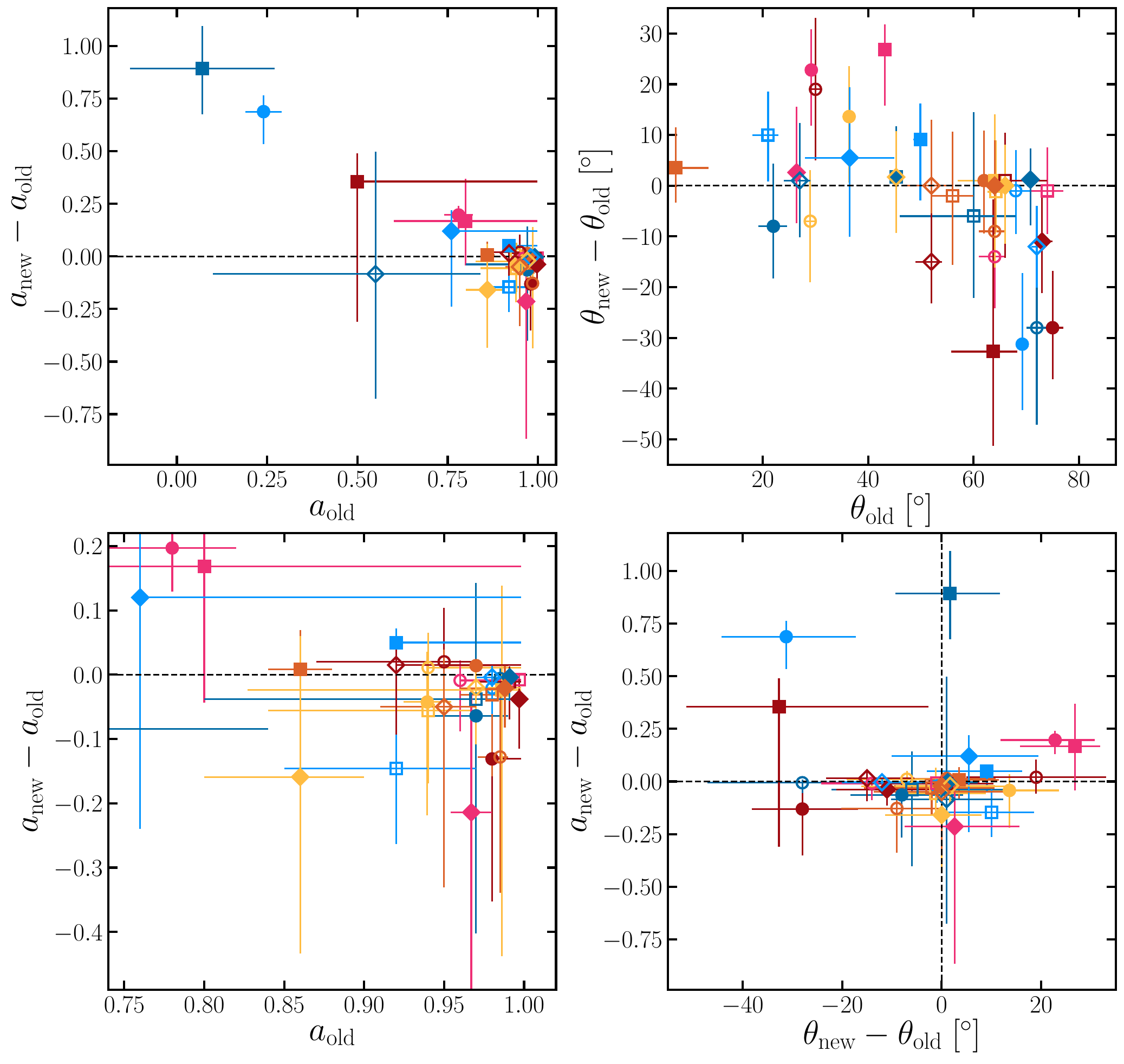}
    \caption{Top: change in spin (left) and inclination (right) when compared to the previously existing measurements. Bottom left: same as the top left, but focused on $a_{\rm old}\geq0.75$, for visual clarity. Bottom right: change in spin vs. change in inclination. The markers represent the same sources as in Figure \ref{fig:old_vs_new_1}, with all values taken from Table \ref{tab:all_spins}.}
    \label{fig:old_vs_new_2}
\end{figure}

In order to further probe whether any trends are highlighted by our analysis, we plot the change in the measured spins and inclinations versus the literature measurements in the top panels of Figure \ref{fig:old_vs_new_2}. The lower-left panel in Figure \ref{fig:old_vs_new_2} shows the same difference in spin, just focusing on values larger than $a=0.75$, where most points are concentrated. Again, for spins, the largest change occurs for previously low values, while originally larger spins are now smaller, but with larger uncertainties. Similar to the previous plot, the trend shown in the top-right panel of Figure \ref{fig:old_vs_new_2} suggests that lower inclinations are now measured to be slightly higher, while previously high inclinations are now slightly lower, all with increased uncertainties. Nevertheless, the measurements are still in good agreement with the previously reported values. Lastly, the bottom-right panel in Figure \ref{fig:old_vs_new_2} shows the change in spin versus the change in inclination. As no trends are obvious, this suggests that the previous uncertain spin measurements were not generally due to uncertain inclination measurements.

%paragraphs about the disk truncaiton and hardness part
Throughout our analysis, we only treated observations in which the sources were at an Eddington fraction between $10^{-3}\leq L/L_{\rm Edd}\leq 0.3$, under the assumption that for this range, the accretion disk extends to the ISCO and is not truncated. If applying the assumption that the inner disk extends to the ISCO while it is in fact truncated at larger radii, our measurements based on the particular spectra should be interpreted as lower limits. However, as the calculation of Eddington fraction can, in principle, be biased by erroneous existing BH mass or distance estimates, another observable that could point to the presence of disk truncation is the spectral hardness, with some studies indicating that during the hard state, the accretion disk should be truncated at large radii (see, e.g. \citealt{2007A&ARv..15....1D}). To test the robustness of our measurements, we recombined the measurements obtained when analyzing the spectra from four sources that have been observed multiple times, while focusing on observations occurring at high Eddington fractions and/or low hardness.

\begin{figure}[ht]
    \centering
    \includegraphics[width= 0.48\textwidth]{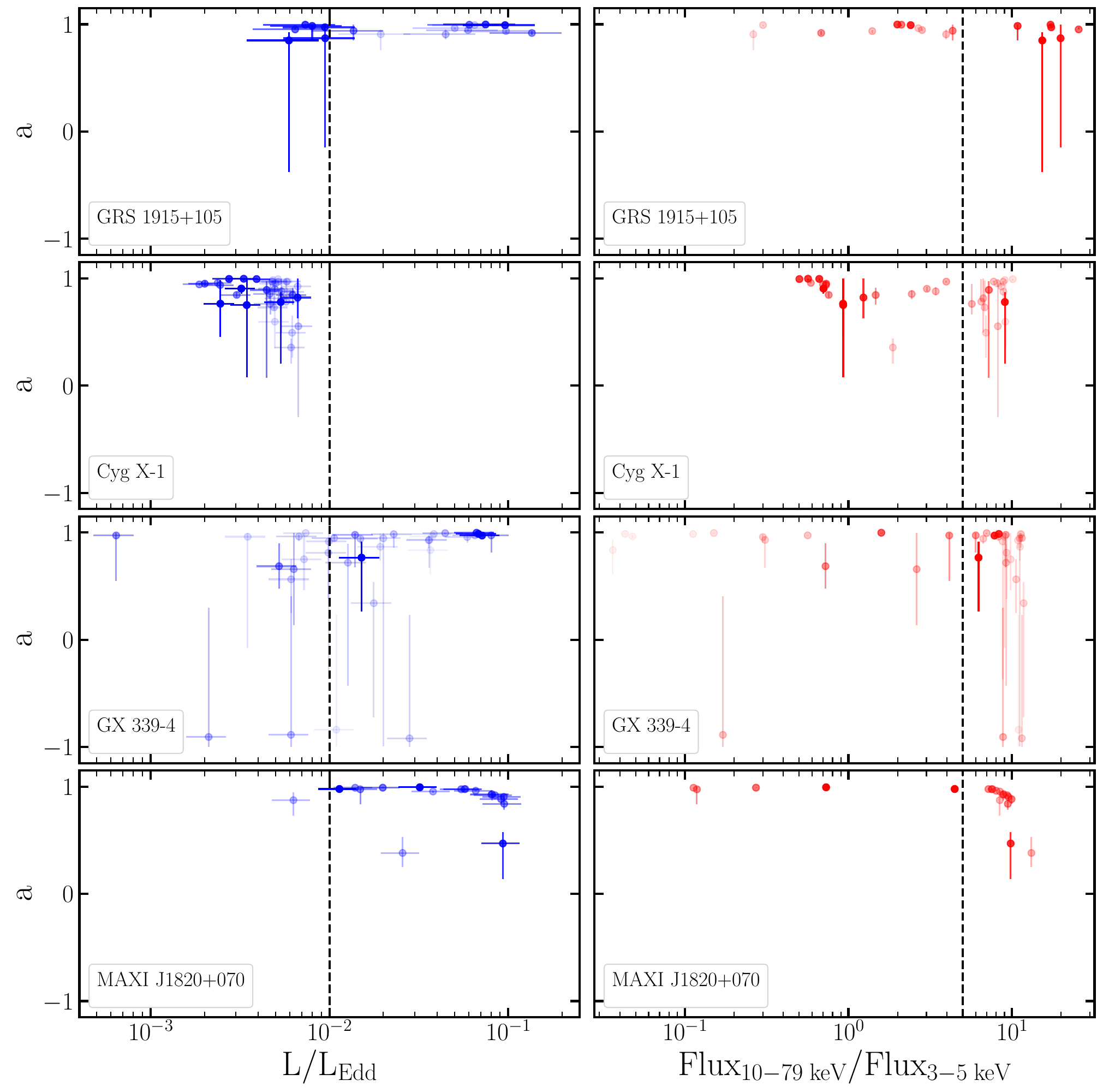}
    \caption{BH spin vs. Eddington fraction (left) and hardness (right), defined as the ratio of the flux in the 10-79~keV band to the flux in the 3-5~keV band, for GRS 1915+105, Cygnus X-1, GX 339-4, and MAXI J1820+070. The vertical dashed lines show an Eddington ratio of $10^{-2}$ in the left panels, and a hardness of $\rm Flux_{\rm 10-79\;keV}/Flux_{\rm 10-79\;keV}=5$. The transparency of the points is proportional to the strength of reflection during the observations.}
    \label{fig:spin_vs_edd_hardness}
\end{figure}

Figure \ref{fig:spin_vs_edd_hardness} shows the measured spins for all the treated observations of GRS 1915+105, Cygnus X-1, GX 339-4, and MAXI J1820+070 vs. the Eddington fraction during the observations (left, blue) and vs. the hardness ratio during the observations (right, red), defined as the ratio of the flux in the 10-70~keV band to the flux in the 3-5~keV band. The transparency of the points is proportional to the strength of reflection during the observations, which was used as weighting while combining the independent measurements into the reported values. The vertical dashed lines in the left and right sets of panels show the boundaries used for this experiment: we only combined observations for which $10^{-2}\leq L/L_{\rm Edd}\leq 0.3$ and for which $\rm Flux_{\rm 10-79\;keV}/Flux_{\rm 3-5\;keV}\leq 5$. In other words, we only combined the points to the right of the vertical dashed line in the left panels, and to the left of the vertical line in the right panels. While these choices are somewhat arbitrary, they clearly highlight the magnitude by which our measurements would be impacted by restricting the observations treated in this analysis to more explicitly agree with prior expectations of accretion disks extending to the ISCO.

When constraining the Eddington fraction, all observations of Cygnus X-1 fall while the source was below $L/L_{\rm Edd}=10^{-2}$, making this experiment irrelevant. For MAXI J1820+070, only one observation was cut out, for which the reflection strength was low and the spin was below the previously determined combined value, leading to a nearly unchanged measurement. For the other two sources (GRS 1915+105 and GX 339-4), by removing the observations for which $L/L_{\rm Edd}\leq10^{-2}$, we filtered out the main contribution to the combined measurement that provides support for a low or negative spin, leading to a decrease in the uncertainty of our spin constraints. When only considering observations for which $\rm Flux_{\rm 10-79\;keV}/Flux_{\rm 3-5\;keV}\leq 5$, all four sources lose support for low spin, with the combined measurements resulting in narrower uncertainties on a higher BH spin. Still, all the measurements obtained through this experiment are in full agreement with the original measurements which incorporate all available data. 

Additionally, we reanalyzed all observations of GX 339-4 to simultaneously allow the inner disk radius and the BH spin to vary. We chose the observations of GX 339-4 for this experiment, as this source spans the broadest range of Eddington fraction of the sources treated in this paper. We note that this treatment does not fix the spin to some value, as similar previous studies do. By allowing the inner disk radius and the spin to vary simultaneously, we properly infer the extent of the variation of the two parameters. If one were to fix the spin to some arbitrarily high value, the inner disk radius would be artificially biased to lower values. A maximally spinning BH with a disk truncated at 6 $r_{\rm g}$ would produce similar gravitational distortions to a nonspinning BH with the disk extending to the ISCO. Therefore, if the spin is erroneously fixed at a high value, for the same spectral shape, the inferred disk truncation would be larger than when allowing the spin to vary freely.

\begin{figure}[ht]
    \centering
    \includegraphics[width= 0.48\textwidth]{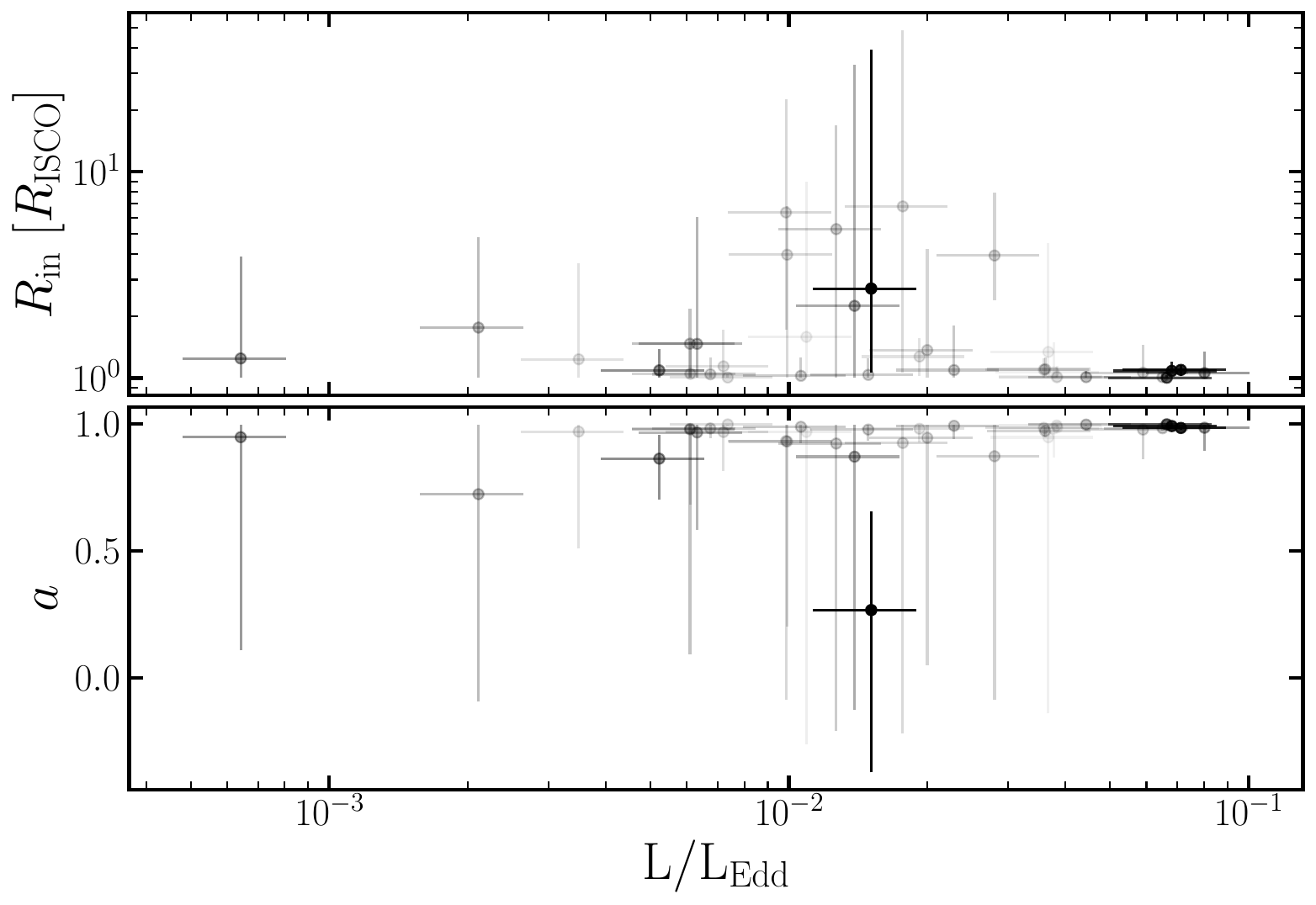}
    \caption{Results of simultaneous fits to the inner disk radius (top) and BH spin (bottom) as a function of Eddington fraction for all the observations of GX 339-4 treated in this paper. The transparency of the points is proportional to the strength of reflection, which was used throughout the paper as weighting for combining independent measurements. No evidence for disk truncation is clear, throughout the entire regime of Eddington fraction.}
    \label{fig:GX_339_free_rin}
\end{figure}

Figure \ref{fig:GX_339_free_rin} shows the inner disk radius in units of the ISCO radius (top) and the BH spin (bottom) measured by running the MCMC analysis of all the treated observations of GX 339-4. The transparency of the points is proportional to the strength of reflection during the observation. The figure indicates that there is no clear sign of disk truncation throughout the entire range of Eddington fractions treated in our analysis. It is important to acknowledge that this experiment was performed for a single source, and that these conclusions do not necessarily translate directly to all sources. The spin of the BH, the degree of misalignment between the BH and binary system, and the mass and metallicity of the donor star, could cause the critical Eddington fraction to differ between sources.  However, even when considering the same source, outbursts are observed to differ significantly; factors including the ratio of coronal and disk flux, mass loss in winds, and the rate of change in the mass accretion rate could cause the critical Eddington fraction to change between outbursts.  The fact that no observations of GX 339-4 show strong evidence of truncation above $ L/L_{\rm Edd} \geq 0.001$ across several outbursts may signal that these considerations are minor. This analysis was performed in order to highlight that our initial assumption, of the accretion disk extending to the ISCO for Eddington fractions between $10^{-3}\leq L/L_{\rm Edd}\leq 0.3$, holds. While this assumption is likely to not be universally true for all sources, or even for all observations of a single source, it is a reasonable assumption given the observable quantities.

Based on the 36 spins reported in this paper, we can place constraints on the distribution of BH spins observed in XB systems. However, it is important to acknowledge that these values do not take into account any possible observational or selection biases, that could lead to a preferential detection of highly spinning BH systems. Within $1\sigma$ uncertainties, $86\%$ of spins are consistent with $a\geq0.95$, $94\%$ are consistent with $a\geq0.9$, and $100\%$ are consistent with $a\geq0.7$, the theoretical upper limit for the spin of a neutron star (NS; \citealt{2011ApJ...728...12L}). This suggests that despite possible selection effects, relativistic reflection could be a pragmatic tool for distinguishing between BH and NS XB systems. Given the reported values, $28\%$ of the spins are definitively $a\geq0.9$ within $1\sigma$ uncertainty (value minus $1\sigma$ error $\geq0.9$). Conversely, looking at the lower limits on the distribution, we find limited support for low BH spins in the BHs in XBs, with $\sim3\%$ of measurements allowing values $a\leq0$ within $1\sigma$, $5.5\%$ allowing $a\leq0.4$, and $\sim17\%$ allowing $a\leq0.6$ within $1\sigma$ (value minus $1\sigma$ error $\leq0.6$). Again, these results strongly suggest that the BHs observed in XBs and those merging in BBH systems observed though GWs indeed come from different distributions, and that their formation paths are determined by the properties of the progenitor stars and by binary interactions both before and after BH formation.

We attempted to illustrate the importance that small-scale source variability between observations can have on inducing changes in the \textit{measured} BH spin. The true value of the BH spin is unlikely to change on timescales comparable to the age of X-ray astronomy, meaning that the variability in BH spin that we see is entirely owed to our incomplete characterization of the physical processes that create the observed spectra. In the future, this uncertainty will likely be mediated by using more complex theoretical models, which can simultaneously explain the spectral, timing, and X-ray polarization measurements of our target sources. The era of high-resolution spectroscopy started by the launch of XRISM will shed light on narrow spectral features present in X-ray spectra that were previously unaccounted for, and the importance of those for spin measurements is yet to be determined. Analyzing XRISM spectra will likely require more complex theoretical models, which will incorporate a more physically accurate description of the mechanisms at play in XB systems.

Lastly, by systematically analyzing such a large sample of BH XB, we have compiled an extensive data set that illustrates the behavior of the models when applied to a variety of observations taken across an array of physical configurations of the sources. This entire data set will be treated in a future paper, in order to assess the existence of possible trends or degeneracies in the entire parameter space. That study will quantify the importance of correlations between parameters, and their impact on the ability to recover the BH spin.

\begin{acknowledgments}
We thank the anonymous reviewer for their comments and suggestions, which have improved the quality of this paper PD acknowledges helpful conversations and advice from Brian Grefenstette, Dominic Walton, and Dan Wilkins.

\textit{Software:} 
\texttt{Astropy} (\citealt{astropy:2013, astropy:2018}), 
\texttt{emcee} (\citealt{2013PASP..125..306F}), 
\texttt{numpy} (\citealt{harris2020array}), 
\texttt{matplotlib} (\citealt{Hunter:2007}), 
\texttt{scipy} (\citealt{2020SciPy-NMeth}), 
\texttt{pandas} (\citealt{2022zndo...6702671R, mckinney-proc-scipy-2010}), 
\texttt{corner} (\citealt{corner}), 
\texttt{iPython} (\citealt{PER-GRA:2007}), 
\texttt{Xspec} (\citealt{1996ASPC..101...17A}), 
\texttt{relxill} (\citealt{2014MNRAS.444L.100D, 2014ApJ...782...76G}).
\end{acknowledgments}

%\newpage

\bibliography{paper}{}

\begin{thebibliography}{}
\expandafter\ifx\csname natexlab\endcsname\relax\def\natexlab#1{#1}\fi
\providecommand{\url}[1]{\href{#1}{#1}}
\providecommand{\dodoi}[1]{doi:~\href{http://doi.org/#1}{\nolinkurl{#1}}}
\providecommand{\doeprint}[1]{\href{http://ascl.net/#1}{\nolinkurl{http://ascl.net/#1}}}
\providecommand{\doarXiv}[1]{\href{https://arxiv.org/abs/#1}{\nolinkurl{https://arxiv.org/abs/#1}}}

\bibitem[{{Abbott} {et~al.}(2023{\natexlab{a}}){Abbott}, {Abbott}, {Acernese}, {Ackley}, {Adams}, {Adhikari}, {Adhikari}, {Adya}, {Affeldt}, {Agarwal}, {Agathos}, {Agatsuma}, {Aggarwal}, {Aguiar}, {Aiello}, {Ain}, {Ajith}, {Akcay}, {Akutsu}, {Albanesi}, {Allocca}, {Altin}, {Amato}, {Anand}, {Anand}, {Ananyeva}, {Anderson}, {Anderson}, {Ando}, {Andrade}, {Andres}, {Andri{\'c}}, {Angelova}, {Ansoldi}, {Antelis}, {Antier}, {Appert}, {Arai}, {Arai}, {Arai}, {Araki}, {Araya}, {Araya}, {Areeda}, {Ar{\`e}ne}, {Aritomi}, {Arnaud}, {Arogeti}, {Aronson}, {Arun}, {Asada}, {Asali}, {Ashton}, {Aso}, {Assiduo}, {Aston}, {Astone}, {Aubin}, {Austin}, {Babak}, {Badaracco}, {Bader}, {Badger}, {Bae}, {Bae}, {Baer}, {Bagnasco}, {Bai}, {Baiotti}, {Baird}, {Bajpai}, {Ball}, {Ballardin}, {Ballmer}, {Balsamo}, {Baltus}, {Banagiri}, {Bankar}, {Barayoga}, {Barbieri}, {Barish}, {Barker}, {Barneo}, {Barone}, {Barr}, {Barsotti}, {Barsuglia}, {Barta}, {Bartlett}, {Barton}, {Bartos}, {Bassiri}, {Basti}, {Bawaj}, {Bayley}, {Baylor},
  {Bazzan}, {B{\'e}csy}, {Bedakihale}, {Bejger}, {Belahcene}, {Benedetto}, {Beniwal}, {Bennett}, {Bentley}, {Benyaala}, {Bergamin}, {Berger}, {Bernuzzi}, {Berry}, {Bersanetti}, {Bertolini}, {Betzwieser}, {Beveridge}, {Bhandare}, {Bhardwaj}, {Bhattacharjee}, {Bhaumik}, {Bilenko}, {Billingsley}, {Bini}, {Birney}, {Birnholtz}, {Biscans}, {Bischi}, {Biscoveanu}, {Bisht}, {Biswas}, {Bitossi}, {Bizouard}, {Blackburn}, {Blair}, {Blair}, {Blair}, {Bobba}, {Bode}, {Boer}, {Bogaert}, {Boldrini}, {Bonavena}, {Bondu}, {Bonilla}, {Bonnand}, {Booker}, {Boom}, {Bork}, {Boschi}, {Bose}, {Bose}, {Bossilkov}, {Boudart}, {Bouffanais}, {Bozzi}, {Bradaschia}, {Brady}, {Bramley}, {Branch}, {Branchesi}, {Brandt}, {Brau}, {Breschi}, {Briant}, {Briggs}, {Brillet}, {Brinkmann}, {Brockill}, {Brooks}, {Brooks}, {Brown}, {Brunett}, {Bruno}, {Bruntz}, {Bryant}, {Bulik}, {Bulten}, {Buonanno}, {Buscicchio}, {Buskulic}, {Buy}, {Byer}, {Davies}, {Cadonati}, {Cagnoli}, {Cahillane}, {Bustillo}, {Callaghan}, {Callister}, {Calloni}, {Cameron},
  {Camp}, {Canepa}, {Canevarolo}, {Cannavacciuolo}, {Cannon}, {Cao}, {Cao}, {Capocasa}, {Capote}, {Carapella}, {Carbognani}, {Carlin}, {Carney}, {Carpinelli}, {Carrillo}, {Carullo}, {Carver}, {Diaz}, {Casentini}, {Castaldi}, {Caudill}, {Cavagli{\`a}}, {Cavalier}, {Cavalieri}, {Ceasar}, {Cella}, {Cerd{\'a}-Dur{\'a}n}, {Cesarini}, {Chaibi}, {Chakravarti}, {Subrahmanya}, {Champion}, {Chan}, {Chan}, {Chan}, {Chan}, {Chan}, {Chandra}, {Chanial}, {Chao}, {Chapman-Bird}, {Charlton}, {Chase}, {Chassande-Mottin}, {Chatterjee}, {Chatterjee}, {Chatterjee}, {Chaturvedi}, {Chaty}, {Chatziioannou}, {Chen}, {Chen}, {Chen}, {Chen}, {Chen}, {Chen}, {Chen}, {Chen}, {Cheng}, {Cheong}, {Cheung}, {Chia}, {Chiadini}, {Chiang}, {Chiarini}, {Chierici}, {Chincarini}, {Chiofalo}, {Chiummo}, {Cho}, {Cho}, {Choudhary}, {Choudhary}, {Christensen}, {Chu}, {Chu}, {Chu}, {Chua}, {Chung}, {Ciani}, {Ciecielag}, {Cie{\'s}lar}, {Cifaldi}, {Ciobanu}, {Ciolfi}, {Cipriano}, {Cirone}, {Clara}, {Clark}, {Clark}, {Clarke}, {Clearwater}, {Clesse},
  {Cleva}, {Coccia}, {Codazzo}, {Cohadon}, {Cohen}, {Cohen}, {Colleoni}, {Collette}, {Colombo}, {Colpi}, {Compton}, {Constancio}, {Conti}, {Cooper}, {Corban}, {Corbitt}, {Cordero-Carri{\'o}n}, {Corezzi}, {Corley}, {Cornish}, {Corre}, {Corsi}, {Cortese}, {Costa}, {Cotesta}, {Coughlin}, {Coulon}, {Countryman}, {Cousins}, {Couvares}, {Coward}, {Cowart}, {Coyne}, {Coyne}, {Creighton}, {Creighton}, {Criswell}, {Croquette}, {Crowder}, {Cudell}, {Cullen}, {Cumming}, {Cummings}, {Cunningham}, {Cuoco}, {Cury{\l}o}, {Dabadie}, {Canton}, {Dall'Osso}, {D{\'a}lya}, {Dana}, {Daneshgaranbajastani}, {D'Angelo}, {Danila}, {Danilishin}, {D'Antonio}, {Danzmann}, {Darsow-Fromm}, {Dasgupta}, {Datrier}, {Dattilo}, {Dave}, {Davier}, {Davis}, {Davis}, {Daw}, {de Alarc{\'o}n}, {Dean}, {Debra}, {Deenadayalan}, {Degallaix}, {de Laurentis}, {Del{\'e}glise}, {Del Favero}, {de Lillo}, {de Lillo}, {Del Pozzo}, {Demarchi}, {de Matteis}, {D'Emilio}, {Demos}, {Dent}, {Depasse}, {de Pietri}, {De Rosa}, {de Rossi}, {Desalvo}, {de Simone},
  {Dhurandhar}, {D{\'\i}az}, {Diaz-Ortiz}, {Didio}, {Dietrich}, {di Fiore}, {di Fronzo}, {di Giorgio}, {di Giovanni}, {di Giovanni}, {di Girolamo}, {di Lieto}, {Ding}, {di Pace}, {di Palma}, {di Renzo}, {Divakarla}, {Dmitriev}, {Doctor}, {D'Onofrio}, {Donovan}, {Dooley}, {Doravari}, {Dorrington}, {Drago}, {Driggers}, {Drori}, {Ducoin}, {Dupej}, {Durante}, {D'Urso}, {Duverne}, {Dwyer}, {Eassa}, {Easter}, {Ebersold}, {Eckhardt}, {Eddolls}, {Edelman}, {Edo}, {Edy}, {Effler}, {Eguchi}, {Eichholz}, {Eikenberry}, {Eisenmann}, {Eisenstein}, {Ejlli}, {Engelby}, {Enomoto}, {Errico}, {Essick}, {Estell{\'e}s}, {Estevez}, {Etienne}, {Etzel}, {Evans}, {Evans}, {Ewing}, {Fafone}, {Fair}, {Fairhurst}, {Farah}, {Farinon}, {Farr}, {Farr}, {Farrow}, {Fauchon-Jones}, {Favaro}, {Favata}, {Fays}, {Fazio}, {Feicht}, {Fejer}, {Fenyvesi}, {Ferguson}, {Fernandez-Galiana}, {Ferrante}, {Ferreira}, {Fidecaro}, {Figura}, {Fiori}, {Fishbach}, {Fisher}, {Fittipaldi}, {Fiumara}, {Flaminio}, {Floden}, {Fong}, {Font}, {Fornal}, {Forsyth},
  {Franke}, {Frasca}, {Frasconi}, {Frederick}, {Freed}, {Frei}, {Freise}, {Frey}, {Fritschel}, {Frolov}, {Fronz{\'e}}, {Fujii}, {Fujikawa}, {Fukunaga}, {Fukushima}, {Fulda}, {Fyffe}, {Gabbard}, {Gabella}, {Gadre}, {Gair}, {Gais}, {Galaudage}, {Gamba}, {Ganapathy}, {Ganguly}, {Gao}, {Gaonkar}, {Garaventa}, {Garc{\'\i}a}, {Garc{\'\i}a-N{\'u}{\~n}ez}, {Garc{\'\i}a-Quir{\'o}s}, {Garufi}, {Gateley}, {Gaudio}, {Gayathri}, {Ge}, {Gemme}, {Gennai}, {George}, {George}, {Gerberding}, {Gergely}, {Gewecke}, {Ghonge}, {Ghosh}, {Ghosh}, {Ghosh}, {Ghosh}, {Giacomazzo}, {Giacoppo}, {Giaime}, {Giardina}, {Gibson}, {Gier}, {Giesler}, {Giri}, {Gissi}, {Glanzer}, {Gleckl}, {Godwin}, {Goetz}, {Goetz}, {Gohlke}, {Golomb}, {Goncharov}, {Gonz{\'a}lez}, {Gopakumar}, {Gosselin}, {Gouaty}, {Gould}, {Grace}, {Grado}, {Granata}, {Granata}, {Grant}, {Gras}, {Grassia}, {Gray}, {Gray}, {Greco}, {Green}, {Green}, {Gretarsson}, {Gretarsson}, {Griffith}, {Griffiths}, {Griggs}, {Grignani}, {Grimaldi}, {Grimm}, {Grote}, {Grunewald}, {Gruning},
  {Guerra}, {Guidi}, {Guimaraes}, {Guix{\'e}}, {Gulati}, {Guo}, {Guo}, {Gupta}, {Gupta}, {Gupta}, {Gustafson}, {Gustafson}, {Guzman}, {Ha}, {Haegel}, {Hagiwara}, {Haino}, {Halim}, {Hall}, {Hamilton}, {Hammond}, {Han}, {Haney}, {Hanks}, {Hanna}, {Hannam}, {Hannuksela}, {Hansen}, {Hansen}, {Hanson}, {Harder}, {Hardwick}, {Haris}, {Harms}, {Harry}, {Harry}, {Hartwig}, {Hasegawa}, {Haskell}, {Hasskew}, {Haster}, {Hattori}, {Haughian}, {Hayakawa}, {Hayama}, {Hayes}, {Healy}, {Heidmann}, {Heidt}, {Heintze}, {Heinze}, {Heinzel}, {Heitmann}, {Hellman}, {Hello}, {Helmling-Cornell}, {Hemming}, {Hendry}, {Heng}, {Hennes}, {Hennig}, {Hennig}, {Hernandez}, {Hernandez Vivanco}, {Heurs}, {Hild}, {Hill}, {Himemoto}, {Hines}, {Hiranuma}, {Hirata}, {Hirose}, {Hochheim}, {Hofman}, {Hohmann}, {Holcomb}, {Holland}, {Holley-Bockelmann}, {Hollows}, {Holmes}, {Holt}, {Holz}, {Hong}, {Hopkins}, {Hough}, {Hourihane}, {Howell}, {Hoy}, {Hoyland}, {Hreibi}, {Hsieh}, {Hsu}, {Huang}, {Huang}, {Huang}, {Huang}, {Huang}, {Huang},
  {H{\"u}bner}, {Huddart}, {Hughey}, {Hui}, {Hui}, {Husa}, {Huttner}, {Huxford}, {Huynh-Dinh}, {Ide}, {Idzkowski}, {Iess}, {Ikenoue}, {Imam}, {Inayoshi}, {Ingram}, {Inoue}, {Ioka}, {Isi}, {Isleif}, {Ito}, {Itoh}, {Iyer}, {Izumi}, {Jaberianhamedan}, {Jacqmin}, {Jadhav}, {Jadhav}, {James}, {Jan}, {Jani}, {Janquart}, {Janssens}, {Janthalur}, {Jaranowski}, {Jariwala}, {Jaume}, {Jenkins}, {Jenner}, {Jeon}, {Jeunon}, {Jia}, {Jin}, {Johns}, {Johnson-McDaniel}, {Jones}, {Jones}, {Jones}, {Jones}, {Jones}, {Jonker}, {Ju}, {Jung}, {Jung}, {Junker}, {Juste}, {Kaihotsu}, {Kajita}, {Kakizaki}, {Kalaghatgi}, {Kalogera}, {Kamai}, {Kamiizumi}, {Kanda}, {Kandhasamy}, {Kang}, {Kanner}, {Kao}, {Kapadia}, {Kapasi}, {Karat}, {Karathanasis}, {Karki}, {Kashyap}, {Kasprzack}, {Kastaun}, {Katsanevas}, {Katsavounidis}, {Katzman}, {Kaur}, {Kawabe}, {Kawaguchi}, {Kawai}, {Kawasaki}, {K{\'e}f{\'e}lian}, {Keitel}, {Key}, {Khadka}, {Khalili}, {Khan}, {Khazanov}, {Khetan}, {Khursheed}, {Kijbunchoo}, {Kim}, {Kim}, {Kim}, {Kim}, {Kim}, {Kim},
  {Kimball}, {Kimura}, {Kinley-Hanlon}, {Kirchhoff}, {Kissel}, {Kita}, {Kitazawa}, {Kleybolte}, {Klimenko}, {Knee}, {Knowles}, {Knyazev}, {Koch}, {Koekoek}, {Kojima}, {Kokeyama}, {Koley}, {Kolitsidou}, {Kolstein}, {Komori}, {Kondrashov}, {Kong}, {Kontos}, {Koper}, {Korobko}, {Kotake}, {Kovalam}, {Kozak}, {Kozakai}, {Kozu}, {Kringel}, {Krishnendu}, {Kr{\'o}lak}, {Kuehn}, {Kuei}, {Kuijer}, {Kulkarni}, {Kumar}, {Kumar}, {Kumar}, {Kumar}, {Kume}, {Kuns}, {Kuo}, {Kuo}, {Kuromiya}, {Kuroyanagi}, {Kusayanagi}, {Kuwahara}, {Kwak}, {Lagabbe}, {Laghi}, {Lalande}, {Lam}, {Lamberts}, {Landry}, {Lane}, {Lang}, {Lange}, {Lantz}, {La Rosa}, {Lartaux-Vollard}, {Lasky}, {Laxen}, {Lazzarini}, {Lazzaro}, {Leaci}, {Leavey}, {Lecoeuche}, {Lee}, {Lee}, {Lee}, {Lee}, {Lee}, {Lee}, {Lehmann}, {Lema{\^\i}tre}, {Leonardi}, {Leroy}, {Letendre}, {Levesque}, {Levin}, {Leviton}, {Leyde}, {Li}, {Li}, {Li}, {Li}, {Li}, {Li}, {Lin}, {Lin}, {Lin}, {Lin}, {Lin}, {Linde}, {Linker}, {Linley}, {Littenberg}, {Liu}, {Liu}, {Liu}, {Liu}, {Llamas},
  {Llorens-Monteagudo}, {Lo}, {Lockwood}, {Loh}, {London}, {Longo}, {Lopez}, {Portilla}, {Lorenzini}, {Loriette}, {Lormand}, {Losurdo}, {Lott}, {Lough}, {Lousto}, {Lovelace}, {Lucaccioni}, {L{\"u}ck}, {Lumaca}, {Lundgren}, {Luo}, {Lynam}, {Macas}, {Macinnis}, {MacLeod}, {MacMillan}, {Macquet}, {Hernandez}, {Magazz{\`u}}, {Magee}, {Maggiore}, {Magnozzi}, {Mahesh}, {Majorana}, {Makarem}, {Maksimovic}, {Maliakal}, {Malik}, {Man}, {Mandic}, {Mangano}, {Mango}, {Mansell}, {Manske}, {Mantovani}, {Mapelli}, {Marchesoni}, {Marchio}, {Marion}, {Mark}, {M{\'a}rka}, {M{\'a}rka}, {Markakis}, {Markosyan}, {Markowitz}, {Maros}, {Marquina}, {Marsat}, {Martelli}, {Martin}, {Martin}, {Martinez}, {Martinez}, {Martinez}, {Martinovic}, {Martynov}, {Marx}, {Masalehdan}, {Mason}, {Massera}, {Masserot}, {Massinger}, {Masso-Reid}, {Mastrogiovanni}, {Matas}, {Mateu-Lucena}, {Matichard}, {Matiushechkina}, {Mavalvala}, {McCann}, {McCarthy}, {McClelland}, {McClincy}, {McCormick}, {McCuller}, {McGhee}, {McGuire}, {McIsaac}, {McIver},
  {McRae}, {McWilliams}, {Meacher}, {Mehmet}, {Mehta}, {Meijer}, {Melatos}, {Melchor}, {Mendell}, {Menendez-Vazquez}, {Menoni}, {Mercer}, {Mereni}, {Merfeld}, {Merilh}, {Merritt}, {Merzougui}, {Meshkov}, {Messenger}, {Messick}, {Meyers}, {Meylahn}, {Mhaske}, {Miani}, {Miao}, {Michaloliakos}, {Michel}, {Michimura}, {Middleton}, {Milano}, {Miller}, {Miller}, {Miller}, {Millhouse}, {Mills}, {Milotti}, {Minazzoli}, {Minenkov}, {Mio}, {Mir}, {Miravet-Ten{\'e}s}, {Mishra}, {Mishra}, {Mistry}, {Mitra}, {Mitrofanov}, {Mitselmakher}, {Mittleman}, {Miyakawa}, {Miyamoto}, {Miyazaki}, {Miyo}, {Miyoki}, {Mo}, {Modafferi}, {Moguel}, {Mogushi}, {Mohapatra}, {Mohite}, {Molina}, {Molina-Ruiz}, {Mondin}, {Montani}, {Moore}, {Moraru}, {Morawski}, {More}, {Moreno}, {Moreno}, {Mori}, {Morisaki}, {Moriwaki}, {Morr{\'a}s}, {Mours}, {Mow-Lowry}, {Mozzon}, {Muciaccia}, {Mukherjee}, {Mukherjee}, {Mukherjee}, {Mukherjee}, {Mukherjee}, {Mukund}, {Mullavey}, {Munch}, {Mu{\~n}iz}, {Murray}, {Musenich}, {Muusse}, {Nadji}, {Nagano},
  {Nagano}, {Nagar}, {Nakamura}, {Nakano}, {Nakano}, {Nakashima}, {Nakayama}, {Napolano}, {Nardecchia}, {Narikawa}, {Naticchioni}, {Nayak}, {Nayak}, {Negishi}, {Neil}, {Neilson}, {Nelemans}, {Nelson}, {Nery}, {Neubauer}, {Neunzert}, {Ng}, {Ng}, {Nguyen}, {Nguyen}, {Nguyen}, {Quynh}, {Ni}, {Nichols}, {Nishizawa}, {Nissanke}, {Nitoglia}, {Nocera}, {Norman}, {North}, {Nozaki}, {Siles}, {Nuttall}, {Oberling}, {O'Brien}, {Obuchi}, {O'Dell}, {Oelker}, {Ogaki}, {Oganesyan}, {Oh}, {Oh}, {Oh}, {Ohashi}, {Ohishi}, {Ohkawa}, {Ohme}, {Ohta}, {Okada}, {Okutani}, {Okutomi}, {Olivetto}, {Oohara}, {Ooi}, {Oram}, {O'Reilly}, {Ormiston}, {Ormsby}, {Ortega}, {O'Shaughnessy}, {O'Shea}, {Oshino}, {Ossokine}, {Osthelder}, {Otabe}, {Ottaway}, {Overmier}, {Pace}, {Pagano}, {Page}, {Pagliaroli}, {Pai}, {Pai}, {Palamos}, {Palashov}, {Palomba}, {Pan}, {Pan}, {Panda}, {Pang}, {Pang}, {Pankow}, {Pannarale}, {Pant}, {Panther}, {Paoletti}, {Paoli}, {Paolone}, {Parisi}, {Park}, {Park}, {Parker}, {Pascucci}, {Pasqualetti}, {Passaquieti},
  {Passuello}, {Patel}, {Pathak}, {Patricelli}, {Patron}, {Paul}, {Payne}, {Pedraza}, {Pegoraro}, {Pele}, {Arellano}, {Penn}, {Perego}, {Pereira}, {Pereira}, {Perez}, {P{\'e}rigois}, {Perkins}, {Perreca}, {Perri{\`e}s}, {Petermann}, {Petterson}, {Pfeiffer}, {Pham}, {Phukon}, {Piccinni}, {Pichot}, {Piendibene}, {Piergiovanni}, {Pierini}, {Pierro}, {Pillant}, {Pillas}, {Pilo}, {Pinard}, {Pinto}, {Pinto}, {Piotrzkowski}, {Piotrzkowski}, {Pirello}, {Pitkin}, {Placidi}, {Planas}, {Plastino}, {Pluchar}, {Poggiani}, {Polini}, {Pong}, {Ponrathnam}, {Popolizio}, {Porter}, {Poulton}, {Powell}, {Pracchia}, {Pradier}, {Prajapati}, {Prasai}, {Prasanna}, {Pratten}, {Principe}, {Prodi}, {Prokhorov}, {Prosposito}, {Prudenzi}, {Puecher}, {Punturo}, {Puosi}, {Puppo}, {P{\"u}rrer}, {Qi}, {Quetschke}, {Quitzow-James}, {Qutob}, {Raab}, {Raaijmakers}, {Radkins}, {Radulesco}, {Raffai}, {Rail}, {Raja}, {Rajan}, {Ramirez}, {Ramirez}, {Ramos-Buades}, {Rana}, {Rapagnani}, {Rapol}, {Ray}, {Raymond}, {Raza}, {Razzano}, {Read}, {Rees},
  {Regimbau}, {Rei}, {Reid}, {Reid}, {Reitze}, {Relton}, {Renzini}, {Rettegno}, {Reza}, {Rezac}, {Ricci}, {Richards}, {Richardson}, {Richardson}, {Riemenschneider}, {Riles}, {Rinaldi}, {Rink}, {Rizzo}, {Robertson}, {Robie}, {Robinet}, {Rocchi}, {Rodriguez}, {Rolland}, {Rollins}, {Romanelli}, {Romano}, {Romel}, {Romero-Rodr{\'\i}guez}, {Romero-Shaw}, {Romie}, {Ronchini}, {Rosa}, {Rose}, {Rosi{\'n}ska}, {Ross}, {Rowan}, {Rowlinson}, {Roy}, {Roy}, {Roy}, {Rozza}, {Ruggi}, {Ruiz-Rocha}, {Ryan}, {Sachdev}, {Sadecki}, {Sadiq}, {Sago}, {Saito}, {Saito}, {Sakai}, {Sakai}, {Sakellariadou}, {Sakuno}, {Salafia}, {Salconi}, {Saleem}, {Salemi}, {Samajdar}, {Sanchez}, {Sanchez}, {Sanchez}, {Sanchis-Gual}, {Sanders}, {Sanuy}, {Saravanan}, {Sarin}, {Sassolas}, {Satari}, {Sathyaprakash}, {Sato}, {Sato}, {Sauter}, {Savage}, {Sawada}, {Sawant}, {Sawant}, {Sayah}, {Schaetzl}, {Scheel}, {Scheuer}, {Schiworski}, {Schmidt}, {Schmidt}, {Schnabel}, {Schneewind}, {Schofield}, {Sch{\"o}nbeck}, {Schulte}, {Schutz}, {Schwartz}, {Scott},
  {Scott}, {Seglar-Arroyo}, {Sekiguchi}, {Sekiguchi}, {Sellers}, {Sengupta}, {Sentenac}, {Seo}, {Sequino}, {Sergeev}, {Setyawati}, {Shaffer}, {Shahriar}, {Shams}, {Shao}, {Sharma}, {Sharma}, {Shawhan}, {Shcheblanov}, {Shibagaki}, {Shikauchi}, {Shimizu}, {Shimoda}, {Shimode}, {Shinkai}, {Shishido}, {Shoda}, {Shoemaker}, {Shoemaker}, {Shyamsundar}, {Sieniawska}, {Sigg}, {Singer}, {Singh}, {Singh}, {Singha}, {Sintes}, {Sipala}, {Skliris}, {Slagmolen}, {Slaven-Blair}, {Smetana}, {Smith}, {Smith}, {Soldateschi}, {Somala}, {Somiya}, {Son}, {Soni}, {Soni}, {Sordini}, {Sorrentino}, {Sorrentino}, {Sotani}, {Soulard}, {Souradeep}, {Sowell}, {Spagnuolo}, {Spencer}, {Spera}, {Srinivasan}, {Srivastava}, {Srivastava}, {Staats}, {Stachie}, {Steer}, {Steinhoff}, {Steinlechner}, {Steinlechner}, {Stevenson}, {Stops}, {Stover}, {Strain}, {Strang}, {Stratta}, {Strunk}, {Sturani}, {Stuver}, {Sudhagar}, {Sudhir}, {Sugimoto}, {Suh}, {Sullivan}, {Sullivan}, {Summerscales}, {Sun}, {Sun}, {Sunil}, {Sur}, {Suresh}, {Sutton}, {Suzuki},
  {Suzuki}, {Swinkels}, {Szczepa{\'n}czyk}, {Szewczyk}, {Tacca}, {Tagoshi}, {Tait}, {Takahashi}, {Takahashi}, {Takamori}, {Takano}, {Takeda}, {Takeda}, {Talbot}, {Talbot}, {Tanaka}, {Tanaka}, {Tanaka}, {Tanaka}, {Tanaka}, {Tanasijczuk}, {Tanioka}, {Tanner}, {Tao}, {Tao}, {Mart{\'\i}n}, {Taranto}, {Tasson}, {Telada}, {Tenorio}, {Terhune}, {Terkowski}, {Thirugnanasambandam}, {Thomas}, {Thomas}, {Thomas}, {Thompson}, {Thondapu}, {Thorne}, {Thrane}, {Tiwari}, {Tiwari}, {Tiwari}, {Toivonen}, {Toland}, {Tolley}, {Tomaru}, {Tomigami}, {Tomura}, {Tonelli}, {Torres-Forn{\'e}}, {Torrie}, {E Melo}, {T{\"o}yr{\"a}}, {Trapananti}, {Travasso}, {Traylor}, {Trevor}, {Tringali}, {Tripathee}, {Troiano}, {Trovato}, {Trozzo}, {Trudeau}, {Tsai}, {Tsai}, {Tsang}, {Tsang}, {Tsao}, {Tse}, {Tso}, {Tsubono}, {Tsuchida}, {Tsukada}, {Tsuna}, {Tsutsui}, {Tsuzuki}, {Turbang}, {Turconi}, {Tuyenbayev}, {Ubhi}, {Uchikata}, {Uchiyama}, {Udall}, {Ueda}, {Uehara}, {Ueno}, {Ueshima}, {Unnikrishnan}, {Uraguchi}, {Urban}, {Ushiba}, {Utina},
  {Vahlbruch}, {Vajente}, {Vajpeyi}, {Valdes}, {Valentini}, {Valsan}, {van Bakel}, {van Beuzekom}, {van den Brand}, {van den Broeck}, {Vander-Hyde}, {van der Schaaf}, {van Heijningen}, {Vanosky}, {van Putten}, {van Remortel}, {Vardaro}, {Vargas}, {Varma}, {Vas{\'u}th}, {Vecchio}, {Vedovato}, {Veitch}, {Veitch}, {Venneberg}, {Venugopalan}, {Verkindt}, {Verma}, {Verma}, {Veske}, {Vetrano}, {Vicer{\'e}}, {Vidyant}, {Viets}, {Vijaykumar}, {Villa-Ortega}, {Vinet}, {Virtuoso}, {Vitale}, {Vo}, {Vocca}, {von Reis}, {von Wrangel}, {Vorvick}, {Vyatchanin}, {Wade}, {Wade}, {Wagner}, {Walet}, {Walker}, {Wallace}, {Wallace}, {Walsh}, {Wang}, {Wang}, {Wang}, {Ward}, {Warner}, {Was}, {Washimi}, {Washington}, {Watchi}, {Weaver}, {Webster}, {Weinert}, {Weinstein}, {Weiss}, {Weller}, {Weller}, {Wellmann}, {Wen}, {We{\ss}els}, {Wette}, {Whelan}, {White}, {Whiting}, {Whittle}, {Wilken}, {Williams}, {Williams}, {Williams}, {Williamson}, {Willis}, {Willke}, {Wilson}, {Winkler}, {Wipf}, {Wlodarczyk}, {Woan}, {Woehler}, {Wofford},
  {Wong}, {Wu}, {Wu}, {Wu}, {Wu}, {Wysocki}, {Xiao}, {Xu}, {Yamada}, {Yamamoto}, {Yamamoto}, {Yamamoto}, {Yamamoto}, {Yamashita}, {Yamazaki}, {Yang}, {Yang}, {Yang}, {Yang}, {Yang}, {Yap}, {Yeeles}, {Yelikar}, {Ying}, {Yokogawa}, {Yokoyama}, {Yokozawa}, {Yoo}, {Yoshioka}, {Yu}, {Yu}, {Yuzurihara}, {Zadro{\.z}ny}, {Zanolin}, {Zeidler}, {Zelenova}, {Zendri}, {Zevin}, {Zhan}, {Zhang}, {Zhang}, {Zhang}, {Zhang}, {Zhang}, {Zhao}, {Zhao}, {Zhao}, {Zhao}, {Zheng}, {Zhou}, {Zhou}, {Zhu}, {Zhu}, {Zimmerman}, {Zlochower}, {Zucker}, {Zweizig}, {Ligo Scientific Collaboration}, {VIRGO Collaboration}, \& {Kagra Collaboration}}]{2023PhRvX..13d1039A}
{Abbott}, R., {Abbott}, T.~D., {Acernese}, F., {et~al.} 2023{\natexlab{a}}, Physical Review X, 13, 041039, \dodoi{10.1103/PhysRevX.13.041039}

\bibitem[{{Abbott} {et~al.}(2023{\natexlab{b}}){Abbott}, {Abbott}, {Acernese}, {Ackley}, {Adams}, {Adhikari}, {Adhikari}, {Adya}, {Affeldt}, {Agarwal}, {Agathos}, {Agatsuma}, {Aggarwal}, {Aguiar}, {Aiello}, {Ain}, {Ajith}, {Akutsu}, {de Alarc{\'o}n}, {Akcay}, {Albanesi}, {Allocca}, {Altin}, {Amato}, {Anand}, {Anand}, {Ananyeva}, {Anderson}, {Anderson}, {Ando}, {Andrade}, {Andres}, {Andri{\'c}}, {Angelova}, {Ansoldi}, {Antelis}, {Antier}, {Antonini}, {Appert}, {Arai}, {Arai}, {Arai}, {Araki}, {Araya}, {Araya}, {Areeda}, {Ar{\`e}ne}, {Aritomi}, {Arnaud}, {Arogeti}, {Aronson}, {Arun}, {Asada}, {Asali}, {Ashton}, {Aso}, {Assiduo}, {Aston}, {Astone}, {Aubin}, {Austin}, {Babak}, {Badaracco}, {Bader}, {Badger}, {Bae}, {Bae}, {Baer}, {Bagnasco}, {Bai}, {Baiotti}, {Baird}, {Bajpai}, {Ball}, {Ballardin}, {Ballmer}, {Balsamo}, {Baltus}, {Banagiri}, {Bankar}, {Barayoga}, {Barbieri}, {Barish}, {Barker}, {Barneo}, {Barone}, {Barr}, {Barsotti}, {Barsuglia}, {Barta}, {Bartlett}, {Barton}, {Bartos}, {Bassiri}, {Basti},
  {Bawaj}, {Bayley}, {Baylor}, {Bazzan}, {B{\'e}csy}, {Bedakihale}, {Bejger}, {Belahcene}, {Benedetto}, {Beniwal}, {Bennett}, {Bentley}, {Benyaala}, {Bergamin}, {Berger}, {Bernuzzi}, {Berry}, {Bersanetti}, {Bertolini}, {Betzwieser}, {Beveridge}, {Bhandare}, {Bhardwaj}, {Bhattacharjee}, {Bhaumik}, {Bilenko}, {Billingsley}, {Bini}, {Birney}, {Birnholtz}, {Biscans}, {Bischi}, {Biscoveanu}, {Bisht}, {Biswas}, {Bitossi}, {Bizouard}, {Blackburn}, {Blair}, {Blair}, {Blair}, {Bobba}, {Bode}, {Boer}, {Bogaert}, {Boldrini}, {Bonavena}, {Bondu}, {Bonilla}, {Bonnand}, {Booker}, {Boom}, {Bork}, {Boschi}, {Bose}, {Bose}, {Bossilkov}, {Boudart}, {Bouffanais}, {Bozzi}, {Bradaschia}, {Brady}, {Bramley}, {Branch}, {Branchesi}, {Brandt}, {Brau}, {Breschi}, {Briant}, {Briggs}, {Brillet}, {Brinkmann}, {Brockill}, {Brooks}, {Brooks}, {Brown}, {Brunett}, {Bruno}, {Bruntz}, {Bryant}, {Bulik}, {Bulten}, {Buonanno}, {Buscicchio}, {Buskulic}, {Buy}, {Byer}, {Cadonati}, {Cagnoli}, {Cahillane}, {Bustillo}, {Callaghan}, {Callister},
  {Calloni}, {Cameron}, {Camp}, {Canepa}, {Canevarolo}, {Cannavacciuolo}, {Cannon}, {Cao}, {Cao}, {Capocasa}, {Capote}, {Carapella}, {Carbognani}, {Carlin}, {Carney}, {Carpinelli}, {Carrillo}, {Carullo}, {Carver}, {Diaz}, {Casentini}, {Castaldi}, {Caudill}, {Cavagli{\`a}}, {Cavalier}, {Cavalieri}, {Ceasar}, {Cella}, {Cerd{\'a}-Dur{\'a}n}, {Cesarini}, {Chaibi}, {Chakravarti}, {Subrahmanya}, {Champion}, {Chan}, {Chan}, {Chan}, {Chan}, {Chan}, {Chandra}, {Chanial}, {Chao}, {Chapman-Bird}, {Charlton}, {Chase}, {Chassande-Mottin}, {Chatterjee}, {Chatterjee}, {Chatterjee}, {Chaturvedi}, {Chaty}, {Chatziioannou}, {Chen}, {Chen}, {Chen}, {Chen}, {Chen}, {Chen}, {Chen}, {Chen}, {Cheng}, {Cheong}, {Cheung}, {Chia}, {Chiadini}, {Chiang}, {Chiarini}, {Chierici}, {Chincarini}, {Chiofalo}, {Chiummo}, {Cho}, {Cho}, {Choudhary}, {Choudhary}, {Christensen}, {Chu}, {Chu}, {Chu}, {Chua}, {Chung}, {Ciani}, {Ciecielag}, {Cie{\'s}lar}, {Cifaldi}, {Ciobanu}, {Ciolfi}, {Cipriano}, {Cirone}, {Clara}, {Clark}, {Clark}, {Clarke},
  {Clearwater}, {Clesse}, {Cleva}, {Coccia}, {Codazzo}, {Cohadon}, {Cohen}, {Cohen}, {Colleoni}, {Collette}, {Colombo}, {Colpi}, {Compton}, {Constancio}, {Conti}, {Cooper}, {Corban}, {Corbitt}, {Cordero-Carri{\'o}n}, {Corezzi}, {Corley}, {Cornish}, {Corre}, {Corsi}, {Cortese}, {Costa}, {Cotesta}, {Coughlin}, {Coulon}, {Countryman}, {Cousins}, {Couvares}, {Coward}, {Cowart}, {Coyne}, {Coyne}, {Creighton}, {Creighton}, {Criswell}, {Croquette}, {Crowder}, {Cudell}, {Cullen}, {Cumming}, {Cummings}, {Cunningham}, {Cuoco}, {Cury{\l}o}, {Dabadie}, {Canton}, {Dall'Osso}, {D{\'a}lya}, {Dana}, {Daneshgaranbajastani}, {D'Angelo}, {Danila}, {Danilishin}, {D'Antonio}, {Danzmann}, {Darsow-Fromm}, {Dasgupta}, {Datrier}, {Datta}, {Dattilo}, {Dave}, {Davier}, {Davies}, {Davis}, {Davis}, {Daw}, {Dean}, {Debra}, {Deenadayalan}, {Degallaix}, {de Laurentis}, {Del{\'e}glise}, {Del Favero}, {de Lillo}, {de Lillo}, {Del Pozzo}, {Demarchi}, {de Matteis}, {D'Emilio}, {Demos}, {Dent}, {Depasse}, {de Pietri}, {De Rosa}, {de Rossi},
  {Desalvo}, {de Simone}, {Dhurandhar}, {D{\'\i}az}, {Diaz-Ortiz}, {Didio}, {Dietrich}, {di Fiore}, {di Fronzo}, {di Giorgio}, {di Giovanni}, {di Giovanni}, {di Girolamo}, {di Lieto}, {Ding}, {di Pace}, {di Palma}, {di Renzo}, {Divakarla}, {Dmitriev}, {Doctor}, {D'Onofrio}, {Donovan}, {Dooley}, {Doravari}, {Dorrington}, {Drago}, {Driggers}, {Drori}, {Ducoin}, {Dupej}, {Durante}, {D'Urso}, {Duverne}, {Dwyer}, {Eassa}, {Easter}, {Ebersold}, {Eckhardt}, {Eddolls}, {Edelman}, {Edo}, {Edy}, {Effler}, {Eguchi}, {Eichholz}, {Eikenberry}, {Eisenmann}, {Eisenstein}, {Ejlli}, {Engelby}, {Enomoto}, {Errico}, {Essick}, {Estell{\'e}s}, {Estevez}, {Etienne}, {Etzel}, {Evans}, {Evans}, {Ewing}, {Fafone}, {Fair}, {Fairhurst}, {Farah}, {Farinon}, {Farr}, {Farr}, {Farrow}, {Fauchon-Jones}, {Favaro}, {Favata}, {Fays}, {Fazio}, {Feicht}, {Fejer}, {Fenyvesi}, {Ferguson}, {Fernandez-Galiana}, {Ferrante}, {Ferreira}, {Fidecaro}, {Figura}, {Fiori}, {Fishbach}, {Fisher}, {Fittipaldi}, {Fiumara}, {Flaminio}, {Floden}, {Fong}, {Font},
  {Fornal}, {Forsyth}, {Franke}, {Frasca}, {Frasconi}, {Frederick}, {Freed}, {Frei}, {Freise}, {Frey}, {Fritschel}, {Frolov}, {Fronz{\'e}}, {Fujii}, {Fujikawa}, {Fukunaga}, {Fukushima}, {Fulda}, {Fyffe}, {Gabbard}, {Gadre}, {Gair}, {Gais}, {Galaudage}, {Gamba}, {Ganapathy}, {Ganguly}, {Gao}, {Gaonkar}, {Garaventa}, {Garc{\'\i}a}, {Garc{\'\i}a-N{\'u}{\~n}ez}, {Garc{\'\i}a-Quir{\'o}s}, {Garufi}, {Gateley}, {Gaudio}, {Gayathri}, {Ge}, {Gemme}, {Gennai}, {George}, {George}, {Gerberding}, {Gergely}, {Gewecke}, {Ghonge}, {Ghosh}, {Ghosh}, {Ghosh}, {Ghosh}, {Giacomazzo}, {Giacoppo}, {Giaime}, {Giardina}, {Gibson}, {Gier}, {Giesler}, {Giri}, {Gissi}, {Glanzer}, {Gleckl}, {Godwin}, {Golomb}, {Goetz}, {Goetz}, {Gohlke}, {Goncharov}, {Gonz{\'a}lez}, {Gopakumar}, {Gosselin}, {Gouaty}, {Gould}, {Grace}, {Grado}, {Granata}, {Granata}, {Grant}, {Gras}, {Grassia}, {Gray}, {Gray}, {Greco}, {Green}, {Green}, {Gretarsson}, {Gretarsson}, {Griffith}, {Griffiths}, {Griggs}, {Grignani}, {Grimaldi}, {Grimm}, {Grote}, {Grunewald},
  {Gruning}, {Guerra}, {Guidi}, {Guimaraes}, {Guix{\'e}}, {Gulati}, {Guo}, {Guo}, {Gupta}, {Gupta}, {Gupta}, {Gustafson}, {Gustafson}, {Guzman}, {Ha}, {Haegel}, {Hagiwara}, {Haino}, {Halim}, {Hall}, {Hamilton}, {Hammond}, {Han}, {Haney}, {Hanks}, {Hanna}, {Hannam}, {Hannuksela}, {Hansen}, {Hansen}, {Hanson}, {Harder}, {Hardwick}, {Haris}, {Harms}, {Harry}, {Harry}, {Hartwig}, {Hasegawa}, {Haskell}, {Hasskew}, {Haster}, {Hattori}, {Haughian}, {Hayakawa}, {Hayama}, {Hayes}, {Healy}, {Heidmann}, {Heidt}, {Heintze}, {Heinze}, {Heinzel}, {Heitmann}, {Hellman}, {Hello}, {Helmling-Cornell}, {Hemming}, {Hendry}, {Heng}, {Hennes}, {Hennig}, {Hennig}, {Hernandez}, {Vivanco}, {Heurs}, {Hild}, {Hill}, {Himemoto}, {Hines}, {Hiranuma}, {Hirata}, {Hirose}, {Hochheim}, {Hofman}, {Hohmann}, {Holcomb}, {Holland}, {Hollows}, {Holmes}, {Holt}, {Holz}, {Hong}, {Hopkins}, {Hough}, {Hourihane}, {Howell}, {Hoy}, {Hoyland}, {Hreibi}, {Hsieh}, {Hsu}, {Huang}, {Huang}, {Huang}, {Huang}, {Huang}, {Huang}, {H{\"u}bner}, {Huddart},
  {Hughey}, {Hui}, {Hui}, {Husa}, {Huttner}, {Huxford}, {Huynh-Dinh}, {Ide}, {Idzkowski}, {Iess}, {Ikenoue}, {Imam}, {Inayoshi}, {Ingram}, {Inoue}, {Ioka}, {Isi}, {Isleif}, {Ito}, {Itoh}, {Iyer}, {Izumi}, {Jaberianhamedan}, {Jacqmin}, {Jadhav}, {Jadhav}, {James}, {Jan}, {Jani}, {Janquart}, {Janssens}, {Janthalur}, {Jaranowski}, {Jariwala}, {Jaume}, {Jenkins}, {Jenner}, {Jeon}, {Jeunon}, {Jia}, {Jin}, {Johns}, {Jones}, {Jones}, {Jones}, {Jones}, {Jones}, {Jonker}, {Ju}, {Jung}, {Jung}, {Junker}, {Juste}, {Kaihotsu}, {Kajita}, {Kakizaki}, {Kalaghatgi}, {Kalogera}, {Kamai}, {Kamiizumi}, {Kanda}, {Kandhasamy}, {Kang}, {Kanner}, {Kao}, {Kapadia}, {Kapasi}, {Karat}, {Karathanasis}, {Karki}, {Kashyap}, {Kasprzack}, {Kastaun}, {Katsanevas}, {Katsavounidis}, {Katzman}, {Kaur}, {Kawabe}, {Kawaguchi}, {Kawai}, {Kawasaki}, {K{\'e}f{\'e}lian}, {Keitel}, {Key}, {Khadka}, {Khalili}, {Khan}, {Khazanov}, {Khetan}, {Khursheed}, {Kijbunchoo}, {Kim}, {Kim}, {Kim}, {Kim}, {Kim}, {Kim}, {Kimball}, {Kimura}, {Kinley-Hanlon},
  {Kirchhoff}, {Kissel}, {Kita}, {Kitazawa}, {Kleybolte}, {Klimenko}, {Knee}, {Knowles}, {Knyazev}, {Koch}, {Koekoek}, {Kojima}, {Kokeyama}, {Koley}, {Kolitsidou}, {Kolstein}, {Komori}, {Kondrashov}, {Kong}, {Kontos}, {Koper}, {Korobko}, {Kotake}, {Kovalam}, {Kozak}, {Kozakai}, {Kozu}, {Kringel}, {Krishnendu}, {Kr{\'o}lak}, {Kuehn}, {Kuei}, {Kuijer}, {Kulkarni}, {Kumar}, {Kumar}, {Kumar}, {Kumar}, {Kume}, {Kuns}, {Kuo}, {Kuo}, {Kuromiya}, {Kuroyanagi}, {Kusayanagi}, {Kuwahara}, {Kwak}, {Lagabbe}, {Laghi}, {Lalande}, {Lam}, {Lamberts}, {Landry}, {Landry}, {Lane}, {Lang}, {Lange}, {Lantz}, {La Rosa}, {Lartaux-Vollard}, {Lasky}, {Laxen}, {Lazzarini}, {Lazzaro}, {Leaci}, {Leavey}, {Lecoeuche}, {Lee}, {Lee}, {Lee}, {Lee}, {Lee}, {Lee}, {Lehmann}, {Lema{\^\i}tre}, {Leonardi}, {Leroy}, {Letendre}, {Levesque}, {Levin}, {Leviton}, {Leyde}, {Li}, {Li}, {Li}, {Li}, {Li}, {Li}, {Lin}, {Lin}, {Lin}, {Lin}, {Lin}, {Linde}, {Linker}, {Linley}, {Littenberg}, {Liu}, {Liu}, {Liu}, {Liu}, {Llamas}, {Llorens-Monteagudo}, {Lo},
  {Lockwood}, {Loh}, {London}, {Longo}, {Lopez}, {Portilla}, {Lorenzini}, {Loriette}, {Lormand}, {Losurdo}, {Lott}, {Lough}, {Lousto}, {Lovelace}, {Lucaccioni}, {L{\"u}ck}, {Lumaca}, {Lundgren}, {Luo}, {Lynam}, {Macas}, {Macinnis}, {MacLeod}, {MacMillan}, {Macquet}, {Hernandez}, {Magazz{\`u}}, {Magee}, {Maggiore}, {Magnozzi}, {Mahesh}, {Majorana}, {Makarem}, {Maksimovic}, {Maliakal}, {Malik}, {Man}, {Mandic}, {Mangano}, {Mango}, {Mansell}, {Manske}, {Mantovani}, {Mapelli}, {Marchesoni}, {Marchio}, {Marion}, {Mark}, {M{\'a}rka}, {M{\'a}rka}, {Markakis}, {Markosyan}, {Markowitz}, {Maros}, {Marquina}, {Marsat}, {Martelli}, {Martin}, {Martin}, {Martinez}, {Martinez}, {Martinez}, {Martinovic}, {Martynov}, {Marx}, {Masalehdan}, {Mason}, {Massera}, {Masserot}, {Massinger}, {Masso-Reid}, {Mastrogiovanni}, {Matas}, {Mateu-Lucena}, {Matichard}, {Matiushechkina}, {Mavalvala}, {McCann}, {McCarthy}, {McClelland}, {McClincy}, {McCormick}, {McCuller}, {McGhee}, {McGuire}, {McIsaac}, {McIver}, {McRae}, {McWilliams},
  {Meacher}, {Mehmet}, {Mehta}, {Meijer}, {Melatos}, {Melchor}, {Mendell}, {Menendez-Vazquez}, {Menoni}, {Mercer}, {Mereni}, {Merfeld}, {Merilh}, {Merritt}, {Merzougui}, {Meshkov}, {Messenger}, {Messick}, {Meyers}, {Meylahn}, {Mhaske}, {Miani}, {Miao}, {Michaloliakos}, {Michel}, {Michimura}, {Middleton}, {Milano}, {Miller}, {Miller}, {Miller}, {Miller}, {Millhouse}, {Mills}, {Milotti}, {Minazzoli}, {Minenkov}, {Mio}, {Mir}, {Miravet-Ten{\'e}s}, {Mishra}, {Mishra}, {Mistry}, {Mitra}, {Mitrofanov}, {Mitselmakher}, {Mittleman}, {Miyakawa}, {Miyamoto}, {Miyazaki}, {Miyo}, {Miyoki}, {Mo}, {Modafferi}, {Moguel}, {Mogushi}, {Mohapatra}, {Mohite}, {Molina}, {Molina-Ruiz}, {Mondin}, {Montani}, {Moore}, {Moraru}, {Morawski}, {More}, {Moreno}, {Moreno}, {Mori}, {Morisaki}, {Moriwaki}, {Morr{\'a}s}, {Mours}, {Mow-Lowry}, {Mozzon}, {Muciaccia}, {Mukherjee}, {Mukherjee}, {Mukherjee}, {Mukherjee}, {Mukherjee}, {Mukund}, {Mullavey}, {Munch}, {Mu{\~n}iz}, {Murray}, {Musenich}, {Muusse}, {Nadji}, {Nagano}, {Nagano}, {Nagar},
  {Nakamura}, {Nakano}, {Nakano}, {Nakashima}, {Nakayama}, {Napolano}, {Nardecchia}, {Narikawa}, {Naticchioni}, {Nayak}, {Nayak}, {Negishi}, {Neil}, {Neilson}, {Nelemans}, {Nelson}, {Nery}, {Neubauer}, {Neunzert}, {Ng}, {Ng}, {Nguyen}, {Nguyen}, {Nguyen}, {Quynh}, {Ni}, {Nichols}, {Nishizawa}, {Nissanke}, {Nitoglia}, {Nocera}, {Norman}, {North}, {Nozaki}, {Siles}, {Nuttall}, {Oberling}, {O'Brien}, {Obuchi}, {O'Dell}, {Oelker}, {Ogaki}, {Oganesyan}, {Oh}, {Oh}, {Oh}, {Ohashi}, {Ohishi}, {Ohkawa}, {Ohme}, {Ohta}, {Okada}, {Okutani}, {Okutomi}, {Olivetto}, {Oohara}, {Ooi}, {Oram}, {O'Reilly}, {Ormiston}, {Ormsby}, {Ortega}, {O'Shaughnessy}, {O'Shea}, {Oshino}, {Ossokine}, {Osthelder}, {Otabe}, {Ottaway}, {Overmier}, {Pace}, {Pagano}, {Page}, {Pagliaroli}, {Pai}, {Pai}, {Palamos}, {Palashov}, {Palomba}, {Pan}, {Pan}, {Panda}, {Pang}, {Pang}, {Pankow}, {Pannarale}, {Pant}, {Panther}, {Paoletti}, {Paoli}, {Paolone}, {Parisi}, {Park}, {Park}, {Parker}, {Pascucci}, {Pasqualetti}, {Passaquieti}, {Passuello}, {Patel},
  {Pathak}, {Patricelli}, {Patron}, {Paul}, {Payne}, {Pedraza}, {Pegoraro}, {Pele}, {Arellano}, {Penn}, {Perego}, {Pereira}, {Pereira}, {Perez}, {P{\'e}rigois}, {Perkins}, {Perreca}, {Perri{\`e}s}, {Petermann}, {Petterson}, {Pfeiffer}, {Pham}, {Phukon}, {Piccinni}, {Pichot}, {Piendibene}, {Piergiovanni}, {Pierini}, {Pierro}, {Pillant}, {Pillas}, {Pilo}, {Pinard}, {Pinto}, {Pinto}, {Piotrzkowski}, {Piotrzkowski}, {Pirello}, {Pitkin}, {Placidi}, {Planas}, {Plastino}, {Pluchar}, {Poggiani}, {Polini}, {Pong}, {Ponrathnam}, {Popolizio}, {Porter}, {Poulton}, {Powell}, {Pracchia}, {Pradier}, {Prajapati}, {Prasai}, {Prasanna}, {Pratten}, {Principe}, {Prodi}, {Prokhorov}, {Prosposito}, {Prudenzi}, {Puecher}, {Punturo}, {Puosi}, {Puppo}, {P{\"u}rrer}, {Qi}, {Quetschke}, {Quitzow-James}, {Raab}, {Raaijmakers}, {Radkins}, {Radulesco}, {Raffai}, {Rail}, {Raja}, {Rajan}, {Ramirez}, {Ramirez}, {Ramos-Buades}, {Rana}, {Rapagnani}, {Rapol}, {Ray}, {Raymond}, {Raza}, {Razzano}, {Read}, {Rees}, {Regimbau}, {Rei}, {Reid},
  {Reid}, {Reitze}, {Relton}, {Renzini}, {Rettegno}, {Reza}, {Rezac}, {Ricci}, {Richards}, {Richardson}, {Richardson}, {Riemenschneider}, {Riles}, {Rinaldi}, {Rink}, {Rizzo}, {Robertson}, {Robie}, {Robinet}, {Rocchi}, {Rodriguez}, {Rolland}, {Rollins}, {Romanelli}, {Romano}, {Romel}, {Romero-Rodr{\'\i}guez}, {Romero-Shaw}, {Romie}, {Ronchini}, {Rosa}, {Rose}, {Rosi{\'n}ska}, {Ross}, {Rowan}, {Rowlinson}, {Roy}, {Roy}, {Roy}, {Rozza}, {Ruggi}, {Ryan}, {Sachdev}, {Sadecki}, {Sadiq}, {Sago}, {Saito}, {Saito}, {Sakai}, {Sakai}, {Sakellariadou}, {Sakuno}, {Salafia}, {Salconi}, {Saleem}, {Salemi}, {Samajdar}, {Sanchez}, {Sanchez}, {Sanchez}, {Sanchis-Gual}, {Sanders}, {Sanuy}, {Saravanan}, {Sarin}, {Sassolas}, {Satari}, {Sathyaprakash}, {Sato}, {Sato}, {Sauter}, {Savage}, {Sawada}, {Sawant}, {Sawant}, {Sayah}, {Schaetzl}, {Scheel}, {Scheuer}, {Schiworski}, {Schmidt}, {Schmidt}, {Schnabel}, {Schneewind}, {Schofield}, {Sch{\"o}nbeck}, {Schulte}, {Schutz}, {Schwartz}, {Scott}, {Scott}, {Seglar-Arroyo}, {Sekiguchi},
  {Sekiguchi}, {Sellers}, {Sengupta}, {Sentenac}, {Seo}, {Sequino}, {Sergeev}, {Setyawati}, {Shaffer}, {Shahriar}, {Shams}, {Shao}, {Sharma}, {Sharma}, {Shawhan}, {Shcheblanov}, {Shibagaki}, {Shikauchi}, {Shimizu}, {Shimoda}, {Shimode}, {Shinkai}, {Shishido}, {Shoda}, {Shoemaker}, {Shoemaker}, {Shyamsundar}, {Sieniawska}, {Sigg}, {Singer}, {Singh}, {Singh}, {Singha}, {Sintes}, {Sipala}, {Skliris}, {Slagmolen}, {Slaven-Blair}, {Smetana}, {Smith}, {Smith}, {Soldateschi}, {Somala}, {Somiya}, {Son}, {Soni}, {Soni}, {Sordini}, {Sorrentino}, {Sorrentino}, {Sotani}, {Soulard}, {Souradeep}, {Sowell}, {Spagnuolo}, {Spencer}, {Spera}, {Srinivasan}, {Srivastava}, {Srivastava}, {Staats}, {Stachie}, {Steer}, {Steinhoff}, {Steinlechner}, {Steinlechner}, {Stevenson}, {Stops}, {Stover}, {Strain}, {Strang}, {Stratta}, {Strunk}, {Sturani}, {Stuver}, {Sudhagar}, {Sudhir}, {Sugimoto}, {Suh}, {Sullivan}, {Summerscales}, {Sun}, {Sun}, {Sunil}, {Sur}, {Suresh}, {Sutton}, {Suzuki}, {Suzuki}, {Swinkels}, {Szczepa{\'n}czyk},
  {Szewczyk}, {Tacca}, {Tagoshi}, {Tait}, {Takahashi}, {Takahashi}, {Takamori}, {Takano}, {Takeda}, {Takeda}, {Talbot}, {Talbot}, {Tanaka}, {Tanaka}, {Tanaka}, {Tanaka}, {Tanaka}, {Tanasijczuk}, {Tanioka}, {Tanner}, {Tao}, {Tao}, {Mart{\'\i}n}, {Taranto}, {Tasson}, {Telada}, {Tenorio}, {Terhune}, {Terkowski}, {Thirugnanasambandam}, {Thomas}, {Thomas}, {Thomas}, {Thompson}, {Thondapu}, {Thorne}, {Thrane}, {Tiwari}, {Tiwari}, {Tiwari}, {Toivonen}, {Toland}, {Tolley}, {Tomaru}, {Tomigami}, {Tomura}, {Tonelli}, {Torres-Forn{\'e}}, {Torrie}, {E Melo}, {T{\"o}yr{\"a}}, {Trapananti}, {Travasso}, {Traylor}, {Trevor}, {Tringali}, {Tripathee}, {Troiano}, {Trovato}, {Trozzo}, {Trudeau}, {Tsai}, {Tsai}, {Tsang}, {Tsang}, {Tsao}, {Tse}, {Tso}, {Tsubono}, {Tsuchida}, {Tsukada}, {Tsuna}, {Tsutsui}, {Tsuzuki}, {Turbang}, {Turconi}, {Tuyenbayev}, {Ubhi}, {Uchikata}, {Uchiyama}, {Udall}, {Ueda}, {Uehara}, {Ueno}, {Ueshima}, {Unnikrishnan}, {Uraguchi}, {Urban}, {Ushiba}, {Utina}, {Vahlbruch}, {Vajente}, {Vajpeyi}, {Valdes},
  {Valentini}, {Valsan}, {van Bakel}, {van Beuzekom}, {van den Brand}, {van den Broeck}, {Vander-Hyde}, {van der Schaaf}, {van Heijningen}, {Vanosky}, {van Putten}, {van Remortel}, {Vardaro}, {Vargas}, {Varma}, {Vas{\'u}th}, {Vecchio}, {Vedovato}, {Veitch}, {Veitch}, {Venneberg}, {Venugopalan}, {Verkindt}, {Verma}, {Verma}, {Veske}, {Vetrano}, {Vicer{\'e}}, {Vidyant}, {Viets}, {Vijaykumar}, {Villa-Ortega}, {Vinet}, {Virtuoso}, {Vitale}, {Vo}, {Vocca}, {von Reis}, {von Wrangel}, {Vorvick}, {Vyatchanin}, {Wade}, {Wade}, {Wagner}, {Walet}, {Walker}, {Wallace}, {Wallace}, {Walsh}, {Wang}, {Wang}, {Wang}, {Ward}, {Warner}, {Was}, {Washimi}, {Washington}, {Watchi}, {Weaver}, {Webster}, {Weinert}, {Weinstein}, {Weiss}, {Weller}, {Wellmann}, {Wen}, {We{\ss}els}, {Wette}, {Whelan}, {White}, {Whiting}, {Whittle}, {Wilken}, {Williams}, {Williams}, {Williamson}, {Willis}, {Willke}, {Wilson}, {Winkler}, {Wipf}, {Wlodarczyk}, {Woan}, {Woehler}, {Wofford}, {Wong}, {Wu}, {Wu}, {Wu}, {Wu}, {Wysocki}, {Xiao}, {Xu}, {Yamada},
  {Yamamoto}, {Yamamoto}, {Yamamoto}, {Yamamoto}, {Yamashita}, {Yamazaki}, {Yang}, {Yang}, {Yang}, {Yang}, {Yang}, {Yap}, {Yeeles}, {Yelikar}, {Ying}, {Yokogawa}, {Yokoyama}, {Yokozawa}, {Yoo}, {Yoshioka}, {Yu}, {Yu}, {Yuzurihara}, {Zadro{\.z}ny}, {Zanolin}, {Zeidler}, {Zelenova}, {Zendri}, {Zevin}, {Zhan}, {Zhang}, {Zhang}, {Zhang}, {Zhang}, {Zhang}, {Zhao}, {Zhao}, {Zhao}, {Zhao}, {Zheng}, {Zhou}, {Zhou}, {Zhu}, {Zhu}, {Zimmerman}, {Zlochower}, {Zucker}, {Zweizig}, {LIGO Scientific Collaboration}, {VIRGO Collaboration}, \& {KAGRA Collaboration}}]{2023PhRvX..13a1048A}
---. 2023{\natexlab{b}}, Physical Review X, 13, 011048, \dodoi{10.1103/PhysRevX.13.011048}

\bibitem[{{Altamirano} {et~al.}(2011){Altamirano}, {Belloni}, {Linares}, {van der Klis}, {Wijnands}, {Curran}, {Kalamkar}, {Stiele}, {Motta}, {Mu{\~n}oz-Darias}, {Casella}, \& {Krimm}}]{2011ApJ...742L..17A}
{Altamirano}, D., {Belloni}, T., {Linares}, M., {et~al.} 2011, \apjl, 742, L17, \dodoi{10.1088/2041-8205/742/2/L17}

\bibitem[{{Arnaud}(1996)}]{1996ASPC..101...17A}
{Arnaud}, K.~A. 1996, in Astronomical Society of the Pacific Conference Series, Vol. 101, Astronomical Data Analysis Software and Systems V, ed. G.~H. {Jacoby} \& J.~{Barnes}, 17

\bibitem[{{Astropy Collaboration} {et~al.}(2013){Astropy Collaboration}, {Robitaille}, {Tollerud}, {Greenfield}, {Droettboom}, {Bray}, {Aldcroft}, {Davis}, {Ginsburg}, {Price-Whelan}, {Kerzendorf}, {Conley}, {Crighton}, {Barbary}, {Muna}, {Ferguson}, {Grollier}, {Parikh}, {Nair}, {Unther}, {Deil}, {Woillez}, {Conseil}, {Kramer}, {Turner}, {Singer}, {Fox}, {Weaver}, {Zabalza}, {Edwards}, {Azalee Bostroem}, {Burke}, {Casey}, {Crawford}, {Dencheva}, {Ely}, {Jenness}, {Labrie}, {Lim}, {Pierfederici}, {Pontzen}, {Ptak}, {Refsdal}, {Servillat}, \& {Streicher}}]{astropy:2013}
{Astropy Collaboration}, {Robitaille}, T.~P., {Tollerud}, E.~J., {et~al.} 2013, \aap, 558, A33, \dodoi{10.1051/0004-6361/201322068}

\bibitem[{{Astropy Collaboration} {et~al.}(2018){Astropy Collaboration}, {Price-Whelan}, {Sip{\H{o}}cz}, {G{\"u}nther}, {Lim}, {Crawford}, {Conseil}, {Shupe}, {Craig}, {Dencheva}, {Ginsburg}, {VanderPlas}, {Bradley}, {P{\'e}rez-Su{\'a}rez}, {de Val-Borro}, {Aldcroft}, {Cruz}, {Robitaille}, {Tollerud}, {Ardelean}, {Babej}, {Bach}, {Bachetti}, {Bakanov}, {Bamford}, {Barentsen}, {Barmby}, {Baumbach}, {Berry}, {Biscani}, {Boquien}, {Bostroem}, {Bouma}, {Brammer}, {Bray}, {Breytenbach}, {Buddelmeijer}, {Burke}, {Calderone}, {Cano Rodr{\'\i}guez}, {Cara}, {Cardoso}, {Cheedella}, {Copin}, {Corrales}, {Crichton}, {D'Avella}, {Deil}, {Depagne}, {Dietrich}, {Donath}, {Droettboom}, {Earl}, {Erben}, {Fabbro}, {Ferreira}, {Finethy}, {Fox}, {Garrison}, {Gibbons}, {Goldstein}, {Gommers}, {Greco}, {Greenfield}, {Groener}, {Grollier}, {Hagen}, {Hirst}, {Homeier}, {Horton}, {Hosseinzadeh}, {Hu}, {Hunkeler}, {Ivezi{\'c}}, {Jain}, {Jenness}, {Kanarek}, {Kendrew}, {Kern}, {Kerzendorf}, {Khvalko}, {King}, {Kirkby}, {Kulkarni},
  {Kumar}, {Lee}, {Lenz}, {Littlefair}, {Ma}, {Macleod}, {Mastropietro}, {McCully}, {Montagnac}, {Morris}, {Mueller}, {Mumford}, {Muna}, {Murphy}, {Nelson}, {Nguyen}, {Ninan}, {N{\"o}the}, {Ogaz}, {Oh}, {Parejko}, {Parley}, {Pascual}, {Patil}, {Patil}, {Plunkett}, {Prochaska}, {Rastogi}, {Reddy Janga}, {Sabater}, {Sakurikar}, {Seifert}, {Sherbert}, {Sherwood-Taylor}, {Shih}, {Sick}, {Silbiger}, {Singanamalla}, {Singer}, {Sladen}, {Sooley}, {Sornarajah}, {Streicher}, {Teuben}, {Thomas}, {Tremblay}, {Turner}, {Terr{\'o}n}, {van Kerkwijk}, {de la Vega}, {Watkins}, {Weaver}, {Whitmore}, {Woillez}, {Zabalza}, \& {Astropy Contributors}}]{astropy:2018}
{Astropy Collaboration}, {Price-Whelan}, A.~M., {Sip{\H{o}}cz}, B.~M., {et~al.} 2018, \aj, 156, 123, \dodoi{10.3847/1538-3881/aabc4f}

\bibitem[{{Bambi} {et~al.}(2021){Bambi}, {Brenneman}, {Dauser}, {Garc{\'\i}a}, {Grinberg}, {Ingram}, {Jiang}, {Liu}, {Lohfink}, {Marinucci}, {Mastroserio}, {Middei}, {Nampalliwar}, {Nied{\'z}wiecki}, {Steiner}, {Tripathi}, \& {Zdziarski}}]{2021SSRv..217...65B}
{Bambi}, C., {Brenneman}, L.~W., {Dauser}, T., {et~al.} 2021, \ssr, 217, 65, \dodoi{10.1007/s11214-021-00841-8}

\bibitem[{{Belloni} {et~al.}(1997){Belloni}, {M{\'e}ndez}, {King}, {van der Klis}, \& {van Paradijs}}]{1997ApJ...479L.145B}
{Belloni}, T., {M{\'e}ndez}, M., {King}, A.~R., {van der Klis}, M., \& {van Paradijs}, J. 1997, \apjl, 479, L145, \dodoi{10.1086/310595}

\bibitem[{{Bhuvana} {et~al.}(2022){Bhuvana}, {Radhika}, \& {Nandi}}]{2022AdSpR..69..483B}
{Bhuvana}, G.~R., {Radhika}, D., \& {Nandi}, A. 2022, Advances in Space Research, 69, 483, \dodoi{10.1016/j.asr.2021.09.036}

\bibitem[{{Brenneman} \& {Reynolds}(2006)}]{2006ApJ...652.1028B}
{Brenneman}, L.~W., \& {Reynolds}, C.~S. 2006, ApJ, 652, 1028, \dodoi{10.1086/508146}

\bibitem[{{Casares} {et~al.}(2009){Casares}, {Orosz}, {Zurita}, {Shahbaz}, {Corral-Santana}, {McClintock}, {Garcia}, {Mart{\'\i}nez-Pais}, {Charles}, {Fender}, \& {Remillard}}]{2009ApJS..181..238C}
{Casares}, J., {Orosz}, J.~A., {Zurita}, C., {et~al.} 2009, \apjs, 181, 238, \dodoi{10.1088/0067-0049/181/1/238}

\bibitem[{{Chauhan} {et~al.}(2019){Chauhan}, {Miller-Jones}, {Anderson}, {Raja}, {Bahramian}, {Hotan}, {Indermuehle}, {Whiting}, {Allison}, {Anderson}, {Bunton}, {Koribalski}, \& {Mahony}}]{2019MNRAS.488L.129C}
{Chauhan}, J., {Miller-Jones}, J.~C.~A., {Anderson}, G.~E., {et~al.} 2019, \mnras, 488, L129, \dodoi{10.1093/mnrasl/slz113}

\bibitem[{{Coughenour} {et~al.}(2023){Coughenour}, {Tomsick}, {Mastroserio}, {Steiner}, {Connors}, {Jiang}, {Hare}, {Shaw}, {Ludlam}, {Fabian}, {Garc{\'\i}a}, \& {Coley}}]{2023ApJ...949...70C}
{Coughenour}, B.~M., {Tomsick}, J.~A., {Mastroserio}, G., {et~al.} 2023, \apj, 949, 70, \dodoi{10.3847/1538-4357/acc65c}

\bibitem[{{Dauser} {et~al.}(2014){Dauser}, {Garcia}, {Parker}, {Fabian}, \& {Wilms}}]{2014MNRAS.444L.100D}
{Dauser}, T., {Garcia}, J., {Parker}, M.~L., {Fabian}, A.~C., \& {Wilms}, J. 2014, MNRAS, 444, L100, \dodoi{10.1093/mnrasl/slu125}

\bibitem[{{della Valle} {et~al.}(1994){della Valle}, {Mirabel}, \& {Rodriguez}}]{1994A&A...290..803D}
{della Valle}, M., {Mirabel}, I.~F., \& {Rodriguez}, L.~F. 1994, \aap, 290, 803

\bibitem[{{Done} {et~al.}(2007){Done}, {Gierli{\'n}ski}, \& {Kubota}}]{2007A&ARv..15....1D}
{Done}, C., {Gierli{\'n}ski}, M., \& {Kubota}, A. 2007, \aapr, 15, 1, \dodoi{10.1007/s00159-007-0006-1}

\bibitem[{{Dong} {et~al.}(2022){Dong}, {Liu}, {Tuo}, {Steiner}, {Ge}, {Garc{\'\i}a}, \& {Cao}}]{2022MNRAS.514.1422D}
{Dong}, Y., {Liu}, Z., {Tuo}, Y., {et~al.} 2022, \mnras, 514, 1422, \dodoi{10.1093/mnras/stac1466}

\bibitem[{{Draghis} {et~al.}(2023{\natexlab{a}}){Draghis}, {Miller}, {Brumback}, {Fabian}, {Tomsick}, \& {Zoghbi}}]{2023ApJ...954...62D}
{Draghis}, P.~A., {Miller}, J.~M., {Brumback}, M.~C., {et~al.} 2023{\natexlab{a}}, \apj, 954, 62, \dodoi{10.3847/1538-4357/ace7b3}

\bibitem[{{Draghis} {et~al.}(2020){Draghis}, {Miller}, {Cackett}, {Kammoun}, {Reynolds}, {Tomsick}, \& {Zoghbi}}]{2020ApJ...900...78D}
{Draghis}, P.~A., {Miller}, J.~M., {Cackett}, E.~M., {et~al.} 2020, \apj, 900, 78, \dodoi{10.3847/1538-4357/aba2ec}

\bibitem[{{Draghis} {et~al.}(2021){Draghis}, {Miller}, {Zoghbi}, {Kammoun}, {Reynolds}, \& {Tomsick}}]{2021ApJ...920...88D}
{Draghis}, P.~A., {Miller}, J.~M., {Zoghbi}, A., {et~al.} 2021, \apj, 920, 88, \dodoi{10.3847/1538-4357/ac1270}

\bibitem[{{Draghis} {et~al.}(2023{\natexlab{b}}){Draghis}, {Miller}, {Zoghbi}, {Reynolds}, {Costantini}, {Gallo}, \& {Tomsick}}]{Draghis23}
---. 2023{\natexlab{b}}, \apj, 946, 19, \dodoi{10.3847/1538-4357/acafe7}

\bibitem[{{Draghis} {et~al.}(2023{\natexlab{c}}){Draghis}, {Balakrishnan}, {Miller}, {Cackett}, {Fabian}, {Miller-Jones}, {Ng}, {Raymond}, {Reynolds}, \& {Zoghbi}}]{2023ApJ...947...39D}
{Draghis}, P.~A., {Balakrishnan}, M., {Miller}, J.~M., {et~al.} 2023{\natexlab{c}}, \apj, 947, 39, \dodoi{10.3847/1538-4357/acc1c8}

\bibitem[{{El-Batal} {et~al.}(2016){El-Batal}, {Miller}, {Reynolds}, {Boggs}, {Chistensen}, {Craig}, {Fuerst}, {Hailey}, {Harrison}, {Stern}, {Tomsick}, {Walton}, \& {Zhang}}]{2016ApJ...826L..12E}
{El-Batal}, A.~M., {Miller}, J.~M., {Reynolds}, M.~T., {et~al.} 2016, \apjl, 826, L12, \dodoi{10.3847/2041-8205/826/1/L12}

\bibitem[{{Fabian} {et~al.}(2000){Fabian}, {Iwasawa}, {Reynolds}, \& {Young}}]{2000PASP..112.1145F}
{Fabian}, A.~C., {Iwasawa}, K., {Reynolds}, C.~S., \& {Young}, A.~J. 2000, \pasp, 112, 1145, \dodoi{10.1086/316610}

\bibitem[{{Fabian} {et~al.}(2014){Fabian}, {Parker}, {Wilkins}, {Miller}, {Kara}, {Reynolds}, \& {Dauser}}]{2014MNRAS.439.2307F}
{Fabian}, A.~C., {Parker}, M.~L., {Wilkins}, D.~R., {et~al.} 2014, \mnras, 439, 2307, \dodoi{10.1093/mnras/stu045}

\bibitem[{{Feng} {et~al.}(2022{\natexlab{a}}){Feng}, {Zhao}, {Li}, {Gou}, {Jia}, {Liao}, \& {Wang}}]{2022MNRAS.516.2074F}
{Feng}, Y., {Zhao}, X., {Li}, Y., {et~al.} 2022{\natexlab{a}}, \mnras, 516, 2074, \dodoi{10.1093/mnras/stac1868}

\bibitem[{{Feng} {et~al.}(2022{\natexlab{b}}){Feng}, {Zhao}, {Gou}, {Li}, {Steiner}, {Garc{\'\i}a}, {Wang}, {Jia}, {Liao}, \& {Li}}]{2022SCPMA..6519512F}
{Feng}, Y., {Zhao}, X., {Gou}, L., {et~al.} 2022{\natexlab{b}}, Science China Physics, Mechanics, and Astronomy, 65, 219512, \dodoi{10.1007/s11433-021-1790-7}

\bibitem[{Foreman-Mackey(2016)}]{corner}
Foreman-Mackey, D. 2016, The Journal of Open Source Software, 1, 24, \dodoi{10.21105/joss.00024}

\bibitem[{{Foreman-Mackey} {et~al.}(2013){Foreman-Mackey}, {Hogg}, {Lang}, \& {Goodman}}]{2013PASP..125..306F}
{Foreman-Mackey}, D., {Hogg}, D.~W., {Lang}, D., \& {Goodman}, J. 2013, \pasp, 125, 306, \dodoi{10.1086/670067}

\bibitem[{{Gallegos-Garcia} {et~al.}(2022){Gallegos-Garcia}, {Fishbach}, {Kalogera}, {L Berry}, \& {Doctor}}]{2022ApJ...938L..19G}
{Gallegos-Garcia}, M., {Fishbach}, M., {Kalogera}, V., {L Berry}, C.~P., \& {Doctor}, Z. 2022, \apjl, 938, L19, \dodoi{10.3847/2041-8213/ac96ef}

\bibitem[{{Gallo} {et~al.}(2015){Gallo}, {Wilkins}, {Bonson}, {Chiang}, {Grupe}, {Parker}, {Zoghbi}, {Fabian}, {Komossa}, \& {Longinotti}}]{2015MNRAS.446..633G}
{Gallo}, L.~C., {Wilkins}, D.~R., {Bonson}, K., {et~al.} 2015, \mnras, 446, 633, \dodoi{10.1093/mnras/stu2108}

\bibitem[{{Garc{\'\i}a} {et~al.}(2014){Garc{\'\i}a}, {Dauser}, {Lohfink}, {Kallman}, {Steiner}, {McClintock}, {Brenneman}, {Wilms}, {Eikmann}, {Reynolds}, \& {Tombesi}}]{2014ApJ...782...76G}
{Garc{\'\i}a}, J., {Dauser}, T., {Lohfink}, A., {et~al.} 2014, ApJ, 782, 76, \dodoi{10.1088/0004-637X/782/2/76}

\bibitem[{{Gendreau} {et~al.}(2016){Gendreau}, {Arzoumanian}, {Adkins}, {Albert}, {Anders}, {Aylward}, {Baker}, {Balsamo}, {Bamford}, {Benegalrao}, {Berry}, {Bhalwani}, {Black}, {Blaurock}, {Bronke}, {Brown}, {Budinoff}, {Cantwell}, {Cazeau}, {Chen}, {Clement}, {Colangelo}, {Coleman}, {Coopersmith}, {Dehaven}, {Doty}, {Egan}, {Enoto}, {Fan}, {Ferro}, {Foster}, {Galassi}, {Gallo}, {Green}, {Grosh}, {Ha}, {Hasouneh}, {Heefner}, {Hestnes}, {Hoge}, {Jacobs}, {J{\o}rgensen}, {Kaiser}, {Kellogg}, {Kenyon}, {Koenecke}, {Kozon}, {LaMarr}, {Lambertson}, {Larson}, {Lentine}, {Lewis}, {Lilly}, {Liu}, {Malonis}, {Manthripragada}, {Markwardt}, {Matonak}, {Mcginnis}, {Miller}, {Mitchell}, {Mitchell}, {Mohammed}, {Monroe}, {Montt de Garcia}, {Mul{\'e}}, {Nagao}, {Ngo}, {Norris}, {Norwood}, {Novotka}, {Okajima}, {Olsen}, {Onyeachu}, {Orosco}, {Peterson}, {Pevear}, {Pham}, {Pollard}, {Pope}, {Powers}, {Powers}, {Price}, {Prigozhin}, {Ramirez}, {Reid}, {Remillard}, {Rogstad}, {Rosecrans}, {Rowe}, {Sager}, {Sanders},
  {Savadkin}, {Saylor}, {Schaeffer}, {Schweiss}, {Semper}, {Serlemitsos}, {Shackelford}, {Soong}, {Struebel}, {Vezie}, {Villasenor}, {Winternitz}, {Wofford}, {Wright}, {Yang}, \& {Yu}}]{2016SPIE.9905E..1HG}
{Gendreau}, K.~C., {Arzoumanian}, Z., {Adkins}, P.~W., {et~al.} 2016, in Society of Photo-Optical Instrumentation Engineers (SPIE) Conference Series, Vol. 9905, Space Telescopes and Instrumentation 2016: Ultraviolet to Gamma Ray, ed. J.-W.~A. {den Herder}, T.~{Takahashi}, \& M.~{Bautz}, 99051H, \dodoi{10.1117/12.2231304}

\bibitem[{{Gou} {et~al.}(2009){Gou}, {McClintock}, {Liu}, {Narayan}, {Steiner}, {Remillard}, {Orosz}, {Davis}, {Ebisawa}, \& {Schlegel}}]{2009ApJ...701.1076G}
{Gou}, L., {McClintock}, J.~E., {Liu}, J., {et~al.} 2009, \apj, 701, 1076, \dodoi{10.1088/0004-637X/701/2/1076}

\bibitem[{{Gou} {et~al.}(2014){Gou}, {McClintock}, {Remillard}, {Steiner}, {Reid}, {Orosz}, {Narayan}, {Hanke}, \& {Garc{\'\i}a}}]{2014ApJ...790...29G}
{Gou}, L., {McClintock}, J.~E., {Remillard}, R.~A., {et~al.} 2014, \apj, 790, 29, \dodoi{10.1088/0004-637X/790/1/29}

\bibitem[{Harris {et~al.}(2020)Harris, Millman, van~der Walt, Gommers, Virtanen, Cournapeau, Wieser, Taylor, Berg, Smith, Kern, Picus, Hoyer, van Kerkwijk, Brett, Haldane, del R{\'{i}}o, Wiebe, Peterson, G{\'{e}}rard-Marchant, Sheppard, Reddy, Weckesser, Abbasi, Gohlke, \& Oliphant}]{harris2020array}
Harris, C.~R., Millman, K.~J., van~der Walt, S.~J., {et~al.} 2020, Nature, 585, 357, \dodoi{10.1038/s41586-020-2649-2}

\bibitem[{{Harrison} {et~al.}(2013){Harrison}, {Craig}, {Christensen}, \& et~al.}]{2013ApJ...770..103H}
{Harrison}, F.~A., {Craig}, W.~W., {Christensen}, F.~E., \& et~al. 2013, ApJ, 770, 103, \dodoi{10.1088/0004-637X/770/2/103}

\bibitem[{Hunter(2007)}]{Hunter:2007}
Hunter, J.~D. 2007, Computing in Science \& Engineering, 9, 90, \dodoi{10.1109/MCSE.2007.55}

\bibitem[{{Iyer} {et~al.}(2015){Iyer}, {Nandi}, \& {Mandal}}]{2015ApJ...807..108I}
{Iyer}, N., {Nandi}, A., \& {Mandal}, S. 2015, \apj, 807, 108, \dodoi{10.1088/0004-637X/807/1/108}

\bibitem[{{Jana} {et~al.}(2021{\natexlab{a}}){Jana}, {Naik}, {Chatterjee}, \& {Jaisawal}}]{2021MNRAS.507.4779J}
{Jana}, A., {Naik}, S., {Chatterjee}, D., \& {Jaisawal}, G.~K. 2021{\natexlab{a}}, \mnras, 507, 4779, \dodoi{10.1093/mnras/stab2448}

\bibitem[{{Jana} {et~al.}(2021{\natexlab{b}}){Jana}, {Jaisawal}, {Naik}, {Kumari}, {Chatterjee}, {Chatterjee}, {Bhowmick}, {Chakrabarti}, {Chang}, \& {Debnath}}]{2021RAA....21..125J}
{Jana}, A., {Jaisawal}, G.~K., {Naik}, S., {et~al.} 2021{\natexlab{b}}, Research in Astronomy and Astrophysics, 21, 125, \dodoi{10.1088/1674-4527/21/5/125}

\bibitem[{{Jia} {et~al.}(2022){Jia}, {Zhao}, {Gou}, {Garc{\'\i}a}, {Liao}, {Feng}, {Li}, {Wang}, {Li}, \& {Wu}}]{2022MNRAS.511.3125J}
{Jia}, N., {Zhao}, X., {Gou}, L., {et~al.} 2022, \mnras, 511, 3125, \dodoi{10.1093/mnras/stac121}

\bibitem[{{Jiang} {et~al.}(2022){Jiang}, {Buisson}, {Dauser}, {Fabian}, {F{\"u}rst}, {Gallo}, {Harrison}, {Parker}, {Steiner}, {Tomsick}, {Ubach}, \& {Walton}}]{2022MNRAS.514.1952J}
{Jiang}, J., {Buisson}, D. J.~K., {Dauser}, T., {et~al.} 2022, \mnras, 514, 1952, \dodoi{10.1093/mnras/stac1401}

\bibitem[{{Kaastra} \& {Bleeker}(2016)}]{2016A&A...587A.151K}
{Kaastra}, J.~S., \& {Bleeker}, J.~A.~M. 2016, A\&A, 587, A151, \dodoi{10.1051/0004-6361/201527395}

\bibitem[{{Keck} {et~al.}(2015){Keck}, {Brenneman}, {Ballantyne}, {Bauer}, {Boggs}, {Christensen}, {Craig}, {Dauser}, {Elvis}, {Fabian}, {Fuerst}, {Garc{\'\i}a}, {Grefenstette}, {Hailey}, {Harrison}, {Madejski}, {Marinucci}, {Matt}, {Reynolds}, {Stern}, {Walton}, \& {Zoghbi}}]{2015ApJ...806..149K}
{Keck}, M.~L., {Brenneman}, L.~W., {Ballantyne}, D.~R., {et~al.} 2015, \apj, 806, 149, \dodoi{10.1088/0004-637X/806/2/149}

\bibitem[{{King} {et~al.}(2015){King}, {Miller}, {Raymond}, {Reynolds}, \& {Morningstar}}]{2015ApJ...813L..37K}
{King}, A.~L., {Miller}, J.~M., {Raymond}, J., {Reynolds}, M.~T., \& {Morningstar}, W. 2015, \apjl, 813, L37, \dodoi{10.1088/2041-8205/813/2/L37}

\bibitem[{{King} {et~al.}(2014){King}, {Walton}, {Miller}, {Barret}, {Boggs}, {Christensen}, {Craig}, {Fabian}, {F{\"u}rst}, {Hailey}, {Harrison}, {Krivonos}, {Mori}, {Natalucci}, {Stern}, {Tomsick}, \& {Zhang}}]{2014ApJ...784L...2K}
{King}, A.~L., {Walton}, D.~J., {Miller}, J.~M., {et~al.} 2014, \apjl, 784, L2, \dodoi{10.1088/2041-8205/784/1/L2}

\bibitem[{{Lamer} {et~al.}(2021){Lamer}, {Schwope}, {Predehl}, {Traulsen}, {Wilms}, \& {Freyberg}}]{2021A&A...647A...7L}
{Lamer}, G., {Schwope}, A.~D., {Predehl}, P., {et~al.} 2021, \aap, 647, A7, \dodoi{10.1051/0004-6361/202039757}

\bibitem[{{Lo} \& {Lin}(2011)}]{2011ApJ...728...12L}
{Lo}, K.-W., \& {Lin}, L.-M. 2011, ApJ, 728, 12, \dodoi{10.1088/0004-637X/728/1/12}

\bibitem[{{Luque-Escamilla} {et~al.}(2015){Luque-Escamilla}, {Mart{\'\i}}, \& {Mart{\'\i}nez-Aroza}}]{2015A&A...584A.122L}
{Luque-Escamilla}, P.~L., {Mart{\'\i}}, J., \& {Mart{\'\i}nez-Aroza}, J. 2015, \aap, 584, A122, \dodoi{10.1051/0004-6361/201527238}

\bibitem[{{Madsen} {et~al.}(2018){Madsen}, {Harrison}, {Broadway}, {Christensen}, {Descalle}, {Ferreira}, {Grefenstette}, {Gurgew}, {Hornschemeier}, {Miyasaka}, {Okajima}, {Pike}, {Pivovaroff}, {Saha}, {Stern}, {Vogel}, {Windt}, \& {Zhang}}]{2018SPIE10699E..6MM}
{Madsen}, K.~K., {Harrison}, F., {Broadway}, D., {et~al.} 2018, in Society of Photo-Optical Instrumentation Engineers (SPIE) Conference Series, Vol. 10699, Space Telescopes and Instrumentation 2018: Ultraviolet to Gamma Ray, ed. J.-W.~A. {den Herder}, S.~{Nikzad}, \& K.~{Nakazawa}, 106996M, \dodoi{10.1117/12.2314117}

\bibitem[{{Majumder} {et~al.}(2024){Majumder}, {Kushwaha}, {Das}, \& {Nandi}}]{2024MNRAS.527L..76M}
{Majumder}, S., {Kushwaha}, A., {Das}, S., \& {Nandi}, A. 2024, \mnras, 527, L76, \dodoi{10.1093/mnrasl/slad148}

\bibitem[{{Miller} {et~al.}(2009){Miller}, {Reynolds}, {Fabian}, {Miniutti}, \& {Gallo}}]{2009ApJ...697..900M}
{Miller}, J.~M., {Reynolds}, C.~S., {Fabian}, A.~C., {Miniutti}, G., \& {Gallo}, L.~C. 2009, \apj, 697, 900, \dodoi{10.1088/0004-637X/697/1/900}

\bibitem[{{Miller} {et~al.}(2013){Miller}, {Parker}, {Fuerst}, {Bachetti}, {Harrison}, {Barret}, {Boggs}, {Chakrabarty}, {Christensen}, {Craig}, {Fabian}, {Grefenstette}, {Hailey}, {King}, {Stern}, {Tomsick}, {Walton}, \& {Zhang}}]{2013ApJ...775L..45M}
{Miller}, J.~M., {Parker}, M.~L., {Fuerst}, F., {et~al.} 2013, \apjl, 775, L45, \dodoi{10.1088/2041-8205/775/2/L45}

\bibitem[{{Miller} {et~al.}(2015){Miller}, {Tomsick}, {Bachetti}, {Wilkins}, {Boggs}, {Christensen}, {Craig}, {Fabian}, {Grefenstette}, {Hailey}, {Harrison}, {Kara}, {King}, {Stern}, \& {Zhang}}]{2015ApJ...799L...6M}
{Miller}, J.~M., {Tomsick}, J.~A., {Bachetti}, M., {et~al.} 2015, \apjl, 799, L6, \dodoi{10.1088/2041-8205/799/1/L6}

\bibitem[{{Miller-Jones} {et~al.}(2009){Miller-Jones}, {Jonker}, {Dhawan}, {Brisken}, {Rupen}, {Nelemans}, \& {Gallo}}]{2009ApJ...706L.230M}
{Miller-Jones}, J.~C.~A., {Jonker}, P.~G., {Dhawan}, V., {et~al.} 2009, \apjl, 706, L230, \dodoi{10.1088/0004-637X/706/2/L230}

\bibitem[{{Miller-Jones} {et~al.}(2021){Miller-Jones}, {Bahramian}, {Orosz}, {Mandel}, {Gou}, {Maccarone}, {Neijssel}, {Zhao}, {Zi{\'o}{\l}kowski}, {Reid}, {Uttley}, {Zheng}, {Byun}, {Dodson}, {Grinberg}, {Jung}, {Kim}, {Marcote}, {Markoff}, {Rioja}, {Rushton}, {Russell}, {Sivakoff}, {Tetarenko}, {Tudose}, \& {Wilms}}]{2021Sci...371.1046M}
{Miller-Jones}, J. C.~A., {Bahramian}, A., {Orosz}, J.~A., {et~al.} 2021, Science, 371, 1046, \dodoi{10.1126/science.abb3363}

\bibitem[{{Mori} {et~al.}(2019){Mori}, {Hailey}, {Mandel}, {Schutt}, {Bachetti}, {Coerver}, {Baganoff}, {Dykaar}, {Grindlay}, {Haggard}, {Heuer}, {Hong}, {Hord}, {Jin}, {Nynka}, {Ponti}, \& {Tomsick}}]{2019ApJ...885..142M}
{Mori}, K., {Hailey}, C.~J., {Mandel}, S., {et~al.} 2019, \apj, 885, 142, \dodoi{10.3847/1538-4357/ab4b47}

\bibitem[{{Motta} \& {Belloni}(2023)}]{2023arXiv230700867M}
{Motta}, S.~E., \& {Belloni}, T.~M. 2023, arXiv e-prints, arXiv:2307.00867, \dodoi{10.48550/arXiv.2307.00867}

\bibitem[{{Mudambi} {et~al.}(2020){Mudambi}, {Rao}, {Gudennavar}, {Misra}, \& {Bubbly}}]{2020MNRAS.498.4404M}
{Mudambi}, S.~P., {Rao}, A., {Gudennavar}, S.~B., {Misra}, R., \& {Bubbly}, S.~G. 2020, \mnras, 498, 4404, \dodoi{10.1093/mnras/staa2656}

\bibitem[{{Mushotzky}(2018)}]{2018SPIE10699E..29M}
{Mushotzky}, R. 2018, in Society of Photo-Optical Instrumentation Engineers (SPIE) Conference Series, Vol. 10699, Space Telescopes and Instrumentation 2018: Ultraviolet to Gamma Ray, ed. J.-W.~A. {den Herder}, S.~{Nikzad}, \& K.~{Nakazawa}, 1069929, \dodoi{10.1117/12.2310003}

\bibitem[{{Neilsen} {et~al.}(2011){Neilsen}, {Remillard}, \& {Lee}}]{2011ApJ...737...69N}
{Neilsen}, J., {Remillard}, R.~A., \& {Lee}, J.~C. 2011, \apj, 737, 69, \dodoi{10.1088/0004-637X/737/2/69}

\bibitem[{{Orosz} {et~al.}(2011){Orosz}, {McClintock}, {Aufdenberg}, {Remillard}, {Reid}, {Narayan}, \& {Gou}}]{2011ApJ...742...84O}
{Orosz}, J.~A., {McClintock}, J.~E., {Aufdenberg}, J.~P., {et~al.} 2011, \apj, 742, 84, \dodoi{10.1088/0004-637X/742/2/84}

\bibitem[{{Orosz} {et~al.}(2014){Orosz}, {Steiner}, {McClintock}, {Buxton}, {Bailyn}, {Steeghs}, {Guberman}, \& {Torres}}]{2014ApJ...794..154O}
{Orosz}, J.~A., {Steiner}, J.~F., {McClintock}, J.~E., {et~al.} 2014, \apj, 794, 154, \dodoi{10.1088/0004-637X/794/2/154}

\bibitem[{{Orosz} {et~al.}(2009){Orosz}, {Steeghs}, {McClintock}, {Torres}, {Bochkov}, {Gou}, {Narayan}, {Blaschak}, {Levine}, {Remillard}, {Bailyn}, {Dwyer}, \& {Buxton}}]{2009ApJ...697..573O}
{Orosz}, J.~A., {Steeghs}, D., {McClintock}, J.~E., {et~al.} 2009, \apj, 697, 573, \dodoi{10.1088/0004-637X/697/1/573}

\bibitem[{{Parker} {et~al.}(2015){Parker}, {Tomsick}, {Miller}, {Yamaoka}, {Lohfink}, {Nowak}, {Fabian}, {Alston}, {Boggs}, {Christensen}, {Craig}, {F{\"u}rst}, {Gandhi}, {Grefenstette}, {Grinberg}, {Hailey}, {Harrison}, {Kara}, {King}, {Stern}, {Walton}, {Wilms}, \& {Zhang}}]{2015ApJ...808....9P}
{Parker}, M.~L., {Tomsick}, J.~A., {Miller}, J.~M., {et~al.} 2015, \apj, 808, 9, \dodoi{10.1088/0004-637X/808/1/9}

\bibitem[{{Parker} {et~al.}(2016){Parker}, {Tomsick}, {Kennea}, {Miller}, {Harrison}, {Barret}, {Boggs}, {Christensen}, {Craig}, {Fabian}, {F{\"u}rst}, {Grinberg}, {Hailey}, {Romano}, {Stern}, {Walton}, \& {Zhang}}]{2016ApJ...821L...6P}
{Parker}, M.~L., {Tomsick}, J.~A., {Kennea}, J.~A., {et~al.} 2016, \apjl, 821, L6, \dodoi{10.3847/2041-8205/821/1/L6}

\bibitem[{P\'erez \& Granger(2007)}]{PER-GRA:2007}
P\'erez, F., \& Granger, B.~E. 2007, Computing in Science and Engineering, 9, 21, \dodoi{10.1109/MCSE.2007.53}

\bibitem[{{Pike} {et~al.}(2022){Pike}, {Negoro}, {Tomsick}, {Bachetti}, {Brumback}, {Connors}, {Garc{\'\i}a}, {Grefenstette}, {Hare}, {Harrison}, {Jaodand}, {Ludlam}, {Mastroserio}, {Mihara}, {Shidatsu}, {Sugizaki}, \& {Takagi}}]{2022ApJ...927..190P}
{Pike}, S.~N., {Negoro}, H., {Tomsick}, J.~A., {et~al.} 2022, \apj, 927, 190, \dodoi{10.3847/1538-4357/ac5258}

\bibitem[{{Ray} {et~al.}(2019){Ray}, {Arzoumanian}, {Ballantyne}, {Bozzo}, {Brandt}, {Brenneman}, {Chakrabarty}, {Christophersen}, {DeRosa}, {Feroci}, {Gendreau}, {Goldstein}, {Hartmann}, {Hernanz}, {Jenke}, {Kara}, {Maccarone}, {McDonald}, {Nowak}, {Phlips}, {Remillard}, {Stevens}, {Tomsick}, {Watts}, {Wilson-Hodge}, {Wood}, {Zane}, {Ajello}, {Alston}, {Altamirano}, {Antoniou}, {Arur}, {Ashton}, {Auchettl}, {Ayres}, {Bachetti}, {Balokovic}, {Baring}, {Baykal}, {Begelman}, {Bhat}, {Bogdanov}, {Briggs}, {Bulbul}, {Bult}, {Burns}, {Cackett}, {Campana}, {Caspi}, {Cavecchi}, {Chenevez}, {Cherry}, {Corbet}, {Corcoran}, {Corsi}, {Degenaar}, {Drake}, {Eikenberry}, {Enoto}, {Fragile}, {Fuerst}, {Gandhi}, {Garcia}, {Goldstein}, {Gonzalez}, {Grefenstette}, {Grinberg}, {Grossan}, {Guillot}, {Guver}, {Haggard}, {Heinke}, {Heinz}, {Hemphill}, {Homan}, {Hui}, {Huppenkothen}, {Ingram}, {Irwin}, {Jaisawal}, {Jaodand}, {Kalemci}, {Kaplan}, {Keek}, {Kennea}, {Kerr}, {van der Klis}, {Kocevski}, {Koss}, {Kowalski}, {Lai},
  {Lamb}, {Laycock}, {Lazio}, {Lazzati}, {Longcope}, {Loewenstein}, {Maitra}, {Majid}, {Maksym}, {Malacaria}, {Margutti}, {Martindale}, {McHardy}, {Meyer}, {Middleton}, {Miller}, {Miller}, {Motta}, {Neilsen}, {Nelson}, {Noble}, {O'Brien}, {Osborne}, {Osten}, {Ozel}, {Palliyaguru}, {Pasham}, {Patruno}, {Pelassa}, {Petropoulou}, {Pilia}, {Pohl}, {Pooley}, {Prescod-Weinstein}, {Psaltis}, {Raaijmakers}, {Reynolds}, {Riley}, {Salvesen}, {Santangelo}, {Scaringi}, {Schanne}, {Schnittman}, {Smith}, {Smith}, {Snios}, {Steiner}, {Steiner}, {Stella}, {Strohmayer}, {Sun}, {Tauris}, {Taylor}, {Tohuvavohu}, {Vacchi}, {Vasilopoulos}, {Veledina}, {Walsh}, {Weinberg}, {Wilkins}, {Willingale}, {Wilms}, {Winter}, {Wolff}, {in 't Zand}, {Zezas}, {Zhang}, \& {Zoghbi}}]{2019arXiv190303035R}
{Ray}, P.~S., {Arzoumanian}, Z., {Ballantyne}, D., {et~al.} 2019, arXiv e-prints, arXiv:1903.03035, \dodoi{10.48550/arXiv.1903.03035}

\bibitem[{{Reback} {et~al.}(2022){Reback}, {Jbrockmendel}, {McKinney}, {Van Den Bossche}, {Roeschke}, {Augspurger}, {Hawkins}, {Cloud}, {Gfyoung}, {Sinhrks}, {Hoefler}, {Klein}, {Petersen}, {Tratner}, {She}, {Ayd}, {Naveh}, {Darbyshire}, {Shadrach}, {Garcia}, {Schendel}, {Hayden}, {Saxton}, {Gorelli}, {Li}, {W{\"o}rtwein}, {Zeitlin}, {Jancauskas}, {McMaster}, \& {Li}}]{2022zndo...6702671R}
{Reback}, J., {Jbrockmendel}, {McKinney}, W., {et~al.} 2022, {pandas-dev/pandas: Pandas 1.4.3}, v1.4.3, Zenodo,  Zenodo, \dodoi{10.5281/zenodo.6702671}

\bibitem[{{Reid} {et~al.}(2014){Reid}, {McClintock}, {Steiner}, {Steeghs}, {Remillard}, {Dhawan}, \& {Narayan}}]{2014ApJ...796....2R}
{Reid}, M.~J., {McClintock}, J.~E., {Steiner}, J.~F., {et~al.} 2014, \apj, 796, 2, \dodoi{10.1088/0004-637X/796/1/2}

\bibitem[{{Reynolds}(2021)}]{2021ARA&A..59..117R}
{Reynolds}, C.~S. 2021, \araa, 59, 117, \dodoi{10.1146/annurev-astro-112420-035022}

\bibitem[{{Russell} {et~al.}(2019){Russell}, {Tetarenko}, {Miller-Jones}, {Sivakoff}, {Parikh}, {Rapisarda}, {Wijnands}, {Corbel}, {Tremou}, {Altamirano}, {Baglio}, {Ceccobello}, {Degenaar}, {van den Eijnden}, {Fender}, {Heywood}, {Krimm}, {Lucchini}, {Markoff}, {Russell}, {Soria}, \& {Woudt}}]{2019ApJ...883..198R}
{Russell}, T.~D., {Tetarenko}, A.~J., {Miller-Jones}, J.~C.~A., {et~al.} 2019, \apj, 883, 198, \dodoi{10.3847/1538-4357/ab3d36}

\bibitem[{{Shahbaz} {et~al.}(1994){Shahbaz}, {Ringwald}, {Bunn}, {Naylor}, {Charles}, \& {Casares}}]{1994MNRAS.271L..10S}
{Shahbaz}, T., {Ringwald}, F.~A., {Bunn}, J.~C., {et~al.} 1994, \mnras, 271, L10, \dodoi{10.1093/mnras/271.1.L10}

\bibitem[{{Shang} {et~al.}(2019){Shang}, {Debnath}, {Chatterjee}, {Jana}, {Chakrabarti}, {Chang}, {Yap}, \& {Chiu}}]{2019ApJ...875....4S}
{Shang}, J.~R., {Debnath}, D., {Chatterjee}, D., {et~al.} 2019, \apj, 875, 4, \dodoi{10.3847/1538-4357/ab0c1e}

\bibitem[{{Stecchini} {et~al.}(2020){Stecchini}, {D'Amico}, {Jablonski}, {Castro}, \& {Braga}}]{2020MNRAS.493.2694S}
{Stecchini}, P.~E., {D'Amico}, F., {Jablonski}, F., {Castro}, M., \& {Braga}, J. 2020, \mnras, 493, 2694, \dodoi{10.1093/mnras/staa417}

\bibitem[{{Steiner} {et~al.}(2014){Steiner}, {McClintock}, {Orosz}, {Remillard}, {Bailyn}, {Kolehmainen}, \& {Straub}}]{2014ApJ...793L..29S}
{Steiner}, J.~F., {McClintock}, J.~E., {Orosz}, J.~A., {et~al.} 2014, \apjl, 793, L29, \dodoi{10.1088/2041-8205/793/2/L29}

\bibitem[{{Steiner} {et~al.}(2012){Steiner}, {Reis}, {Fabian}, {Remillard}, {McClintock}, {Gou}, {Cooke}, {Brenneman}, \& {Sanders}}]{2012MNRAS.427.2552S}
{Steiner}, J.~F., {Reis}, R.~C., {Fabian}, A.~C., {et~al.} 2012, \mnras, 427, 2552, \dodoi{10.1111/j.1365-2966.2012.22128.x}

\bibitem[{{Svoboda} {et~al.}(2024){Svoboda}, {Dov{\v{c}}iak}, {Steiner}, {Muleri}, {Ingram}, {Yilmaz}, {Rodriguez Cavero}, {Marra}, {Poutanen}, {Veledina}, {Mojaver}, {Bianchi}, {Garc{\'\i}a}, {Kaaret}, {Krawczynski}, {Matt}, {Podgorn{\'y}}, {Weisskopf}, {Kislat}, {Petrucci}, {Brigitte}, {Bursa}, {Fabiani}, {Hu}, {Chun}, {Mastroserio}, {Mikus̆incov{\'a}}, {Ratheesh}, {Romani}, {Soffitta}, {Ursini}, {Zane}, {Agudo}, {Antonelli}, {Bachetti}, {Baldini}, {Baumgartner}, {Bellazzini}, {Bongiorno}, {Bonino}, {Brez}, {Bucciantini}, {Capitanio}, {Castellano}, {Cavazzuti}, {Chen}, {Ciprini}, {Costa}, {De Rosa}, {Del Monte}, {Di Gesu}, {Di Lalla}, {Di Marco}, {Donnarumma}, {Doroshenko}, {Ehlert}, {Enoto}, {Evangelista}, {Ferrazzoli}, {Gunji}, {Hayashida}, {Heyl}, {Iwakiri}, {Jorstad}, {Karas}, {Kitaguchi}, {Kolodziejczak}, {La Monaca}, {Latronico}, {Liodakis}, {Maldera}, {Manfreda}, {Marin}, {Marinucci}, {Marscher}, {Marshall}, {Massaro}, {Mitsuishi}, {Mizuno}, {Negro}, {Ng}, {O'Dell}, {Omodei}, {Oppedisano},
  {Papitto}, {Pavlov}, {Peirson}, {Perri}, {Pesce-Rollins}, {Pilia}, {Possenti}, {Puccetti}, {Ramsey}, {Rankin}, {Roberts}, {Sgr{\`o}}, {Slane}, {Spandre}, {Swartz}, {Tamagawa}, {Tavecchio}, {Taverna}, {Tawara}, {Tennant}, {Thomas}, {Tombesi}, {Trois}, {Tsygankov}, {Turolla}, {Vink}, {Wu}, \& {Xie}}]{2024ApJ...960....3S}
{Svoboda}, J., {Dov{\v{c}}iak}, M., {Steiner}, J.~F., {et~al.} 2024, \apj, 960, 3, \dodoi{10.3847/1538-4357/ad0842}

\bibitem[{{Tao} {et~al.}(2019){Tao}, {Tomsick}, {Qu}, {Zhang}, {Zhang}, \& {Bu}}]{2019ApJ...887..184T}
{Tao}, L., {Tomsick}, J.~A., {Qu}, J., {et~al.} 2019, \apj, 887, 184, \dodoi{10.3847/1538-4357/ab5282}

\bibitem[{{Tashiro} {et~al.}(2020){Tashiro}, {Maejima}, {Toda}, {Kelley}, {Reichenthal}, {Hartz}, {Petre}, {Williams}, {Guainazzi}, {Costantini}, {Fujimoto}, {Hayashida}, {Henegar-Leon}, {Holland}, {Ishisaki}, {Kilbourne}, {Loewenstein}, {Matsushita}, {Mori}, {Okajima}, {Porter}, {Sneiderman}, {Takei}, {Terada}, {Tomida}, {Yamaguchi}, {Watanabe}, {Akamatsu}, {Arai}, {Audard}, {Awaki}, {Babyk}, {Bamba}, {Bando}, {Behar}, {Bialas}, {Boissay-Malaquin}, {Brenneman}, {Brown}, {Canavan}, {Chiao}, {Comber}, {Corrales}, {Cumbee}, {de Vries}, {den Herder}, {Dercksen}, {Diaz-Trigo}, {DiPirro}, {Done}, {Dotani}, {Ebisawa}, {Eckart}, {Eckert}, {Eguchi}, {Enoto}, {Ezoe}, {Ferrigno}, {Fujita}, {Fukazawa}, {Furuzawa}, {Gallo}, {Gorter}, {Grim}, {Gu}, {Hagino}, {Hamaguchi}, {Hatsukade}, {Hawthorn}, {Hayashi}, {Hell}, {Hiraga}, {Hodges-Kluck}, {Horiuchi}, {Hornschemeier}, {Hoshino}, {Ichinohe}, {Iga}, {Iizuka}, {Ishida}, {Ishihama}, {Ishikawa}, {Ishimura}, {Jaffe}, {Kaastra}, {Kallman}, {Kara}, {Katsuda}, {Kenyon}, {Kimball},
  {Kitaguchi}, {Kitamoto}, {Kobayashi}, {Kobayashi}, {Kohmura}, {Kubota}, {Leutenegger}, {Li}, {Lockard}, {Maeda}, {Markevitch}, {Martz}, {Matsumoto}, {Matsuzaki}, {McCammon}, {McLaughlin}, {McNamara}, {Miko}, {Miller}, {Miller}, {Minesugi}, {Mitani}, {Mitsuishi}, {Mizumoto}, {Mizuno}, {Mukai}, {Murakami}, {Mushotzky}, {Nakajima}, {Nakamura}, {Nakazawa}, {Natsukari}, {Nigo}, {Nishioka}, {Nobukawa}, {Nobukawa}, {Noda}, {Odaka}, {Ogawa}, {Ohashi}, {Ohno}, {Ohta}, {Okamoto}, {Ota}, {Ozaki}, {Paltani}, {Plucinsky}, {Pottschmidt}, {Sampson}, {Sasaki}, {Sato}, {Sato}, {Sato}, {Sawada}, {Seta}, {Shibano}, {Shida}, {Shidatsu}, {Shigeto}, {Shinozaki}, {Shirron}, {Simionescu}, {Smith}, {Someya}, {Soong}, {Sugawara}, {Sugawara}, {Szymkowiak}, {Takahashi}, {Takeshima}, {Tamagawa}, {Tamura}, {Tanaka}, {Tanimoto}, {Terashima}, {Tsuboi}, {Tsujimoto}, {Tsunemi}, {Tsuru}, {Uchida}, {Uchida}, {Uchiyama}, {Ueda}, {Uno}, {Vink}, {Watanabe}, {Witthoeft}, {Wolfs}, {Yamada}, {Yamaoka}, {Yamasaki}, {Yamauchi}, {Yamauchi},
  {Yanagase}, {Yaqoob}, {Yasuda}, {Yoshida}, {Yoshioka}, \& {Zhuravleva}}]{2020SPIE11444E..22T}
{Tashiro}, M., {Maejima}, H., {Toda}, K., {et~al.} 2020, in Society of Photo-Optical Instrumentation Engineers (SPIE) Conference Series, Vol. 11444, Space Telescopes and Instrumentation 2020: Ultraviolet to Gamma Ray, ed. J.-W.~A. {den Herder}, S.~{Nikzad}, \& K.~{Nakazawa}, 1144422, \dodoi{10.1117/12.2565812}

\bibitem[{{Verner} {et~al.}(1996){Verner}, {Ferland}, {Korista}, \& {Yakovlev}}]{1996ApJ...465..487V}
{Verner}, D.~A., {Ferland}, G.~J., {Korista}, K.~T., \& {Yakovlev}, D.~G. 1996, \apj, 465, 487, \dodoi{10.1086/177435}

\bibitem[{Virtanen {et~al.}(2020)Virtanen, Gommers, Oliphant, Haberland, Reddy, Cournapeau, Burovski, Peterson, Weckesser, Bright, {van der Walt}, Brett, Wilson, Millman, Mayorov, Nelson, Jones, Kern, Larson, Carey, Polat, Feng, Moore, {VanderPlas}, Laxalde, Perktold, Cimrman, Henriksen, Quintero, Harris, Archibald, Ribeiro, Pedregosa, {van Mulbregt}, \& {SciPy 1.0 Contributors}}]{2020SciPy-NMeth}
Virtanen, P., Gommers, R., Oliphant, T.~E., {et~al.} 2020, Nature Methods, 17, 261, \dodoi{10.1038/s41592-019-0686-2}

\bibitem[{{Walton} {et~al.}(2017){Walton}, {Mooley}, {King}, {Tomsick}, {Miller}, {Dauser}, {Garc{\'\i}a}, {Bachetti}, {Brightman}, {Fabian}, {Forster}, {F{\"u}rst}, {Gandhi}, {Grefenstette}, {Harrison}, {Madsen}, {Meier}, {Middleton}, {Natalucci}, {Rahoui}, {Rana}, \& {Stern}}]{2017ApJ...839..110W}
{Walton}, D.~J., {Mooley}, K., {King}, A.~L., {et~al.} 2017, \apj, 839, 110, \dodoi{10.3847/1538-4357/aa67e8}

\bibitem[{{Wang} {et~al.}(2018{\natexlab{a}}){Wang}, {Kawai}, {Shidatsu}, {Tachibana}, {Yoshii}, {Sudo}, \& {Kubota}}]{2018PASJ...70...67W}
{Wang}, S., {Kawai}, N., {Shidatsu}, M., {et~al.} 2018{\natexlab{a}}, \pasj, 70, 67, \dodoi{10.1093/pasj/psy058}

\bibitem[{{Wang} {et~al.}(2018{\natexlab{b}}){Wang}, {M{\'e}ndez}, {Altamirano}, {Court}, {Beri}, \& {Cheng}}]{2018MNRAS.478.4837W}
{Wang}, Y., {M{\'e}ndez}, M., {Altamirano}, D., {et~al.} 2018{\natexlab{b}}, \mnras, 478, 4837, \dodoi{10.1093/mnras/sty1372}

\bibitem[{{W}es {M}c{K}inney(2010)}]{mckinney-proc-scipy-2010}
{W}es {M}c{K}inney. 2010, in {P}roceedings of the 9th {P}ython in {S}cience {C}onference, ed. {S}t\'efan van~der {W}alt \& {J}arrod {M}illman, 56 -- 61, \dodoi{10.25080/Majora-92bf1922-00a}

\bibitem[{{Wilms} {et~al.}(2000){Wilms}, {Allen}, \& {McCray}}]{2000ApJ...542..914W}
{Wilms}, J., {Allen}, A., \& {McCray}, R. 2000, \apj, 542, 914, \dodoi{10.1086/317016}

\bibitem[{{Xu} {et~al.}(2020){Xu}, {Harrison}, {Tomsick}, {Walton}, {Barret}, {Garc{\'\i}a}, {Hare}, \& {Parker}}]{2020ApJ...893...30X}
{Xu}, Y., {Harrison}, F.~A., {Tomsick}, J.~A., {et~al.} 2020, \apj, 893, 30, \dodoi{10.3847/1538-4357/ab7dc0}

\bibitem[{{Xu} {et~al.}(2018){Xu}, {Harrison}, {Kennea}, {Walton}, {Tomsick}, {Miller}, {Barret}, {Fabian}, {Forster}, {F{\"u}rst}, {Gandhi}, \& {Garc{\'\i}a}}]{2018ApJ...865...18X}
{Xu}, Y., {Harrison}, F.~A., {Kennea}, J.~A., {et~al.} 2018, \apj, 865, 18, \dodoi{10.3847/1538-4357/aada03}

\bibitem[{{Yao} {et~al.}(2021){Yao}, {Kulkarni}, {Burdge}, {Caiazzo}, {De}, {Dong}, {Fremling}, {Kasliwal}, {Kupfer}, {van Roestel}, {Sollerman}, {Bagdasaryan}, {Bellm}, {Cenko}, {Drake}, {Duev}, {Graham}, {Kaye}, {Masci}, {Miranda}, {Prince}, {Riddle}, {Rusholme}, \& {Soumagnac}}]{2021ApJ...920..120Y}
{Yao}, Y., {Kulkarni}, S.~R., {Burdge}, K.~B., {et~al.} 2021, \apj, 920, 120, \dodoi{10.3847/1538-4357/ac15f9}

\bibitem[{{Zdziarski} {et~al.}(2024){Zdziarski}, {Banerjee}, {Chand}, {Dewangan}, {Misra}, {Szanecki}, \& {Nied{\'z}wiecki}}]{2024ApJ...962..101Z}
{Zdziarski}, A.~A., {Banerjee}, S., {Chand}, S., {et~al.} 2024, \apj, 962, 101, \dodoi{10.3847/1538-4357/ad1b60}

\bibitem[{{Zhu} {et~al.}(2023){Zhu}, {Chen}, \& {Wang}}]{2023MNRAS.523.4394Z}
{Zhu}, H., {Chen}, X., \& {Wang}, W. 2023, \mnras, 523, 4394, \dodoi{10.1093/mnras/stad1656}

\end{thebibliography}
\bibliographystyle{aasjournal}

\appendix
\section{Observations Used} \label{sec:obsids}
For IGR J17091-3624  we analyzed ObsIDs 80001041002, 80202014002, 80202014004, 80202014006, 80202015002, 80202015004, 80702315002, 80702315004, 80702315006, 80801324002, 80801324004, 80802321002, 80802321003, and 80802321005.

For MAXI J1535-571, we fit the spectra from ObsIDs 80302309002, 80302309004, 80302309006, 80302309008, 80302309010, 80302309012, 80302309014, 80402302002, 80402302004, 80402302006, 80402302008, 80402302009, 80402302010, 90301013002, 90301327002, and 90301327004.

For GRS 1915+105, we fit the spectra from ObsIDs 10002004001, 30302018002, 30302020006, 30302020008, 30502008002, 30502008004, 30502008006, 30602008006, 80401312002, 90001001002, 90201053002, 90202045004, 90301001002, 90501321002, 90501346002, 90701323002, and 90701332002.

For Cygnus X-1, the ObsIDs for the individual observations analyzed are 00001011001, 00001011002, 10002003001, 10014001001, 10102001002, 30001011002, 30001011003, 30001011005, 30001011007, 30001011009, 30001011011, 30002150002, 30002150004, 30002150006, 30002150008, 30101022002, 30202032002, 30302019002, 30302019004, 30302019006, 30302019008, 30302019010, 30302019012, 30702017002, 30702017004, 30702017006, 80502335002, 80502335004, 80502335006, 90101019002, 90101020002, 90802013002, 90802013004, and 90802013006.

For GX 339-4, we fit ObsIDs 80001013002, 80001013004, 80001013006, 80001013008, 80001013010, 80001015001, 80001015003, 80102011002, 80102011004, 80102011006, 80102011008, 80102011010, 80302304002, 80302304004, 80302304005, 80302304007, 80502325002, 80502325004, 80502325006, 80502325008, 80601302002, 80601302004, 80601302006, 80702316002, 80702316005, 80801341002, 90401369002, 90401369004, 90502356004, 90702303001, 90702303003, 90702303005, 90702303007, 90702303009, 90702303011, and 90702303013.

\section{Sources wherein no measurement was possible}\label{sec:no_measurement}
\subsection{MAXI J1810-222}\label{sec:MAXI_J1810-222}
There are two NuSTAR observations of BH candidate MAXI J1810-222 (ObsIDs 90402367002 and 90410345001). During both observations, Mode 1 scientific data were unavailable, so we analyzed the NuSTAR Mode 6 data. The spectra show no signs of relativistic reflection. We fit the spectra from both observations in the 3-10 keV band.

The best fit for the spectra from both observations was achieved using the \texttt{diskbb+powerlaw} model, with Galactic absorption through \texttt{TBabs} not being statistically required by the data. For ObsID 90402367002, we obtain $\chi^2/\nu=67.97/51$, with a measured disk temperature of $\sim0.5$ keV and a power-law index $\Gamma\sim3.9\pm0.2$. For ObsID 90410345001, we obtain $\chi^2/\nu=37.49/34$, with a measured disk temperature of $\sim0.6$ keV and an unconstrained power-law index $\Gamma$. We conclude that no spin measurement of this source is possible given the available archival NuSTAR observations.

\section{Figures}\label{sec:figures_and_tables}
This appendix contains all the figures that show the fit residuals and the posterior probability distributions resulting from the MCMC analysis for all the observations of each source.

\begin{figure}[ht]
    \centering
    \includegraphics[width= 0.95\textwidth]{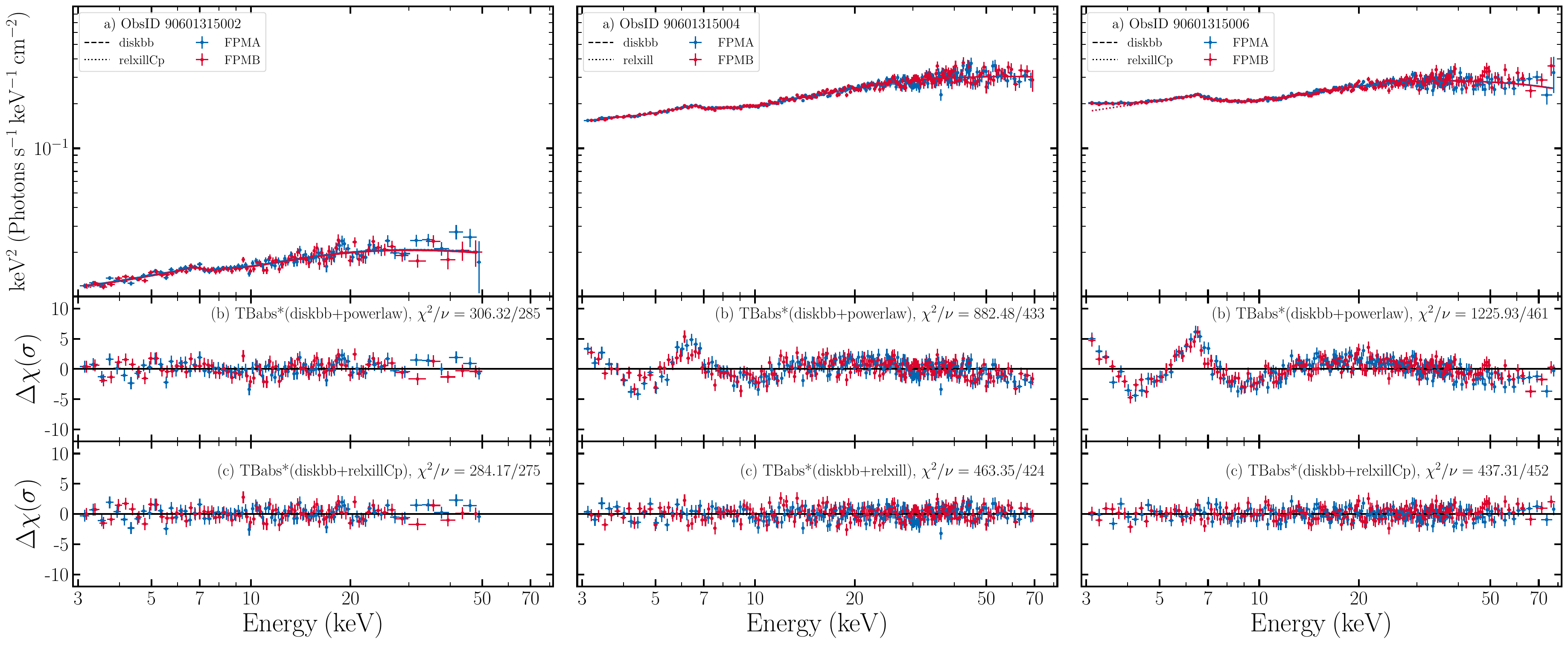}
    \caption{The three panels represent the three observations of AT 2019wey analyzed. The top panels show the unfolded spectra, with the NuSTAR FPMA spectra shown in blue and FPMB spectra in red. The reported best-fit models are shown by the solid lines, while the contributions of the \texttt{relxill} components are shown by the dotted lines and the contributions of the \texttt{diskbb} components are shown through the dashed lines. The middle and bottom panels show the residuals in terms of $\sigma$ produced when fitting the spectra with \texttt{TBabs*(diskbb+powerlaw)} (middle) and the best-performing reflection models (bottom), respectively, together with the statistic produced. Figure discussed in Section \ref{sec:AT_2019wey}.}
    \label{fig:AT_2019wey_delchi}
\end{figure}

\begin{figure}[ht]
    \centering
    \includegraphics[width= 0.95\textwidth]{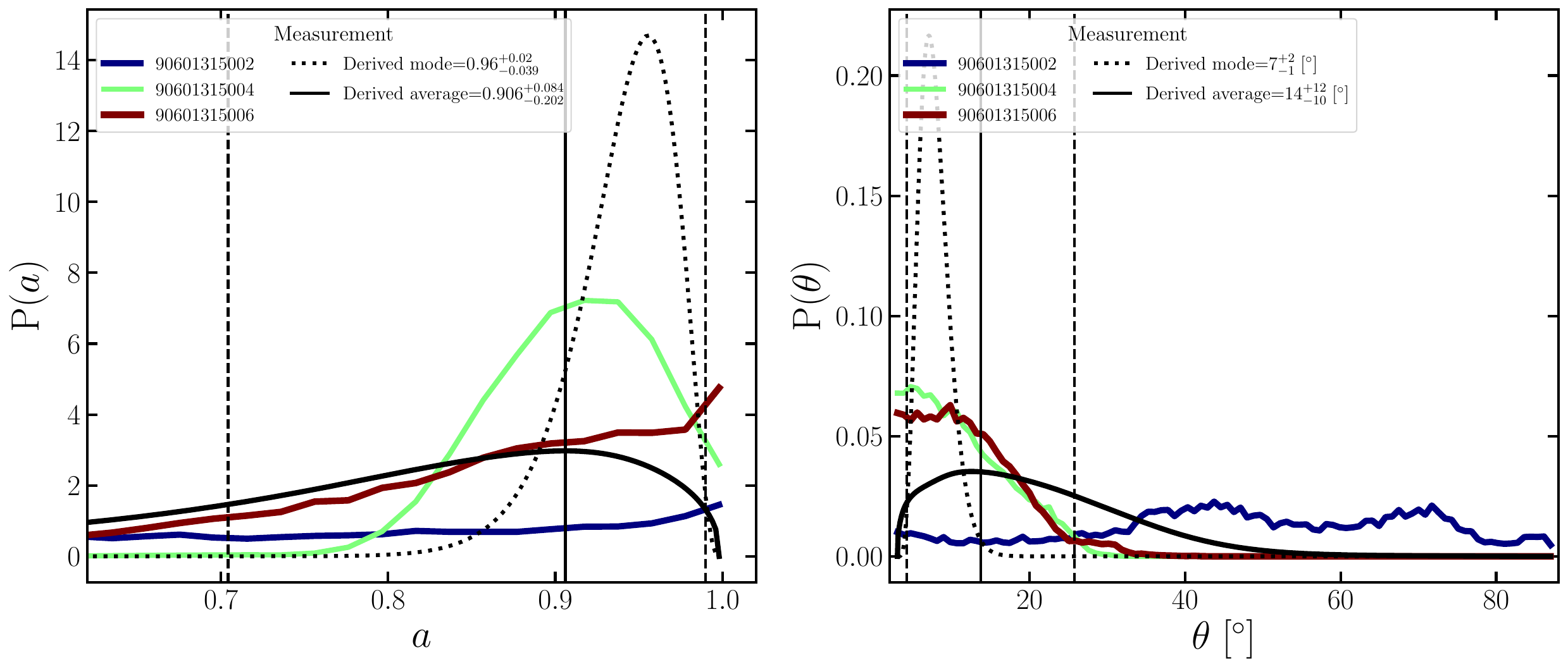}
    \caption{The left panel shows the posterior distributions resulting from the MCMC analysis of AT 2019wey for spin, while the right panel shows the posterior distributions for the inclination of the inner accretion disk in the model. The different colored lines represent the three observations analyzed, and the black curves represent the combined inferred distribution, with the thickness of the lines being proportional to the ratio of the reflected flux to the total flux in the 3-79~keV band, which were used as weighting when combining the posterior distributions. The solid vertical black lines represent the modes of the combined distribution, and the dashed vertical black lines represent the $1\sigma$ credible intervals of the measurements. Figure discussed in Section \ref{sec:AT_2019wey}.}
    \label{fig:AT_2019wey_combined}
\end{figure}

\begin{figure}[ht]
    \centering
    \includegraphics[width= 0.8\textwidth]{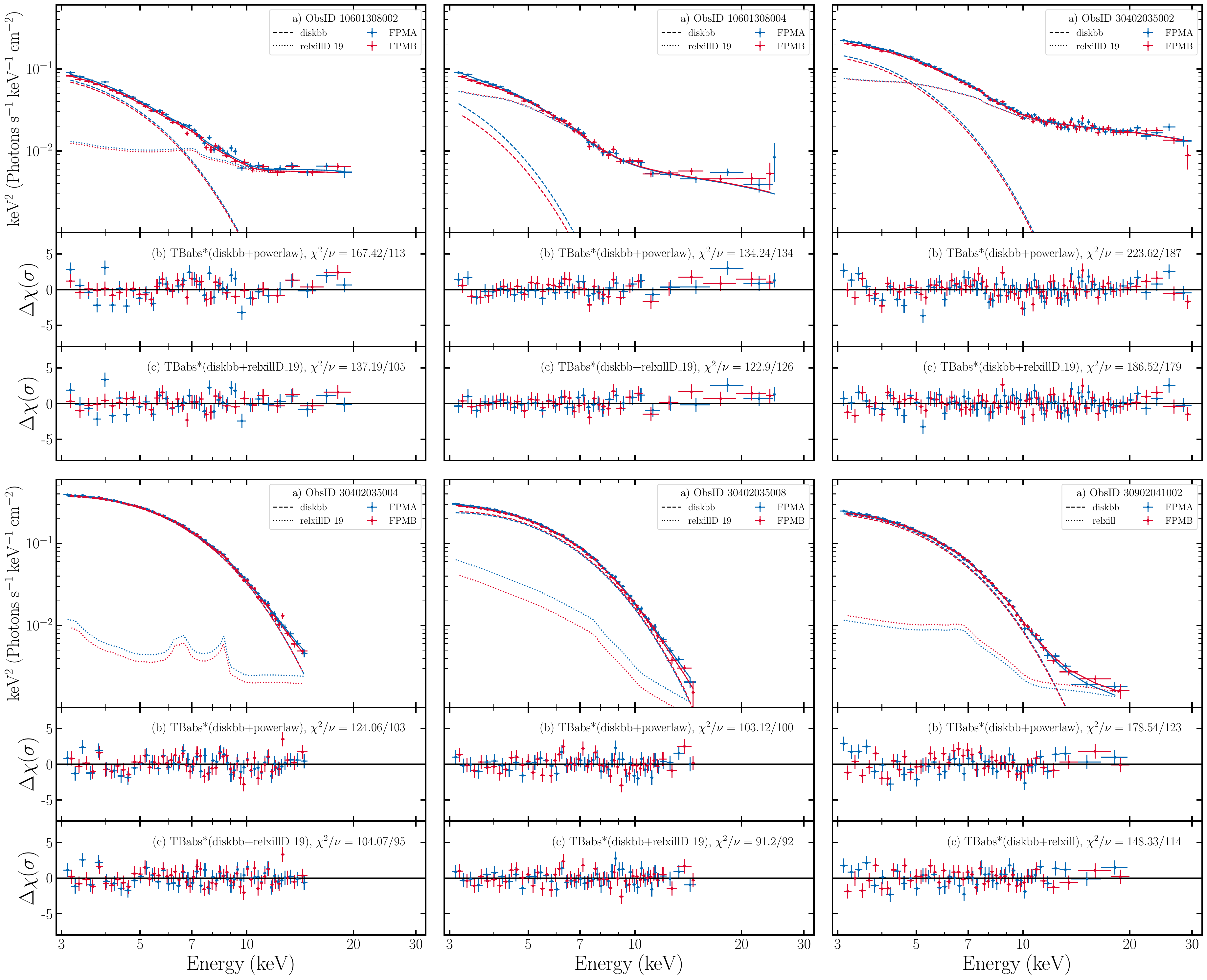}
    \caption{Unfolded spectra and fit residuals for the LMC X-3 data. Explanations are analogous to Figure \ref{fig:AT_2019wey_delchi}. Figure discussed in Section \ref{sec:LMC_X-3}.}
    \label{fig:LMC_X3_delchi}
\end{figure}
\begin{figure}[ht]
    \centering
    \includegraphics[width= 0.95\textwidth]{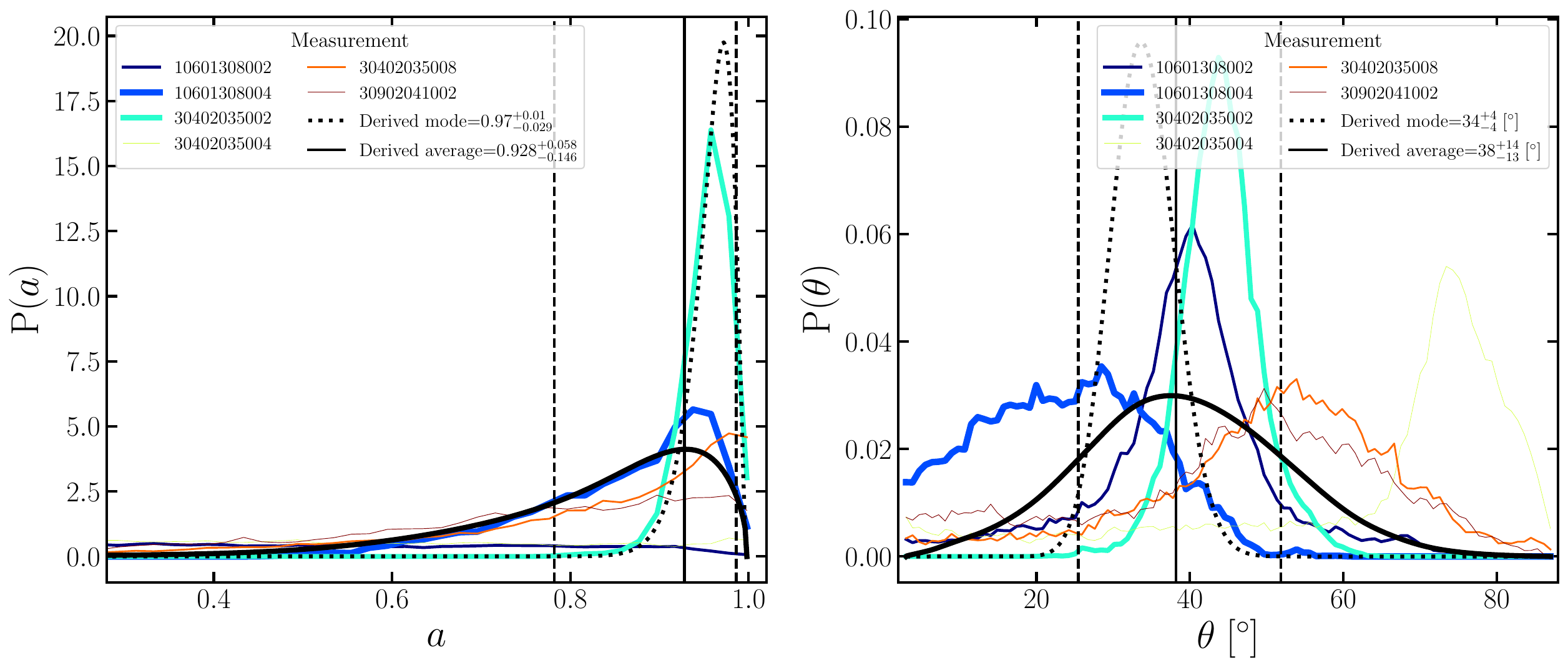}
    \caption{Posterior distributions for the analysis of LMC X-3 data. Explanations are analogous to Figure \ref{fig:AT_2019wey_combined}. Figure discussed in Section \ref{sec:LMC_X-3}.}
    \label{fig:LMC_X3_combined}
\end{figure}

\begin{figure}[ht]
    \centering
    \includegraphics[width= 0.6\textwidth]{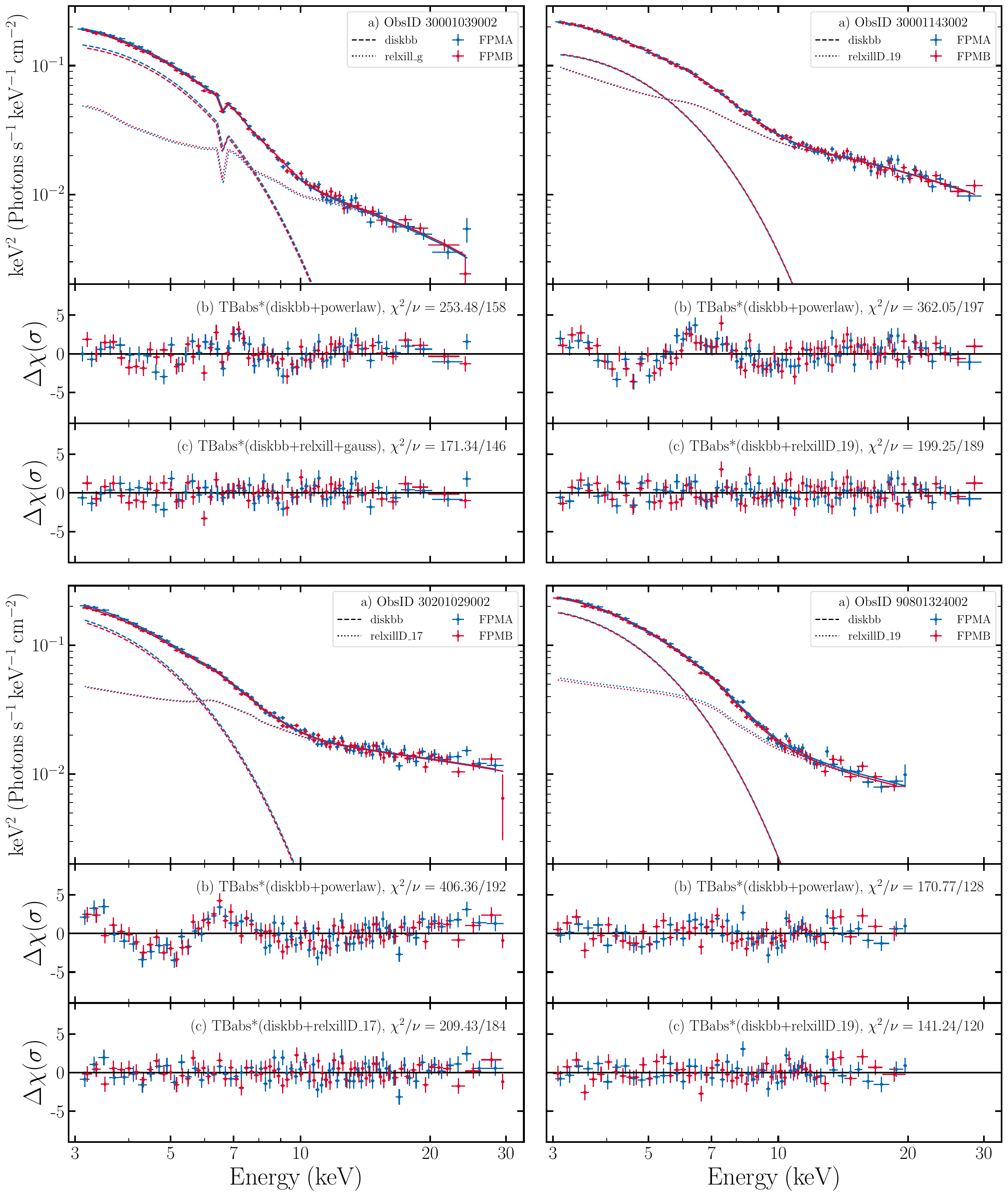}
    \caption{Unfolded spectra and fit residuals for the LMC X-1 data. Explanations are analogous to Figure \ref{fig:AT_2019wey_delchi}. Figure discussed in Section \ref{sec:LMC_X-1}.}
    \label{fig:LMC_X1_delchi}
\end{figure}

\begin{figure}[ht]
    \centering
    \includegraphics[width= 0.95\textwidth]{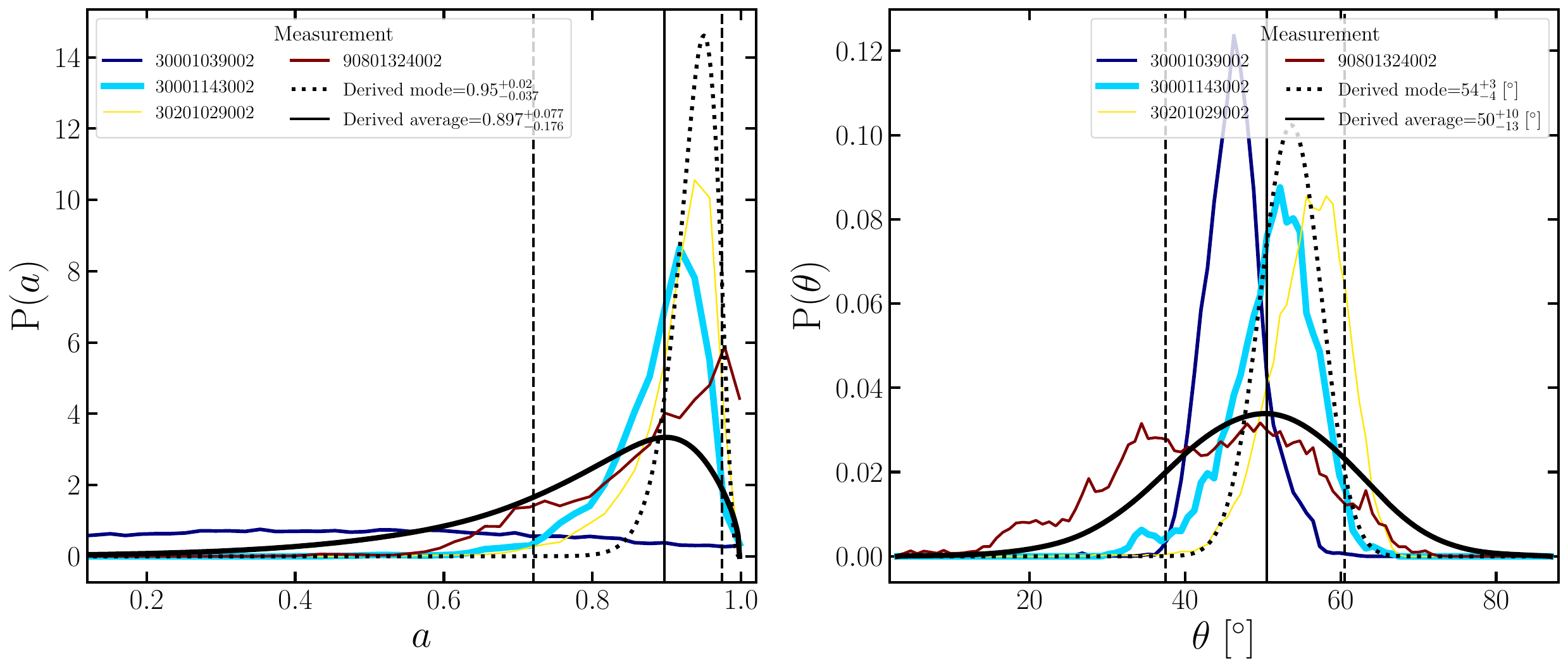}
    \caption{Posterior distributions for the analysis of LMC X-1 data. Explanations are analogous to Figure \ref{fig:AT_2019wey_combined}. Figure discussed in Section \ref{sec:LMC_X-1}.}
    \label{fig:LMC_X1_combined}
\end{figure}

\begin{figure}[ht]
    \centering
    \includegraphics[width= 0.7\textwidth]{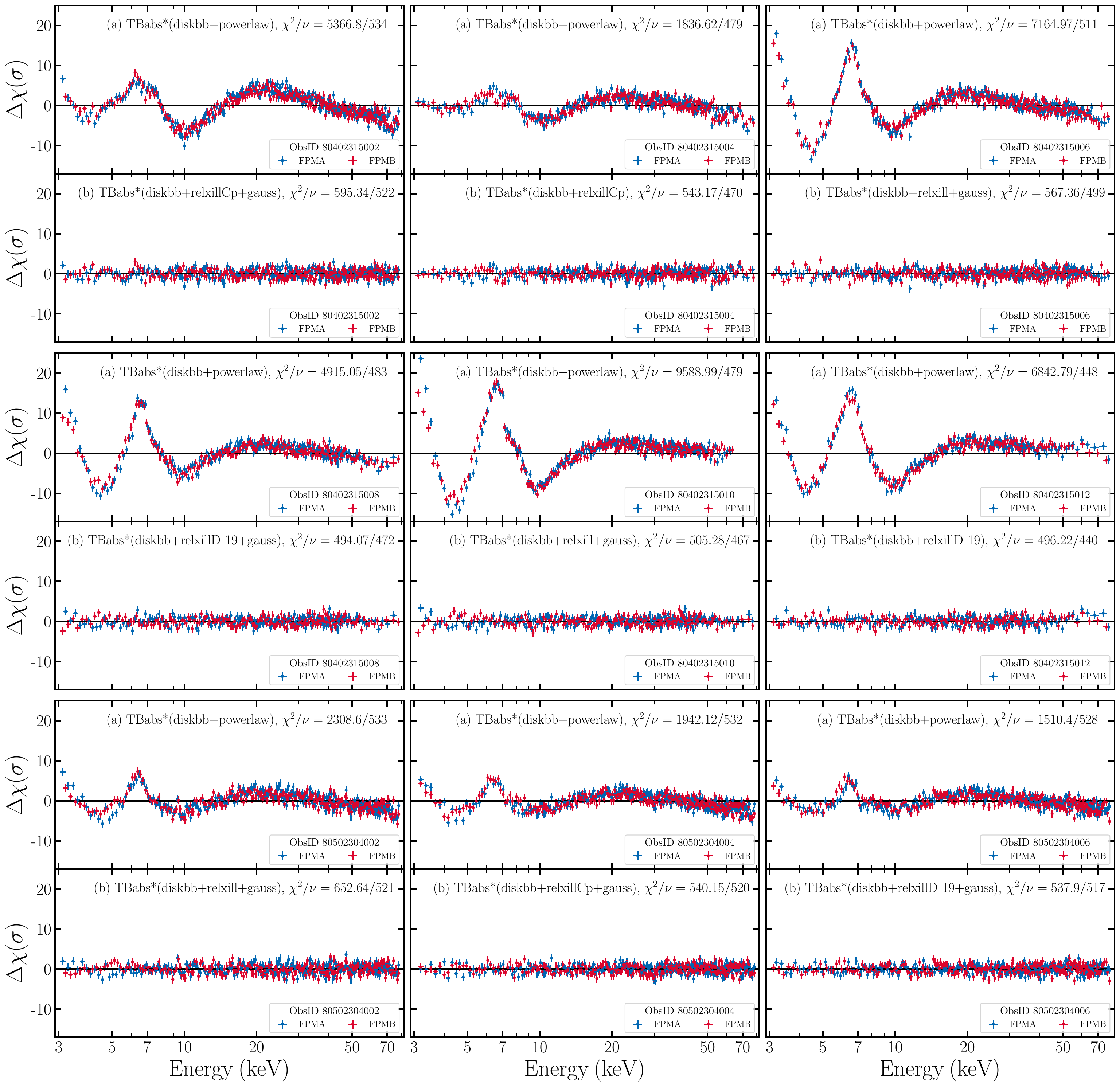}
    \caption{Fit residuals for the MAXI J1348-630 data. Explanations are analogous to Figure \ref{fig:AT_2019wey_delchi}. Figure discussed in Section \ref{sec:MAXI_J1348-630}.}
    \label{fig:MAXI_J1348-630_delchi}
\end{figure}

\begin{figure}[ht]
    \centering
    \includegraphics[width= 0.95\textwidth]{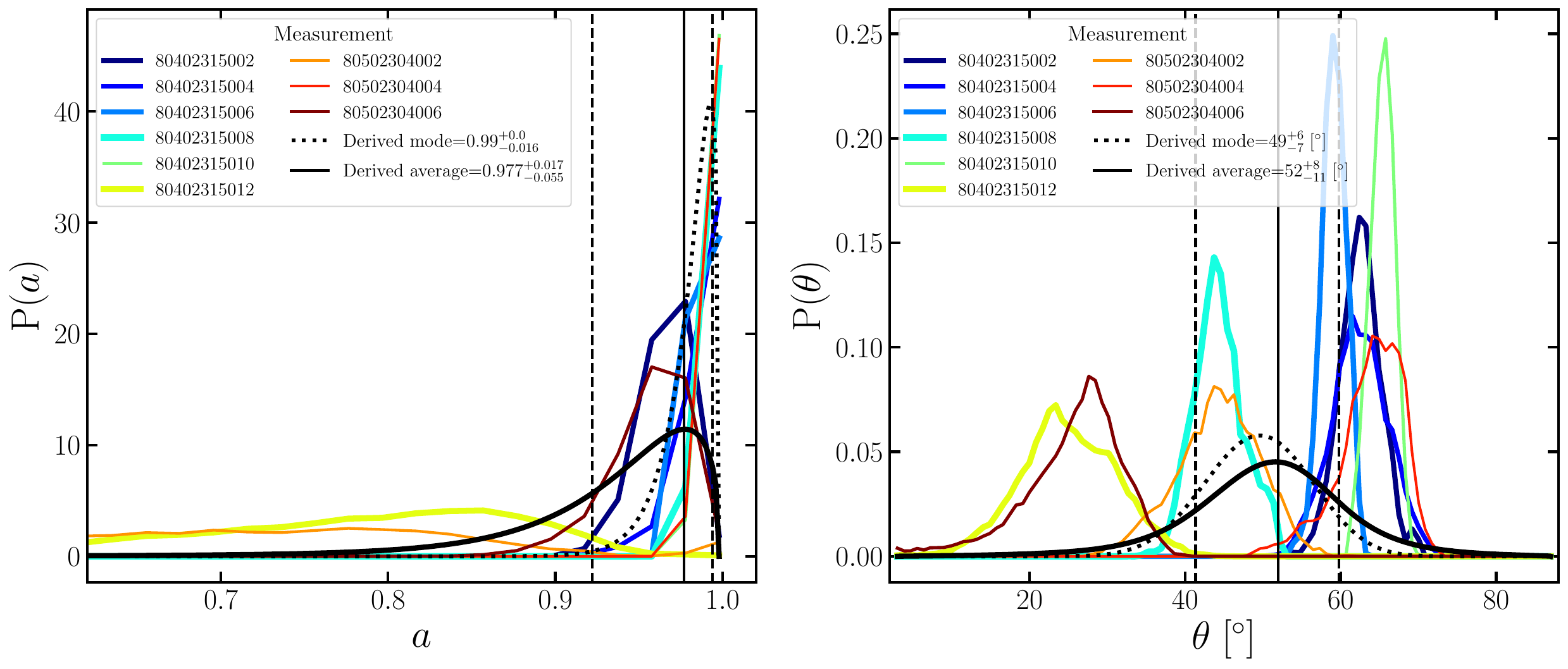}
    \caption{Posterior distributions for the analysis of MAXI J1348-630 data. Explanations are analogous to Figure \ref{fig:AT_2019wey_combined}. Figure discussed in Section \ref{sec:MAXI_J1348-630}.}
    \label{fig:MAXI_J1348-630_combined}
\end{figure}

\begin{figure}[ht]
    \centering
    \includegraphics[width= 0.95\textwidth]{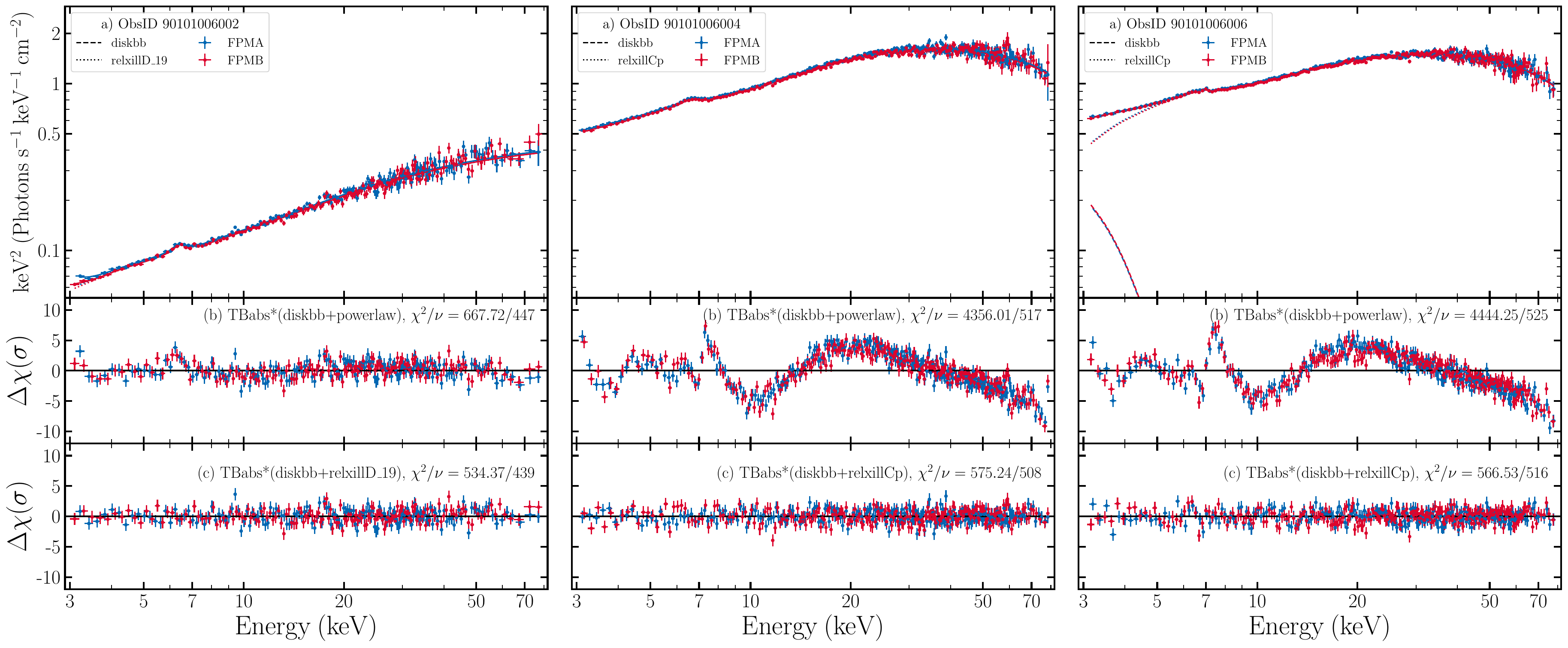}
    \caption{Unfolded spectra and fit residuals for the GS 1354-645 data. Explanations are analogous to Figure \ref{fig:AT_2019wey_delchi}. Figure discussed in Section \ref{sec:GS_1354-645}.}
    \label{fig:GS_1354_delchi}
\end{figure}

\begin{figure}[ht]
    \centering
    \includegraphics[width= 0.95\textwidth]{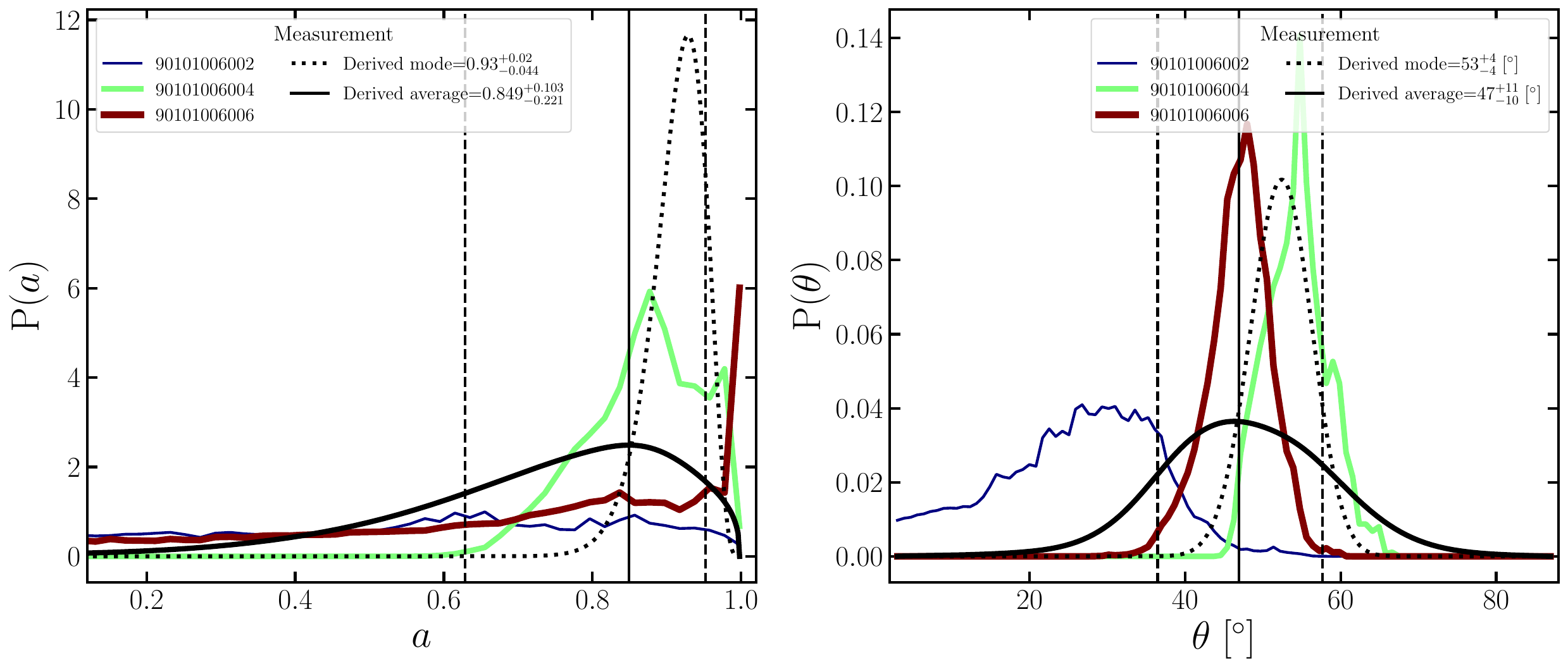}
    \caption{Posterior distributions for the analysis of GS 1354-645 data. Explanations are analogous to Figure \ref{fig:AT_2019wey_combined}. Figure discussed in Section \ref{sec:GS_1354-645}.}
    \label{fig:GS_1354_combined}
\end{figure}

\begin{figure}[ht]
    \centering
    \includegraphics[width= 0.7\textwidth]{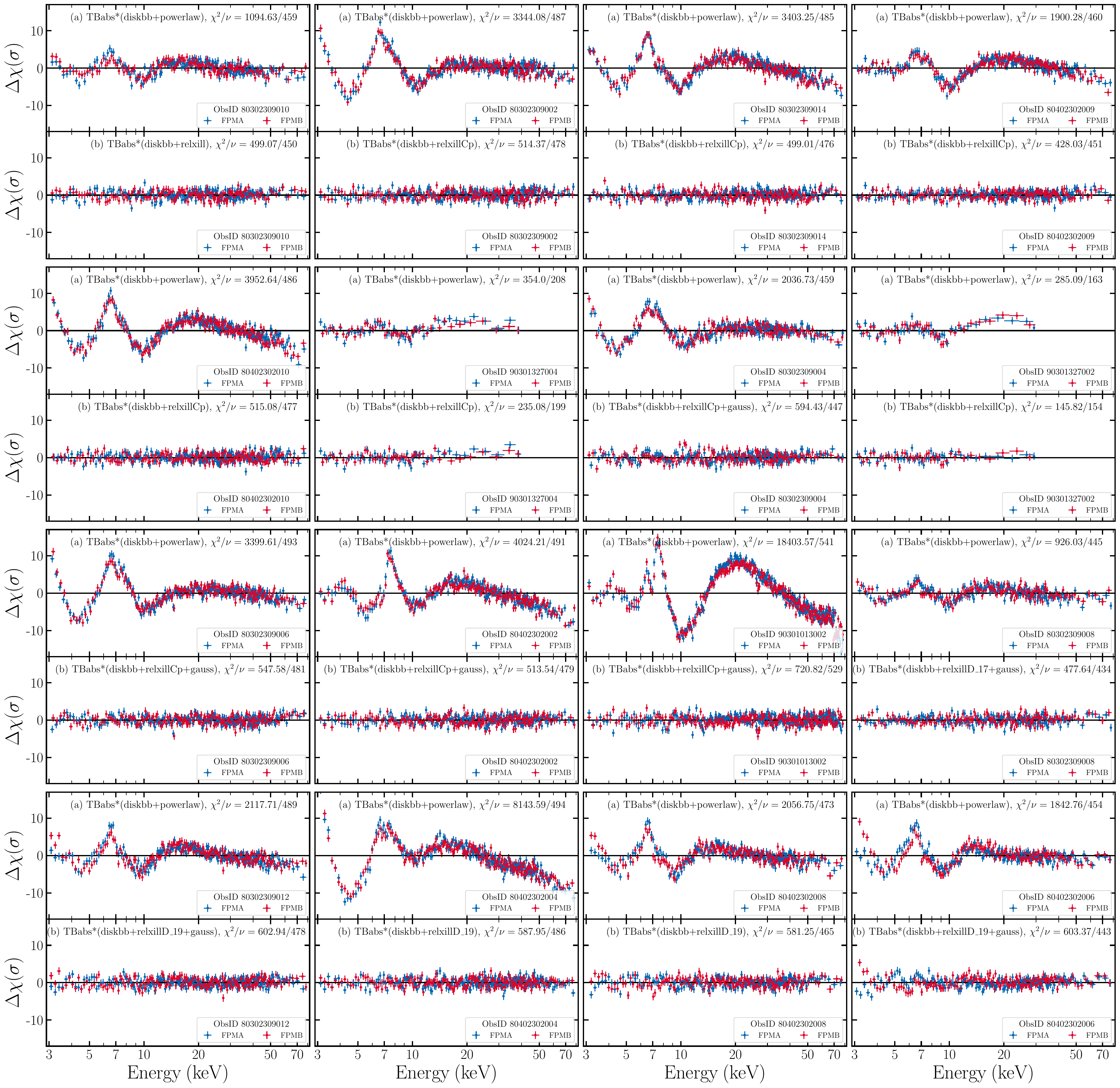}
    \caption{Fit residuals for the MAXI J1535-571 data. Explanations are analogous to Figure \ref{fig:AT_2019wey_delchi}. Figure discussed in Section \ref{sec:MAXI_J1535-571}.}
    \label{fig:MAXI_J1535_delchi}
\end{figure}
\begin{figure}[ht]
    \centering
    \includegraphics[width= 0.95\textwidth]{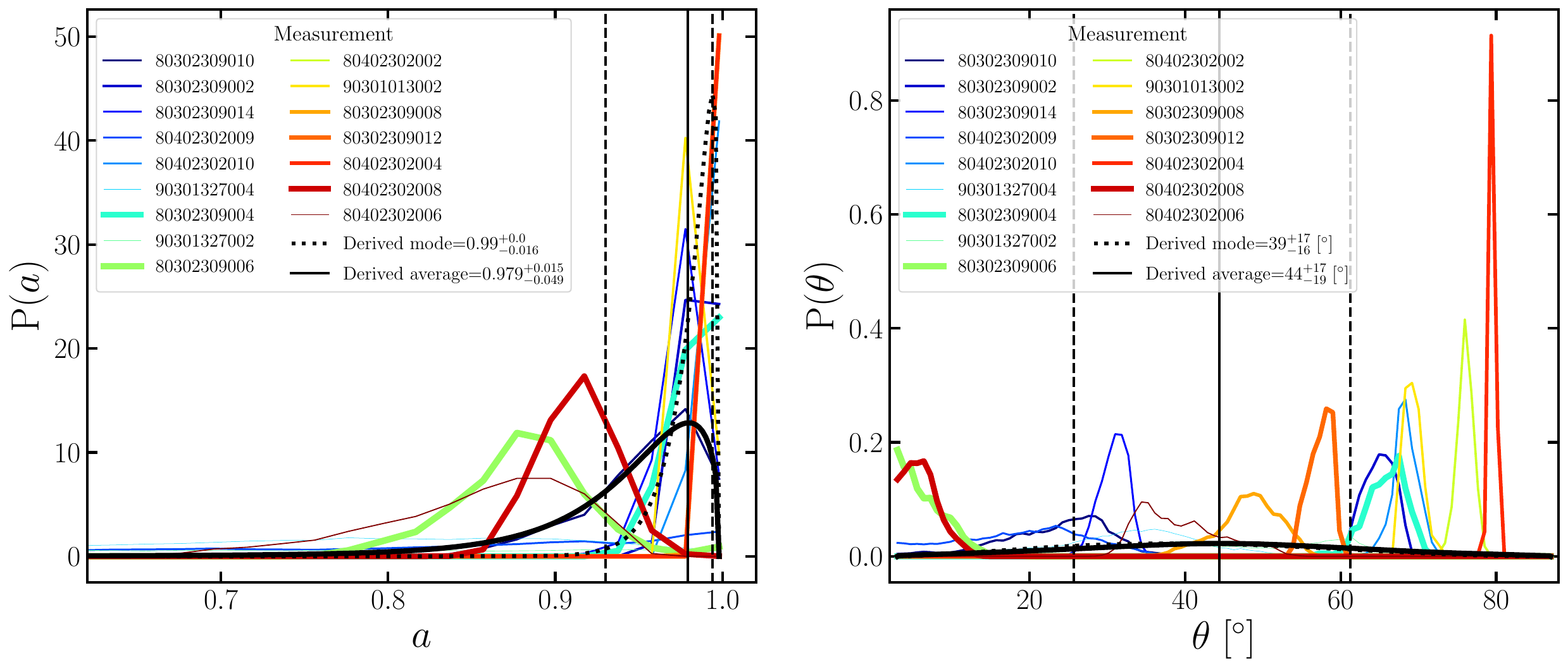}
    \caption{Posterior distributions for the analysis of MAXI J1535-571 data. Explanations are analogous to Figure \ref{fig:AT_2019wey_combined}. Figure discussed in Section \ref{sec:MAXI_J1535-571}.}
    \label{fig:MAXI_J1535_combined}
\end{figure}

\begin{figure}[ht]
    \centering
    \includegraphics[width= 0.6\textwidth]{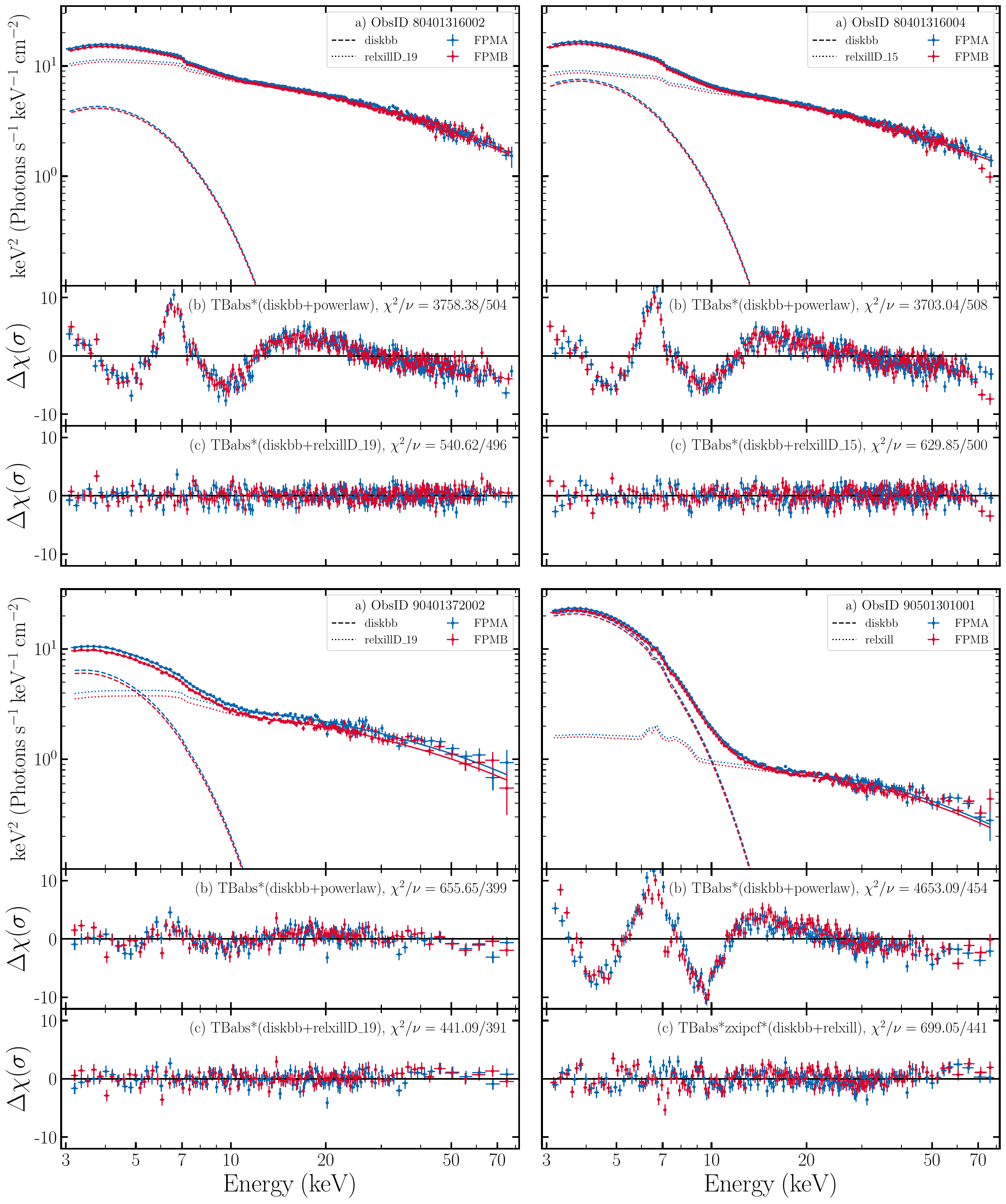}
    \caption{Unfolded spectra and fit residuals for the MAXI J1631-479 data. Explanations are analogous to Figure \ref{fig:AT_2019wey_delchi}. Figure discussed in Section \ref{sec:MAXI_J1631-479}.}
    \label{fig:MAXI_J1631_delchi}
\end{figure}

\begin{figure}[ht]
    \centering
    \includegraphics[width= 0.85\textwidth]{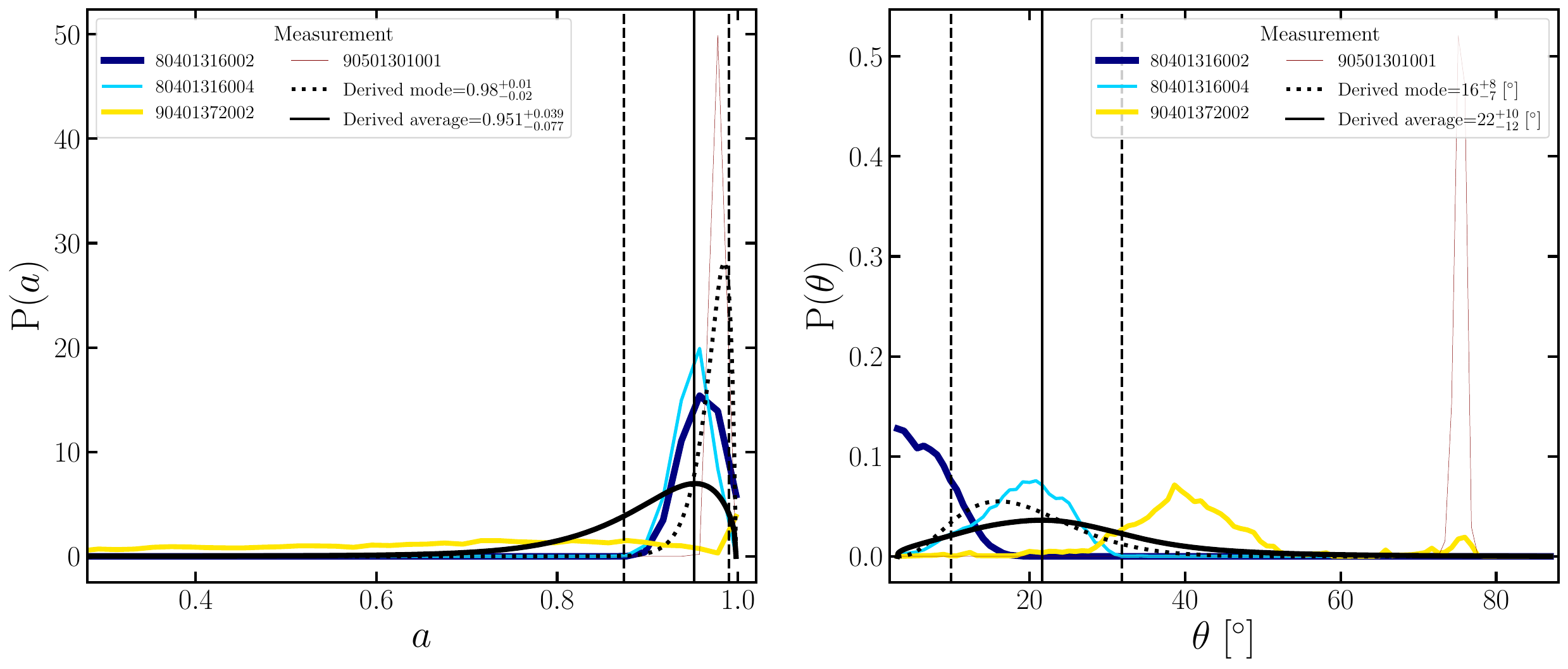}
    \caption{Posterior distributions for the analysis of MAXI J1631-479 data. Explanations are analogous to Figure \ref{fig:AT_2019wey_combined}. Figure discussed in Section \ref{sec:MAXI_J1631-479}.}
    \label{fig:MAXI_J1631_combined}
\end{figure}

\begin{figure}[ht]
    \centering
    \includegraphics[width= 0.8\textwidth]{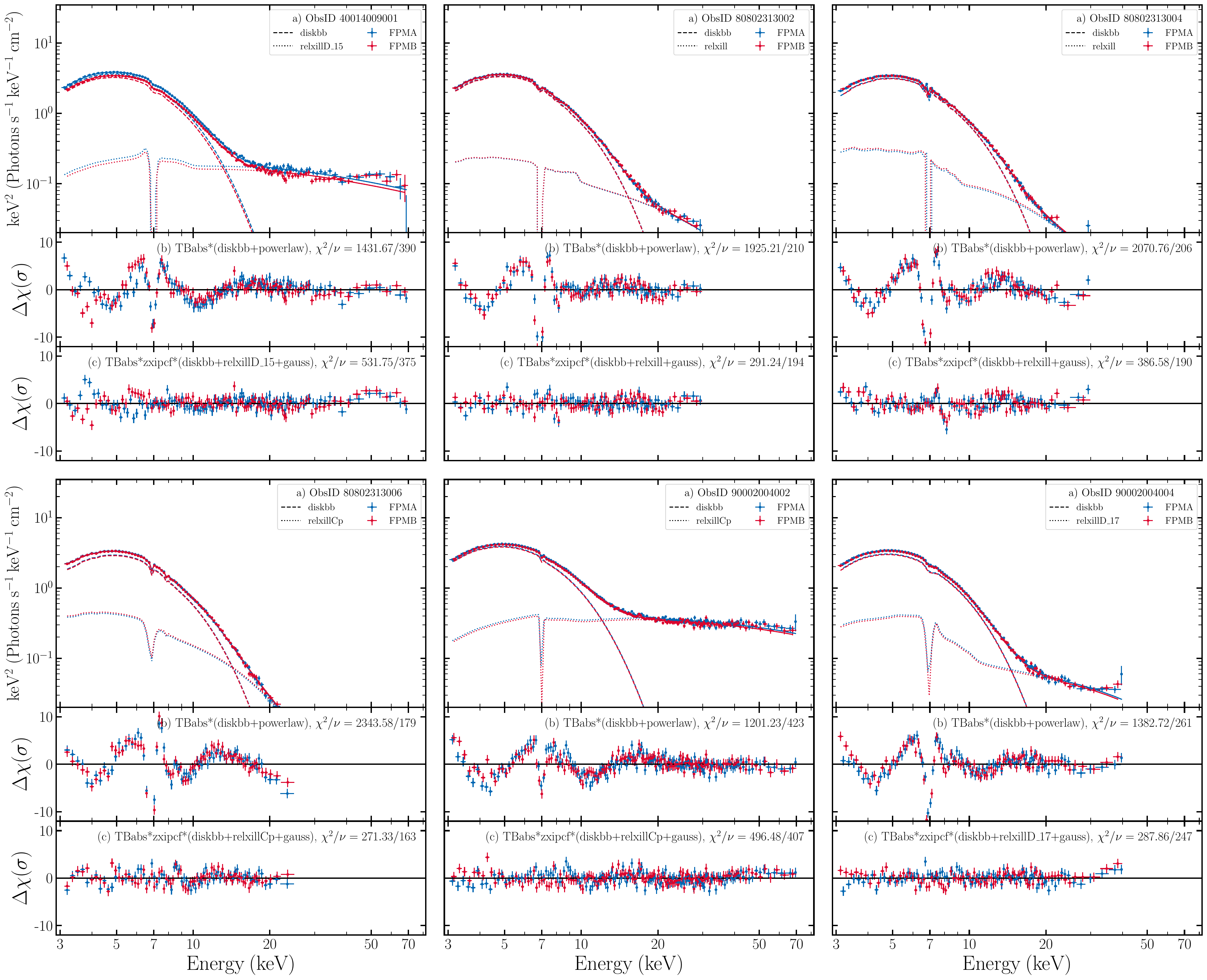}
    \caption{Unfolded spectra and fit residuals for the 4U 1630-472 data. Explanations are analogous to Figure \ref{fig:AT_2019wey_delchi}. Figure discussed in Section \ref{sec:4U_1630-472}.}
    \label{fig:4U_1630_delchi}
\end{figure}
\begin{figure}[ht]
    \centering
    \includegraphics[width= 0.95\textwidth]{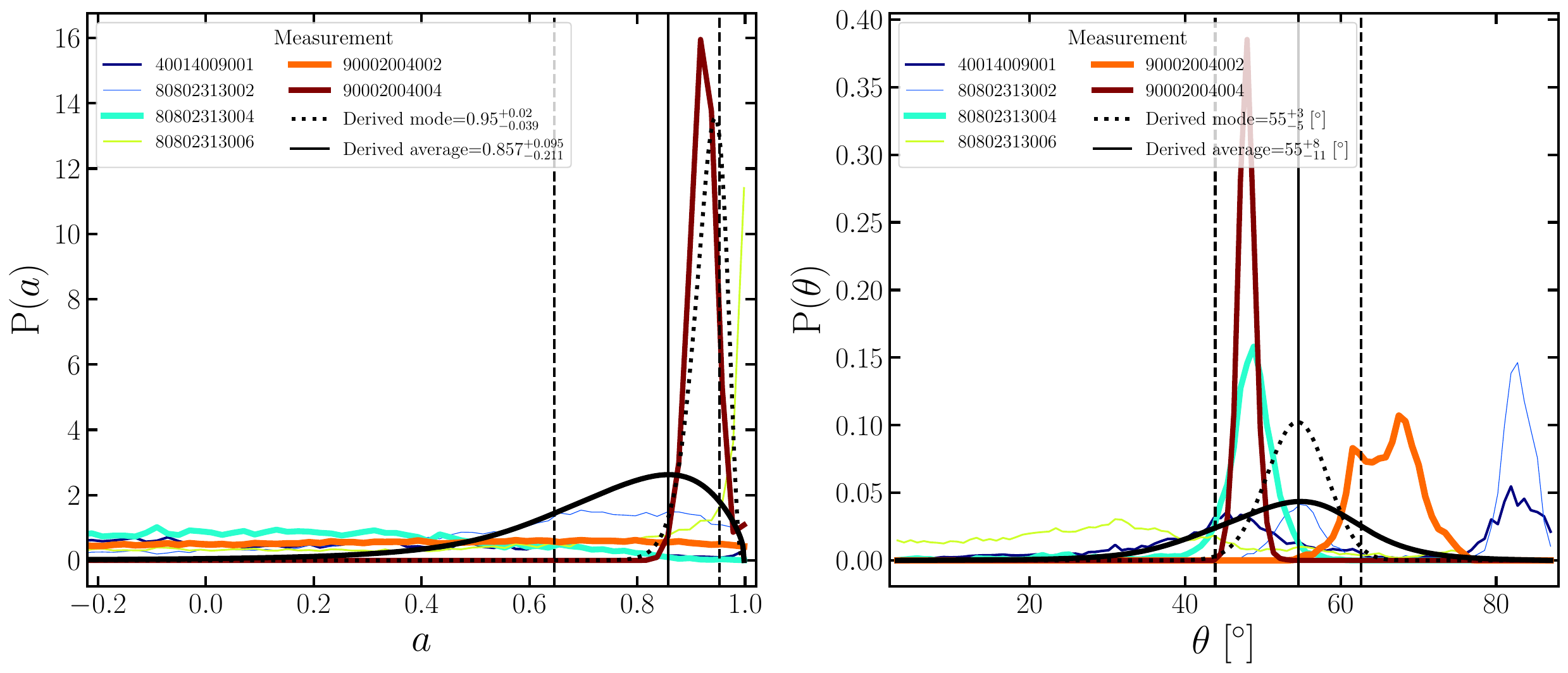}
    \caption{Posterior distributions for the analysis of 4U 1630-472 data. Explanations are analogous to Figure \ref{fig:AT_2019wey_combined}. Figure discussed in Section \ref{sec:4U_1630-472}.}
    \label{fig:4U_1630_combined}
\end{figure}

\begin{figure}[ht]
    \centering
    \includegraphics[width= 0.7\textwidth]{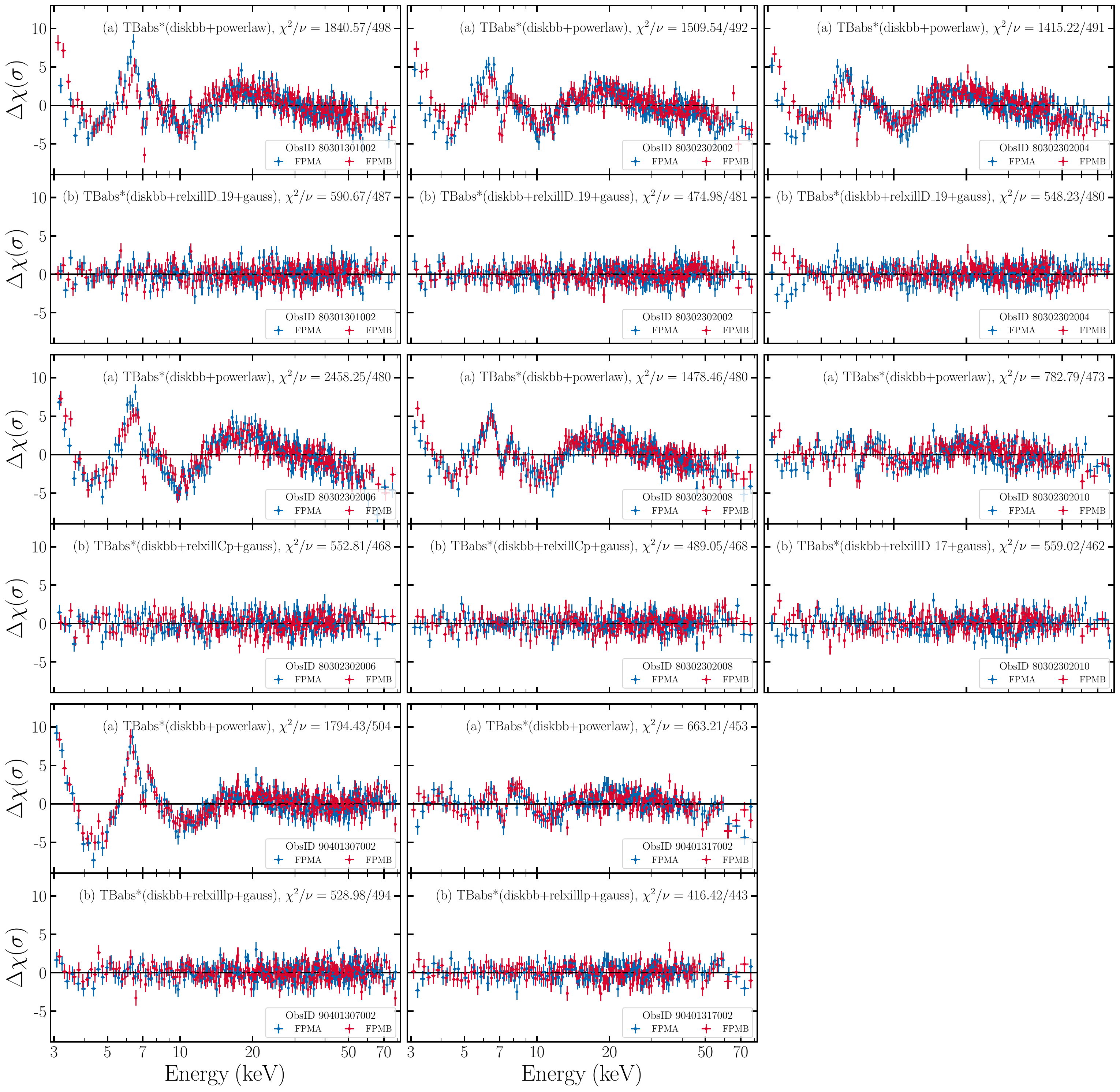}
    \caption{Fit residuals for the Swift J1658-4242 data. Explanations are analogous to Figure \ref{fig:AT_2019wey_delchi}. Figure discussed in Section \ref{sec:Swift_J1658.2-4242}.}
    \label{fig:Swift_J1658-4242_delchi}
\end{figure}
\begin{figure}[ht]
    \centering
    \includegraphics[width= 0.95\textwidth]{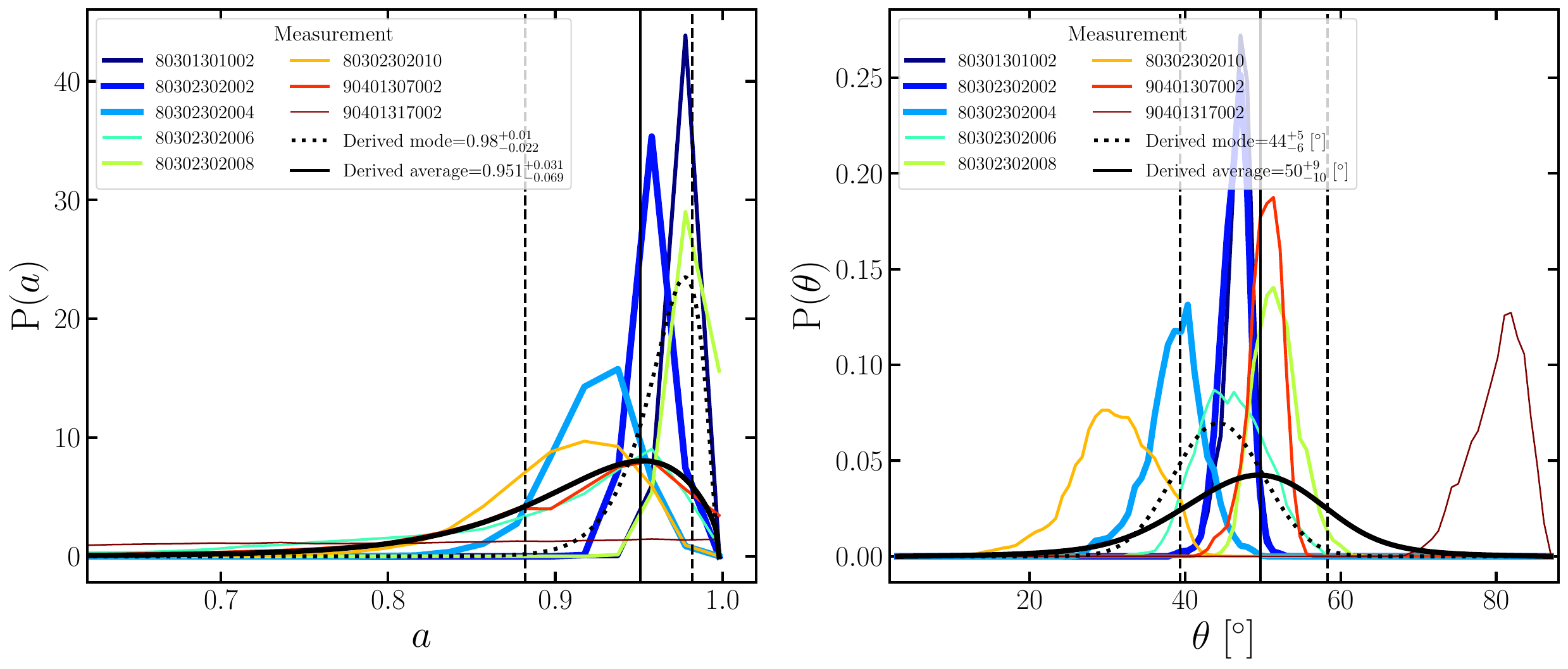}
    \caption{Posterior distributions for the analysis of Swift J1658-4242 data. Explanations are analogous to Figure \ref{fig:AT_2019wey_combined}. Figure discussed in Section \ref{sec:Swift_J1658.2-4242}.}
    \label{fig:Swift_J1658-4242_combined}
\end{figure}

\begin{figure}[ht]
    \centering
    \includegraphics[width= 0.95\textwidth]{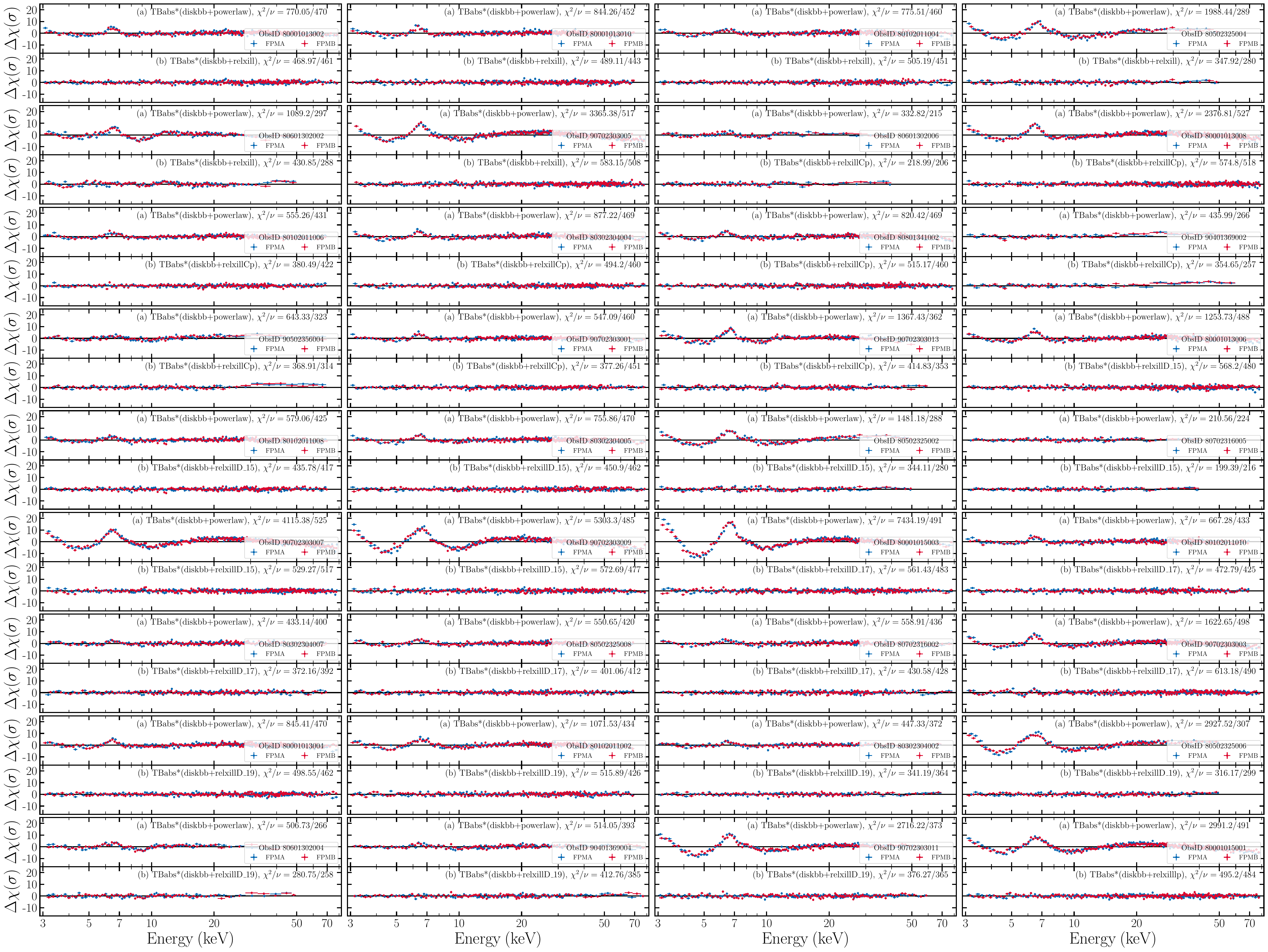}
    \caption{Fit residuals for the GX 339-4 data. Explanations are analogous to Figure \ref{fig:AT_2019wey_delchi}. Figure discussed in Section \ref{sec:GX_339-4}.}
    \label{fig:GX_339_delchi}
\end{figure}
\begin{figure}[ht]
    \centering
    \includegraphics[width= 0.95\textwidth]{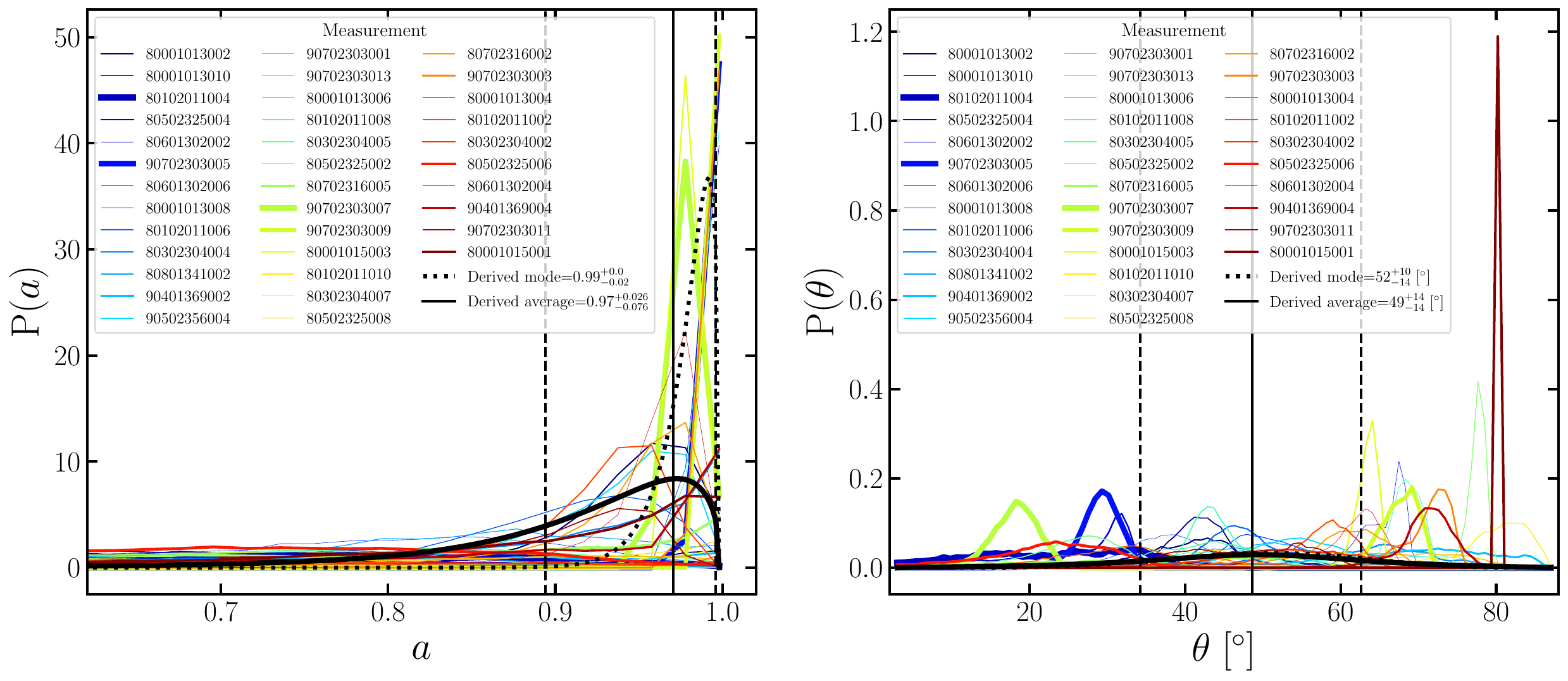}
    \caption{Posterior distributions for the analysis of GX 339-4 data. Explanations are analogous to Figure \ref{fig:AT_2019wey_combined}. Figure discussed in Section \ref{sec:GX_339-4}.}
    \label{fig:GX_339_combined}
\end{figure}

\begin{figure}[ht]
    \centering
    \includegraphics[width= 0.7\textwidth]{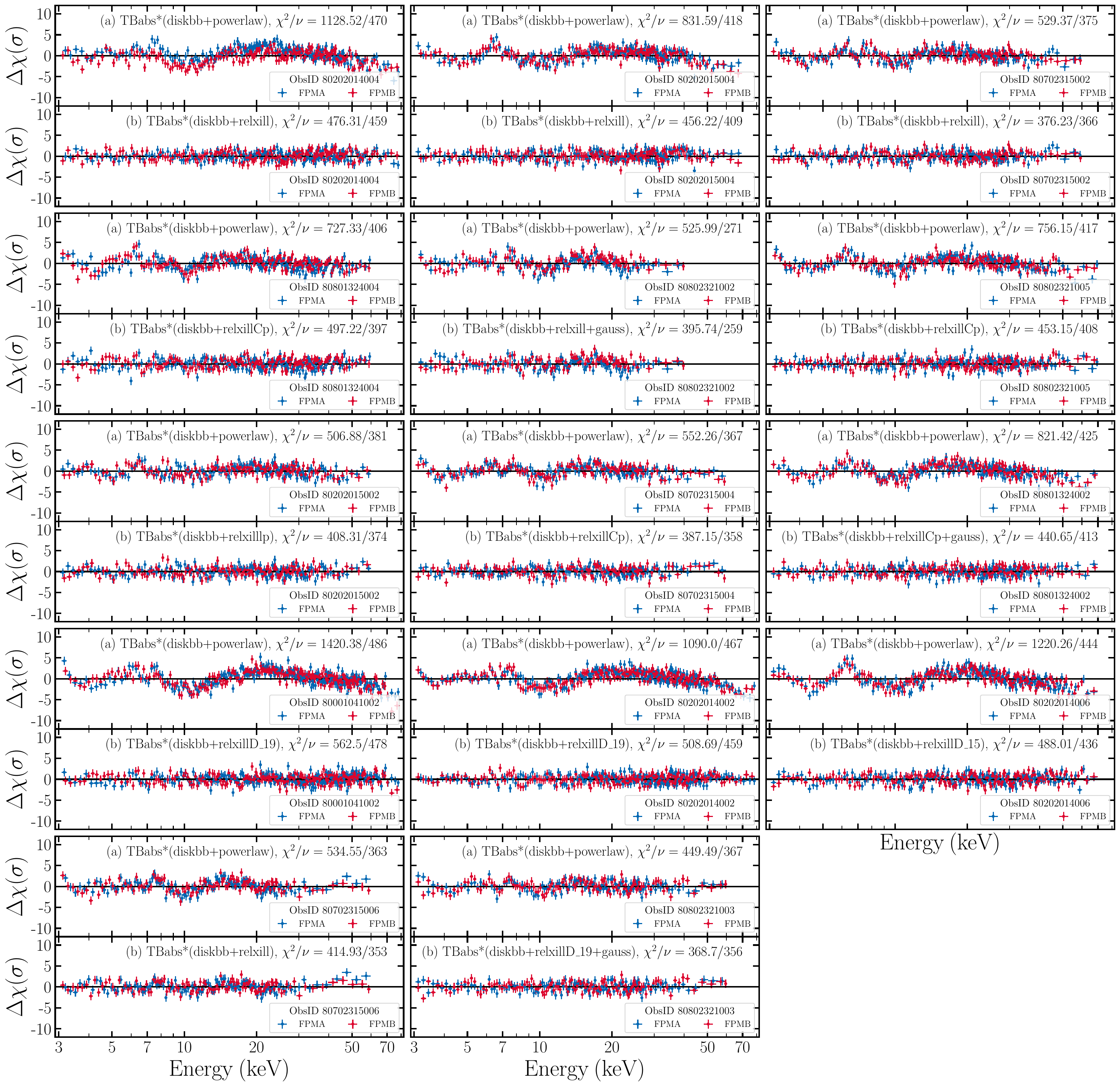}
    \caption{Fit residuals for the IGR J17091-3624 data. Explanations are analogous to Figure \ref{fig:AT_2019wey_delchi}. Figure discussed in Section \ref{sec:IGR_J17091-3624}.}
    \label{fig:IGR_J17091_delchi}
\end{figure}
\begin{figure}[ht]
    \centering
    \includegraphics[width= 0.85\textwidth]{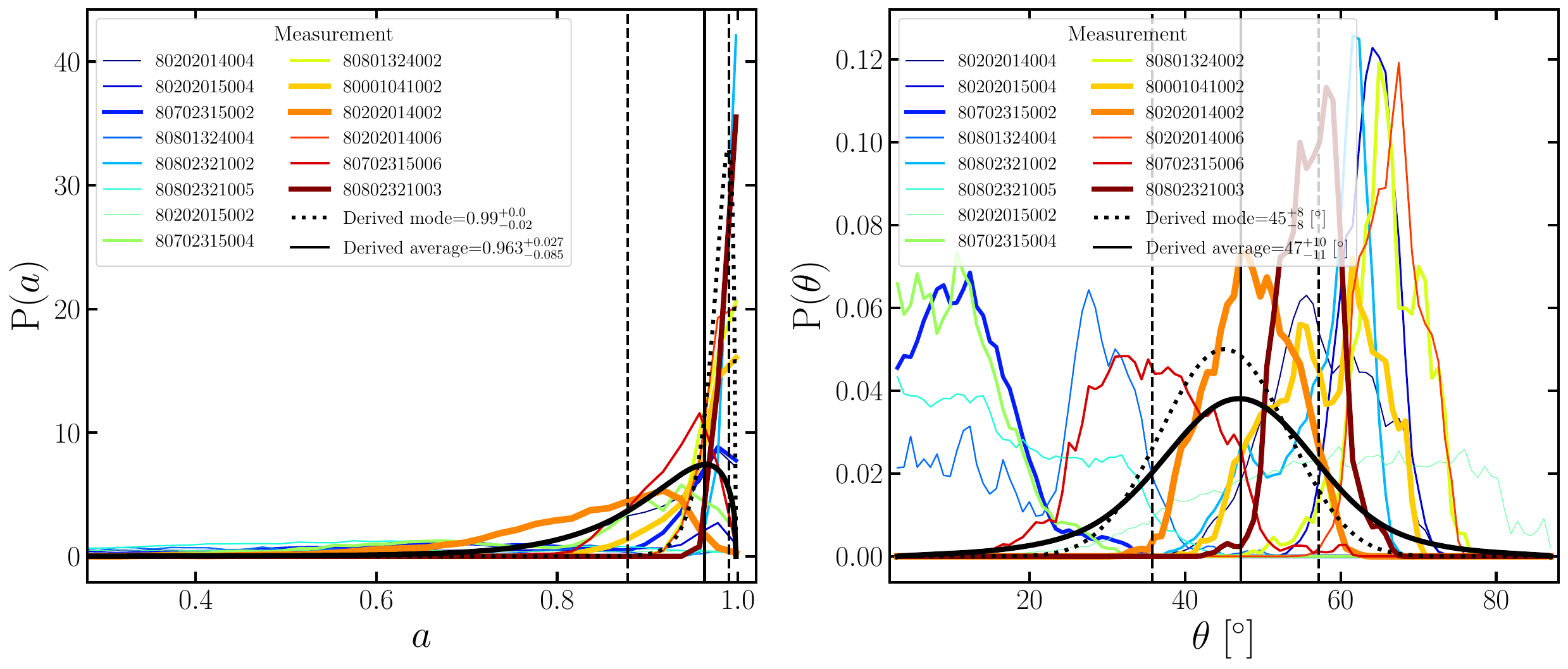}
    \caption{Posterior distributions for the analysis of IGR J17091-3624 data. Explanations are analogous to Figure \ref{fig:AT_2019wey_combined}. Figure discussed in Section \ref{sec:IGR_J17091-3624}.}
    \label{fig:IGR_J17091_combined}
\end{figure}

\begin{figure}[ht]
    \centering
    \includegraphics[width= 0.8\textwidth]{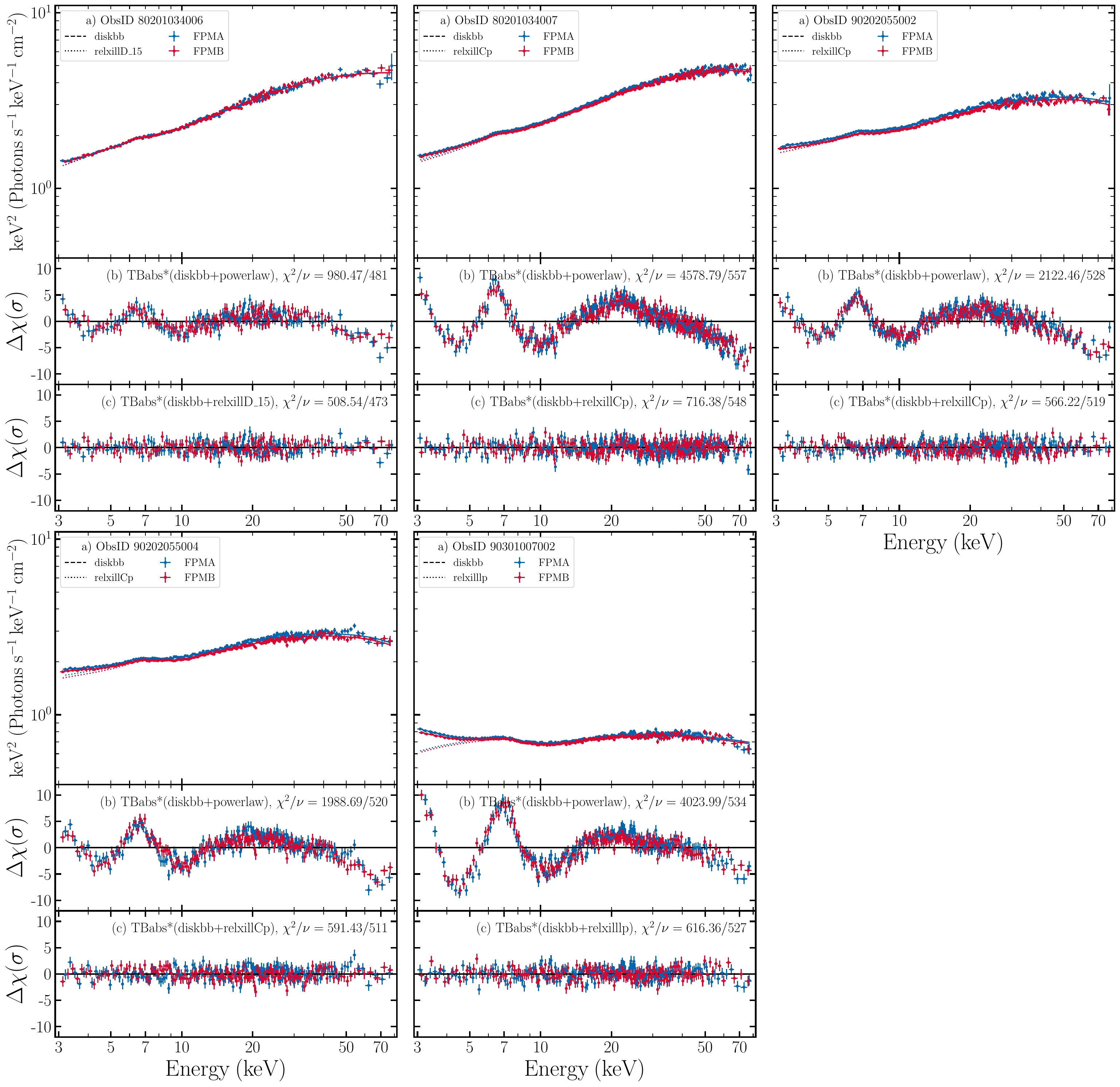}
    \caption{Unfolded spectra and fit residuals for the GRS 1716-249 data. Explanations are analogous to Figure \ref{fig:AT_2019wey_delchi}. Figure discussed in Section \ref{sec:GRS_1716-249}.}
    \label{fig:GRS_1716_delchi}
\end{figure}

\begin{figure}[ht]
    \centering
    \includegraphics[width= 0.85\textwidth]{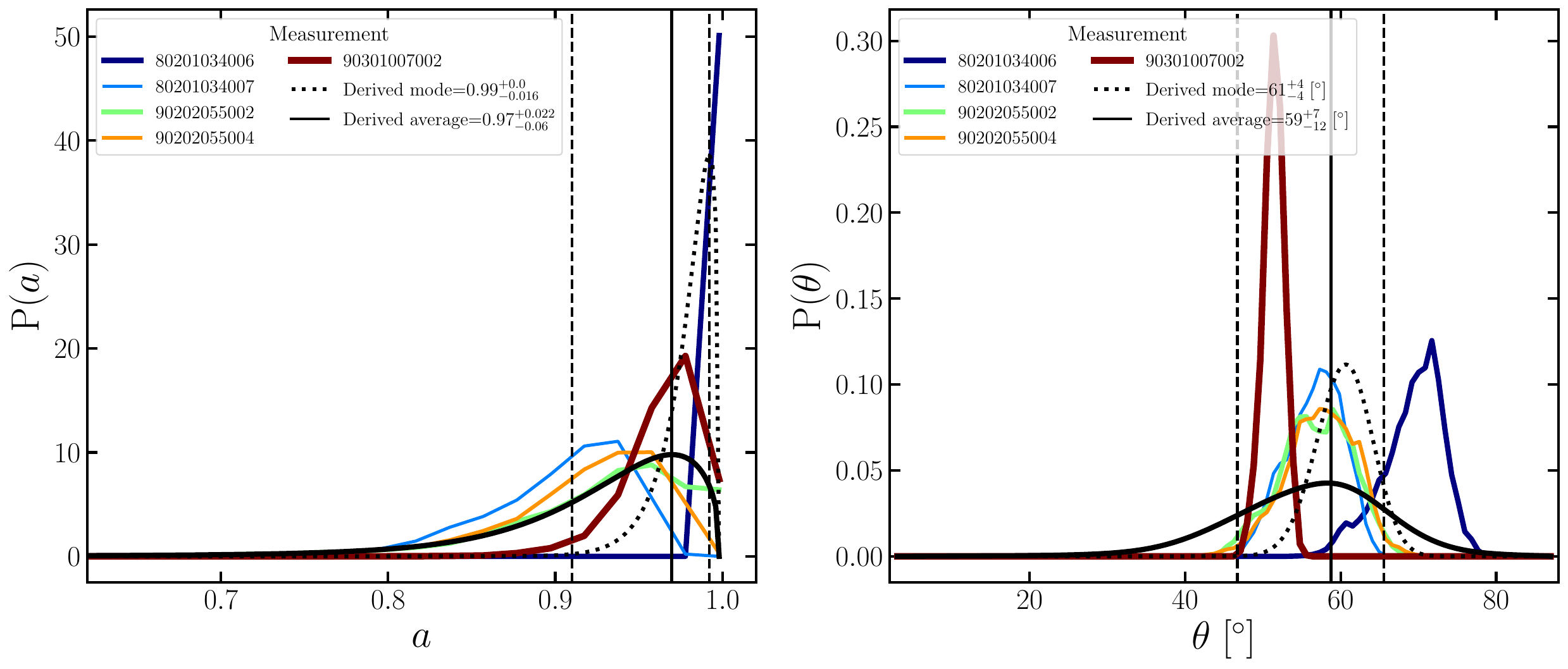}
    \caption{Posterior distributions for the analysis of GRS 1716-249 data. Explanations are analogous to Figure \ref{fig:AT_2019wey_combined}. Figure discussed in Section \ref{sec:GRS_1716-249}.}
    \label{fig:GRS_1716_combined}
\end{figure}

\begin{figure}[ht]
    \centering
    \includegraphics[width= 0.95\textwidth]{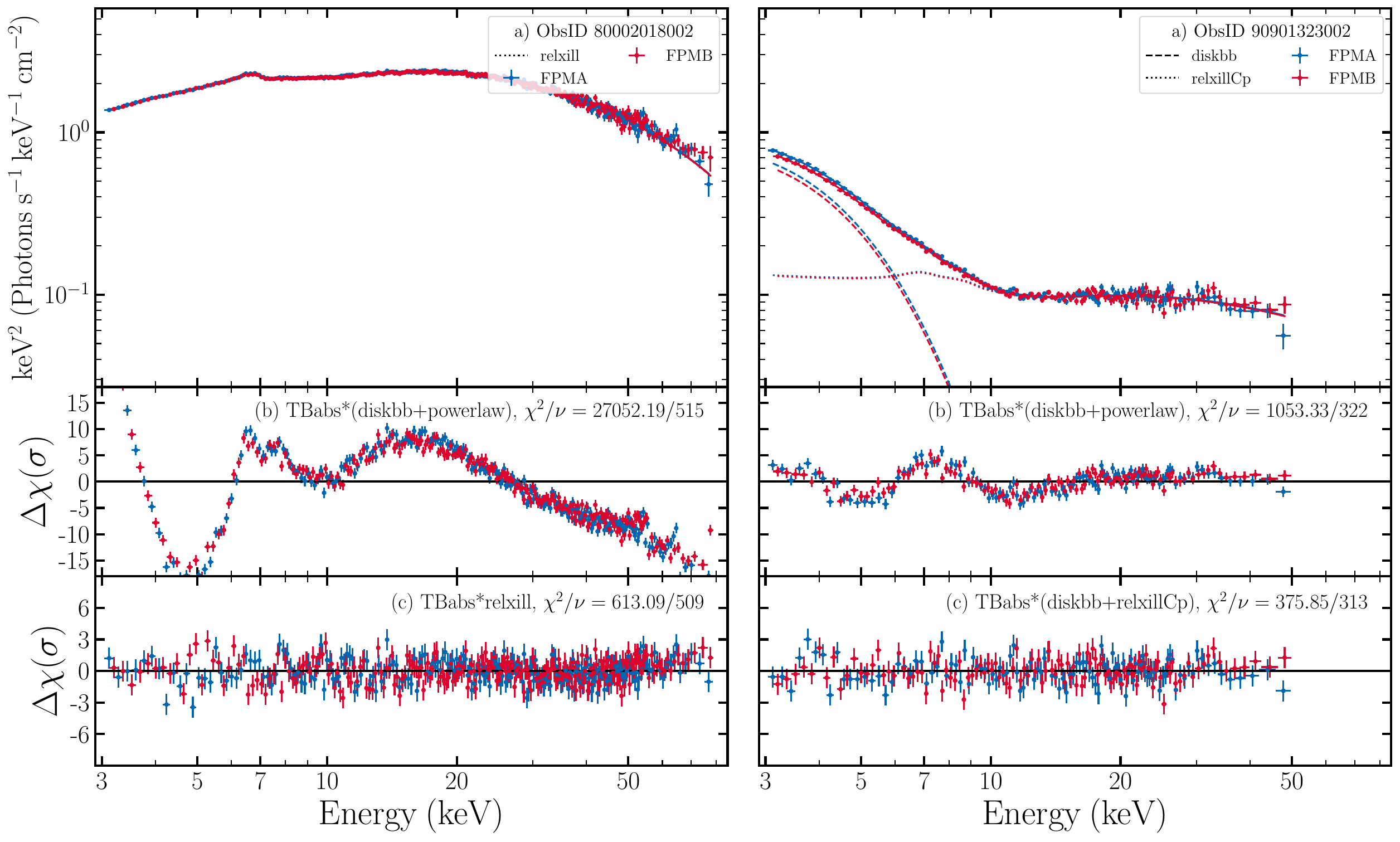}
    \caption{Unfolded spectra and fit residuals for the GRS 1739-278 data. Explanations are analogous to Figure \ref{fig:AT_2019wey_delchi}. Figure discussed in Section \ref{sec:GRS_1739-278}.}
    \label{fig:GRS_1739_delchi}
\end{figure}
\begin{figure}[ht]
    \centering
    \includegraphics[width= 0.95\textwidth]{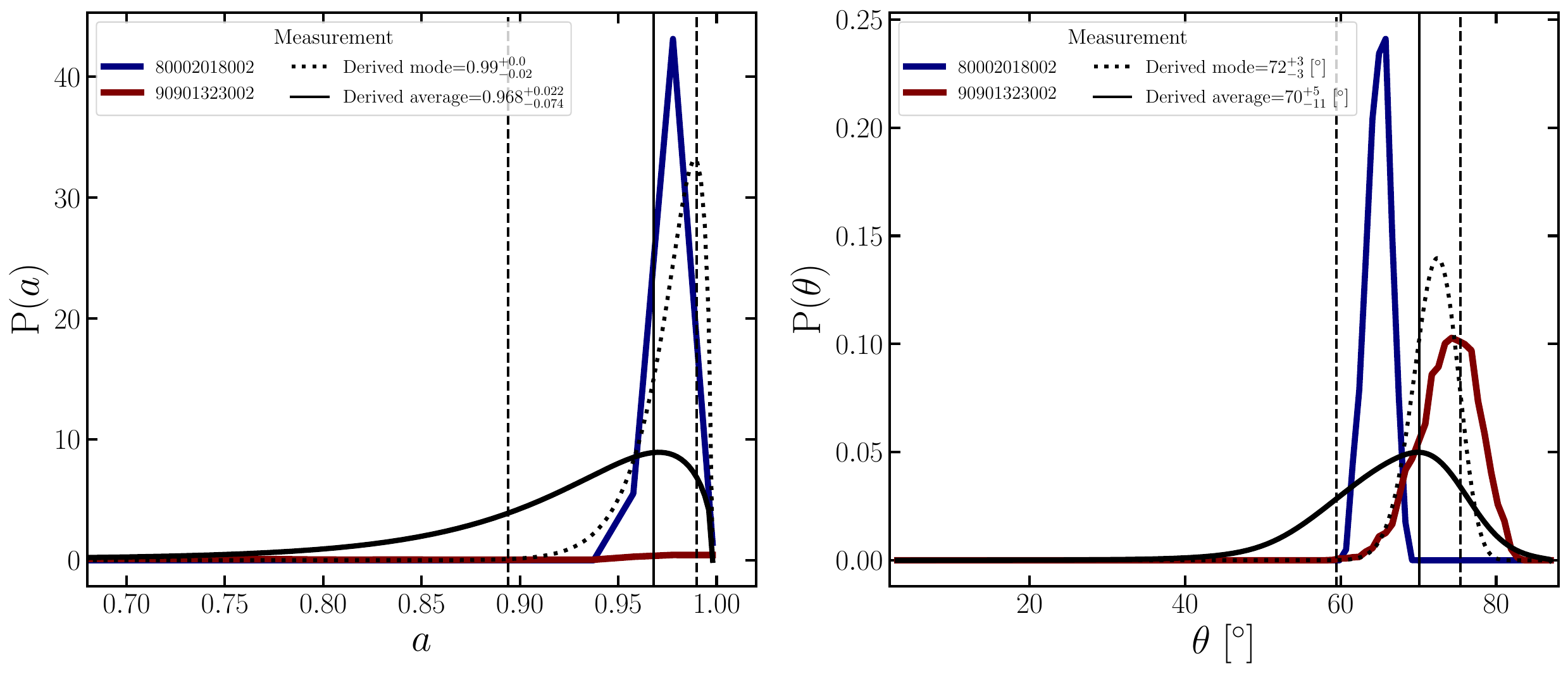}
    \caption{Posterior distributions for the analysis of GRS 1739-278 data. Explanations are analogous to Figure \ref{fig:AT_2019wey_combined}. Figure discussed in Section \ref{sec:GRS_1739-278}.}
    \label{fig:GRS_1739_combined}
\end{figure}

\begin{figure}[ht]
    \centering
    \includegraphics[width= 0.6\textwidth]{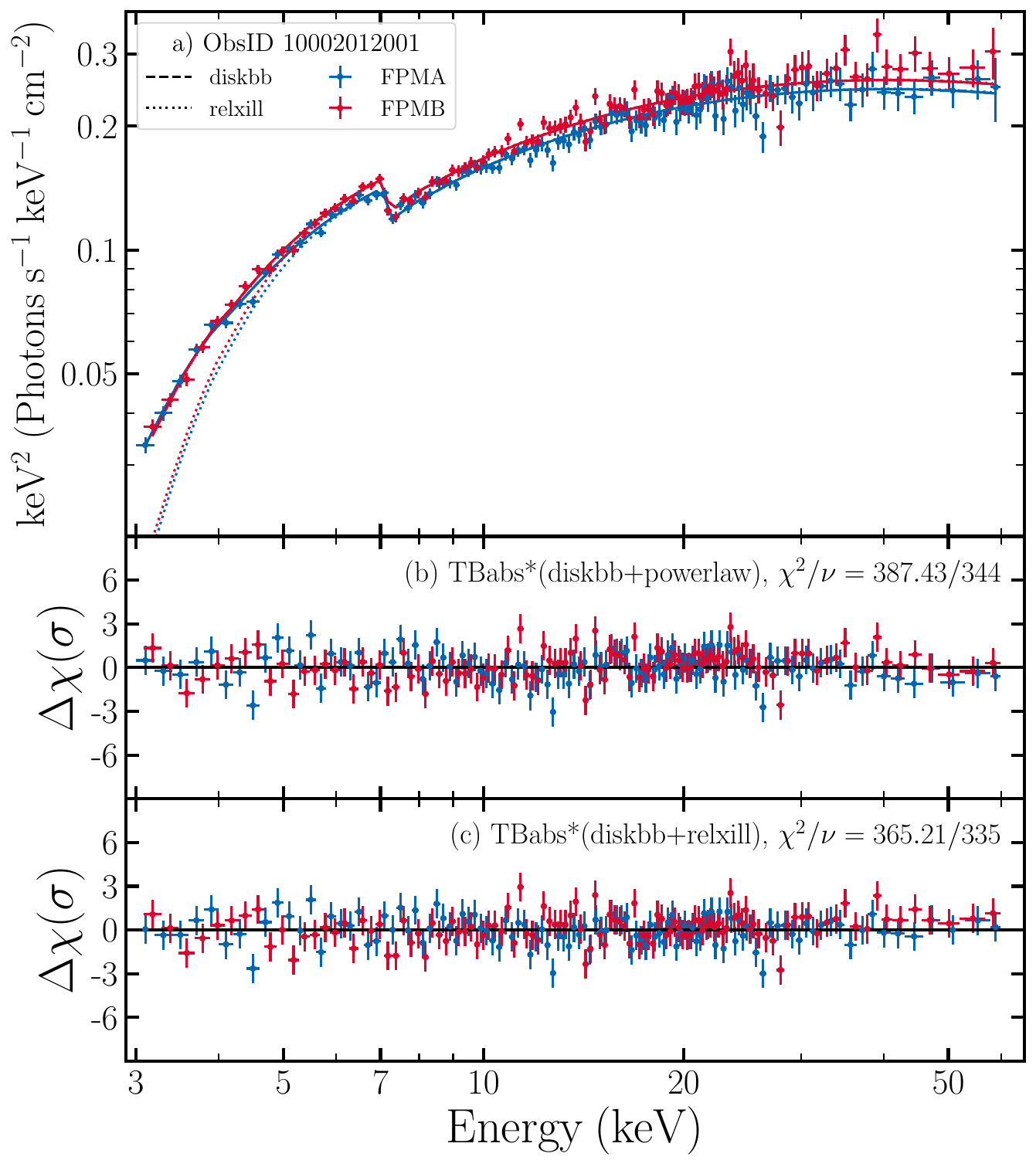}
    \caption{Unfolded spectra and fit residuals for the 1E 1740.7-2942 data. Explanations are analogous to Figure \ref{fig:AT_2019wey_delchi}. Figure discussed in Section \ref{sec:1E_1740.7-2942}.}
    \label{fig:1E_1740_delchi}
\end{figure}
\begin{figure}[ht]
    \centering
    \includegraphics[width= 0.8\textwidth]{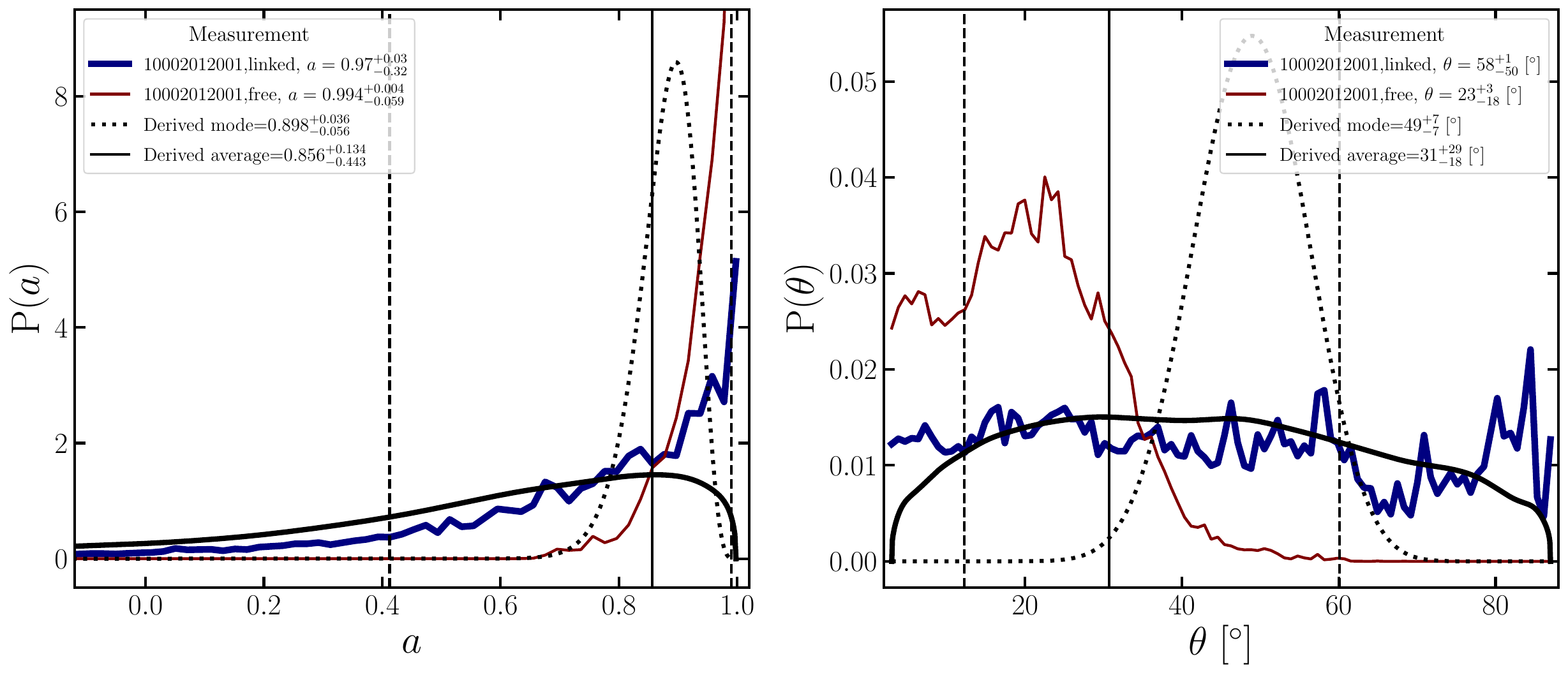}
    \caption{Posterior distributions for the analysis of 1E 1740.7-294 data. Explanations are analogous to Figure \ref{fig:AT_2019wey_combined}. The black curves represent the distributions inferred using our posterior combining algorithm on the individual posterior distribution from the MCMC run on ObsID 10002012001. The red curves show the posterior distributions for spin and inclination obtained when running the MCMC analysis on fits to the NuSTAR spectra from ObsID 10002012001 when allowing the power law index to vary freely between the spectra from the two NuSTAR FPM detectors. Figure discussed in Section \ref{sec:1E_1740.7-2942}.}
    \label{fig:1E_1740_combined}
\end{figure}

\begin{figure}[ht]
    \centering
    \includegraphics[width= 0.6\textwidth]{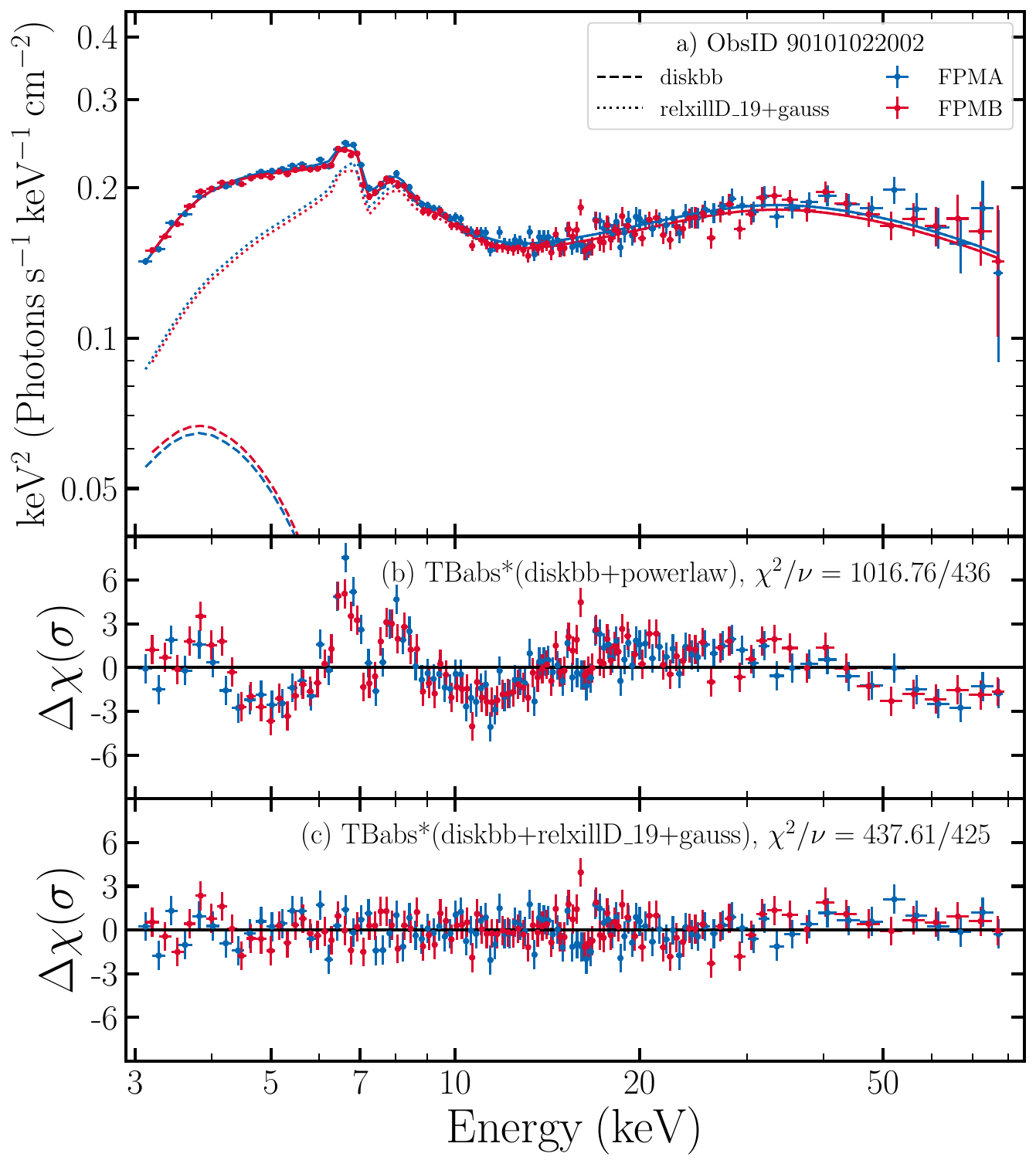}
    \caption{Unfolded spectra and fit residuals for the T15 data. Explanations are analogous to Figure \ref{fig:AT_2019wey_delchi}. Figure discussed in Section \ref{sec:T15}.}
    \label{fig:T15_delchi}
\end{figure}
\begin{figure}[ht]
    \centering
    \includegraphics[width= 0.8\textwidth]{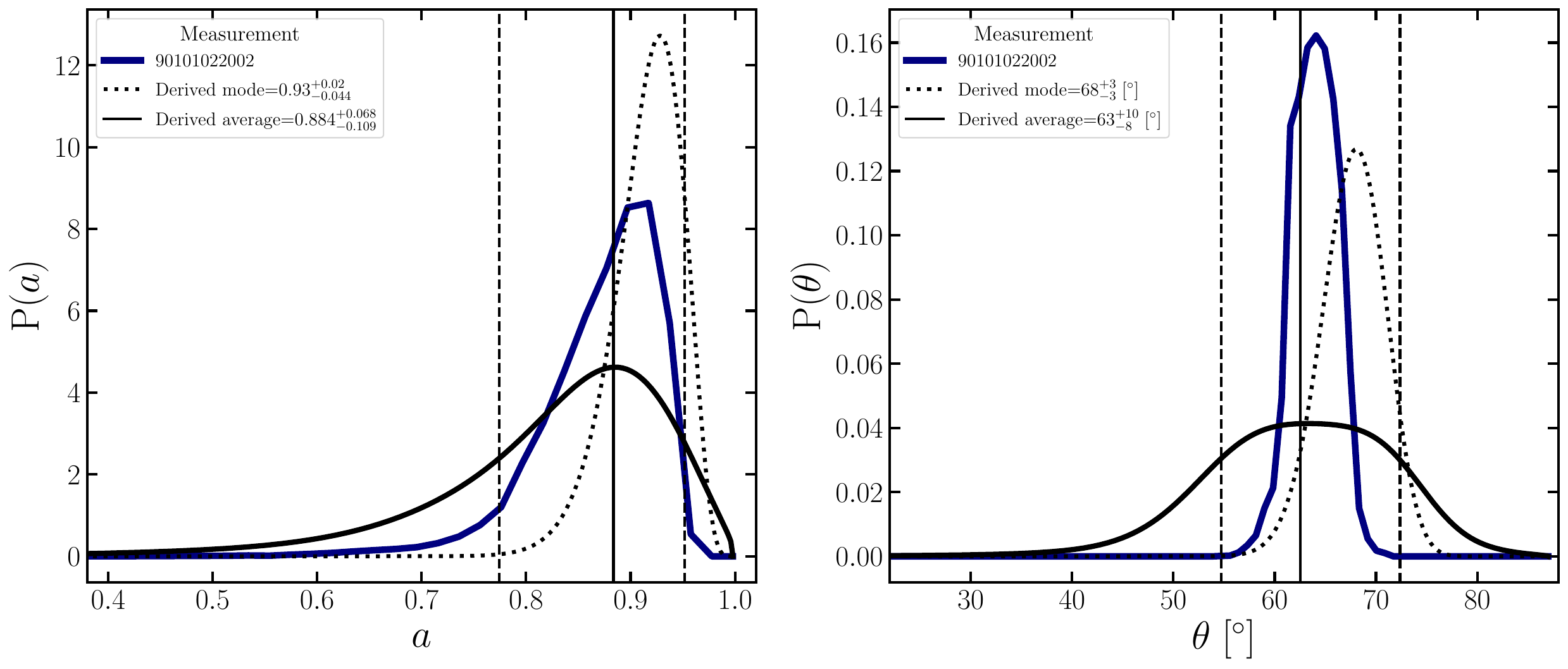}
    \caption{Posterior distributions for the analysis of T15 data. Explanations are analogous to Figure \ref{fig:AT_2019wey_combined}. The black curves represent the distributions inferred using our posterior combining algorithm on the individual posterior distribution from the MCMC run on ObsID 90101022002.  Figure discussed in Section \ref{sec:T15}.}
    \label{fig:T15_combined}
\end{figure}

\begin{figure}[ht]
    \centering
    \includegraphics[width= 0.6\textwidth]{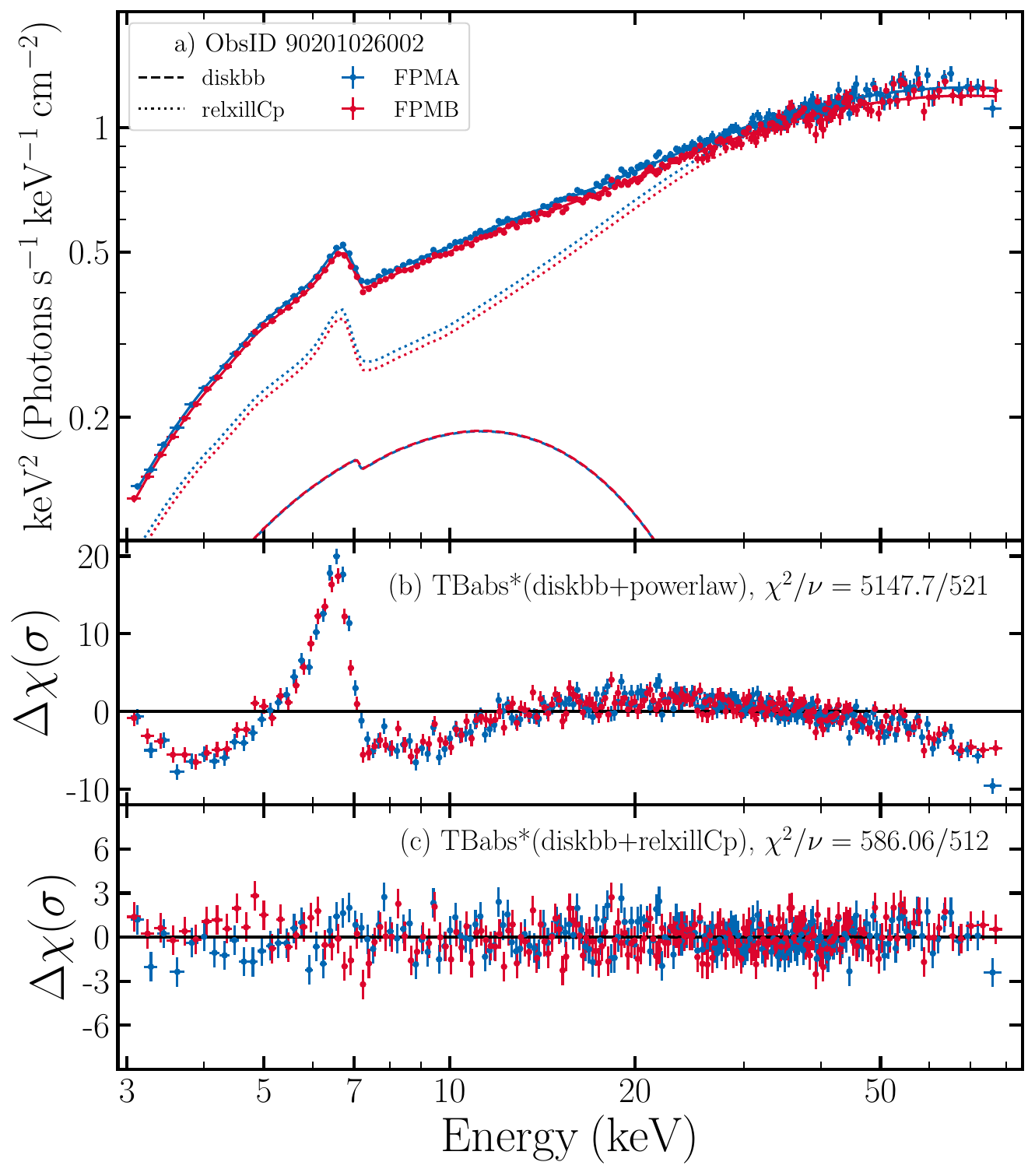}
    \caption{Unfolded spectra and fit residuals for the T37 data. Explanations are analogous to Figure \ref{fig:AT_2019wey_delchi}. Figure discussed in Section \ref{sec:T37}.}
    \label{fig:T37_delchi}
\end{figure}
\begin{figure}[ht]
    \centering
    \includegraphics[width= 0.8\textwidth]{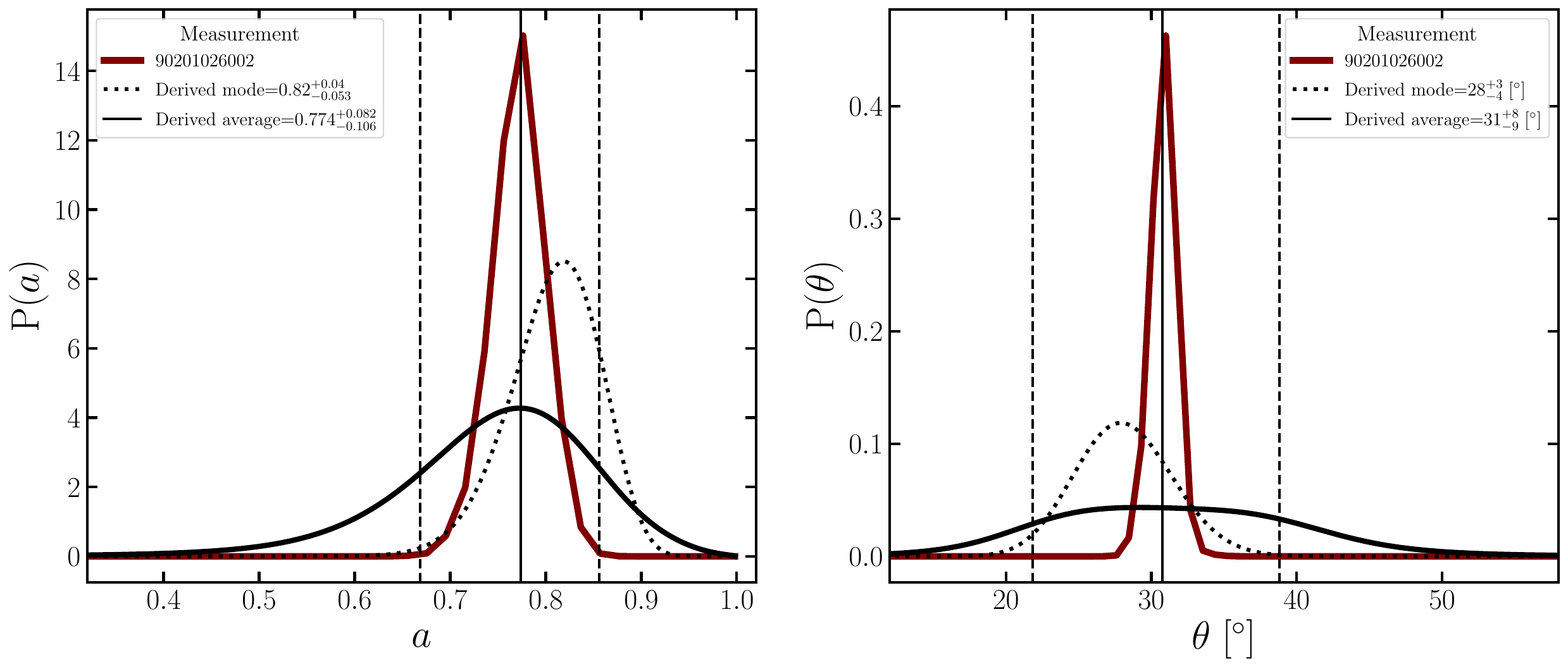}
    \caption{Posterior distributions for the analysis of T37 data. Explanations are analogous to Figure \ref{fig:AT_2019wey_combined}. The black curves represent the distributions inferred using our posterior combining algorithm on the individual posterior distribution from the MCMC run on ObsID 90201026002. Figure discussed in Section \ref{sec:T37}.}
    \label{fig:T37_combined}
\end{figure}

\begin{figure}[ht]
    \centering
    \includegraphics[width= 0.65\textwidth]{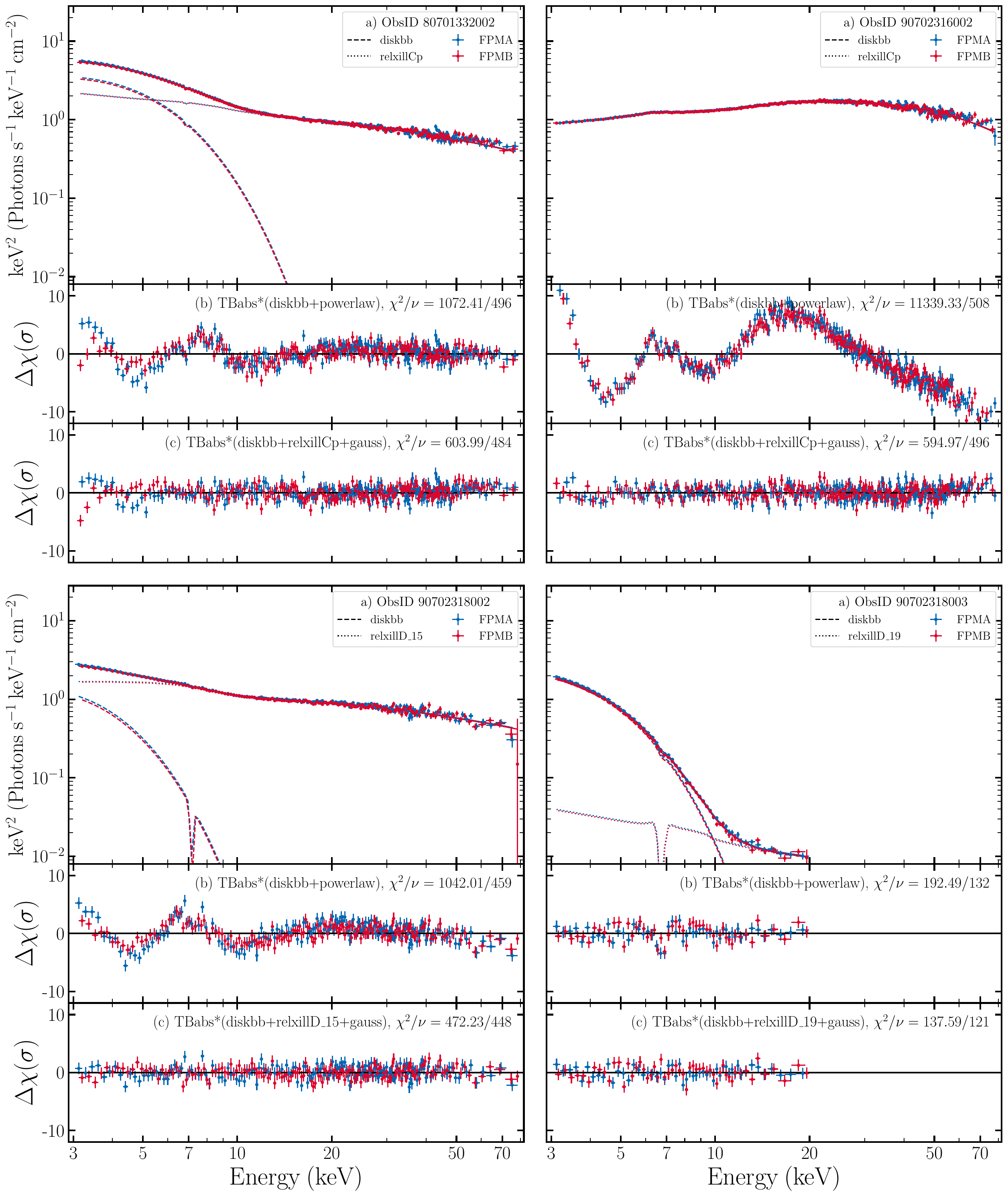}
    \caption{Unfolded spectra and fit residuals for the MAXI J1803-298 data. Explanations are analogous to Figure \ref{fig:AT_2019wey_delchi}. Figure discussed in Section \ref{sec:MAXI_J1803-298}.}
    \label{fig:MAXI_J1803_delchi}
\end{figure}
\begin{figure}[ht]
    \centering
    \includegraphics[width= 0.8\textwidth]{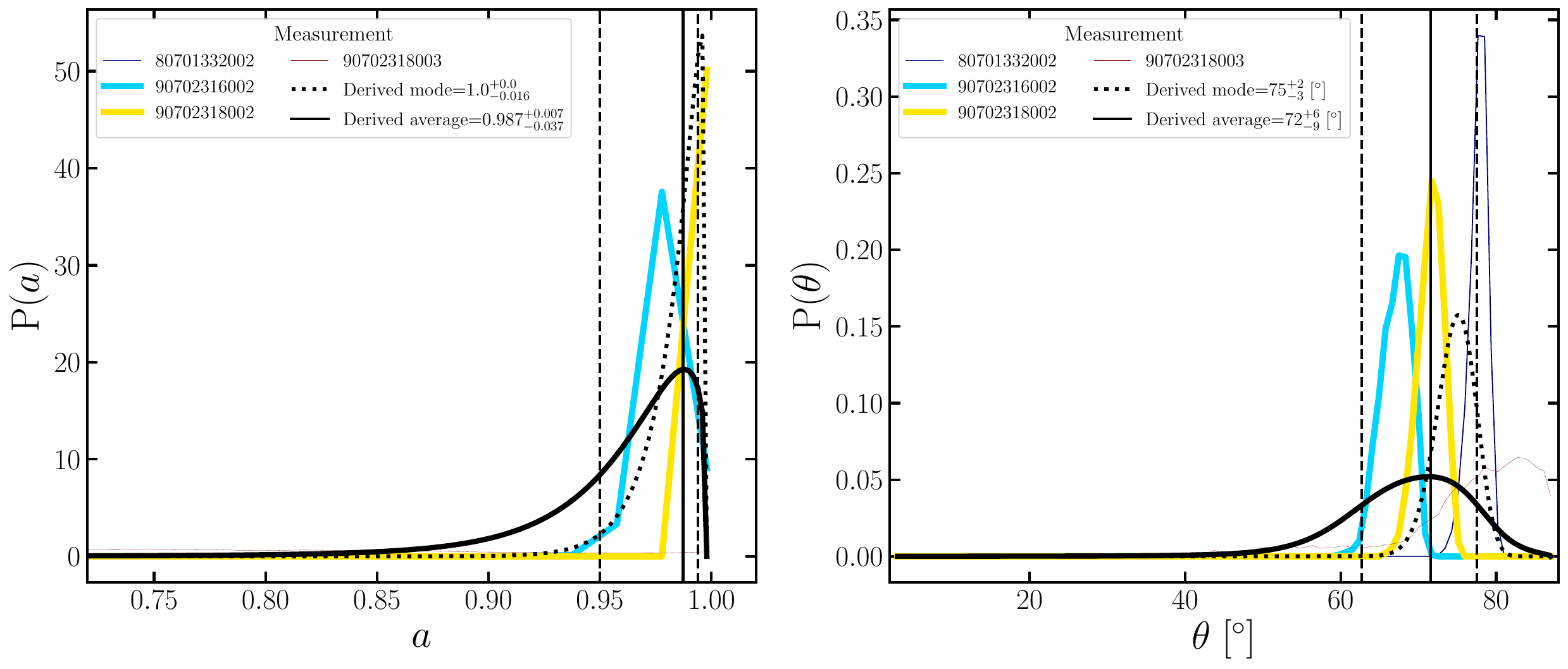}
    \caption{Posterior distributions for the analysis of MAXI J1803-298 data. Explanations are analogous to Figure \ref{fig:AT_2019wey_combined}. Figure discussed in Section \ref{sec:MAXI_J1803-298}.}
    \label{fig:MAXI_J1803_combined}
\end{figure}

\begin{figure}[ht]
    \centering
    \includegraphics[width= 0.95\textwidth]{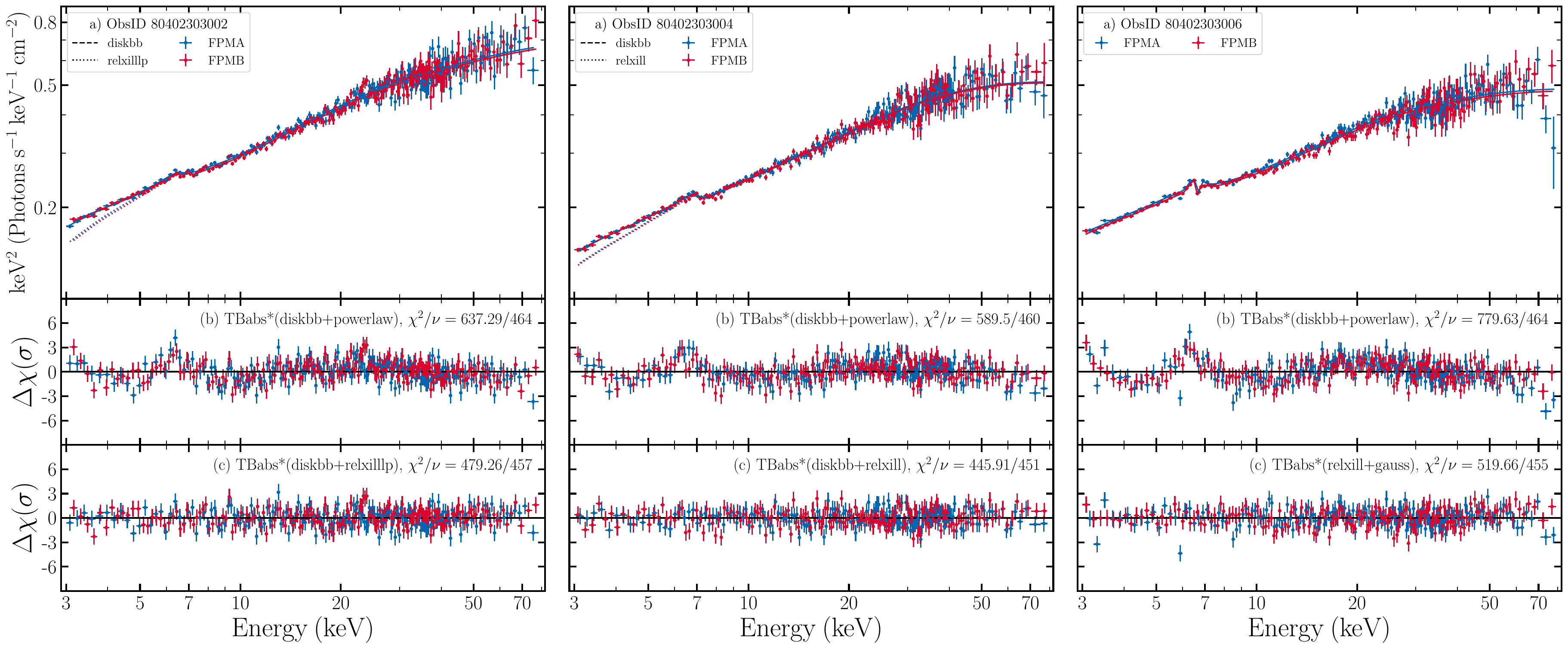}
    \caption{Unfolded spectra and fit residuals for the MAXI J1813-095 data. Explanations are analogous to Figure \ref{fig:AT_2019wey_delchi}. Figure discussed in Section \ref{sec:MAXI_J1813-095}.}
    \label{fig:MAXI_J1813-095_delchi}
\end{figure}
\begin{figure}[ht]
    \centering
    \includegraphics[width= 0.95\textwidth]{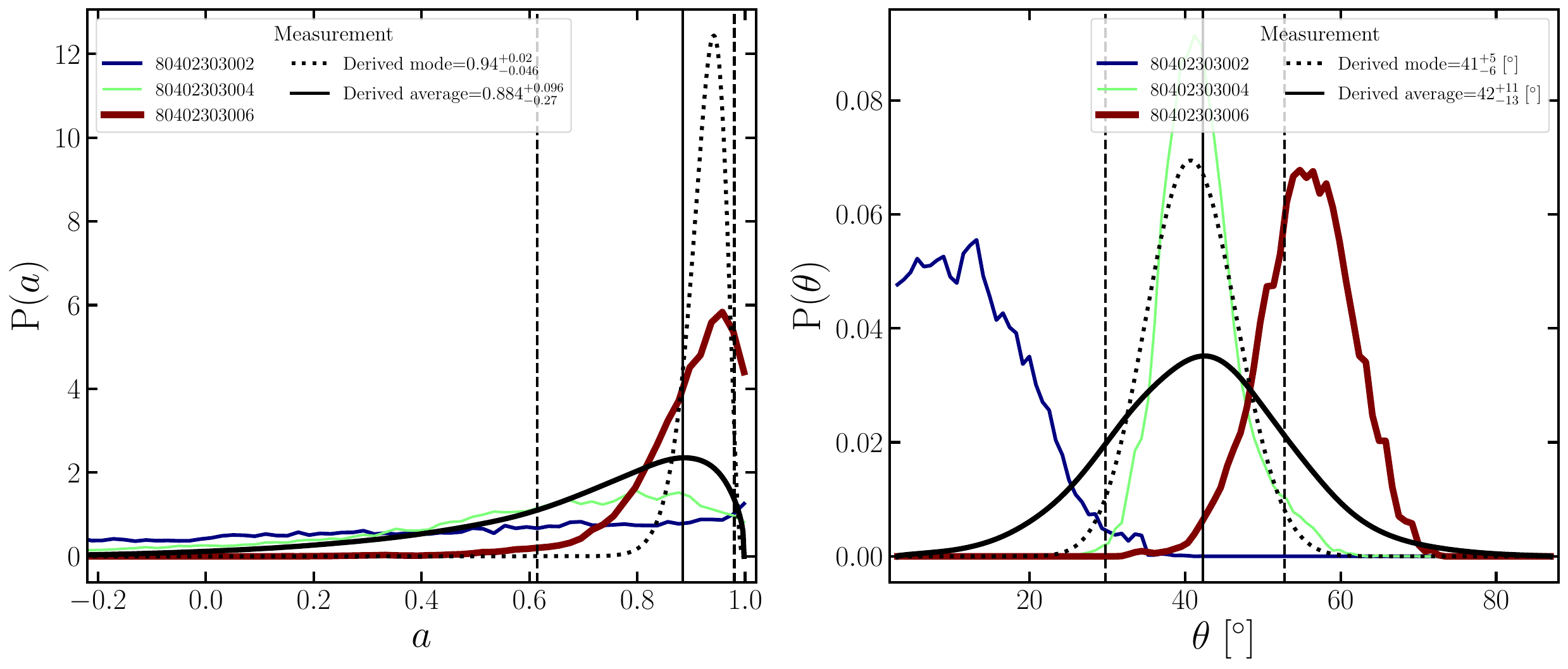}
    \caption{Posterior distributions for the analysis of MAXI J1813-095 data. Explanations are analogous to Figure \ref{fig:AT_2019wey_combined}. Figure discussed in Section \ref{sec:MAXI_J1813-095}.}
    \label{fig:MAXI_J1813-095_combined}
\end{figure}

\begin{figure}[ht]
    \centering
    \includegraphics[width= 0.95\textwidth]{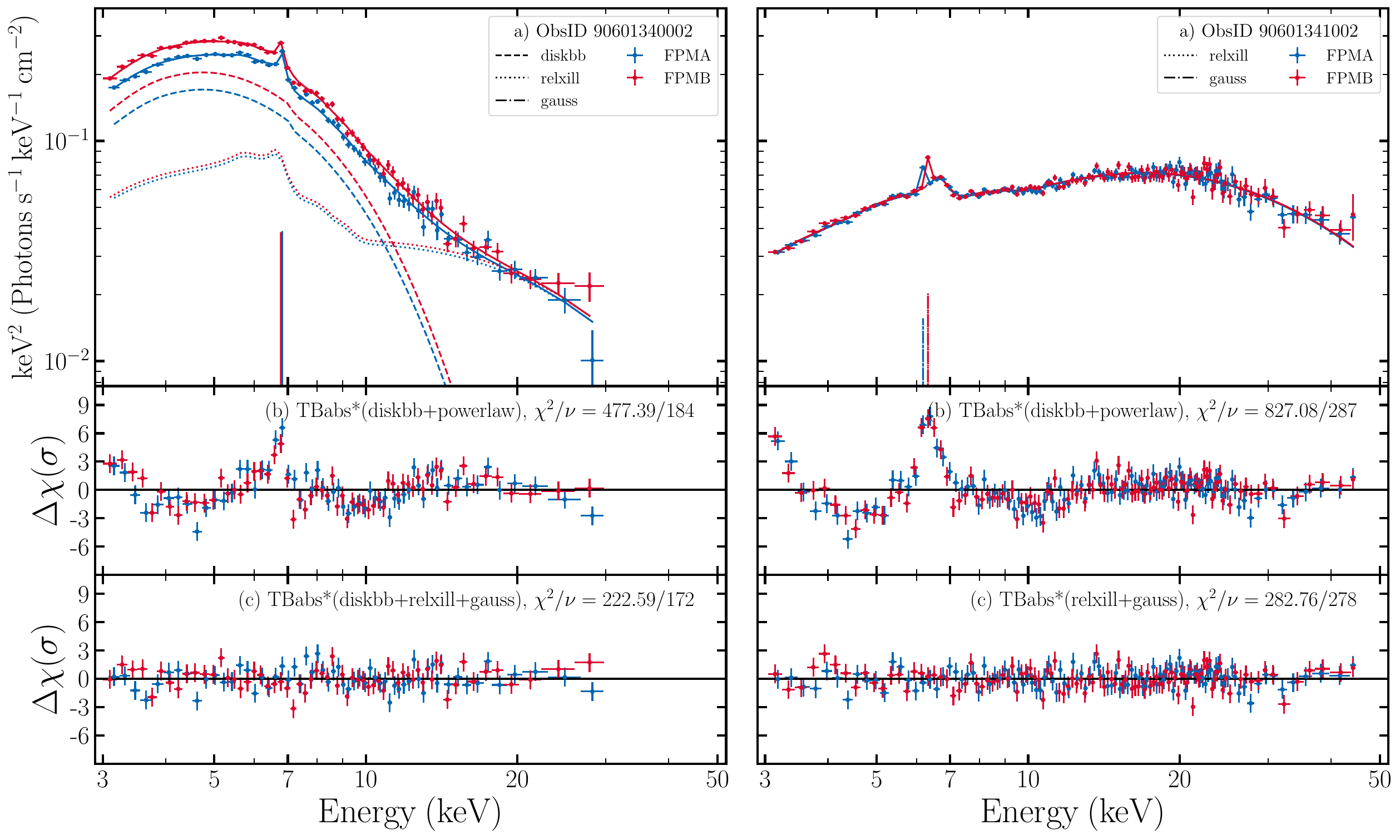}
    \caption{Unfolded spectra and fit residuals for the MAXI J1848-015 data. Explanations are analogous to Figure \ref{fig:AT_2019wey_delchi}. Figure discussed in Section \ref{sec:MAXI_J1848-015}.}
    \label{fig:MAXI_J1848_delchi}
\end{figure}
\begin{figure}[ht]
    \centering
    \includegraphics[width= 0.95\textwidth]{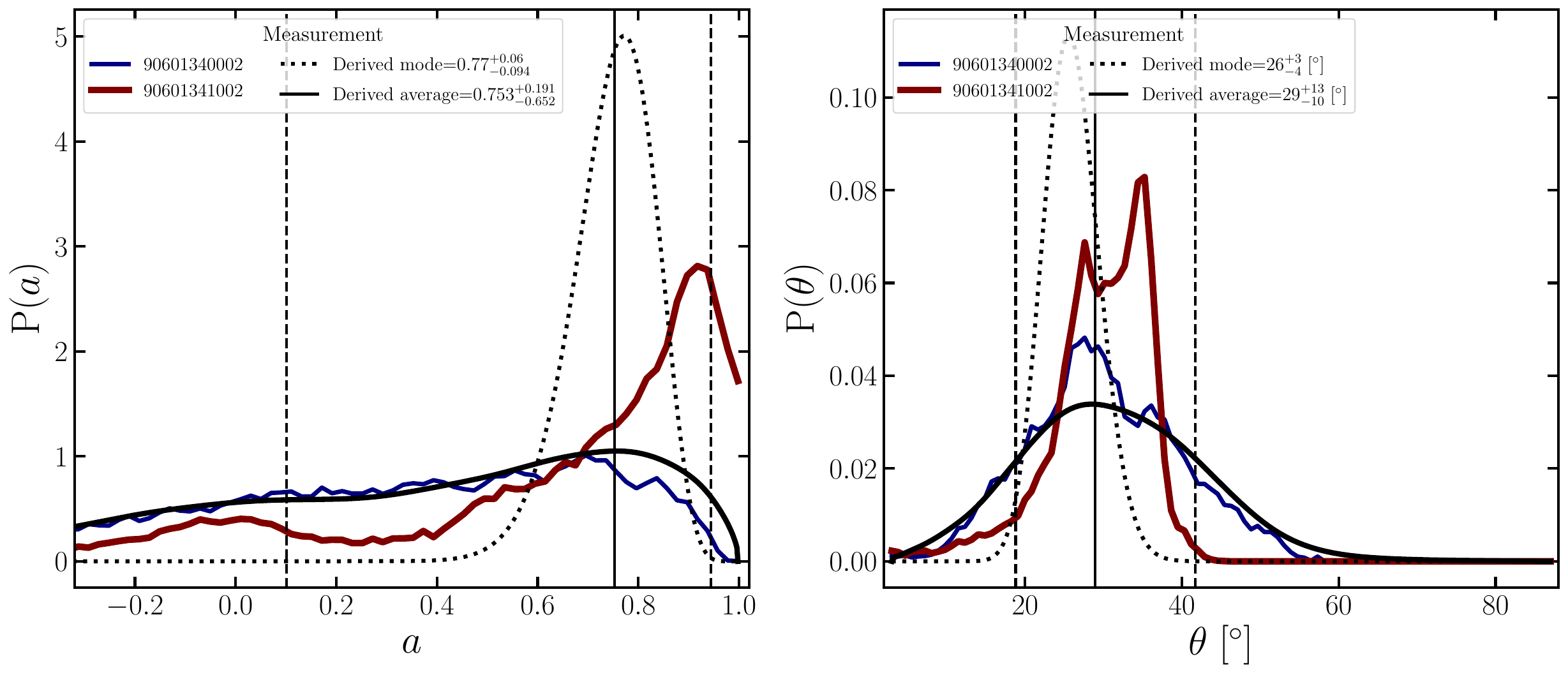}
    \caption{Posterior distributions for the analysis of MAXI J1848-015 data. Explanations are analogous to Figure \ref{fig:AT_2019wey_combined}.. Figure discussed in Section \ref{sec:MAXI_J1848-015}.}
    \label{fig:MAXI_J1848_combined}
\end{figure}

\begin{figure}[ht]
    \centering
    \includegraphics[width= 0.8\textwidth]{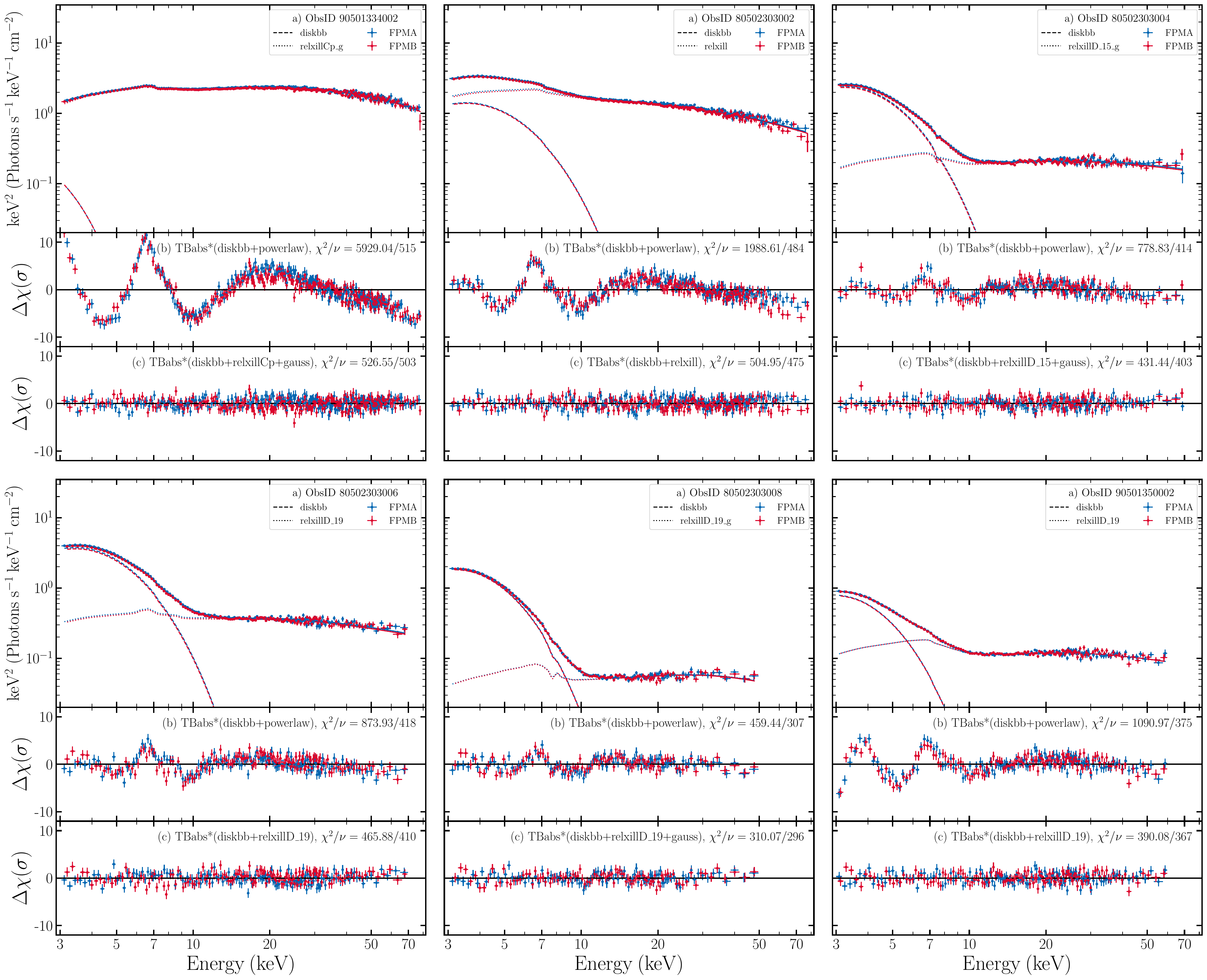}
    \caption{Unfolded spectra and fit residuals for the EXO 1846-031 data. Explanations are analogous to Figure \ref{fig:AT_2019wey_delchi}. Figure discussed in Section \ref{sec:EXO_1846-031}.}
    \label{fig:EXO_1846_delchi}
\end{figure}
\begin{figure}[ht]
    \centering
    \includegraphics[width= 0.95\textwidth]{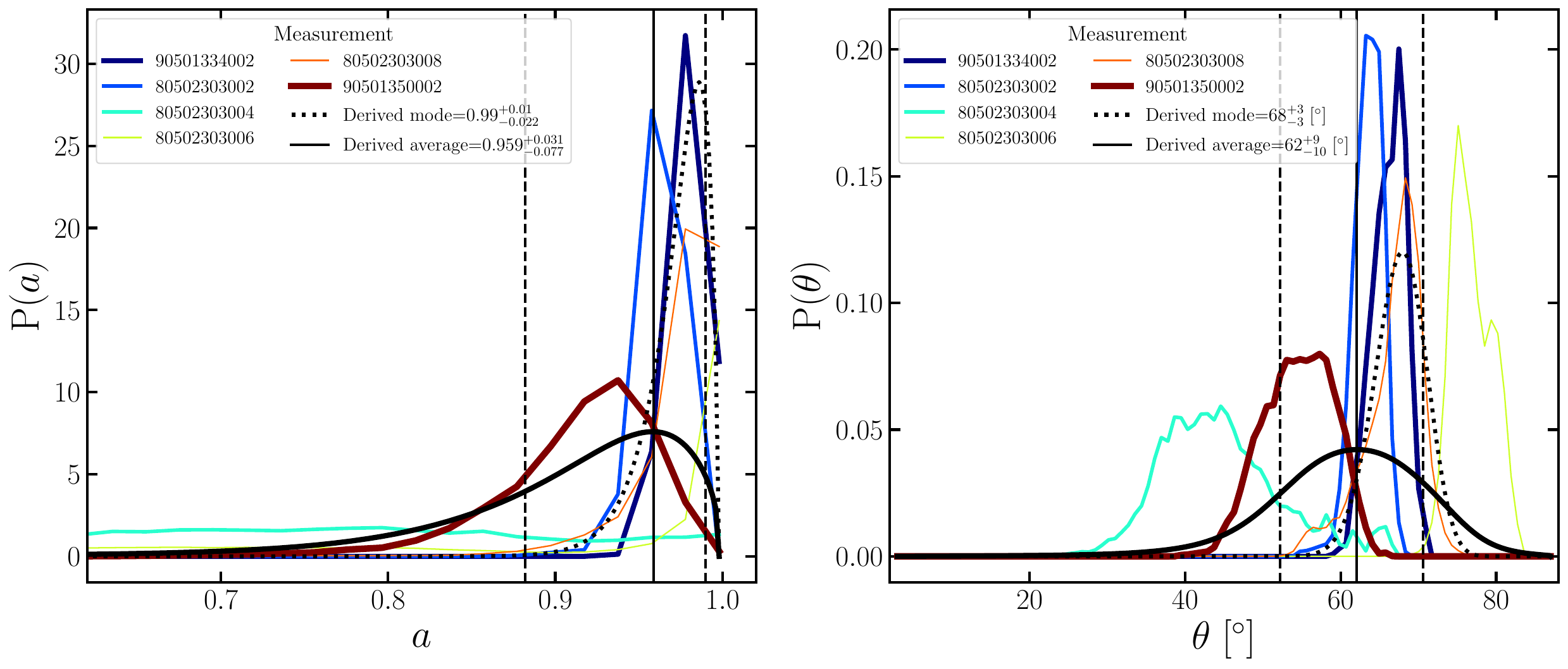}
    \caption{Posterior distributions for the analysis of EXO 1846-031 data. Explanations are analogous to Figure \ref{fig:AT_2019wey_combined}. Figure discussed in Section \ref{sec:EXO_1846-031}.}
    \label{fig:EXO_1846_combined}
\end{figure}

\begin{figure}[ht]
    \centering
    \includegraphics[width= 0.95\textwidth]{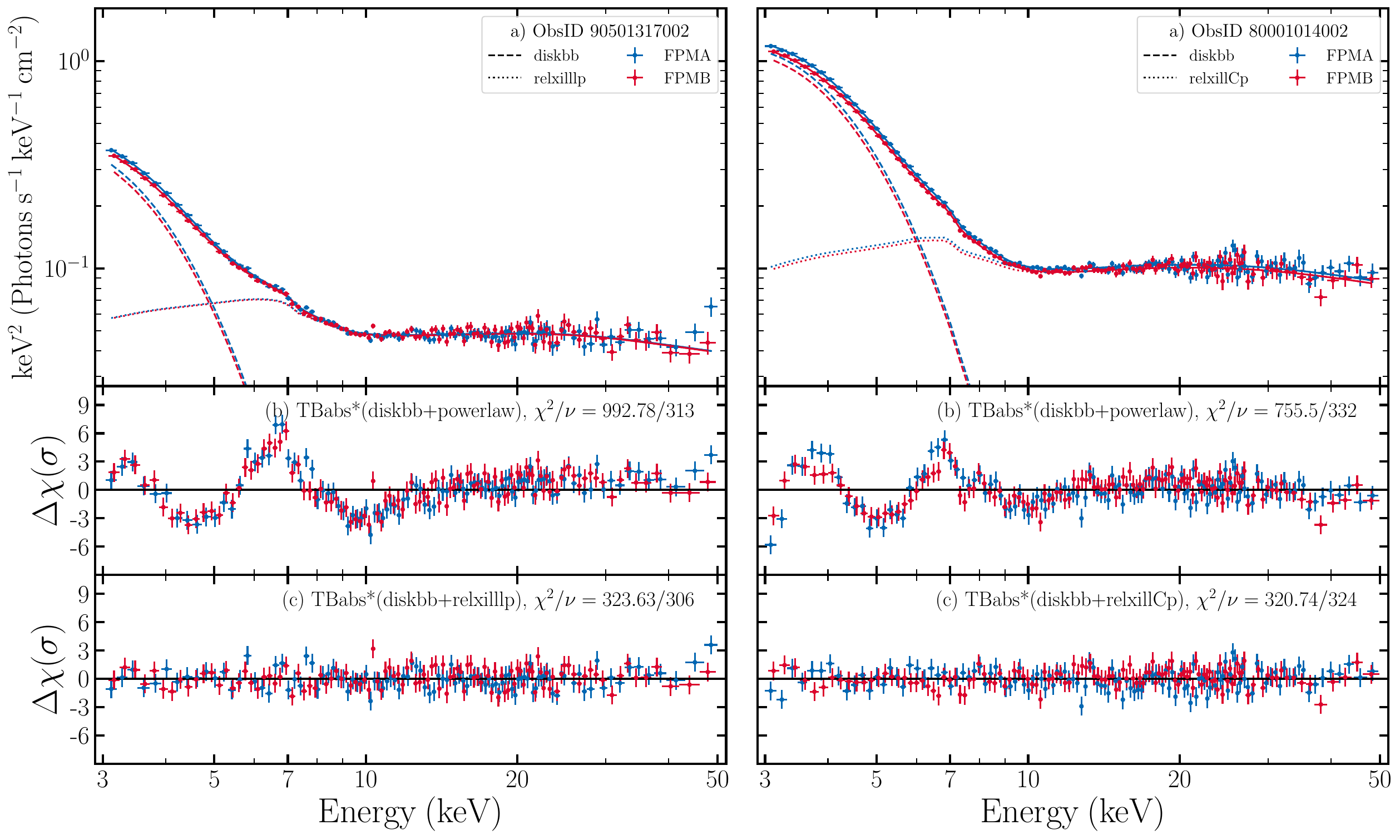}
    \caption{Unfolded spectra and fit residuals for the XTE J1908+094 data. Explanations are analogous to Figure \ref{fig:AT_2019wey_delchi}. Figure discussed in Section \ref{sec:XTE_J1908+094}.}
    \label{fig:XTE_J1908_delchi}
\end{figure}
\begin{figure}[ht]
    \centering
    \includegraphics[width= 0.95\textwidth]{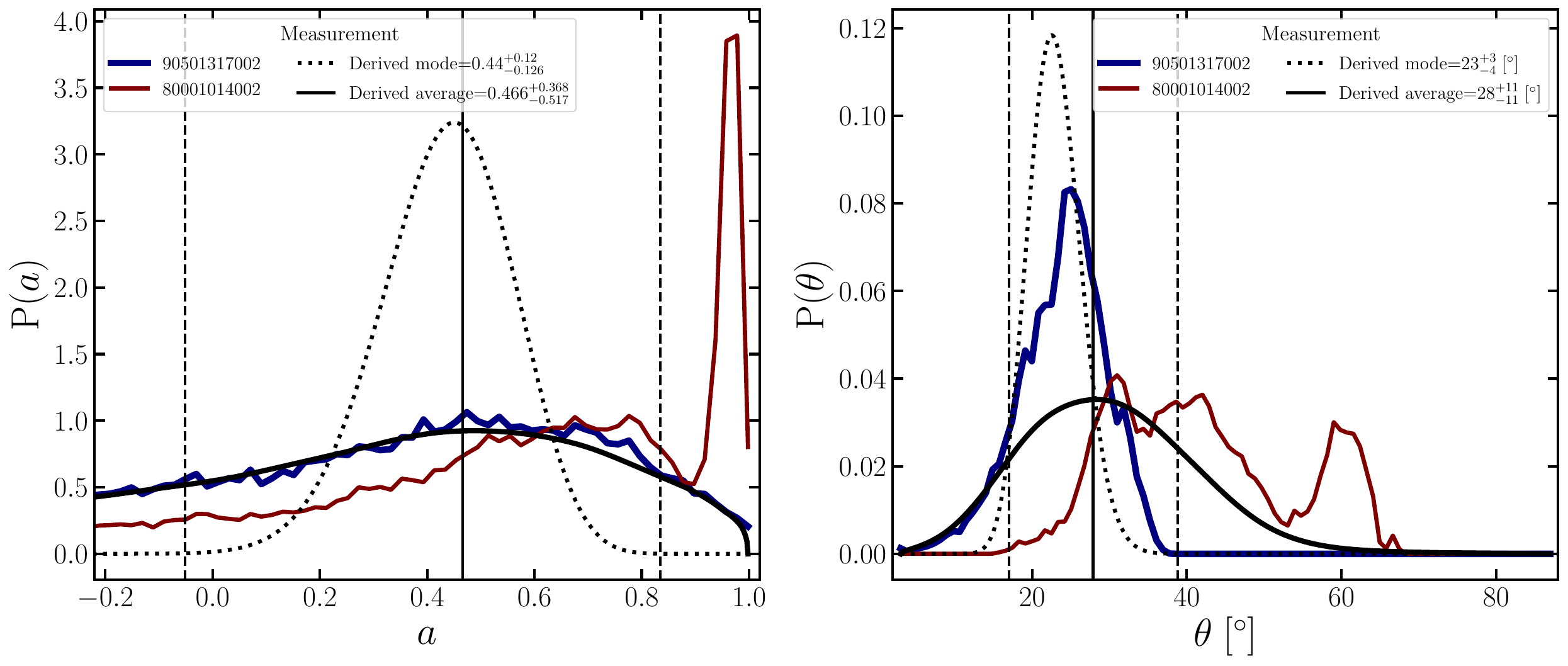}
    \caption{Posterior distributions for the analysis of XTE J1908+094 data. Explanations are analogous to Figure \ref{fig:AT_2019wey_combined}. Figure discussed in Section \ref{sec:XTE_J1908+094}.}
    \label{fig:XTE_J1908_combined}
\end{figure}

\begin{figure}[ht]
    \centering
    \includegraphics[width= 0.8\textwidth]{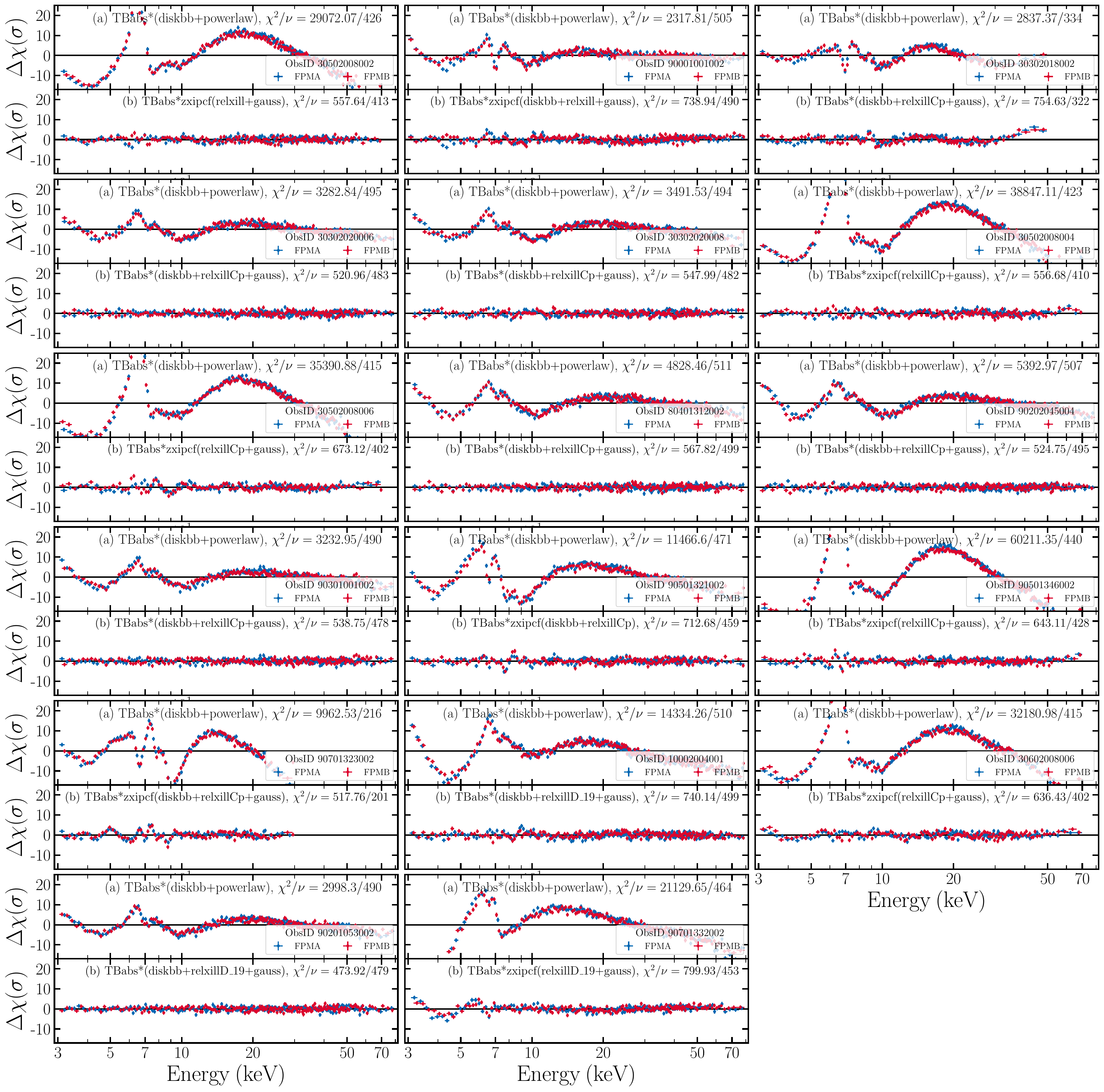}
    \caption{Fit residuals for the GRS 1915+105 data. Explanations are analogous to Figure \ref{fig:AT_2019wey_delchi}. Figure discussed in Section \ref{sec:GRS_1915+105}.}
    \label{fig:GRS_1915_delchi}
\end{figure}
\begin{figure}[ht]
    \centering
    \includegraphics[width= 0.8\textwidth]{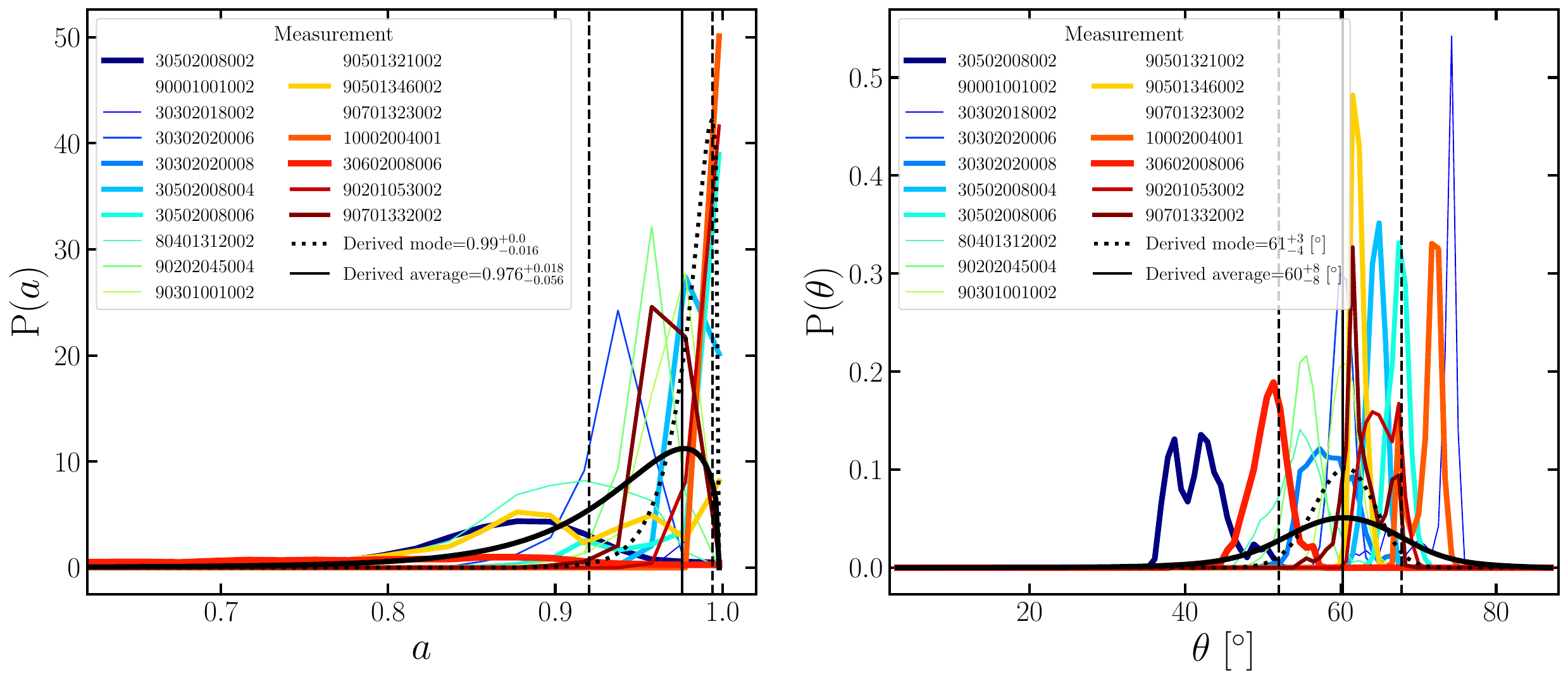}
    \caption{Posterior distributions for the analysis of GRS 1915+105 data. Explanations are analogous to Figure \ref{fig:AT_2019wey_combined}. Figure discussed in Section \ref{sec:GRS_1915+105}.}
    \label{fig:GRS_1915_combined}
\end{figure}

\begin{figure}[ht]
    \centering
    \includegraphics[width= 0.95\textwidth]{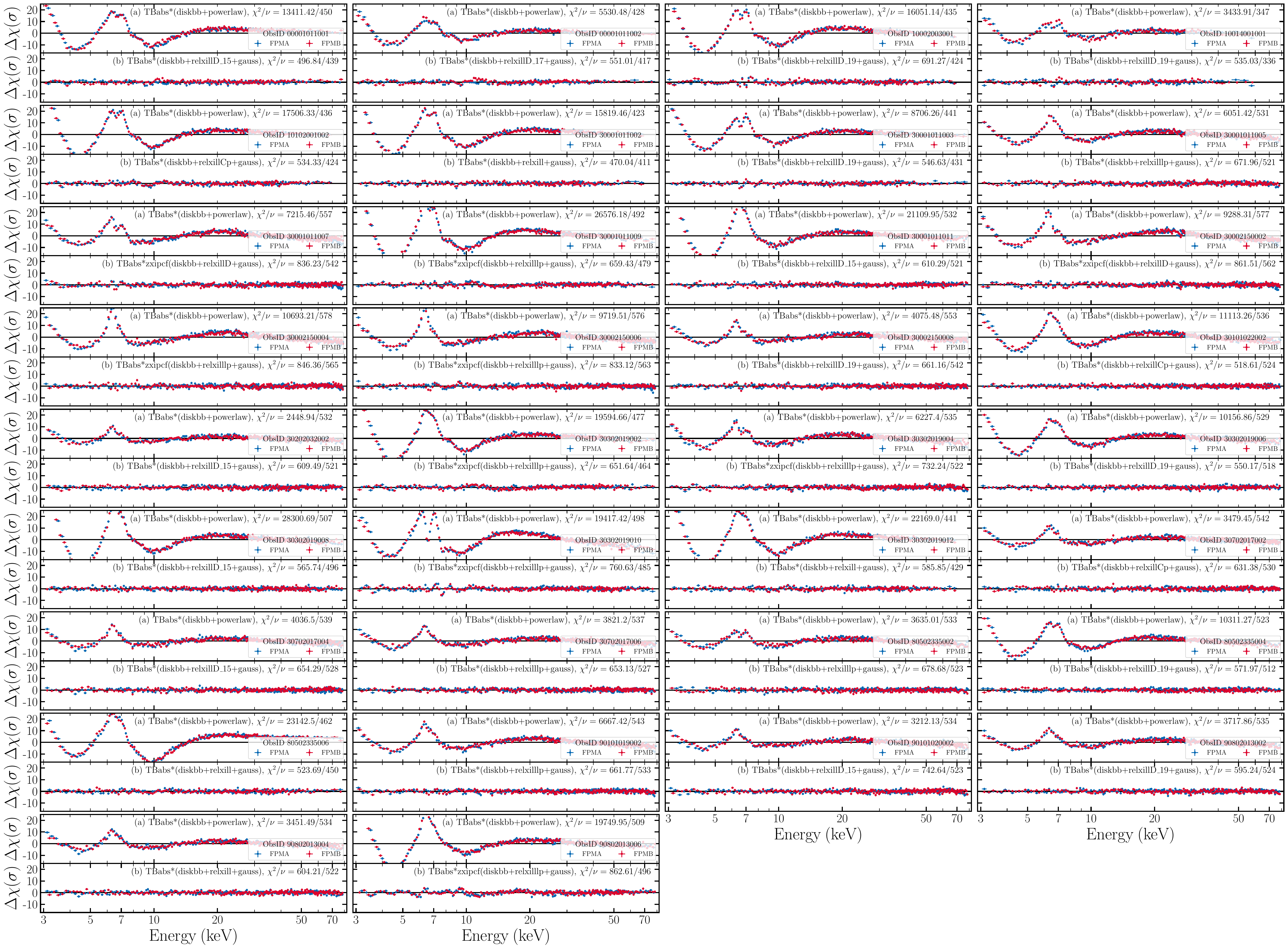}
    \caption{Fit residuals for the Cygnus X-1 data. Explanations are analogous to Figure \ref{fig:AT_2019wey_delchi}. Figure discussed in Section \ref{sec:Cyg_X-1}.}
    \label{fig:Cyg_X-1_delchi}
\end{figure}
\begin{figure}[ht]
    \centering
    \includegraphics[width= 0.95\textwidth]{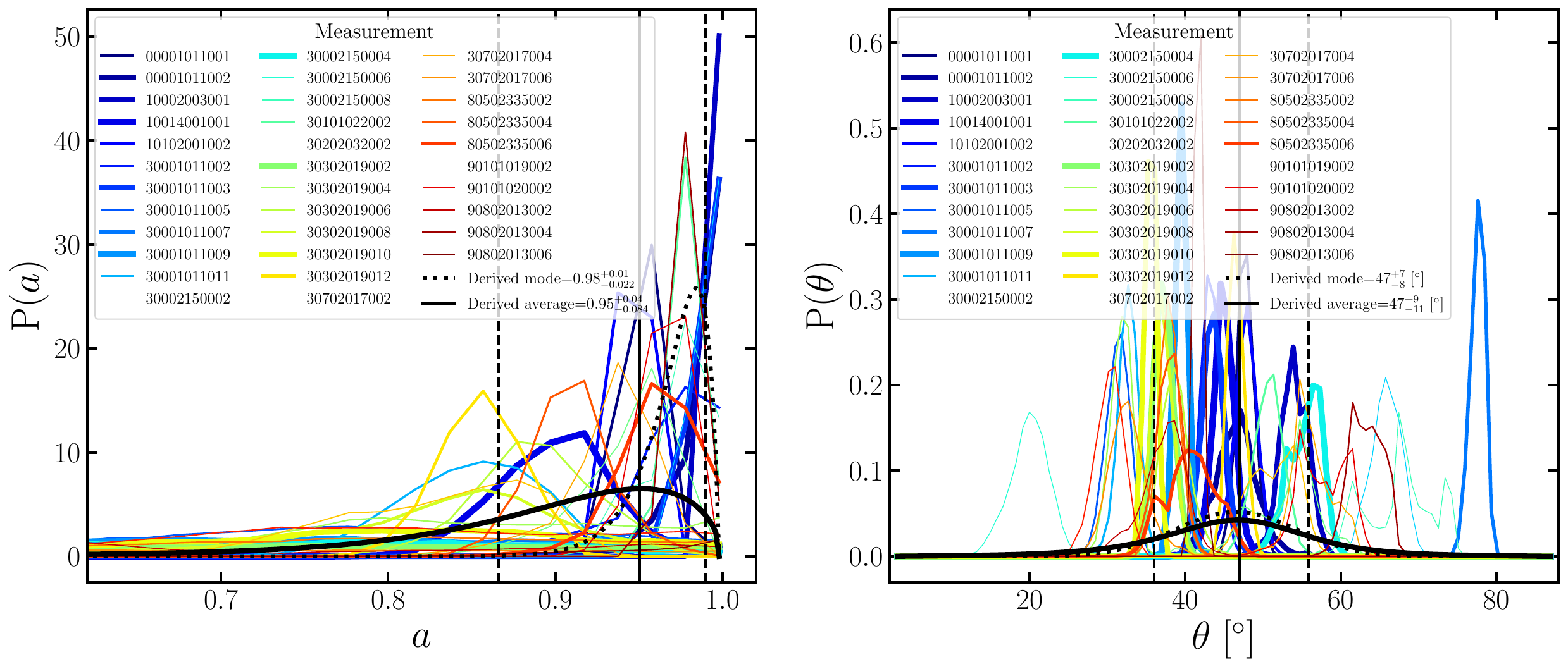}
    \caption{Posterior distributions for the analysis of Cygnus X-1 data. Explanations are analogous to Figure \ref{fig:AT_2019wey_combined}. Figure discussed in Section \ref{sec:Cyg_X-1}.}
    \label{fig:Cyg_X-1_combined}
\end{figure}

\begin{figure}[ht]
    \centering
    \includegraphics[width= 0.6\textwidth]{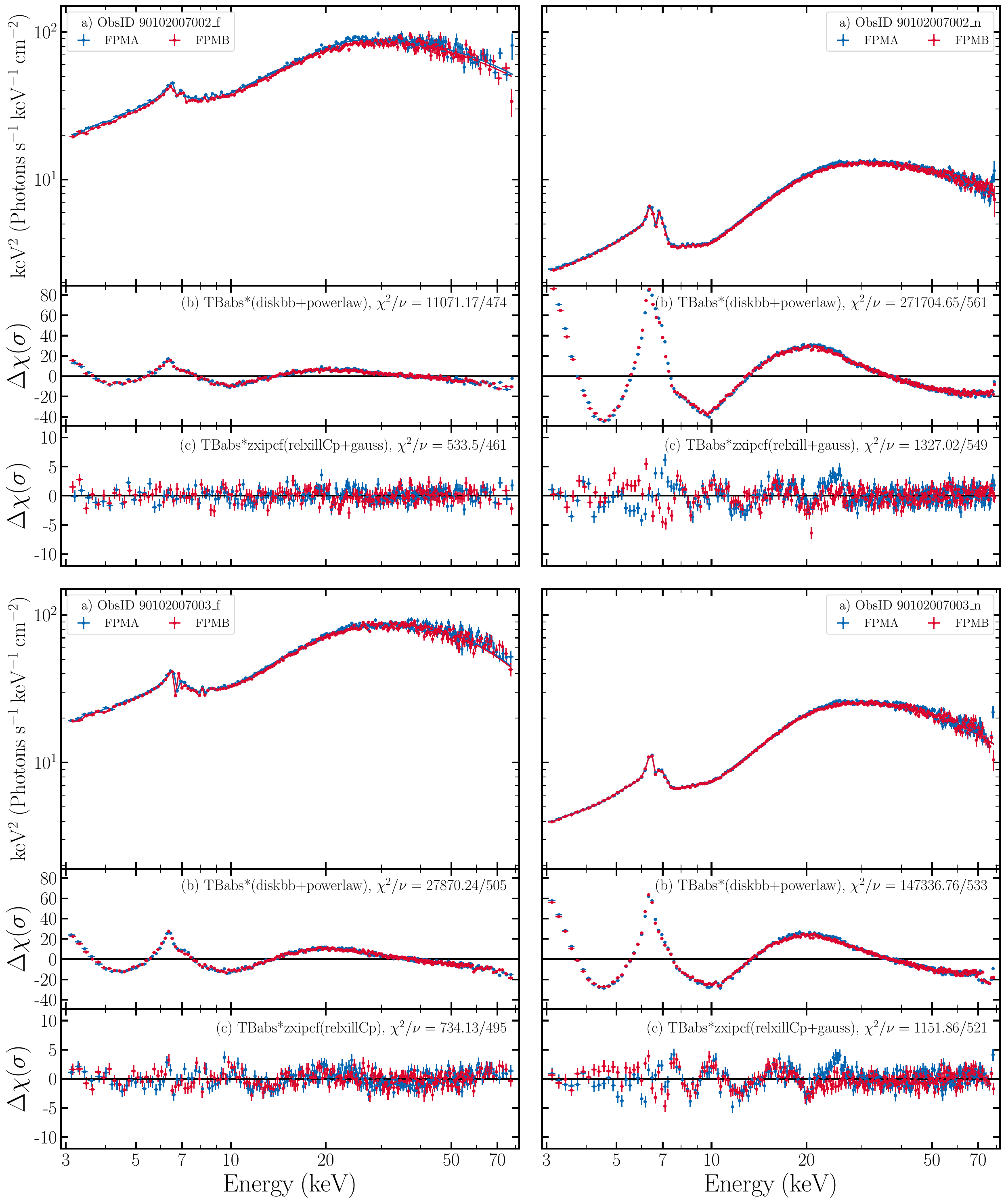}
    \caption{Unfolded spectra and fit residuals for the V404 Cygnus data. Explanations are analogous to Figure \ref{fig:AT_2019wey_delchi}. The flaring state of the source is shown in the left panels, and is indicated by ``$\_$f," while the non-flaring state is shown in the right panels, indicated by ``$\_$n." Figure discussed in Section \ref{sec:V404_Cyg}.}
    \label{fig:V404_Cyg_delchi}
\end{figure}
\begin{figure}[ht]
    \centering
    \includegraphics[width= 0.8\textwidth]{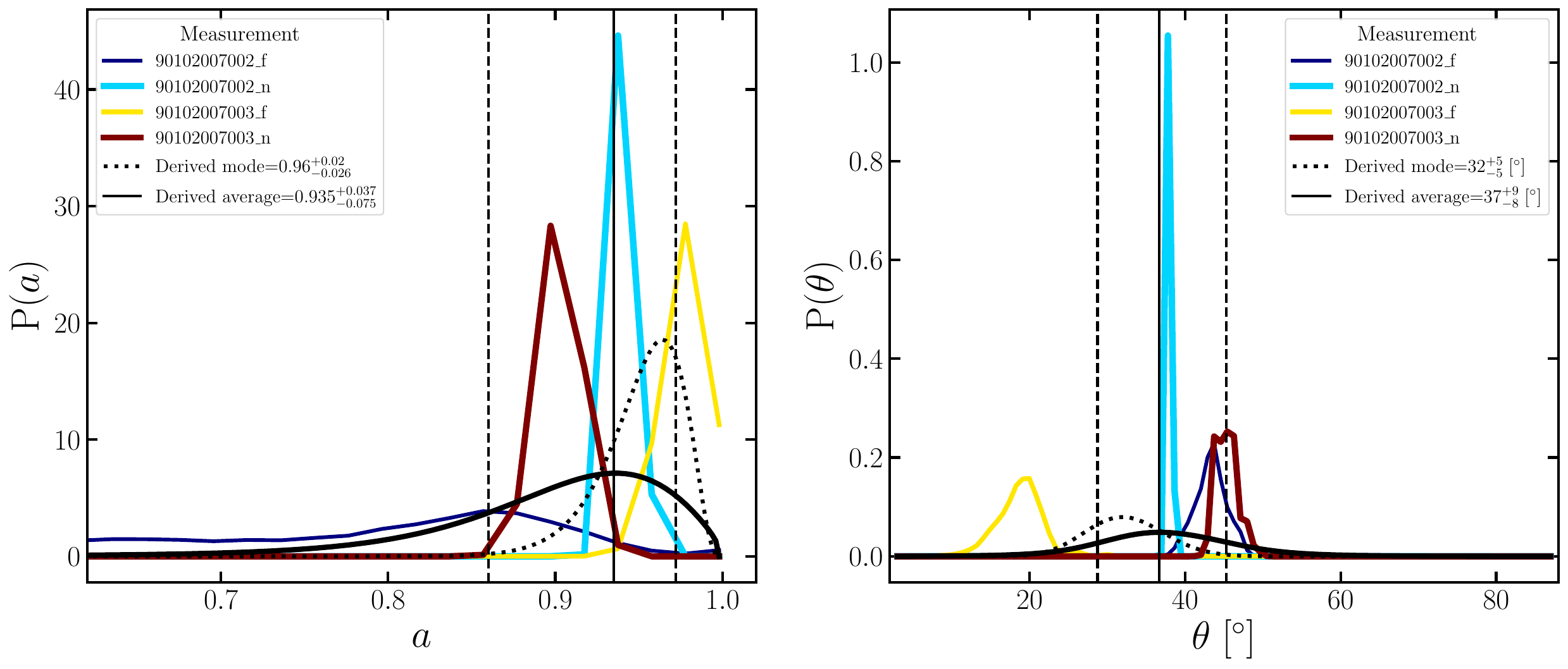}
    \caption{Posterior distributions for the analysis of V404 Cygnus data. Explanations are analogous to Figure \ref{fig:AT_2019wey_combined}. Figure discussed in Section \ref{sec:V404_Cyg}.}
    \label{fig:V404_Cyg_combined}
\end{figure}

\end{document}